\pdfoutput=1
\documentclass[rmp,aps,nofootinbib,twocolumn]{revtex4}
\usepackage{graphicx}

\def\inbar{\,\vrule height1.5ex width.4pt depth0pt}
\def\IR{\relax{\rm I\kern-.18em R}}
\def\IC{\relax\hbox{$\inbar\kern-.3em{\rm C}$}}

\usepackage{amsmath,amssymb,amsbsy}

\newcommand{\txt}[1]{\mathrm{#1}}

\newcommand{\kB}{k_\txt{B}}

\newcommand{\unit}[1]{\ \mathrm{#1}}

\renewcommand{\vec}[1]{\boldsymbol{#1}}

\renewcommand{\Re}{\mathrm{Re}}

\newcommand{\be}{\begin{equation}} \newcommand{\ee}{\end{equation}}
\newcommand{\ba}{\begin{eqnarray}} \newcommand{\ea}{\end{eqnarray}}

\newcommand{\ie}{i.\,e.}
\newcommand{\eg}{e.\,g.}

\begin{document}
\title{Single-electron current sources: towards a refined definition of ampere}

\author{Jukka P.~Pekola}
\email{jukka.pekola@aalto.fi}
\affiliation{Low Temperature Laboratory (OVLL), Aalto University, P.O. Box 13500, FI-00076 AALTO, Finland}
\author{Olli-Pentti Saira}
\affiliation{Low Temperature Laboratory (OVLL), Aalto University, P.O. Box 13500, FI-00076 AALTO, Finland}
\author{Ville F.~Maisi}
\affiliation{Low Temperature Laboratory (OVLL), Aalto University, P.O. Box 13500, FI-00076 AALTO, Finland}
\affiliation{Centre for Metrology and Accreditation (MIKES), P.O. Box 9, 02151 Espoo, Finland}
\author{Antti Kemppinen}
\affiliation{Centre for Metrology and Accreditation (MIKES), P.O. Box 9, 02151 Espoo, Finland}
\author{Mikko M\"ott\"onen}
\affiliation{Low Temperature Laboratory (OVLL), Aalto University, P.O. Box 13500, FI-00076 AALTO, Finland}
\affiliation{COMP Centre of Excellence, Department of Applied Physics, Aalto University,
P.O. Box 13500, FI-00076 AALTO, Finland}
\author{Yuri A.~Pashkin}
\altaffiliation[On leave from ]{Lebedev Physical Institute, Moscow 119991, Russia}
\affiliation{NEC Smart Energy Research Laboratories and RIKEN Advanced Science Institute, 34 Miyukigaoka, Tsukuba, Ibaraki 305-8501, Japan and Department of Physics, Lancaster University, Lancaster, LA1 4YB, UK}
\author{Dmitri V.~Averin}
\affiliation{Department of Physics and Astronomy, Stony Brook University, SUNY, Stony Brook, NY 11794-3800, USA}

\begin{abstract}
Controlling electrons at the level of elementary charge $e$ has been demonstrated experimentally already in the 1980's. Ever since, producing an electrical current $ef$, or its integer multiple, at a drive frequency $f$ has been in a focus of research for metrological purposes. In this review we first discuss the generic physical phenomena and technical constraints that influence charge transport. We then present the broad variety of proposed realizations. Some of them have already proven experimentally to nearly fulfill the demanding needs, in terms of transfer errors and transfer rate, of quantum metrology of electrical quantities, whereas some others are currently "just" wild ideas, still often potentially competitive if technical constraints can be lifted. We also discuss the important issues of read-out of single-electron events and potential error correction schemes based on them. Finally, we give an account of the status of single-electron current sources in the bigger framework of electric quantum standards and of the future international SI system of units, and briefly discuss the applications and uses of single-electron devices outside the metrological context.
\end{abstract}

%\date{May 2001}
\maketitle
\tableofcontents

%
% Chapter 1
%

\section{INTRODUCTION}
\label{sec:intro}

Future definition of the ampere is foreseen to be based on fixing the elementary
charge $e$. Its most direct realization would be the transport of a known number of
electrons. Over the past quarter of a century, we have witnessed progress towards ever better control of individual electrons.  
Since single-electron tunneling is by now a well established subject, several reviews of its different aspects exist in literature \cite{Averin1991,Averin1992,Durrani2009,Sohn1997,vanderWiel2002}. 

Several milestones have been achieved in the progress towards a single-electron current source since the initial proposals of the metrological triangle in the mid 80's~\cite{Likharev1985}. The single-electron ampere would be based on transporting an electron with charge $e$, or rather a known number $N$ of electrons, $Ne$, in each operation of a control parameter that is cyclically repeated at frequency $f$, so that the output dc current is ideally equal to $Nef$. The needs of precision metrology generally state that this operation has to be performed at a relative error level not larger than $10^{-8}$ and at the same time the current level needs to be several hundreds of pico-amperes~\cite{Feltin2009}. Just a few years after the initial theoretical proposals, the first metallic \cite{Geerligs1990,Pothier1991a,Pothier1992} and semiconducting \cite{Kouwenhoven1991} single-electron turnstiles and pumps demonstrated currents $I=Nef$ with an error of a few percent, still orders of magnitude away from what is needed. Like often in precision metrology, the pursuit of higher accuracy has been a pacemaker for understanding new physics, since the errors that need to be suppressed are often a result of interesting physical phenomena. For instance, quantum multi-electron processes and non-equilibrium phenomena have been intensively studied in order to improve the performance of single-electron sources. In five years, the accuracy of single-electron pumps was remarkably improved by another five to six orders of magnitude \cite{Keller1996} by effectively suppressing the so-called co-tunneling current, however, at the expense of significantly increased complexity of the device and reduced overall magnitude of the output current (a few pico-amperes) of the pump. Alternative ideas were to be found. At the same time, single-electron conveyors in semiconducting channels using surface-acoustic wave driving yielded promising results, in particular in terms of significantly increased current level~\cite{Shilton1996}. Yet likely due to overheating effects in the channel, it may turn out to be difficult to suppress thermal errors to the desired level using this technique.

Interestingly there was a decade of reduced progress in the field, until in the second half of the previous decade several new proposals and implementations were put forward. The most promising of these devices are undeniably the sources based on a quantum dot~\cite{Blumenthal2007}, down to a single-parameter ac control \cite{Kaestner2008}, and a SINIS turnstile~\cite{Pekola2008}, which is a basic single-electron transistor with superconducting leads and normal metal island. These simple devices promise high accuracy, and a possibility to run many of them in parallel~\cite{Maisi2009}. At around the same time, other promising ideas came out, for example
a quantum phase slip based superconducting current standard~\cite{Mooij2006}). Quantum phase slips provide the mechanism for the
existence of the Coulomb blockade effects in superconducting wires without tunnel barriers~\cite{Astafiev2012}, , and could potentially lead to current standards
producing larger currents. At the time of writing this review, we are definitely witnessing a period of intense activity in the field in a well-founded atmosphere of optimism.

%
% Chapter 2
%

\section{PRINCIPLES OF MANIPULATING INDIVIDUAL ELECTRONS}
\label{sec:principles}

\subsection{Charge quantization on mesoscopic conductors}

In this section, we briefly summarize the essential concepts of single-electron device physics,
with the emphasis on the topics needed for the subsequent discussion of the quantized current sources. 
We focus mostly on metallic devices since those have an elaborated theory based in first principles. 
The main concepts can be however adapted when considering semiconducting systems as well~\cite{Zimmerman2004}.

As is well known from the elementary treatments of the Bohr's model in quantum mechanics,
electrostatic energy of an electron in a hydrogen atom is roughly equal to the kinetic energy
of its confinement in the atomic orbital of the radius. The fact that the characteristic
energy separation of levels in the confinement energy spectrum decreases much more rapidly
than the electrostatic energy with the size of the confining region ensures then that in
mesoscopic conductors which are large on the atomic scale, the electrostatic energy of
individual electrons can be large even in the regime where the separation of the individual
energy levels associated with quantum confinement of electrons is negligible. As a characteristic
estimate, electrostatic energy of charge $e$ of one electron on a micrometer-size conductor
is on the order of a milli-electron volt, or 10 K in temperature units, and is many orders of
magnitude larger than the energy separation $\delta E$ of electron confinement levels in the same
conductor, which should be about nano-electron volt, well below all practical temperatures. As a
result, at low but easily reachable temperatures in the kelvin and sub-kelvin range, the properties
of mesoscopic conducting islands are dominated by electrostatic energy of individual electrons,
while small $\delta E$ provides one of the conditions that makes it possible to use macroscopic
capacitances to quantitatively describe electrostatics of these conductors even in this
"single-electron" regime. The charging energy $U$ of a system of such conductors can be expressed
then as usual in terms of the numbers $n_j$ of excess electrons charging each conductor and the
capacitance matrix $C$, see, e.g., \cite{Landau1980_Sec2}
\begin{equation}
U(\{n_j\})= \frac{e^2}{2} \sum_{i,j} [C^{-1}]_{i,j} n_i n_j \, ,
\label{a1} \end{equation}
where the sum runs over all conductors in the structure.

The electrostatic energy (\ref{a1}) creates energy gaps separating different charge configurations
$\{n_j\}$ which provide the possibility to distinguish and manipulate these charge configurations.
Historically, one of the first observations of distinct individual electron charges were the
Millikan's experiments on motion of charged micrometer-scale droplets of oil, which produced the
evidence that ``all electrical charges, however produced, are exact multiples of one definite,
elementary, electrical charge'' \cite{Millikan1911}. In those experiments, the oil droplets were,
however, charged randomly by an uncontrollable process of absorption of ions which exist normally
in air. By contrast, in mesoscopic conductors, the charge states $n_j$ can be changed in a
controllable way. Besides the charging energy (\ref{a1}), such a process of controlled manipulation
of individual charges in mesoscopic conductors requires two additional elements. First, the tunnel
junctions formed between the nearest-neighbor electrodes of the structure enable
the electron transfer between these electrodes, and second, the possibility to control the
electrostatic energy gaps by continuous variation of charges on the junctions \cite{Averin1986}.
The simplest way of varying the charges on the tunnel junctions continuously is by placing the
electrodes in external electrical fields \cite{Buttiker1987} that create continuously-varying
potential differences between the electrodes of the structure. Externally-controlled
continuous transport and gate voltages produced in this way can be used then to transfer
 individual electrons in the system of mesoscopic conductors.

%\begin{figure}[ht]
%\includegraphics[width=0.4\textwidth]{a_fig1}
%\setlength{\unitlength}{1.0in}
%\begin{picture}(2.1,1.9)
%\put(0.1,-.05){\epsfxsize1.5in \epsfbox{figures/a_fig1}}
%\end{picture}
%\caption{Schematic diagram of the basic circuit for manipulating individual electrons,
%single-electron box (SEB): conducting island carrying electric charge $en$, and an
%electrostatically coupled external electrode with the charge $e(N-n)$ producing the
%gate voltage $V_g$.
%\label{aa1} } \end{figure}

\begin{figure}[ht]
\centering
\includegraphics[width=0.27\textwidth]{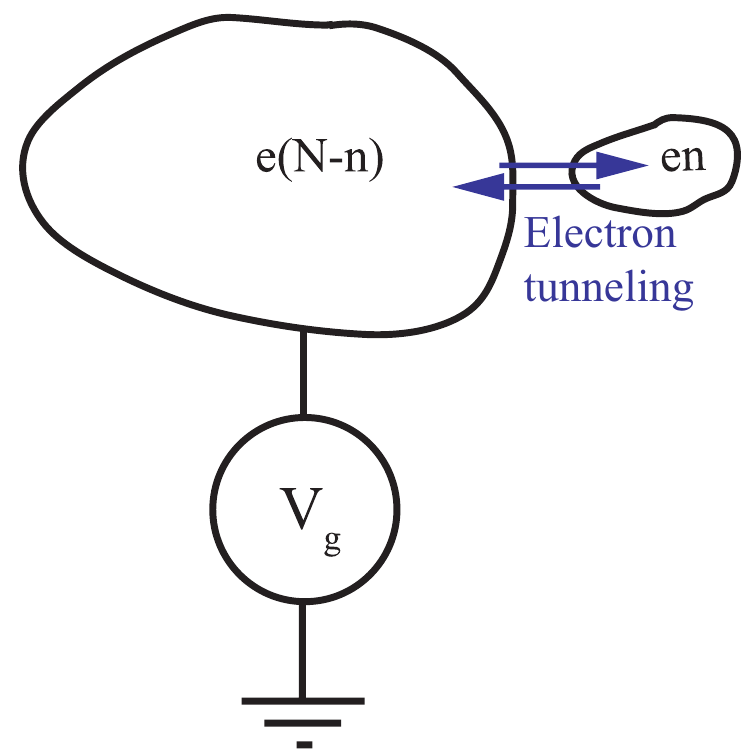}
\caption{Schematic diagram of the basic circuit for manipulating individual electrons,
single-electron box (SEB): conducting island carrying electric charge $en$, and an
electrostatically coupled external electrode with the charge $e(N-n)$ producing the
gate voltage $V_g$.
\label{aa1} }
\end{figure}

A simple model of the sources of continuously-varying external voltages is obtained by taking
some of the electrodes of the structure described by the energy (\ref{a1}) to have very large
self-capacitance and carry large charge, so that the tunneling of a few electrons does not
affect the potentials created by them. For instance, the most basic single-electron structure,
single-electron box (SEB) \cite{Lafarge1991}, can be simplified to the two electrodes, one main
island carrying the charge $en$, and the electrode with the charge $e(N-n)$ creating the gate
voltage $V_g$ (Fig.~\ref{aa1}(a)). Quantitatively, the structure in Fig.~\ref{aa1} is characterized
by the capacitance matrix
\begin{equation}
C = \left( \!   \begin{array}{cc} C_\Sigma^g &-C_g \\ -C_g & C_\Sigma  \end{array} \! \right) \, ,
\label{a2} \end{equation}
where $C_\Sigma^g$, $C_\Sigma > C_g$ are the total capacitances of the gate electrode and the
island, respectively. In the limit $N,\, C_0^g \rightarrow \infty$, with $eN/C_0^g=V_g$, the
energy (\ref{a1}) of the charges shown in Fig.~\ref{aa1} reduces for the capacitance matrix (\ref{a2})
to the following expression:
\begin{equation}
U=U_0 + E_C n^2-e^2nn_g/C_{\Sigma}\, .
\label{a3} \end{equation}
Above, $U_0$ is an $n$-independent energy of creating the source of the gate voltage,
$U_0=e^2N^2/2C_0$ in this case, $E_C\equiv e^2/2C_{\Sigma}$ is the charging energy of one
electron on the main electrode of the box, and $en_g\equiv C_gV_g$ is the charge induced on
this electrode by the gate voltage $V_g$ through the gate capacitance $C_g$. As one can
see from  Eq.~(\ref{a3}), the gate voltage $V_g$ indeed controls the energy gaps separating
the different charge states $n$ of the main island and therefore makes it possible to
manipulate individual electron transitions changing the island charge $en$.

\begin{figure}[t]
\includegraphics[width=0.49 \textwidth]{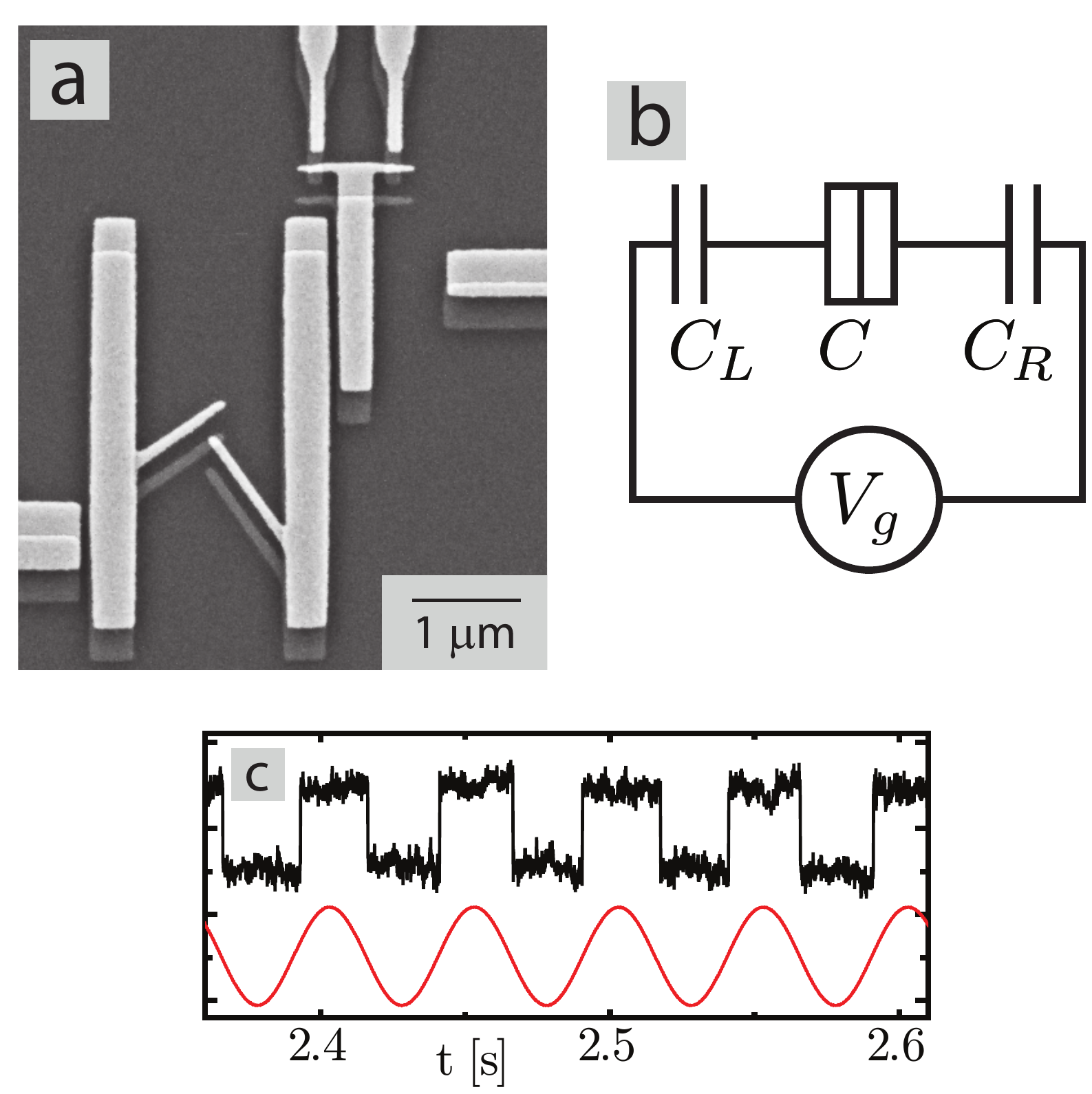}
\caption{Practical SEB: (a) scanning electron micrograph of a realistic box structure,
(b) its equivalent electric circuit, and (c) single-electron transitions in the box
illustrating the ``charge quantization'': a time-dependent gate-voltage $V_g(t)$ (sinusoidal
curve) of an appropriate amplitude drives individual electron transitions changing the box
state between the two discrete charge configurations, electron on the left or on the right island.
These two charge states are detected via the shown two-level detector output
current synchronous with the oscillating $V_g(t)$. (Figure adapted from \protect Refs.~\onlinecite{Saira2010} and \onlinecite{Saira2012b}).
\label{aa2} } \end{figure}

Figure \ref{aa2}(a) shows a scanning electron micrograph of an example of a realistic box
structure, in which, in contrast to the schematic diagram of Fig.~\ref{aa1}, one pays attention
to satisfying several quantitative requirements on the box parameters. First of all, the capacitance
$C_{\Sigma}$ needs to be sufficiently small to have significant charging energy $E_C$, while one
keeps the gate capacitance $C_g$ not very small in comparison to $C_{\Sigma}$, to be able to
manipulate the charge $en$ more easily and also to measure it. The box in Fig.~\ref{aa2}(a) consists
of the similar-size islands, and its equivalent electric circuit is shown in Fig.~\ref{aa2}(b).
The charging energy of the box is described by the same expression (\ref{a3}), with $en$ being the
charge transferred from the left to the right island, and $C_{\Sigma}$ - the total mutual
capacitance between the two islands, $C_{\Sigma}=C+C_g$, where $C_g^{-1} =C_L^{-1}+C_R^{-1}$.
Connecting the box islands to the source of gate voltage $V_g$ through capacitances $C_{L,R}$ on
both sides reduces coupling to parasitic voltage fluctuations in the electrodes of the structure,
responsible for environment-induced tunneling discussed below. Generally, a practical geometric structure
of the box islands is also determined by the fact that the main contribution to the capacitance
$C_{\Sigma}$ comes from the tunnel junction formed in the area where the ``arms'' of the islands
[Fig.~\ref{aa2}(a)] overlap. The size of this area should be minimized to increase $E_C$. At the same
time, the islands themselves can be made much larger than the junctions, to increase the gate
capacitance $C_g$ without affecting strongly the total capacitance $C_{\Sigma}$. Besides increasing
the coupling to the gate voltage created by the two outside horizontal electrodes in the box
structure shown in Fig.~\ref{aa2}(a), larger size of the box islands also increases the coupling to
the single-electron transistor (discussed in more details below) which measures
the charge of the box and can be seen in the upper right corner of Fig.~\ref{aa2}(a).

As was mentioned above, the main qualitative property of the SEB is that it allows one to manipulate
individual electrons through variation of the gate voltage $V_g$. Indeed, at low temperatures,
$T\ll E_C/k_B$, the box occupies the ground state of the charging energy (\ref{a3}). For a
given gate-voltage-induced charge $n_g$, the minimum is achieved when the number $n$ of extra electrons
on the island equals $n_g$ rounded to the nearest integer. This dependence of $n$ on $n_g$ means that one
electron is added or removed from the box island,
changing $n$ by $\pm 1$, whenever $n_g$ passes through a degeneracy point, \ie, $n_g=1/2$ modulo
an integer, at which point the charging energies (\ref{a3}) of the two charge states that differ by
one electron transition, $\delta n=1$, are equal. If the gate voltage increases monotonically,
the dependence $n(n_g)$ has the shape of the ``Coulomb staircase'' \cite{Lafarge1991}, with
each step of the staircase corresponding to the addition of one electron with gate voltage
increase by $\delta n_g=1$. If the gate voltage oscillates in time around the degeneracy point
$n_g=1/2$, as in Fig.~\ref{aa2}(c), with an appropriate amplitude ($\delta n_g \sim 1$), it induces
back-and-forth electron transitions between the two charge states separated by one electron
charge, $\delta n=1$, which can be seen in Fig.~\ref{aa2}(c) as the two-level telegraph signal
of the detector measuring the box charge. Thus, Fig.~\ref{aa2}(c) gives a practical example
of manipulation of an individual electron transition in the SEB.

One of the most interesting dynamic manifestations of the manipulation of individual electrons in a
system of mesoscopic conductors is the possibility to arrange the system dynamics in such a way
that electrons are transferred through it one by one, in a correlated fashion. This can be
achieved, for instance, if the gate voltage $V_g$ of the SEB grows in time at a constant rate
such that effectively a constant dc ``displacement'' current $I=e\dot{n}_g$ is injected in the
box junction. The same dynamic would be obtained if the real dc current $I$ flows into a
mesoscopic tunnel junction. In this case, correlated successive transfer of electrons one by
one through the junction gives rise to the ``single-electron tunneling'' oscillations \cite{Averin1986,Bylander2005}
of voltage on the junction, $\partial U/\partial (en)=e(n-n_g)/C_{\Sigma}$, with frequency $f$
related to the current by the fundamental equation
\begin{equation}
I=ef \, .
\label{a5} \end{equation}
More complex structures than SEB or an individual tunnel junction, like single-electron turnstile \cite{Geerligs1990}
and pump \cite{Pothier1992,Keller1996} discussed below, make it possible to ``invert'' this relation and
transfer one electron per period of the applied gate voltage oscillation with frequency $f$. The
above discussion of the manipulation of individual electrons in the SEB shows that the charge states
$n$, while controlled by the gate voltage $V_g$, remain the same in a range of variation
of $V_g$. Physically, such ``quantization of charge'' results from the fact that an isolated
conductor can contain only an integer total number of electrons, with the charging energy producing
energy gaps separating different electron number states. Charge quantization enables one to
make the accuracy of manipulation of individual electron charges in structures like SEB
very high, in principle approaching the metrological level. Such potentially metrological
accuracy also extends to the transport in turnstiles and pumps, making the current
sources based on single-electron tunneling promising candidates for creation of the quantum standard of
electrical current.

\subsection{Sequential single-electron tunneling}
\label{sec:seqtunneling}

\begin{figure}[ht]
\centering
\includegraphics[width=0.16\textwidth]{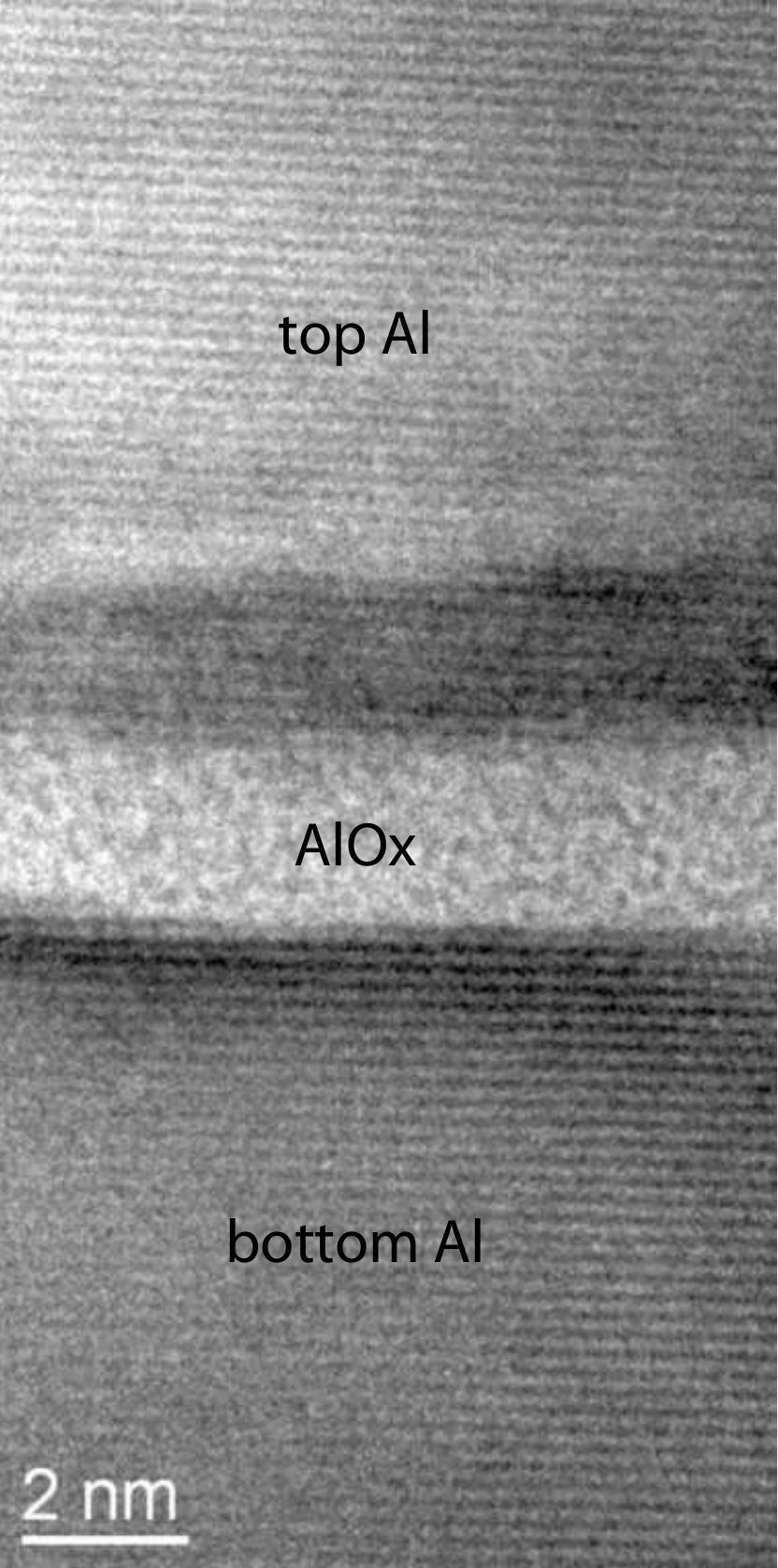}
\caption{High resolution TEM image of a cross-section of an aluminium oxide tunnel junction
(from \protect Ref.~\onlinecite{Prunnila2010}).
\label{aa3} }
\end{figure}

As discussed above, one of the key elements in manipulating individual electrons in
systems of mesoscopic conductors is a tunnel junction, which provides the means to transfer
electrons along the system thus creating the dc current $I$ through it. A tunnel junction \cite{Giaever1960}
is a system of two conductors separated by a layer of insulator that is sufficiently thin to allow
electrons to tunnel between the conductors (see Fig.~\ref{aa3}). For normal conductors, the current
through the junction depends linearly on the applied transport voltage, and is characterized by the 
tunnel conductance $G_T\equiv 1/R_T$. In the single-electron devices, $G_T$ should satisfy two
contrasting requirements. To increase the current $I$ driven through the structure, e.g., to increase
the allowed range of frequencies $f$ for which Eq.~(\ref{a5}) is satisfied accurately, one should
maximize $G_T$. On the other hand, charge quantization on the electrodes of the structure requires
that they are well-isolated from each other, i.e., $G_T$ should be small. The latter condition can
be formulated more quantitatively requiring that the characteristic charging energy $E_C$ of the
localized charge states is well-defined despite the finite lifetime of these states, $\sim G_T/C$,
where $C$ is the typical junction capacitance in the structure: $E_C\gg \hbar G_T/C$. This condition
can be expressed as $G_T \ll 1/R_Q$, where $R_Q \equiv \pi \hbar /e^2 \simeq 13.9$ k$\Omega$ is the
characteristic ``quantum'' resistance. When this condition is satisfied, the localized charge states
provide an appropriate starting point for the description of a single-electron structure, while electron
tunneling can be treated as perturbation. In what follows, we mostly concentrate on such a regime
of ``strong Coulomb blockade'' which is necessary for implementation of precise transport of individual 
electrons, as required for quantized current sources.

%\begin{figure}[ht]
%\includegraphics[width=0.4\textwidth]{a_fig3}
%\setlength{\unitlength}{1.0in}
%\begin{picture}(2.2,1.8)
%\put(0.6,-.05){\epsfxsize1.1in \epsfbox{figures/a_fig3}}
%\end{picture}
%\caption{High resolution TEM image of a cross-section of an aluminium oxide tunnel junction
%(from \protect Ref.~\onlinecite{Prunnila2010}).
%\label{aa3} } \end{figure}

Majority of practical metallic structures employ tunnel junctions based on
barriers formed by either thermal or plasma oxidation of aluminum. The main reason for this are the
superior properties of the aluminum oxide layer, in terms of
its uniformity and electrical and noise properties. A typical barrier structure is shown in
Fig.~\ref{aa3} that includes a high resolution transmission-electron-microscopy (TEM) image of
a cross-section of an aluminum-based junction with amorphous Al$_x$O$_y$ tunnel barrier.
From the point of view of the Landauer-B\"{u}ttiker formula for electric conductance of a mesoscopic
conductor, the junction tunnel conductance can be expressed as $G_T=(1/R_Q)\sum_j T_j$, where the
sum is taken over spin-degenerate electron transport channels propagating across the junction, and
$T_j$ is the quantum mechanical transmission coefficient of the insulator barrier for electrons in
the $j$th channel. Condition of strong Coulomb blockade, $G_T \ll 1/R_Q$, implies that all
individual transmission coefficients are small, $T_j\ll 1$. Although the transmission coefficients
$T_j$ are sensitive to the atomic-scale structure of the junction, the fact that the aluminum oxide
layer is relatively uniform on an intermediate space scale larger than the individual atoms~\cite{Greibe2011} 
allows transport properties to be estimated semi-quantitatively from the ``bulk'' properties of the barrier.

Since the tunnel current depends exponentially on the barrier parameters, the measured electron
tunneling rates in high-resistance junctions and in the large voltage range allow one to estimate
parameters of the aluminum oxide barrier (see, e.g., \onlinecite{Tan2008}): they yield a barrier height 
$U \simeq 2$ eV, and effective electron mass $m_{eff}\simeq 0.5\, m_e$ in terms of the free electron mass 
$m_e$. While the dimensions of the typical tunnel junctions need to be small -- of the order of $\sim 100$ nm, 
also cf. Fig.~\ref{aa2}(a) -- in order to make the junction capacitance sufficiently low, they are still 
quite large on the atomic scale. In this regime, discreteness of the spectrum of the transverse modes $j$ 
is negligible, and the tunnel conductance $G_T$ is proportional to the junction area $A$. For the value of 
specific junction resistance $A/G_T \sim 10$ k$\Omega\, \mu$m$^2$ typical for the tunnel junctions, estimates
using the barrier parameters and the simplest assumption of ballistic transport in the junction give
for the barrier transparency $T\sim 10^{-6}$ corresponding to barrier thickness close to 2 nm (cf.
Fig.~\ref{aa3}). Barrier with this thickness transmits effectively only the electrons impinging on
it orthogonally. This ``focusing'' effect means that the tunnel conductance can be expressed in
terms of one maximum value of the transmission coefficient, $G_T=NT/R_Q$, where the effective number
$N$ of the transport channels in the junction is not determined directly by the density of states
in the electrodes, but depends also on the characteristic ``traversal energy'' $\epsilon_0$ of
the barrier which gives the energy scale on which the barrier transparency changes with energy,
$N\simeq A m \epsilon_0/2 \pi \hbar^2 $. For the parameters of the aluminum oxide barrier mentioned
above, this gives for the area per transport channel, $A/N\simeq 1$ nm$^2$. As will be discussed in
the next subsection, some of the higher-order transitions in the single-electron structures, e.g. Andreev
reflection, depend explicitly on the barrier transmission coefficients $T_j$. In contrast to this,
the lowest-order electron tunneling depends only on the total junction conductance $G_T$.

The most straightforward approach to the description of tunneling in the single-electron structures in the strong
Coulomb blockade regime is based on the tunnel Hamiltonian method \cite{Cohen1962}, in which the junction
is modeled with the following Hamiltonian:
\begin{equation}
H=H_1+H_2+H_T\, , \;\;\; H_T= \sum_{k,p} [T_{kp}c_k^{\dagger} c_p+ h.c.] \, .
\label{a6} \end{equation}
Here $H_{1,2}$ are the Hamiltonians of the junction electrodes, $H_T$ is responsible for
 tunneling, with $c_k$ and $c_p$ denoting electron destruction operator in the two electrodes,
respectively, and $T_{kp}$ are the tunneling amplitudes. As was discussed in the previous section,
discreteness $\delta E$ of the single-particle electron states is completely negligible in a
typical metallic mesoscopic conductor, so these states form a continuum with some density of states
which in a normal metal, is constant on a small energy scale of interest in the structures.
In this case, one can treat $H_T$ using Fermi's golden rule to obtain the rate $\Gamma (E)$ of a
tunneling process that changes the charge configuration $\{n_j\}$ on the system of mesoscopic
conductors by transferring one electron through a tunnel junction between the two conductors.
As a result, changing the electrostatic energy (\ref{a1}) by an amount $E=U(\{n_{j,\mathrm{in}}\} )
-U(\{n_{j,\mathrm{fin}}\} )$, where $\{n_{j,\mathrm{in}}\}$ is the initial and $\{n_{j,\mathrm{fin}}\}$ 
the final charge configuration, we obtain
\begin{equation}
\Gamma (E) = \frac{G_T}{e^2}\int d \epsilon f (\epsilon) [1-f (\epsilon+E)] \nu_1 (\epsilon)
\nu_2 (\epsilon+E) \, .
\label{a7} \end{equation}
In this expression, $f(\epsilon)$ is the equilibrium Fermi distribution function, and $\nu_j(\epsilon)$ 
is the density of the single-particle states in the $j$th electrode of the junction, $j=1,2$, in units of
the normal density of states $\rho_j$, which together with the average of the squares of the
tunneling amplitudes determine the tunnel conductance, $G_T=4\pi e^2 \langle |T_{kp}|^2 \rangle
\rho_1 \rho_2$. Equation (\ref{a7}) assumes that the energy $E\sim E_C$ is much smaller than all
internal energies of the junction in the normal state, in particular the traversal energy
$\epsilon_0$, the condition very well satisfied for practical metallic structures in
which $E_C \sim 1$ meV, while $\epsilon_0 \sim 1$ eV. Using the standard properties of the Fermi
distribution functions, one can see directly that the rate (\ref{a7}) of tunneling between the two
equilibrium electrodes satisfies the necessary detailed balance condition
$\Gamma_{1\rightarrow 2} (-E)=e^{-E/k_BT} \Gamma_{2\rightarrow 1} (E)$. If, in addition, the densities
of states are symmetric with respect to the chemical potentials of the electrodes, the tunneling rate 
is also symmetric, $\Gamma_{1\rightarrow 2} (E)=\Gamma_{2\rightarrow 1} (E)$, and the detailed balance 
condition simplifies to $\Gamma (-E)=e^{-E/k_BT} \Gamma (E)$. The detailed balance condition makes it 
possible to express the tunneling rate (\ref{a7}) in terms of the current-voltage characteristic $I(V)$ of the 
junction at fixed bias voltage $V$:
\begin{equation}
\Gamma (E) = I(E/e)/[e(1-e^{-E/k_BT})] \, .
\label{a7*} \end{equation}

For the NIN junctions, when both electrodes are in the normal (N) states, $\nu_j(\epsilon) \equiv 1$, 
Eq.~(\ref{a7}) gives, in agreement with Eq.~(\ref{a7*}), for the tunneling rate
\begin{equation}
\Gamma (E) = \frac{G_T}{e^2} \frac{E}{1-e^{-E/k_BT}} \, .
\label{a8} \end{equation}
Tunneling of individual electrons with the rate (\ref{a8}) is an irreversible dissipative process
which converts the electrostatic energy change $E$ into internal energy of the electron gas
inside the junction electrodes. In accordance with this understanding, at small temperatures $T$,
the rate (\ref{a8}) vanishes as $e^{E/k_BT}$ for energetically unfavorable transitions, $E<0$, when
the energy for the transition is taken from the thermal fluctuations of the electron reservoirs.
In the regime of allowed transitions, $E\sim E_C>0$, the magnitude of the typical transition rate
$\Gamma \sim G_T/C$ for the realistic values of the parameters, $G_T\sim 1$ M$\Omega$, $C\sim
10^{-16} \ldots 10^{-15}\unit{F}$, is quite high, in the GHz range.

\begin{figure}[ht]
\includegraphics[width=0.49\textwidth]{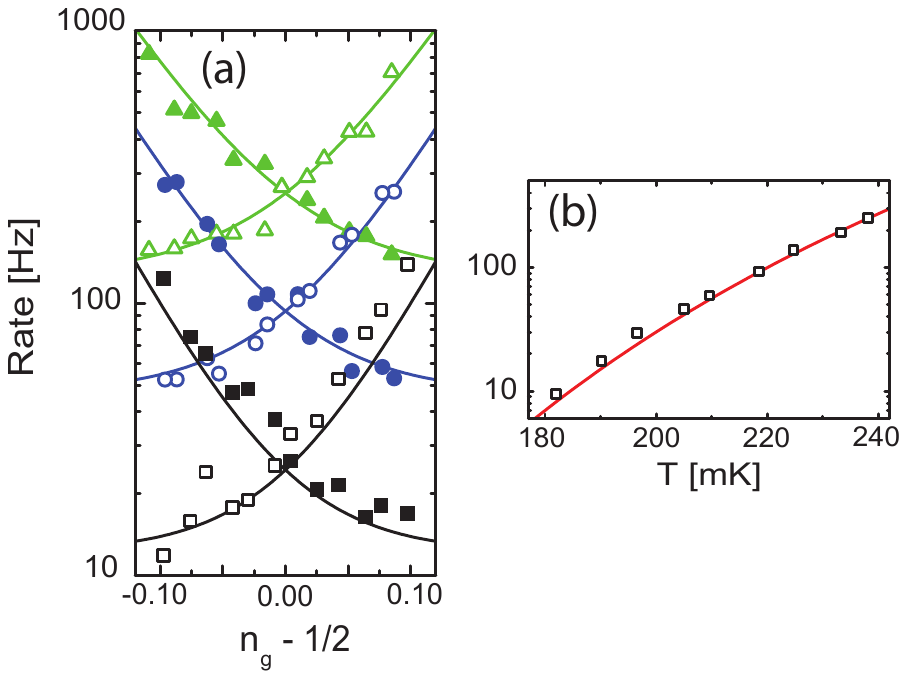}
\caption{(a) Measured thermally activated rates of forward $\Gamma (E)$ and backward $\Gamma (-E)$
tunneling in a ``hybrid'' SIN single-electron box at different temperatures, as a function of the
gate voltage offset from the degeneracy point related to the energy change $E$ in tunneling as
$E=2E_C(n_g-1/2)$. \protect Solid lines are the theory prediction according to Eq.~(\ref{a9}) with
fitted parameters: $E_C=157$ $\mu$eV, $\Delta=218$ $\mu$eV, $1/G_T=100$ M$\Omega$.  (b) The
tunneling rate at degeneracy, $E=0$, as a function of temperature (square markers), and best fit
(solid line) to Eq.~(\ref{a9}). (Figure adapted from \protect Ref.~\onlinecite{Saira2012b}.)
\label{aa4} } \end{figure}

In the SIN junctions, when one of the junction electrodes is a superconductor (S), the BCS density of
states: $\nu_1(\epsilon)=|\epsilon|/(\epsilon^2-\Delta^2)^{1/2}$ for $|\epsilon|>\Delta$, and
vanishing otherwise, implies that at temperatures well below the superconducting energy gap
$\Delta$, the tunneling rate (\ref{a7}) is strongly suppressed and can be reduced into the kHz and even
Hz range. Indeed, evaluating the integral in Eq.~(\ref{a7}) for the SIN junction assuming
$k_BT,E\ll \Delta$, one gets
\begin{equation}
\Gamma (E) = \frac{G_T}{e^2} \sqrt{2\pi \Delta k_BT} e^{-\Delta /k_BT}\, \frac{\sinh(E/k_BT)}{1-e^{-E/k_BT}} \, .
\label{a9} \end{equation}
Figure \ref{aa4} shows the tunneling rate (\ref{a9}) measured in an SIN junction in the
configuration of a ``hybrid'' SEB [see Fig.~\ref{aa2}(a)], in which one of the islands of the box is a
superconductor (aluminum), the other one being normal metal (copper). The electrostatic energy
change $E$ in the case of the box follows from
Eq.~(\ref{a3}) as $E=U(n=0)-U(n=1)=2E_C(n_g-1/2)$, is proportional to the deviation of the gate
voltage of the box from the degeneracy point $n_g=1/2$. The measurements can be described well by
Eq.~(\ref{a9}) with reasonable values of parameters including the superconducting
energy gap $\Delta$ of aluminum.

\begin{figure}[ht]
\includegraphics[width=0.49\textwidth]{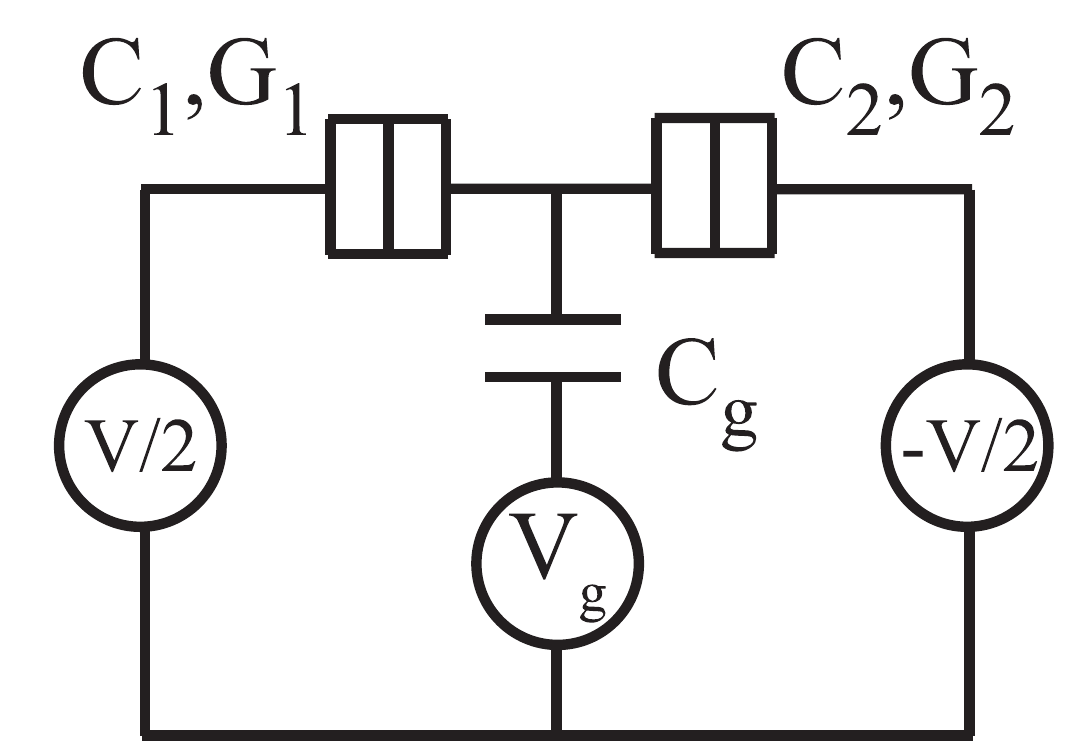}
\caption{Equivalent electric circuit of an SET.
\label{aa5} } \end{figure}

Since the tunneling transitions described quantitatively by the rates (\ref{a7})-(\ref{a9}) are
inherently random stochastic processes, dynamics of the structures in the strong Coulomb blockade
regime and electron transport properties including the dc current $I$,
current noise, or even full statistical distribution of the transferred charge, can be obtained from the
time evolution of the probabilities $p(\{n_j\})$ of various charge configurations $\{n_j\}$ governed by
the standard rate equation for the balance of the probability fluxes. The most basic single-electron system that
allows for the flow of dc current through it, and gives an example of such an equation is the single-electron transistor
(SET)~\cite{Averin1986,Fulton1987,Likharev1987}. The transistor can be viewed as a generalization of 
the SEB, and consists of a mesoscopic conducting island connected by two tunnel junctions to the bulk 
electrodes that provide the transport voltage $V$ across it. The island is also coupled capacitively to 
the source of the gate voltage $V_g$ which controls the flow of current $I$ through the transistor between 
the two electrodes. Equivalent circuit of the transistor is shown in Fig.~\ref{aa5}, and an example of its
geometric structure can be seen in the upper right corner of Fig.~\ref{aa2}(a), where it is used to
measure the charge state of the SEB. The charge configuration of the transistor is characterized simply
by the number $n$ of extra electrons on its central island, and accordingly, the rate equation describing
its dynamics is:
\begin{equation}
\dot{p}(n)=\sum_{n,j,\pm} [ p(n\pm1)\Gamma_j^{(\mp)}(n\pm 1)-p(n)\Gamma_j^{(\pm)}(n) ]\, ,
\label{a10} \end{equation}
where $p(n)$ is the probability distribution of the charge $en$ on the central island of the transistor,
and the rates $\Gamma_j^{(\pm)}(n)$ describe the tunneling processes in junction $j$ with the tunnel
conductance $G_j$ out of the state $n$ in the direction that increases $(+)$ or decreases $(-)$ $n$ by one.
The rates are given by Eqs.~(\ref{a8}) or (\ref{a9}), or their generalizations, depending on the nature of
the transistor electrodes. They depend on the indices of $\Gamma$'s in Eq.~(\ref{a10}) through the change
of $E$ of the charging energy $U$ of the transistor, which is a function of all these indices.
The transistor energy $U$ consists of two parts, one that coincides with the charging energy (\ref{a3})
of the SEB in which $C_{\Sigma}=C_1+C_2+C_g$, and the other, $U_V$, is created by the transport voltage $V$:
\begin{equation}
U_V= -eNV -enV(C_2+C_g/2)/C_{\Sigma} \, .
\label{a11} \end{equation}
Here $N$ is the number of electrons that have been transferred through the transistor. Both the dc
current $I$ through the transistor \cite{Averin1991} and the current noise \cite{Korotkov1994} can be 
calculated starting from Eq.~(\ref{a10}). The main physical property of the transistor transport 
characteristics is that they depend periodically on the gate voltage, in particular $I(n_g+1)=I(n_g)$. 
This dependence of the transistor current on the charge $en_g$ induced on its central island makes the 
SET a charge detector, with sub-electron sensitivity approaching ($10^{-5} \ldots 10^{-6}$)
e/Hz$^{1/2}$ \cite{Zimmerli1992,Krupenin1998,Roschier2001}. As a result, the SET is the most 
standard charge detector for measurements of, e.g., individual electron dynamics in other single-electron structures
[cf. Fig.~\ref{aa2}(a)].

The hybrid SINIS or NISIN transistors have an additional important feature that distinguishes them
from the SETs with normal electrodes. They provide the possibility to realize the regime of
the quantized current $I$ (\ref{a5}), when driven by the ac gate voltage $V_g(t)$ of frequency $f$
\cite{Pekola2008}. This property of the hybrid SETs is one of the main topics of this review
discussed below.

The basic expression (\ref{a7}) for the tunneling rates assumes that the electrodes of the tunnel
junction are in equilibrium at temperature $T$, with the implied assumption that this temperature
coincides with fixed temperature of the whole sample. Since each electron tunneling event deposits an 
amount of heat $\Delta U$ into the electron system of the electrodes, this condition requires that the
relaxation processes in the electrodes are sufficiently effective to maintain the equilibrium. The
relaxation rates decrease rapidly with decreasing temperature, e.g., $\propto T^5$ for electron-phonon
relaxation in an ordinary metal -- see, e.g. \cite{Giazotto2006}. This makes the relaxation insufficient 
and causes the overheating effects to appear at some low temperature, in practice around 0.1 K. Therefore, 
the overheating sets a lower limit to the effective temperature of the transitions, in this way 
limiting the accuracy of control over the individual electron transport.

One more assumption underlying the expression (\ref{a9}) for the tunneling rate in the SIN
junctions is that the electron distribution function is given by the Fermi distribution function
$f(\epsilon)$. As known from the statistical mechanics, even in equilibrium, this requires that
the total effective number of particles that participate in forming this distribution is large. In
normal-metal islands, this requirement is satisfied at temperatures much larger than the single-particle
level spacing, $T\gg \delta E/k_B$, as is the case for practically all metallic tunnel junctions.
In contrast to this, in superconducting islands, this condition can be violated at temperatures below
the superconducting energy gap, $T\ll \Delta/k_B$, when the total number of the quasiparticle excitations
in the electrode is no longer large. The temperature scale of the onset of this ``individual
quasiparticle'' regime can be estimated from the relation $\int_{\Delta}^{\infty} d \epsilon
f (\epsilon) \nu (\epsilon)/\delta E \sim 1$. The main qualitative feature of this regime is the
sensitivity of the electron transport properties of a superconducting island to the parity of the total
number of electrons on it \cite{Averin1992b, Tuominen1992}. In particular, the charge tunneling rate (\ref{a9}) 
in the SIN junction should be modified in this case into the rates of tunneling of individual quasiparticles.
For $T\gg \delta E/k_B$, these rates are still determined by the same average $\langle |T_{kp}|^2 \rangle$
over many single-particle states of the squares of the tunneling amplitudes (\ref{a6}) which give the
tunnel conductance $G_T$, and therefore can be expressed through $G_T$. In the regime of ``strong''
parity effects, when $T\ll T^* \equiv \Delta/k_B\ln N_{eff}$, where $N_{eff}=(2\pi k_BT\Delta)^{1/2}/\delta E$
is the effective number of states for the quasiparticle excitations \cite{Tuominen1992}, an ideal BCS
superconductor should reach the state with no quasiparticles, if the total number $N_0$ of electrons
in the superconductor is even, and precisely one unpaired quasiparticle if $N_0$ is odd. Although many
non-equilibrium processes in realistic superconductors lead to creation of a finite density of
quasiparticle excitations which do not ``freeze out'' at low temperatures (see, e.g., \cite{deVisser2011}),
one can realize the situation with the number of quasiparticles controlled as in an ideal BCS
superconductor, as e.g. in \cite{Tuominen1992,Lafarge1993,Saira2012}. In this regime, the rates of sequential 
charge tunneling between the normal metal electrode and a superconducting island depend on the parity of
the total number $N_0$ of electrons in the island \cite{Schon1994}. For $T\ll T^*$, the tunneling rates
both to and from the island are dominated by the one quasiparticle that exists on the island
for odd $N_0$. When this quasiparticle is equilibrated to the edge of the quasiparticle spectrum at energy
$\Delta$, the rates of tunneling to and from the island (i.e., increasing and decreasing the charge $en$ of 
the island) coincide, and for $|E| \ll \Delta$ are independent of the electrostatic energy change $E$:
\begin{equation}
\Gamma^{odd} = \frac{G_T \delta E}{4e^2} \, .
\label{a14} \end{equation}
For even $N_0$, when there are no quasiparticles on the island, tunneling necessarily involves the
process of creation of a quasiparticle making the tunneling rates energy dependent on the energy
change $E$:
\begin{equation}
\Gamma^{even} (E) = \Gamma^{odd} N_{eff} e^{-(\Delta-E) /k_BT} \, .
\label{a15} \end{equation}
In the hybrid superconductor/normal metal structures, these tunneling rates determine the electron
transport properties through the rate equation (\ref{a10}).

\subsection{Co-tunneling, Andreev reflection, and other higher-order processes}
\label{sec:cotunneling}

Sequential tunneling discussed above represents only the first non-vanishing order of the perturbation
theory in the tunnel Hamiltonian $H_T$ (\ref{a6}). In the strong Coulomb blockade regime $G_T\ll 1/R_Q$,
this approximation provides an excellent starting point for the description of electron transport, 
accounting quantitatively for main observed properties of these structures. However,
more detailed picture of the transport should include also the tunneling processes of higher order
in $H_T$, which involve transfers of more than one electron in one or several tunnel junctions. Although for
$G_T\ll 1/R_Q$, the rates of these more complex multi-step electron "co-tunneling" processes are small in
comparison with the rates of the single-step sequential electron tunneling, they are frequently important
either because they provide the only energetically allowed transport mechanism, or because they limit the
accuracy of the control of the basic sequential single-electron transitions. The simplest example of the co-tunneling
is the current leakage in the SET in the CB regime \cite{Averin1989,Geerligs1990c}, when the bias voltage $V$
is smaller than the CB threshold and any single-step electron transfer that changes the charge $en$ on the
transistor island by $\pm e$ (see Fig.~\ref{aa5}) would increase the charging energy (\ref{a3}) and is
suppressed. In this regime, only the two-step co-tunneling process that consists of electron transfers in
both junctions of the transistor in the same direction gains the bias energy (\ref{a11}). It achieves this
by changing the number $N$ of electrons transported through the transistor by 1 without changing the charge
$en$ on the island. Qualitatively, this process represents a quantum tunneling through the energy barrier
created by the charging energy. Because of the discrete nature of charge transfer in each step of the
co-tunneling, its rate is not suppressed exponentially as for the usual quantum tunneling, and is smaller
only by a factor $G_T R_Q\ll 1$ than the rate of sequential tunneling processes.

In a hybrid SIN junction, in addition to the charging energy, superconducting energy gap $\Delta$
provides an extra energy barrier to tunneling of individual electrons, suppressing the sequential tunneling
rate (\ref{a9}) at low temperatures $T\ll \Delta/k_B$. The gap $\Delta$ exists only for individual electrons,
while pairs of electrons with zero total energy and momentum can enter a superconductor as a Cooper pair,
in the process called Andreev reflection \cite{Andreev1964}. In tunnel junctions, Andreev reflection (AR) 
can be described similarly to the co-tunneling, as a perturbative two-step tunneling process, in which the
transfer of the first electron is virtual and only the second electron transfer makes the process
energetically favorable and real. Quantitatively, the rates of such multi-step transitions can be
determined through their higher-order transition amplitudes constructed according to the standard rules
of the perturbation theory (see, e.g., \cite{Landau1980_Sec43}). For instance, in the simplest example of a 
two-step AR process in a hybrid single-electron box, the elementary amplitude $A(\epsilon_k,\epsilon_l)$
of the process that takes two electrons in the normal electrode with energies $\epsilon_k$, $\epsilon_l$
and transfers them into the superconductor as a Cooper pair can be written as:
\begin{equation}
A (\epsilon_k,\epsilon_l) = \sum_p u_p v_p T_{kp}T_{lp} (\frac{1}{\Omega_p+E_i-\epsilon_k}+
\frac{1}{\Omega_p +E_i -\epsilon_l}) \, .
\label{a20} \end{equation}
The two-step process goes through an intermediate state obtained as a result of the first step of the
process. The intermediate states differ by the order of transfer of the two electrons and by the
single-particle state of energy $\epsilon_p$ in the superconductor in which the virtual quasiparticle
with excitation energy $\Omega_p= (\Delta^2+ \epsilon_p^2 )^{1/2}$ is created. In addition to $\Omega_p$,
the energy of the intermediate state includes the charging energy barrier $E_i$ to the transfer of one
electron from the normal electrode to the superconductor. The standard BCS factors
$v_p=[(1-\epsilon_p/\Omega_p)/2]^{1/2}$ and $u_p=[(1 + \epsilon_p/\Omega_p)/2]^{1/2}$ enter
Eq.~(\ref{a20}) because $v_p$ is the amplitude of state $p$ being empty in the BCS ground state, thus
allowing the first electron transfer, while $u_p$ is the overlap of the doubly occupied orbital state
$p$ with the BCS ground state, which gives the amplitude of return to the ground state after the second
electron transfer. Since no trace of the intermediate states is left in the final state obtained after
the whole AR process is complete, they should be summed over coherently, at the level of the amplitude
$A(\epsilon_k,\epsilon_l)$. By contrast, the initial states $\epsilon_k$, $\epsilon_l$ of the electrons
in the transition are left empty in the final state, and can be used to distinguish between different 
transition processes. This means that they should be summed over incoherently, in the expression for the 
tunneling rate $\Gamma_{AR}$. At small temperatures $k_BT\ll \Delta$, one can neglect thermal excitations 
in the superconductor obtaining the total AR tunneling rate as
\begin{equation}
\Gamma_{AR}= \frac{2\pi}{\hbar}  \sum_{k,l} |A (\epsilon_k,\epsilon_l) |^2 f(\epsilon_k)f(\epsilon_l)
\delta (\epsilon_k+\epsilon_l+E)\, ,
\label{a21} \end{equation}
where $E$ is the electrostatic energy change due to the complete AR tunneling process. The sum over all
states $p$ in the superconductor in Eq.~(\ref{a20}) implies that the contribution of the individual
quasiparticles (which is important in the parity-dependent transition rates (\ref{a14}) and (\ref{a15}))
is negligible in the amplitude $A$, and individual quasiparticles affect $\Gamma_{AR}$ only through the
change of the charging conditions of tunneling.

Result of summation over different single-particle states in Eqs.~(\ref{a20}) and (\ref{a21})
depends on the geometric structure of the SIN junction. For instance, quadratic dependence of the
AR amplitude $A$ (\ref{a20}) on the tunneling amplitudes makes the magnitude of Andreev reflection
sensitive not only to the total tunnel conductance $G_T$ but also to the distribution of the
barrier transmission probabilities. Two main qualitative features of the aluminium oxide tunnel
junctions (Fig.~\ref{aa3}), which are the focus of the main part of this review, are the relatively
thick insulator barrier characterized by the focusing effect on the tunneling electrons, and low
resistance of the junction electrodes. The simplest junction model that takes into account both
features assumes ballistic electron motion that can be separated into different transport channels
throughout the junction. In this case, the states $k,l$ in Eqs.~(\ref{a21}) belong to the same
transport channel, and summation over different channels can be done directly and gives the
effective number $N$ of the channels which, as discussed in the section~\ref{sec:seqtunneling}, is 
limited by the angular dependence of the barrier transmission probabilities 
(see, e.g., \onlinecite{Averin1995}). In the ballistic approximation, Eqs.~(\ref{a20}) and (\ref{a21}) 
give for the AR tunneling rate \cite{Averin2008}:
\begin{eqnarray}
\Gamma_{AR}=\frac{ \hbar G_T^2 \Delta^2}{16\pi e^4 N} \int d\epsilon f(\epsilon -E/2)
f(-\epsilon -E/2) \nonumber \\ \times  | \sum_{\pm} a(\pm \epsilon-E_i-E/2) |^2,
\label{a22} \end{eqnarray}
where
\[ a(\epsilon) \equiv (\epsilon^2-\Delta^2)^{-1/2}  \ln \left[ \frac{\Delta-\epsilon +
(\epsilon^2 -\Delta^2)^{1/2}}{\Delta-\epsilon - (\epsilon^2-\Delta^2)^{1/2}} \right]\, . \]
Equation (\ref{a22}) is well defined if the relevant energies in the amplitude $a(\epsilon)$
do not approach the edge of the superconducting energy gap, $\epsilon \simeq  \Delta$, which gives
logarithmically divergent contribution to $\Gamma_{AR}$. This singularity can be smeared by many
mechanisms, e.g., the non-uniformity of the gap $\Delta$ or finite transmission probability of the
barrier. In the single-electron tunneling regime, the main broadening mechanism should be the lifetime of the intermediate
charge state in the AR process, and can be accounted for by replacing in Eq.~(\ref{a20}) the energy
$E_i$ with $E_i-i\gamma/2$, where $\gamma$ is the rate of sequential lowest-order tunneling out of
the intermediate charge configuration.

\begin{figure}[t]
    \includegraphics[width=8.5cm]{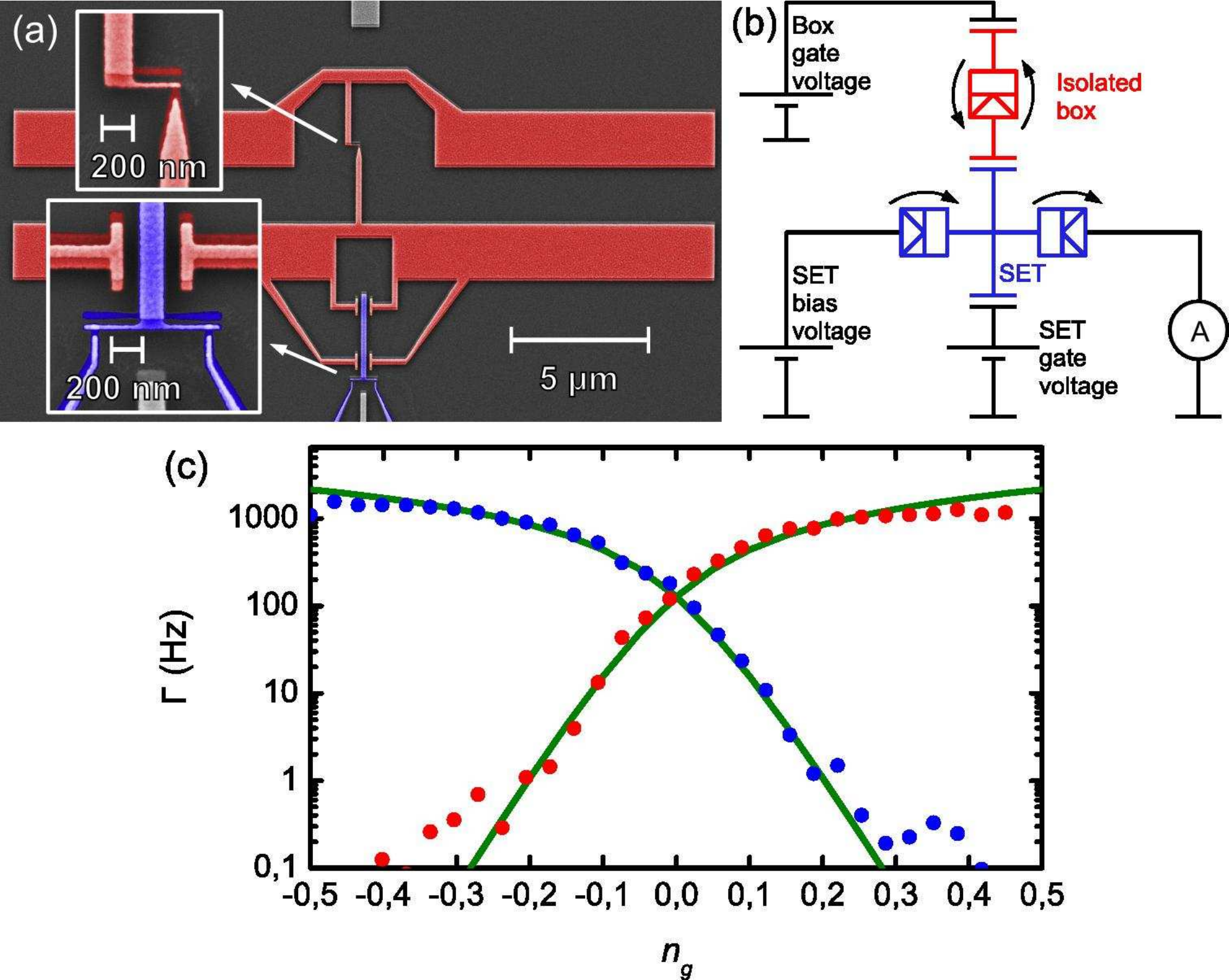}
    \caption{Real time detection of Andreev tunneling in an isolated SEB shown as red in the scanning 
    electron micrograph of (a) and its schematic in panel (b). The electrometer, shown in blue is used for 
    counting the single-electron and Andreev tunneling rates. Panel (c) shows the tunneling rate for AR as 
    blue and red dots for forward and backward directions. The green lines are theoretical calculations 
    where the non-uniformity of the tunnel barrier is taken into account. 
    Figure adapted from Ref.~\cite{Maisi2011}.}
    \label{fig:ARexp}
\end{figure}

Experimentally, individual AR processes can be observed directly in time domain in the hybrid SEB
\cite{Maisi2011}. Such observation allows one to extract the rates of AR tunneling shown in Fig.~\ref{fig:ARexp}
as a function of the normalized gate voltage $n_g$, which determines the energies $E_i$ and $E$ of
the transition, together with the theoretical fit based on Eq.~(\ref{a22}). One can see that
Eq.~(\ref{a22}) describes very well the shape of the curves. It requires, however, considerably,
roughly by a factor 15, smaller effective number $N$ of the transport channels. In practice, the
magnitude of AR tunneling rates that is larger by roughly a factor of 10 than the theoretical
expectation for a given tunnel conductance $G_T$, is the usual feature of the tunnel junctions
(see also, e.g., \cite{Pothier1994,Greibe2011}), and in principle could be accounted by the variation 
of the barrier thickness over the junction area. Unfortunately, there is so far no experimental or 
theoretical evidence that the barrier non-uniformity is indeed the reason for the discrepancy between 
the magnitude of the lowest-order and AR tunneling.

In the structures without superconducting electrodes, multi-step electron transitions, in contrast to
the AR processes, involve electron transfers in different directions and/or across different tunnel
junctions, since a transfer of the two electrons in the same junction and the same direction can not
make the process energetically favorable in the absence of the pairing gap $\Delta$. In the simplest
example of the NININ SET, the two-step co-tunneling process in the CB regime discussed
qualitatively at the beginning of this Section, consists of two electron transfers across the two 
junctions of the transistor. Quantitatively, the rate of this process is dominated by {\em inelastic} 
contribution $\Gamma_{in}$, in 
which the single-particle states $\epsilon_p$, $\epsilon_q$ of electrons in the central island of the
transistor involved in the transfers are different \cite{Averin1989}. As a result, the occupation factors 
of these states are changed, i.e. electron-hole excitations are created after the process is completed,
the fact that implies that contributions to the tunneling rates from different $\epsilon_p$, $\epsilon_q$
should be summed over incoherently. The elementary amplitude $A$ of this process consists then only of
a sum over the two possibilities, one in which an electron is first transferred onto the island
increasing the charging energy of the intermediate state by $E^{(+)}$, and the other, in which an
electron tunnels first from the island still increasing the charging energy but by a different amount
$E^{(-)}$:
\begin{equation}
A = T_{kp}T_{ql} (\frac{1}{\epsilon_p+E^{(+)}-\epsilon_k}+ \frac{1}{\epsilon_l +E^{(-)} -\epsilon_q})
\, .
\label{a25} \end{equation}
The total rate $\Gamma_{in}$ is given then by the sum of $|A|^{2}$ over all single-particle states
involved with the appropriate equilibrium occupation factors, and can be expressed directly through
the junction conductances as:
\begin{eqnarray}
\Gamma_{in} (E)=\frac{ \hbar G_1G_2 }{2\pi e^4} \int d\epsilon_k d\epsilon_p d\epsilon_q d\epsilon_l
f(\epsilon_k) f(\epsilon_q)  \nonumber \\ \times [1-f(\epsilon_p)][1-f(\epsilon_l)]
\delta(E-\epsilon_p+\epsilon_k-\epsilon_l+\epsilon_q) \nonumber \\
\cdot |\frac{1}{\epsilon_p+E^{(+)}-\epsilon_k}+ \frac{1}{\epsilon_l +E^{(-)} -\epsilon_q}|^2,
\label{a26} \end{eqnarray}
where $E=eV$ (see Fig.~\ref{aa5}) is the energy gain due to the transfer of electron charge $e$ through
both junctions of the transistor. Equation (\ref{a26}) shown of the second-order electron co-tunneling
that involves one virtual intermediate stage is indeed smaller than the rate (\ref{a7}) of sequential
single-electron tunneling roughly by a factor $R_QG_T\ll 1$. The derivation above also makes it clear
that the rate of the multi-step electron transitions that go through $n$ virtual intermediate stages
with larger $n$ would be suppressed much more strongly, by a factor $(R_QG_T)^n$.

If the energy gain $E$ and thermal energy $k_BT$ are smaller than the charging energy barriers $E^{(\pm)}$,
Eq.~(\ref{a26}) for the inelastic co-tunneling rate can be simplified to
\begin{equation}
\Gamma_{in} (E) = \frac{ \hbar G_1G_2 }{12\pi e^4}(\frac{1}{E^{(+)}}+ \frac{1}{E^{(-)}})^2
 \frac{E[E^2+(2\pi k_BT)^2]}{1-e^{-E/k_BT}} \, .
\label{a27} \end{equation}
This equation shows that, as a result of creation of excitations in the process of inelastic co-tunneling,
its rate decreases rapidly with decreasing $E$ and $T$. At very low energies, the process of co-tunneling
in the NININ transistor will be dominated by the {\em elastic} contribution, in which an electron is added
to and removed from the same single-particle state of the transistor island, without creating excitations
on the island. Because of the restriction on the involved single-particle states, the rate of such elastic 
contribution contains an additional factor on the order of $\delta E/E_C$  \cite{Averin1990} and can win 
over $\Gamma_{in}$ only at very low temperatures, practically negligible for the structures based on 
the $\mu$m-scale metallic islands considered in this review.

The approach to multi-step electron transitions in the single-electron structures illustrated in this Section
with the examples of Andreev reflection and electron co-tunneling can be directly extended to
other higher-order tunneling processes, e.g. co-tunneling of a Cooper pair and an electron \cite{Averin2008},
which together with Andreev reflection and electron co-tunneling limit in general the accuracy of control
over sequential single-electron transitions.

\subsection{Coulomb blockade of Cooper pair tunneling}
\label{sec:CBCP}

In contrast to the tunneling processes considered above, which involve electrons in the normal-metal
electrodes, tunneling of Cooper pairs in a junction between two superconductors is intrinsically
a dissipationless process \cite{Josephson1962}. As such, it should not be characterized by a tunneling
rate but a tunneling amplitude. Quantitative form of the corresponding term in the junction Hamiltonian
can be written most directly at low energies, $k_BT\, ,E_C \ll \Delta$, when the quasiparticles can not
be excited in the superconducting electrodes of the junction,
and the tunneling of Cooper pair is well-separated from the tunneling of individual electrons.
In this regime, a superconductor can be thought of as a Bose-Einstein condensate of a "mesoscopically" large
number of Cooper pairs which all occupy one quantum state. Transfer of one pair between two such condensates in
the electrodes of a tunnel junction does not have any non-negligible effects on the condensates besides changing
the charge $Q=2en$ on the junction capacitance by $\pm 2e$. Therefore, the part of the Hamiltonian describing
the tunneling of Cooper pairs should contain the terms accounting for the changes of the charge $Q$. Using the
standard notation $-E_J/2$ for the amplitude of Cooper-pair tunneling and including the charging energy
(\ref{a3}), one obtains the Hamiltonian of an SIS tunnel junction or, equivalently, Cooper-pair box in the
following form \cite{Averin1985,Buttiker1987}:
\begin{equation}
H=4 E_C (n-n_g)^2 -\frac{E_J}{2}\sum_{\pm} |n\rangle\langle n\pm 1| \, .
\label{a30} \end{equation}
Here $n$ is the number of Cooper pairs charging the total junction capacitance, and $n_g$ is the continuous
(e.g., gate-voltage-induced) charge on this capacitance normalized now to the Cooper pair charge $2e$. Similarly
to the sequential tunneling rates, the Cooper-pair tunneling amplitude in the Hamiltonian (\ref{a30}) is a
macroscopic parameter which receives contributions from all Cooper pairs in the condensate, and can be expressed
directly through the tunnel conductance $G_T$ of the junction, $E_J=\pi G_T \Delta/2e$ \cite{Ambegaokar1963},
in agreement with the simple fact that the {\em amplitude} of the two-electron tunneling should have the same
dependence on the barrier transparency as the {\em rate} of tunneling of one electron. In the situation of
the junction (\ref{a30}) realized with the actual Bose-Einstein condensates of atoms, such ``Bose Josephson
junction'' can contain relatively small total number of particles, and the tunnel amplitude varies then with
the difference $n$ of the number of particles in the two condensates - see, e.g.,
\cite{Folling2007,Cheinet2008,Averin2008b}.

Dependence on the ground state of the Hamiltonian (\ref{a30}) on the induced charge $n_g$ allows for qualitatively
similar control of the individual Cooper pairs as individual electrons in the normal-state SEB discussed in
Sec.~\ref{sec:seqtunneling}. If $E_J\ll E_C$, precisely one Cooper pair is transferred through the junction,
changing $n$ by $\pm 1$, whenever $n_g$ passes adiabatically through a degeneracy point $n_g=1/2$ modulo an integer.
This leads to the same staircase-like dependence $n(n_g)$ as in the normal case, but with each step corresponding
to the transfer of one more Cooper pair with the increase of $n_g$ by $1$. The main new element of the
superconducting situation is that the SIS junction is intrinsically a coherent quantum system without dissipation,
and if extrinsic sources of decoherence can be made sufficiently weak, should exhibit reversible dynamics of a
simple quantum system. For instance, close to the degeneracy point $n_g=n+1/2$ the two charge states with the same
electrostatic energy, $n$ and $n+1$, are coupled by coherent quantum mechanical tunneling of a Cooper pair, and
the junction behaves as the very basic quantum two-state system \cite{Bouchiat1998,Nakamura1999}. Such two-state
dynamics and general coherent quantum dynamics of the Hamiltonian (\ref{a30}) serve as the basis for the
development of superconducting quantum information devices -- for reviews, see, e.g.,
\cite{Averin2000,Makhlin2001}.

Superconducting junctions also exhibit the dynamics similar to the single-electron tunneling oscillations. If the induced charge $n_g$
grows in time at a constant rate, so that effectively a dc ``displacement'' current $I=2e\dot{n}_g$ is injected
into the junction, Cooper pairs are transferred Cooper pairs are transferred through it in a correlated manner,
one by one, giving rise to the ``Bloch'' oscillations \cite{Averin1985} of voltage across the junction, with
frequency $f$ related to the current $I$:
\begin{equation}
I=2ef \, .
\label{a31} \end{equation}
The Hamiltonian (\ref{a30}) and its extensions to the multi-junction systems can be used to design time-dependent
periodic dynamics with frequency $f$ which would to transfer precisely one Cooper pair per period and therefore
produce the dc current quantized according to Eq.~(\ref{a31}). Although the system dynamics employed for such
Cooper-pair pumping can be of different kinds - see, e.g., \cite{Hoehne2012}, the most typical is the adiabatic
dynamics \cite{Geerligs1991}, in which the pumped charge is related \cite{Pekola1999,Aunola2003,Mottonen2008}
to Berry's phase or \cite{Faoro2003} its non-abelian extensions.

\subsection{Influence of environment on tunneling}
\label{sec:environment induced tunneling}

\begin{figure}[t]
    \includegraphics[width=5cm]{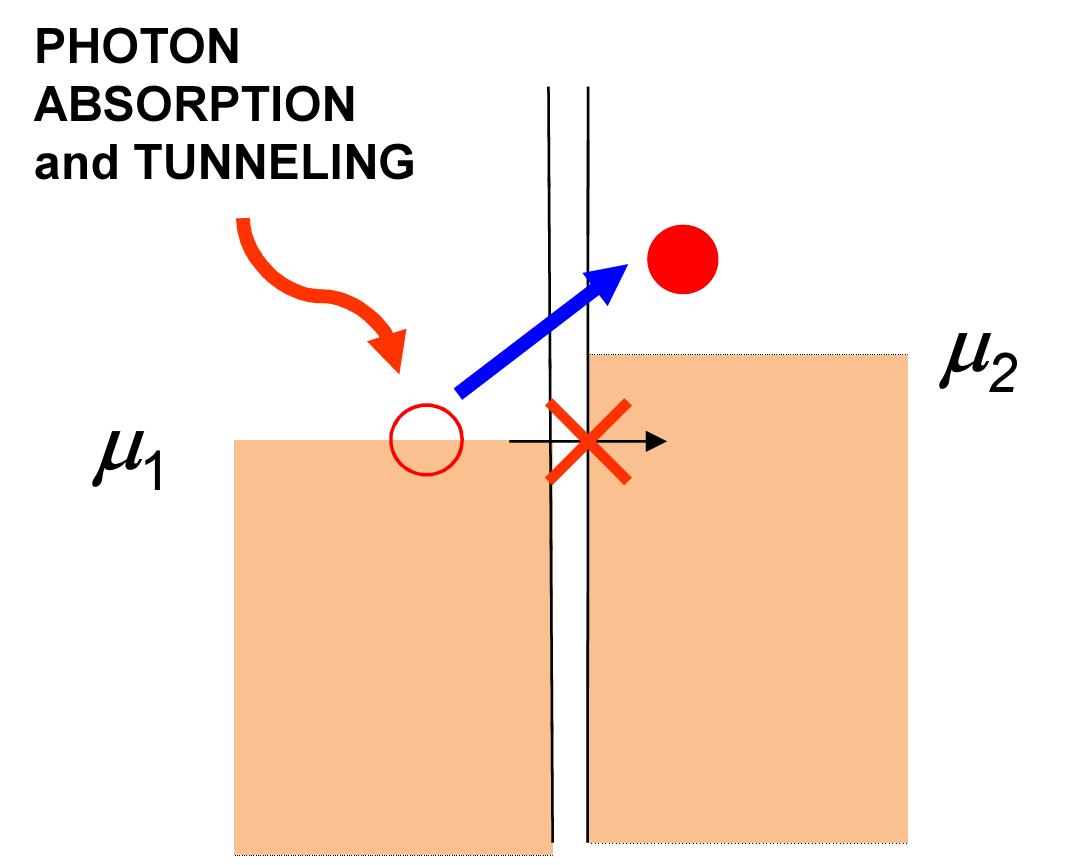}
    \caption{A simple schematic showing photon absorption by a generic tunnel junction, and the inelastic electron tunneling from the left side of the barrier to right.}
    \label{fig:PhotAbs}
\end{figure}

Tunneling in small junctions is influenced by electromagnetic environment. The tunneling rates are modified by photon absorption or emission; Figure~\ref{fig:PhotAbs} depicts schematically a process where tunneling rate in a generic junction is enhanced by absorption of a photon from the environment.

The general theoretical framework of how this happens was put forward in seminal works by Devoret {\sl et al.}~\cite{Devoret1990} and Girvin {\sl et al.}~\cite{Girvin1990}, and later expanded in the review by Ingold and Nazarov~\cite{Ingold1992}. The golden rule type tunneling rates discussed in the earlier sections get modified as
\begin{eqnarray} \label{env1}
\Gamma =
&&\frac{1}{e^2R_T}\int^{\infty}_{-\infty} \int^{\infty}_{-\infty} dE
dE'\, \nu_1(E-\delta E)\nu_2(E')\times\nonumber \\ && f_1(E-\delta E)[1-f_2(E')]P(E-E'),
\end{eqnarray}
where $\nu_i(E)$, $i=1,2$ are the normalized densities of states (DOS) in the two electrodes, $f_i(E)$ are the corresponding energy distributions in the electrodes, and $\delta E$ is the energy cost in the tunneling event. The function $P(E)$ can be interpreted as
the probability density to emit energy $E$ to the environment, which becomes a delta-function in the special case of a junction with perfect voltage bias. The $P(E)$ can be calculated as the following transformation using the phase-phase correlation function $J(t)$:
\begin{equation} \label{env2}
P(E) = \frac{1}{2 \pi \hbar} \int^{\infty}_{-\infty}
\exp \big (J(t) + \frac{i}{\hbar} Et \big) dt.
\end{equation}
By modeling the environment by a frequency $\omega/2\pi$ dependent impedance $Z(\omega)$ in
thermal equilibrium at temperature
$T_\textrm{{\footnotesize{env}}}$, one obtains%~\cite{Ingold92}
\begin{eqnarray}
\label{env3}
J(t) = && 2 \int^{\infty}_{0} \frac{d \omega}{\omega}
\frac{\Re [Z(\omega)]}{R_K}\times\nonumber\\&&\big[ \coth ( \frac{\hbar
\omega}{2k_B T_{\rm env}} ) ( \cos
\omega t - 1 ) - i \sin \omega t \big],
\end{eqnarray}
where $R_K = h/e^2$ is the resistance quantum.

\begin{figure}[t]
    \includegraphics[width=7.5cm]{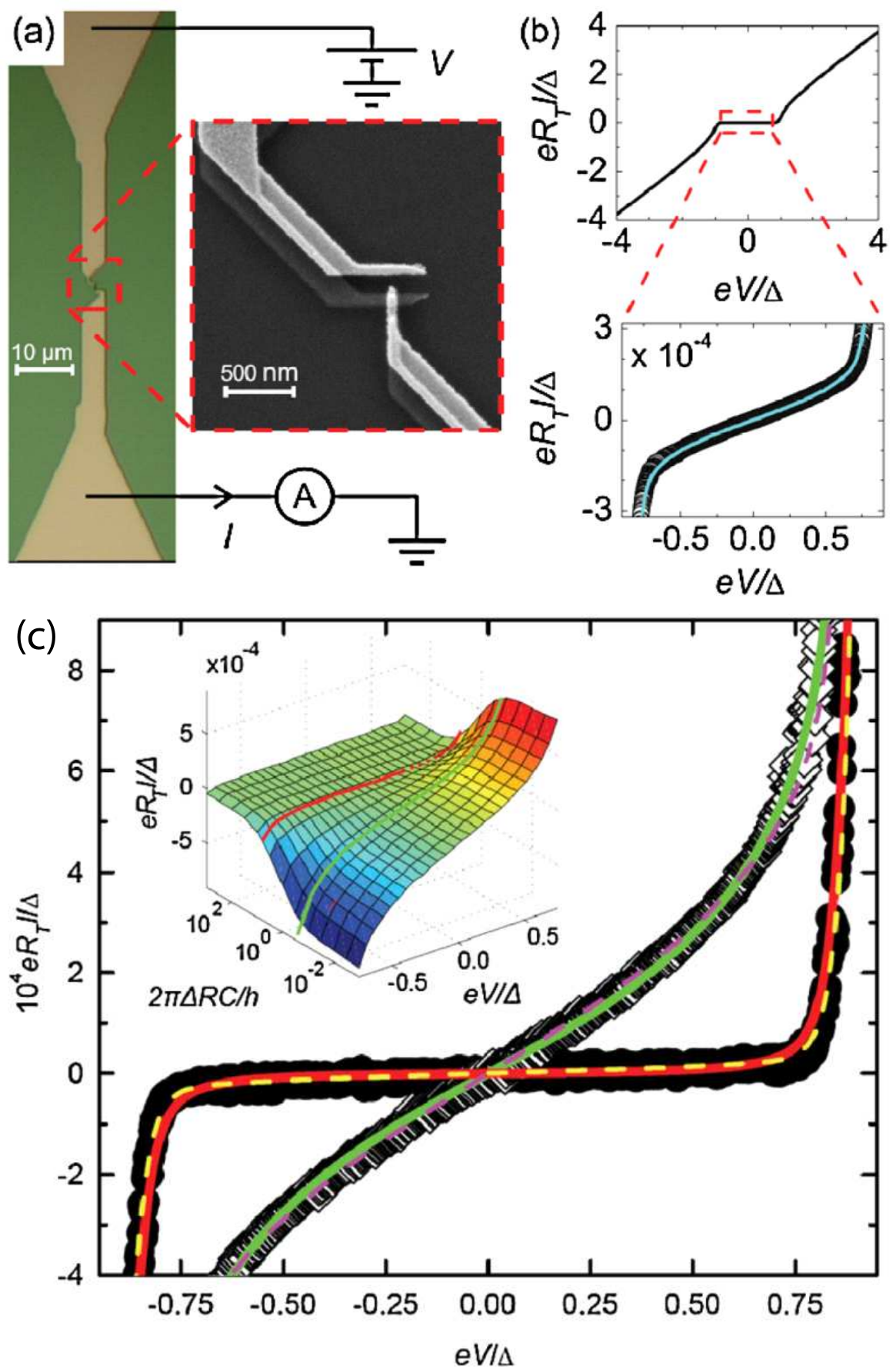}
    \caption{NIS junctions influenced by hot environment. (a) Geometry of a NIS junction made of aluminum (low contrast) as the superconductor
and copper (high contrast) as the normal metal. The tapered ends lead to large pads. (b) Typical IV
characteristics, measured at 50 mK for a junction with $R_T = 30$ k$\Omega$. Linear leakage, i.e. non-vanishing sub-gap current due to coupling to the environment, can be observed. The cyan line is the corresponding theoretical line from the $P(E)$ theory and $RC$ environment with dissipation $R$ at $T_{\rm env}=4.2$ K. (c) Measured IV curves of an NIS junction
with $R_T = 761$ k$\Omega$ on a ground plane providing a large protecting capacitance against thermal fluctuations (solid symbols) and of a
similar junction with $R_T = 627$ k$\Omega$  without the ground plane
(open symbols). Solid lines present the theoretical results for capacitance $C = 10$ pF (red line) and $C = 0.3$ pF
(green line). The resistance and the temperature of the environment
are set to $R=2$ $\Omega$ and $T_{\rm env}=4.2$ K, respectively. The inset shows IV curves
based on the full $P(E)$ calculation as functions of the shunt
capacitance $C$. The red and green lines are reproduced on this
graph from the main figure. Figure adapted from Ref.~\cite{Pekola2010}.}
    \label{fig:NIS}
\end{figure}

Often one can assume that the unintentional environment can be modeled as a wide-band dissipative source in form of and $RC$ circuit. For a purely resistive and capacitive environment
\begin{equation}
\label{env4}
\Re[Z(\omega)] = R/[1 + (
\omega RC)^2 ],
\end{equation}
where $R$ is the resistance of the
environment and $C$ is the total capacitance
including the junction capacitance and parallel shunt capacitors.
This rather simple model has been
successfully applied to explain several experimental
observations, see, e.g., ~\cite{Martinis1993,Hergenrother1995}. For a system with intentionally enhanced capacitance, it could be used to account for experimental improvement of the characteristics of an NIS junction and of a single-electron turnstile~\cite{Pekola2010}, and further improvements have been obtained in Refs.~\cite{Saira2010,Saira2012}, see Sec. 3.2. We show in Fig.~\ref{fig:NIS} an NIS-junction and its current-voltage characteristics under different experimental conditions.

Focusing to the single-electron sources, the environment has at least two effects to be considered. (i) The coupling of the black-body radiation of the hot surrounding environment can induce photon-assisted tunneling. (ii) The intentionally fabricated on-chip environment in the immediate vicinity of the single-electron circuit serves as a filter against the external noise. Moreover, it can influence the tunneling rates in a way improving the performance~\cite{Zorin2000,Lotkhov2001,Bubanja2011}. Detailed discussion on the error processes in pumps, including those due to coupling to the environment, is given further in this review.

\subsection{Heating of single-electron devices}
\label{sec:heating of single-electron devices}

Single-electron circuits operate optimally at low temperatures. The standard condition is that $k_BT \ll E_C$, where $E_C$ is the characteristic charging energy scale. Another condition in superconductor based devices is that $k_BT \ll \Delta$, where $\Delta$ is the energy gap of the superconductor. Since thermal errors in synchronized transfer of electrons are typically proportional to $e^{-E_{\rm CR}/k_BT}$, where $E_{\rm CR}$ is the characteristic energy (in the previous examples $E_C$ or $\Delta$), it is obvious that temperature needs to be more than an order of magnitude below $E_{\rm CR}/k_B$. At low temperatures the overheating becomes a critical issue~\cite{Giazotto2006}. The energy relaxation between the electron system and the bath, typically formed by the phonons, becomes increasingly slow towards low temperatures. Moreover, the various heating rates are typically not scaling down similarly with decreasing temperature.

%Figure xx shows a simplified thermal model of a single-electron device.
Heat is injected to the electron system, first and foremost, as Joule heating due to current in a biased circuit. Other sources of heat include application of dissipative gate voltages or magnetic flux injection, thermal radiation discussed in Section~\ref{sec:environment induced tunneling} and shot-noise induced dissipation by back-action from a charge or current detector. The steady state temperature of the electron system is determined by the balance between the input powers and the heat currents via different relaxation channels. The injected energy relaxes to phonons via electron-phonon relaxation, to the leads by heat transport through the tunnel junctions, and by radiation to other dissipative elements in the cold circuit. We will discuss these processes in more detail below.

{\sl Joule heating and cooling:} In a biased circuit, the total Joule power is $P=IV$ where $V$ is the overall voltage and $I$ the current. This power can, however, be distributed very un-evenly in the different parts of the circuit: in an extreme example, some parts may cool down whereas the others are heavily overheated. Let us focus on dissipation in biased tunnel junctions. The basic example is a tunneling process in a junction between two conductors with essentially constant density of states, which is the case presented by normal metals. At the finite bias voltage $V$ the tunneling electron leaves behind a hole-like excitation and it creates an excited electron in the other electrode, i.e., both electrodes tend to heat up. Quantitatively we can write the expression of the power deposited in the, say, right electrode as
\begin{equation} \label{jh1}
P_R = \frac{1}{e^2R_T}\int dE \, E [f_L(E-eV)-f_R(E)].
\end{equation}
Here, $f_{L,R}$ refer to the energy distributions on the left (L) and right (R) side of the junction, respectively. $P_R=\frac{V^2}{2R_T}$ when $f_L =f_R$, i.e., when the temperatures of the two sides are equal.  By symmetry, or by direct calculation we can verify that the same amount of power is deposited to the left electrode in this situation. Thus the total power dissipation equals $P = P_L + P_R = V^2/R_T = IV$, as it should.

If one of the conductors is superconducting, the current-voltage characteristics are non-linear and the power deposited to each electrode is given by
\begin{equation} \label{jh2}
P_{N,S} = \frac{1}{e^2R_T}\int dE \, \tilde E_{N/S}\, \nu_S(E)[f_N(E-eV)-f_S(E)].
\end{equation}
Here $\tilde E_N = E-eV$, and $\tilde E_S=E$, where $N,S$ now refer to the normal and superconducting leads, respectively. The overall heating is again given by $IV$, but in this case, under bias conditions $eV\simeq \Delta$, $P_N$ on the normal side can become negative (NIS cooling, \cite{Giazotto2006}) and $P_S$ on the superconductor side is always positive, i.e. it is heated up.

Finally, if both sides are superconducting, the current-voltage characteristics are highly non-linear, but due to symmetry $P_L =P_R =IV/2=P/2$.

{\sl Other heating sources:} Overheating of a single-electron circuit can be caused by various other sources. Ac gate-voltages or ac magnetic fluxes can induce dissipative currents and heating due to dielectric losses, and single-electron electrometry or electrometry by a quantum point contact detector can cause effective heating due to the shot-noise back-action coupling to the single-electron circuit, just to mention a few possibilities.

{\sl Energy relaxation by conduction to leads:} If a difference between the electronic temperatures $T_L$ and $T_R$ of the left and right leads exists, $\Delta T \equiv T_L-T_R$, heat $P_{L\rightarrow R}$ can flow electronically through the tunnel barrier. In the case of a normal-normal junction, we have
\begin{eqnarray} \label{jh3}
P_{L\rightarrow R}  && =\frac{1}{e^2R_T}\int dE \, E [f_L(E)-f_R(E)]\nonumber\\&&=\frac{\pi^2k_B^2}{6e^2R_T}(T_L^2-T_R^2).
\end{eqnarray}
where in the last step we assume that the junction is not biased. For a small temperature difference $\Delta T$ about the mean $T=(T_L+T_R)/2$ of the two temperatures, we may then write the thermal conductance $G_{\rm TH}\equiv P_{L\rightarrow R}/\Delta T$ of a NIN tunnel junction as
\begin{equation} \label{jh4}
G_{\rm TH}=\frac{\pi^2k_B^2T}{3e^2R_T},
\end{equation}
which is the Wiedemann-Franz law for a conductor with resistance $R_T$.
For either an NIS or SIS junction, heat conductance is exponentially small at low temperatures due to $\Delta$. Another mechanism for the heat flow is the diffusion in the leads. It is discussed, in particular in superconducting ones, in Section~\ref{sec:sinis_qp}.

{\sl Electron-phonon relaxation:} Electron phonon relaxation is one of the dominant and in many systems one of the best understood relaxation mechanisms. For a normal metal conductor with a uniform temperature $T$ that differs from the bath phonon temperature $T_0$, one may write quite generally~\cite{Wellstood1994}
\begin{equation} \label{jh5}
P_{\rm e-p} =\Sigma \mathcal{V} (T^5-T_0^5),
\end{equation}
where $\Sigma$ is a material constant of order $10^9$ WK$^{-5}$m$^{-3}$~\cite{Giazotto2006}, and $\mathcal{V}$ is the volume of the conductor. This equation holds amazingly well at sub-kelvin temperatures for various metals, irrespective of their dimensions.
In single-electron devices, we typically consider dissipation in a small Coulomb-blockaded region, whose volume and thus, according to Eq.~(\ref{jh5}), the coupling to the phonon bath is weak. Due to the small dimensions, one can typically assume a spatially uniform energy distribution on the conductor, moreover, assumption of overheating with a well defined electron temperature is also justified quite generally.

In some cases these assumptions are not necessarily valid. An important exception is given by superconductors where energy relaxation via phonon emission becomes extremely weak due to the energy gap. At low temperatures the relaxation is limited by the emission of $2\Delta$ phonons corresponding to the recombination of quasiparticles into Cooper pairs~\cite{Rothwarf1967}. In the past few years, several experiments have measured the relaxation rate in this context, see, e.g., \cite{Barends2008}, and the corresponding energy release rate was measured recently in~\cite{Timofeev2009}. According to the latter measurement the recombination-related heat flux is strongly suppressed from that given in Eq.~(\ref{jh5}), being about two orders of magnitude weaker than in the normal state at the temperature $T=0.3T_C$, where $T_C$ is the critical temperature of aluminium. At even lower temperatures the heat current is further suppressed, eventually exponentially as $\propto e^{-\Delta/k_BT}$. Besides recombination, also the diffusive heat conduction is strongly suppressed in a superconductor at $T\ll T_C$. This means that a superconductor is a poor material as a lead of a single-electron source, where non-equilibrium quasiparticles are injected at the rate $f$. The situation can be improved by inserting so-called quasiparticle traps into the circuit, discussed in section~\ref{sec:sinis_qp}.  Yet a fully superconducting Cooper-pair pump can be dissipationless ideally.

\begin{figure}[t]
    \includegraphics[width=8.5cm]{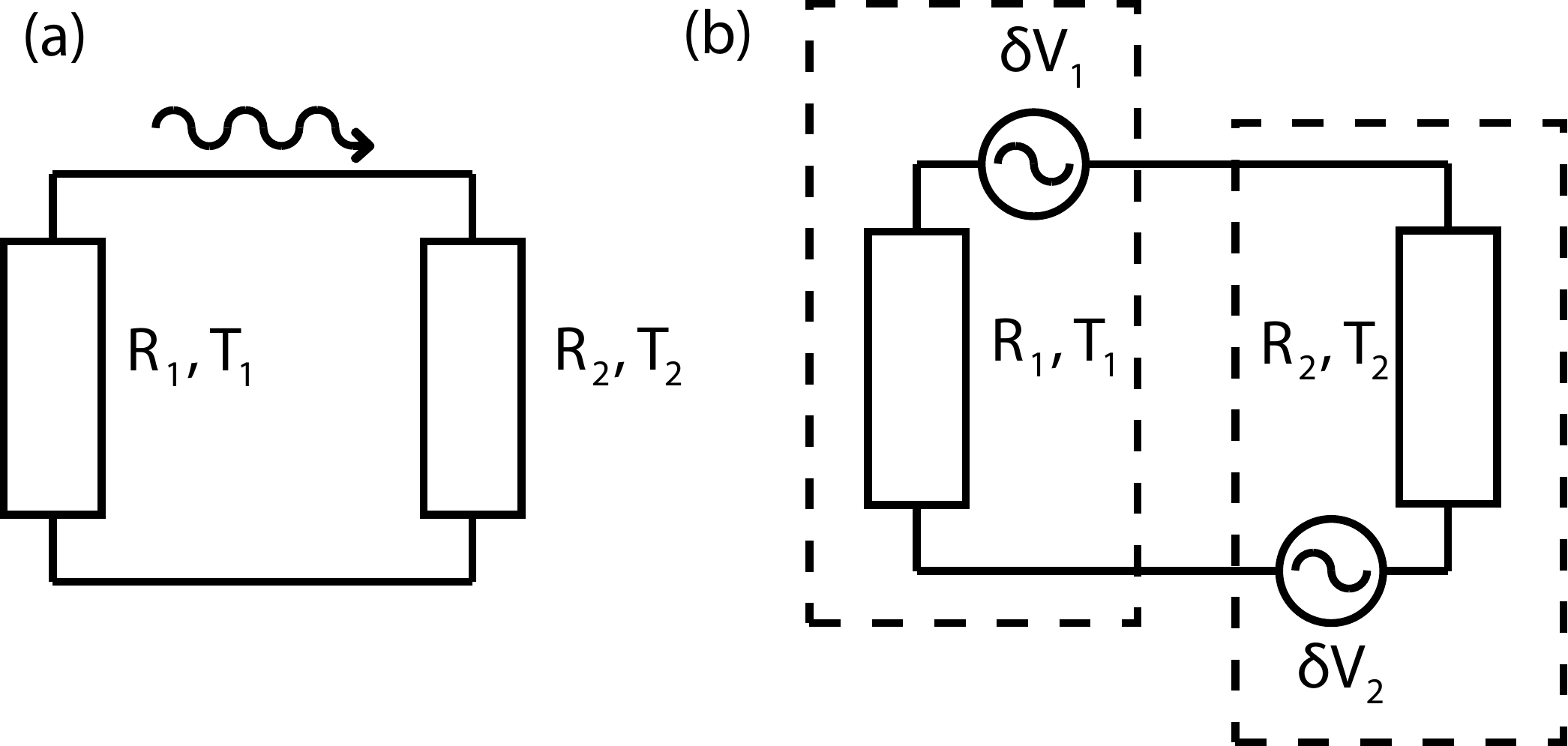}
    \caption{(a) Radiative heat flow is caused by the photons which carry energy between resistors $R_1$ and $R_2$ at temperatures $T_1$ and $T_2$ respectively. The heat transport can be modelled by having voltage fluctuations $\delta V_i$ as shown in (b). Here we have assumed total transmission. The assumption can be relaxed by adding a non-zero impedance to the loop.}
    \label{fig:radHeat}
\end{figure}

{\sl Heating and cooling by radiation:} Coupling of a junction to the electromagnetic environment is associated with heat exchange. Hot environment can induce photon assisted tunneling as discussed in subsection \ref{sec:environment induced tunneling} above. The basic concept of radiative heat transport in an electric circuit has been known since the experiments of Johnson and Nyquist \cite{Johnson1928,Nyquist1928} more than eighty years ago. Electromagnetic radiation on a chip has recently turned out to be an important channel of heat transport at low temperatures~\cite{Schmidt2004,Meschke2006,Timofeev2009a}. If two resistors $R_1$ and $R_2$ at temperatures $T_1$ and $T_2$ are connected directly to each other in a loop the heat exchange between them can be modeled by Langevin type circuit analysis, as indicated in Fig.~\ref{fig:radHeat} by the voltage sources producing thermal noise. Assuming an  idealized quantum limit, where the circuit transmits all frequencies up to the thermal cut-off at $\omega_{\rm th}=k_BT_i/\hbar$, the net heat current between the two resistors is given by
\begin{equation} \label{jh6}
P_{\gamma} =\frac{R_1R_2}{(R_1+R_2)^2}\frac{\pi k_B^2}{12\hbar}(T_1^2-T_2^2).
\end{equation}
This is an interesting limit which applies for circuits on a chip where the stray capacitances and inductances are small enough such that the circuit low-pass cut-off frequency exceeds $\omega_{\rm th}$. Equation~(\ref{jh6}) has an important limit for the maximum coupling with $R_1=R_2$ and for small temperature differences $|T_1-T_2|$, namely $G_{\rm TH}=P_\gamma/(T_1-T_2)=(\pi k_B^2 T)/(6\hbar)\equiv G_Q$, the so-called quantum of thermal conductance~\cite{Pendry1983}. Equation (\ref{jh6}) can be applied also in the case where one of the resistors is replaced by a NIN tunnel junction with the corresponding resistance. Another important case is that of a hot resistor $R$ at temperature $T_{\rm env}$, discussed in the previous subsection. In this limit, with $RC$ cut-off as discussed in  \ref{sec:environment induced tunneling}, one finds that the heat absorption rate by a resistor or a normal tunnel junction (at $T \ll T_{\rm env}$) is given approximately by~\cite{Pekola2007a}
\begin{equation} \label{jh7}
P_{\gamma} =\frac{k_BT_{\rm env}}{R_TC}.
\end{equation}

%
% Chapter 3
%

\section{REALIZATIONS}
\label{sec:realizations}

\subsection{Normal metal devices}
\label{sec:metalpumps}

Single-electron tunneling effects provide means to transport electrons controllably one by one. In this respect the obvious choices are metallic single-electron circuits and semiconducting quantum dots. The metallic ones can be either in their normal state or superconducting, or as hybrids of the two. The quantum dots, metallic hybrids and superconducting circuits will be discussed in later sections. The first single-electron source was a metallic (non-superconducting) turnstile with four tunnel junctions and one active gate~\cite{Geerligs1990}. The word "turnstile" refers to a device that is voltage biased at $V_{\rm b}$ between the external leads, but where the transport of electrons is impeded under idle conditions because of an energy gap. Under the active gate operation, electrons are transported synchronously one at a time. The finite voltage determines the direction of charge transport, at the expense that the device is also dissipative. We will discuss a more recent version of a turnstile in the subsection \ref{sec:hybridpumps}.

\begin{figure}[t]
    \includegraphics[width=0.49\textwidth]{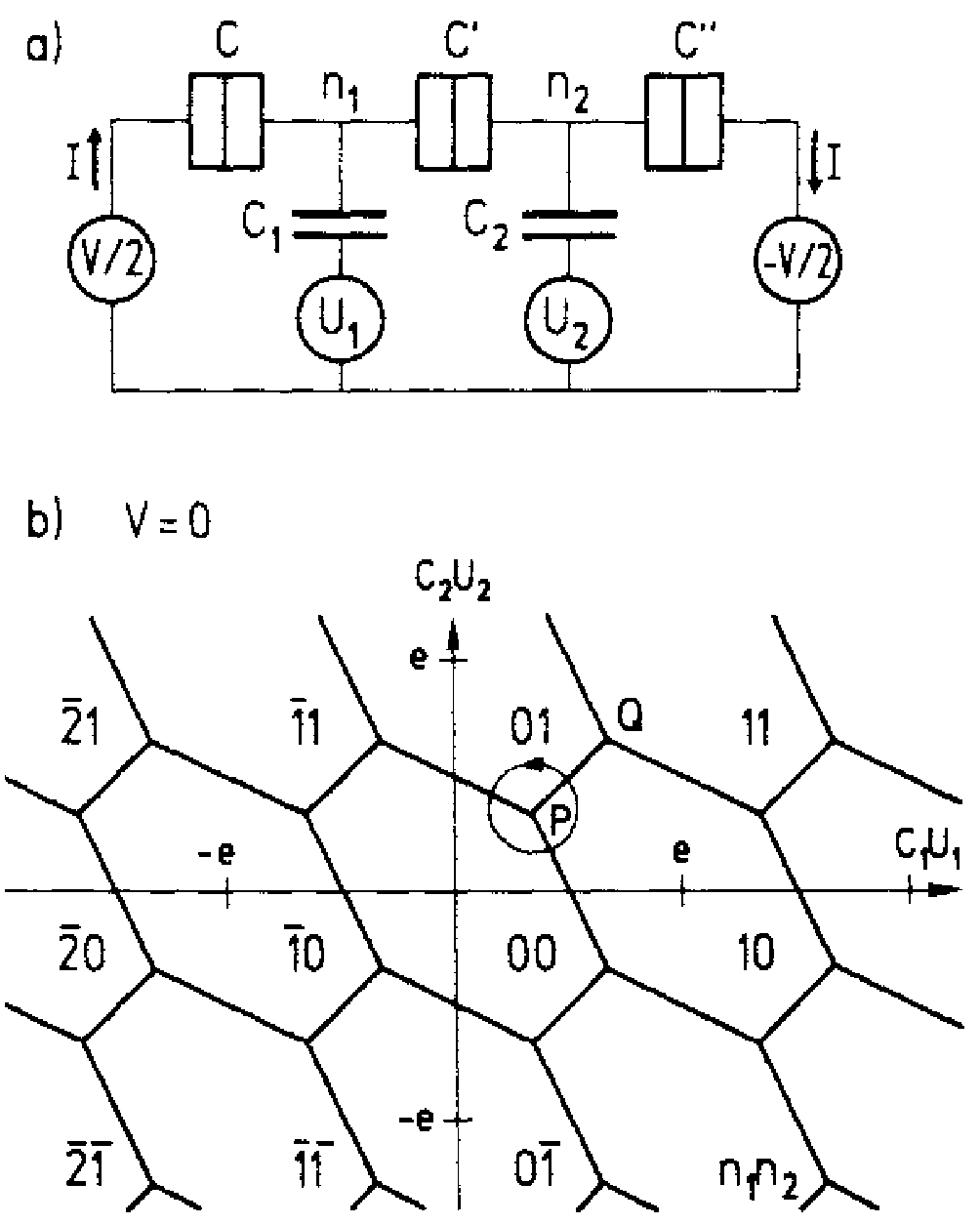}
    \caption{A three-junction pump. Schematics shown in (a), where the pump is biased by voltage $V$ and with gate voltages $U_1$ and $U_2$. (b) The stability diagram of the three-junction pump on the plane of the gate voltages at zero bias voltage. For operation of the pump, see text. Figure from Ref.~\cite{Pothier1991a}.}
    \label{fig:threejunctionpump}
\end{figure}

The most impressive results at the early days of single-electron sources were obtained by metallic multi-junction pumps, operating in a non-superconducting state. A prototype of them, featuring the main principle is the three-junction pump, with two islands and a gate to each of them, see Fig.~\ref{fig:threejunctionpump} (a). This kind of a pump was successfully operated in 1991 by Pothier at al.~\cite{Pothier1991,Pothier1991a,Pothier1992}. Figure~\ref{fig:threejunctionpump} (b) demonstrates the stability diagram of a three-junction pump, which is essentially the same as that of the more common double-island quantum-dot circuit. The two axes here are the two gate voltages, $n_{\rm g1},n_{\rm g2}$, normalized by the voltage corresponding to charge displacement of one electron, i.e., $n_{{\rm g}i}=C_{{\rm g}i}V_{{\rm g},i}/e$, where $C_{{\rm g}i}$ is the gate capacitance of island $i$. The stability diagram consists of lines separating different stable charge states on the islands, indicated by indices $(n_1,n_2)$ in the figure. The important property of this stability diagram is the existence of the nodes where three different charge states become degenerate, being the three states with the lowest energy. The pump is operated around such a node, setting the working point at this node by applying dc voltages to the two gates. Let us focus on one such node, that at $n_{\rm g1}=n_{\rm g2}=1/3$. Now, if the temporally varying gate voltages with frequency $f$ added to these dc gate biases are such that the cyclic trajectory encircles the node at $n_{\rm g1}=n_{\rm g2}=1/3$ counterclockwise, one electron is transported through the pump from left to right. The simplest implementation of such a cyclic trajectory is a circle around the node, which is represented by two equal amplitude (in $n_{{\rm g}i}$) sinusoidal voltages applied to the two gates, phase-shifted by 90 degrees. Let us take point A as the starting point of the cycle. There the system is in the charge state $(0,0)$. Upon crossing the first degeneracy line, the new stable charge state is $(1,0)$, meaning that if an electron has to tunnel from the left lead to island 1, while moving in this part of the stability diagram. Reversible pumping is achieved when $f$ is so slow that the transition occurs right at the degeneracy line. If, however, the pumping frequency is too fast, the tunneling is not occuring before meeting the next degeneracy line, and the pumping fails. Roughly speaking, the tunneling process is stochastic, where the decay time of the Poisson process is determined by the junction resistance (see section~\ref{sec:seqtunneling}), and if the pumping frequency becomes comparable to the inverse decay time for tunneling, the desired event can be missed. In the successful cycle, on the contrary, the system next crosses the degeneracy line between charge states $(1,0)$ and $(0,1)$, and under the same conditions, the system transits to the new stable charge state by an electron tunneling from the left island to the right one. In the remaining part of the cycle, on crossing the last degeneracy line, an electron tunnels from the right island to the right lead completing the cycle where charge $e$ (one electron) has been transported from the left lead to the right one. By cyclically repeating this path at frequency $f$, an average current $I=ef$ runs from right to left, and this current can be read, for instance, by a regular transimpedance amplifier.

One of the advantages of the three junction pump over the early turnstiles is that the device can be operated, in principle, reversibly, since no external bias voltage is needed. It can pump even against moderate bias. Another difference between the pump and the turnstile above is that in the pump there are no unattended islands, on which the charge would be poorly controlled. Yet early realizations using fully normal metal conductors in both the turnstiles and pumps suffered from other error sources which made these devices relatively inaccurate, on the level of 1\%, even at low operation frequencies. A fundamental error source in this case is co-tunneling, discussed in Section~\ref{sec:cotunneling}. To circumvent this problem, a pump with a longer array of junctions is desirable: the error rate due to co-tunneling is effectively suppressed by increasing the number of junctions in the array.

\begin{figure}[t]
    \includegraphics[width=0.49\textwidth]{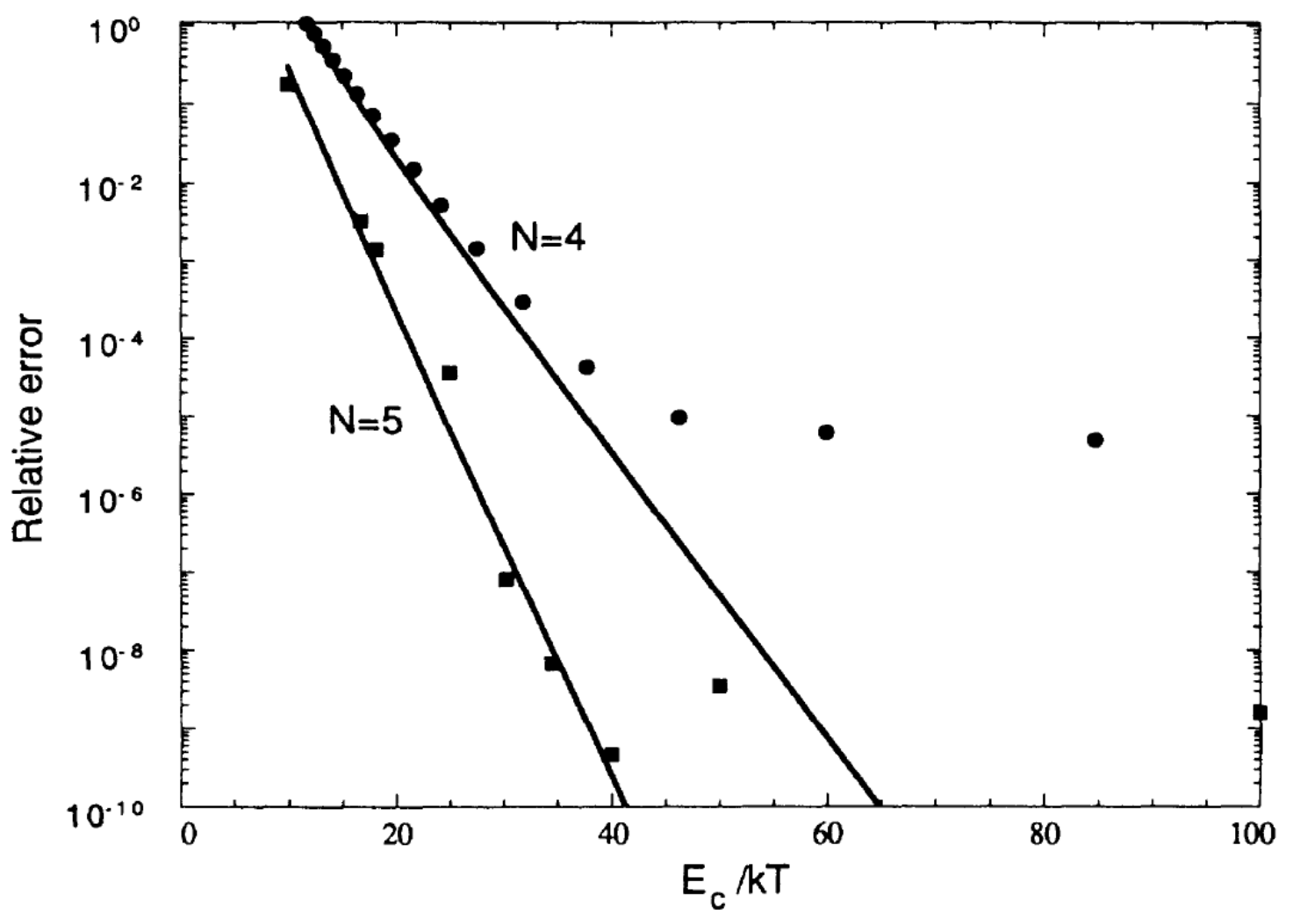}
    \caption{Predicted relative co-tunneling induced error versus inverse temperature for the multi-junction pumps, with $N=4$ (circles) and $N=5$ (squares). The computer simulations (points) and the predictions of analytic results (lines) are shown. Parameters are $R_T=20R_K$, $f =4\cdot 10^{-4} /(R_TC)$, and $CV/e=-0.15$. Figure adapted from Ref.~\cite{Jensen1992}.}
    \label{fig:JensenMartinis92}
\end{figure}

\begin{figure}[t]
    \includegraphics[width=0.49\textwidth]{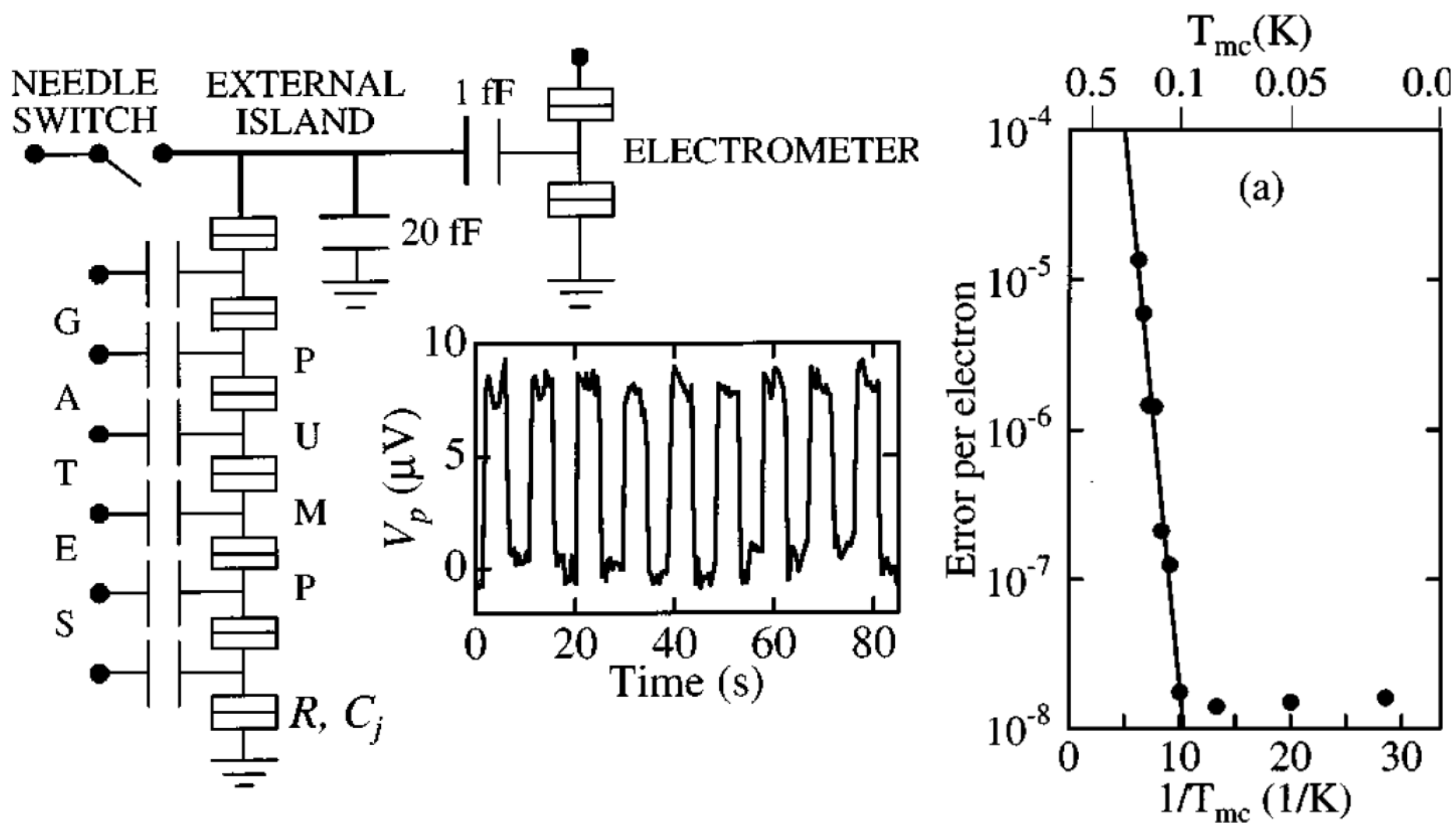}
    \caption{The seven junction pump. The plot on the left shows the schematic of the pump, with six islands, each with a gate. The electrons are pumped to/from the external island on the top, and the charge on the island is detected by a single-electron electrometer. The plot in the middle shows the voltage $V_p$ on the external island vs time when pumping $\pm e$ with a
wait time of 4.5 s in between. The graph on the right shows the pumping error vs temperature of the measurement, demonstrating the 15 ppb accuracy at temperatures below 100 mK. Figure adapted from Ref.~\cite{Keller1996}.}
    \label{fig:KellerAPL}
\end{figure}

Theoretical analysis of co-tunneling in multi-junction pumps in form of $N$ junctions in series with non-superconducting electrodes was performed in Ref.~\cite{Jensen1992,Averin1993}. Thermal co-tunneling errors were analyzed with focus on $N=4$ and $N=5$. The conclusion of the analysis was that under realistic experimental conditions, $N=4$ pump fails to produce accuracy better than about $10^{-5}$, insufficient for metrology, whereas $N=5$ should be sufficiently good at low operation frequencies, as far as co-tunneling is concerned. This is demonstrated in Fig.~\ref{fig:JensenMartinis92}, where relative error of $10^{-8}$ was predicted for an $N=5$ pump at the operation frequency of $f=1.3$ MHz, for a pump with junctions having $R_T=500$ k$\Omega$ and $C=0.6$ fF, if we assume that the working temperature is $T=50$ mK. All these parameters are quite realistic. However, one notes already here that the frequency at which such a multi-junction pump can be operated is very low. In the subsequent experiments~\cite{Martinis1994} the error rate of about $0.5$ ppm could be achieved, but still orders of magnitude above the prediction based on co-tunneling for their circuit parameters and experimental conditions. The authors conjectured this discrepancy to arise from photon-assisted tunneling, see Section~\ref{sec:environment induced tunneling}, considered in this context theoretically, e.g., in Refs.~\cite{White1993,Martinis1993}. After the experiment in~\cite{Martinis1994}, focus was turned into $N=7$ pump where further improved results, 15 pbb, could be obtained at pumping frequencies of about $10$ MHz~\cite{Keller1996}. This impressive result, depicted in Fig.~\ref{fig:KellerAPL}, however of limited applicability in metrology of redefining ampere due to the small magnitude of the current, was proposed to present a capacitance standard based on electron counting~\cite{Keller1999}. Further analysis and experiments on errors in multi-junction pumps can be found, e.g., in Refs.~\cite{Kautz1999,Covington2000,Kautz2000,Jehl2003}.

Another succesful line of metallic single-electron pumps relies on a smaller number of junctions with the help of dissipative on-chip environment used to suppress harmful cotunneling and photon assisted tunneling~\cite{Lotkhov2001,Camarota2012}. These devices will be discussed further in Sections~\ref{subsec:ecs} and \ref{sec:indirectqmt}.

\subsection{Hybrid superconducting--normal metal devices}
\label{sec:hybridpumps}

\subsubsection{Operating principles}

\begin{figure}[t]
    \includegraphics[width=0.49\textwidth]{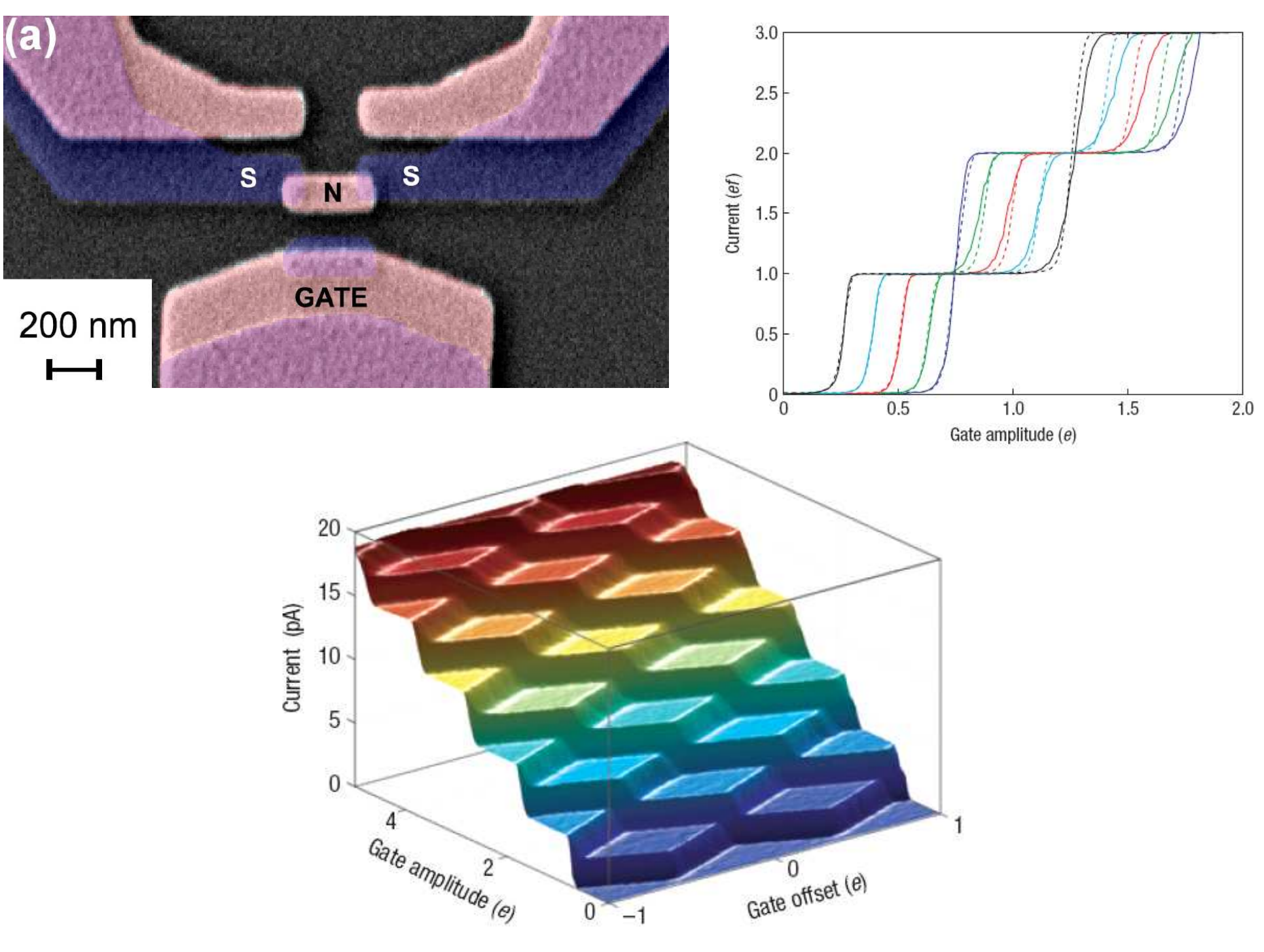}
    \caption{The hybrid NIS turnstile. Top left: A scanning electron micrograph of a SINIS turnstile, which is a hybrid single-electron transistor with superconducting leads and a normal metal island. Top right and bottom: current of a turnstile under rf drive on the gate at different operation points with respect to the dc gate position and rf amplitude of the gate. Figure adapted from Refs.~\cite{Pekola2008,Kemppinen2009a}.}
    \label{fig:SINIS1}
\end{figure}

The hybrid turnstile, originally proposed and demonstrated in  Ref.~\cite{Pekola2008}, is based on a single-electron transistor where the tunnel junctions are formed between a superconductor and a normal metal, see Fig.~\ref{fig:SINIS1} on top left. In principle, it can be realized either in a SINIS or NISIN configuration~\cite{Averin2008,Kemppinen2009}. However, it has turned out for several reasons that the former one is the only potential choice out of the two for accurate synchronized electron transport purposes~\cite{Averin2008}. One reason is that in the NISIN structure, tunneling strongly heats the island due to Joule power and weak energy relaxation in the small superconducting island, whereas in the SINIS case the island is of normal metal, better thermalized to the bath, and under proper operation, it can be cooled, too~\cite{Kafanov2009}. The NISIN turnstile
may also suffer from unpredictable $1e/2e$ periodicity issues. Furthermore, a detailed analysis of the higher-order
tunneling processes shows that cotunneling limits the fundamental accuracy of the NISIN
turnstile, whereas uncertainties below $10^{-8}$ are predicted for the SINIS
version~\cite{Averin2008}. Hence we focus on the SINIS turnstile here.

\begin{figure}[t]
    \includegraphics[width=0.49\textwidth]{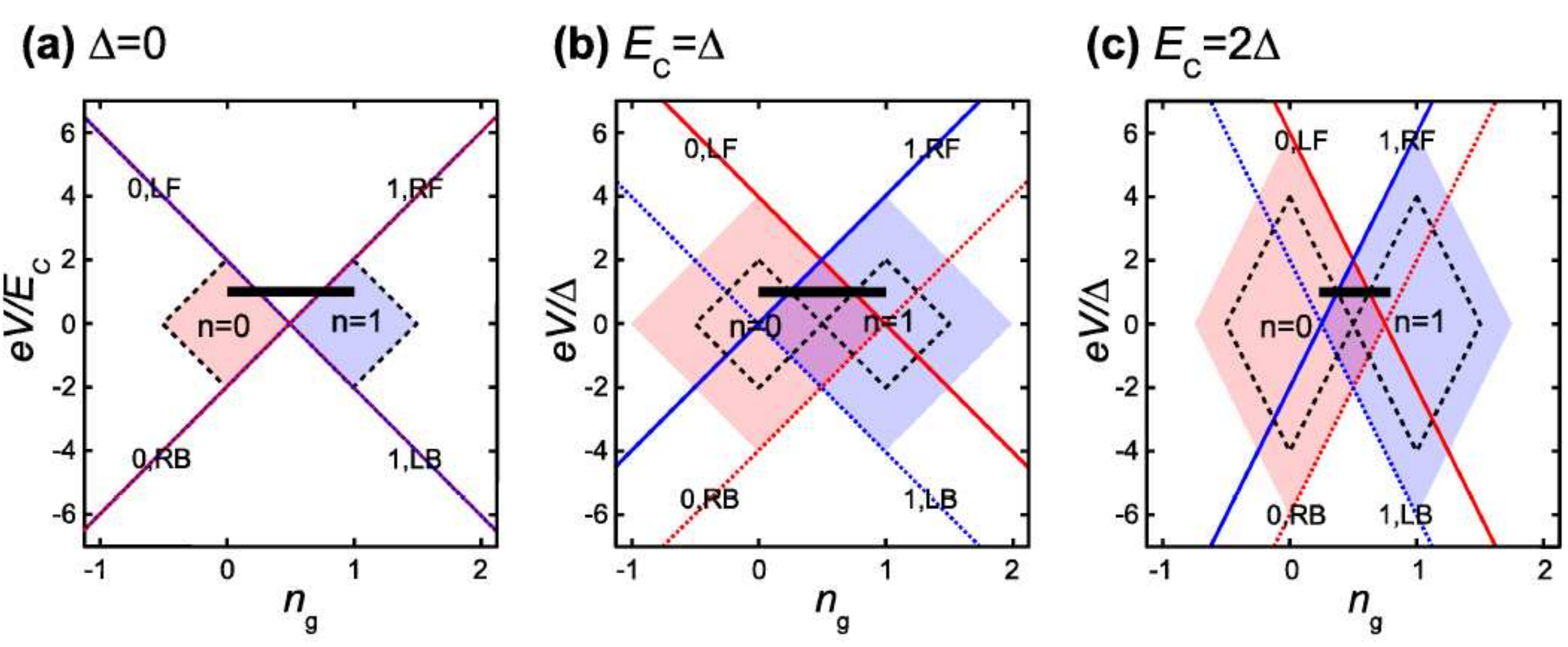}
    \caption{Schematic picture of pumping (a) with a normal SET, (b) with a
   hybrid SET with $E_C=\Delta$, and (c) with a hybrid SET with $E_C=2\Delta$.
   The shaded areas are the stability regions of the charge states $n=0$ and $n=1$. The edges of the
   normal SET stability regions are drawn to all figures with dashed black lines. The long coloured
   lines represent the transition thresholds from states $n=0$ and $n=1$ by tunneling through the
   left (L) or the right (R) junction in the wanted forward (F, solid line) or unwanted backward
   (B, dashed line) direction. The thick black line corresponds to pumping with constant bias voltage $eV_\mathrm{b}/\Delta=1$ and
   a varying gate voltage. Figure from Ref.~\cite{Kemppinen2009a}.}
    \label{fig:timantit}
\end{figure}

The stability diagram of a conventional single-electron transistor is composed of Coulomb diamonds on the gate voltage $V_{\rm g}$ - drain-source voltage $V_{\rm b}$ plane, see Fig. \ref{fig:timantit}. Gate voltages $V_{\rm g}$ are again written in dimensionless form, normalized by the voltage corresponding to charge displacement of one electron, $n_{\rm g}$. In this case the adjacent diamonds touch each other at a single point at $V_{\rm b}=0$, implying that the charge state is not locked for all gate voltage values. The operation of the SINIS-turnstile, on the contrary, is based on the combined effect of the two gaps: the superconducting BCS gap expands the stability
regions of the charge states and the neighboring regions overlap. The principle of operation of the
turnstile is illustrated in Fig.~\ref{fig:timantit}. When the gate charge $n_\mathrm{g}(t)$
is alternated between two neighboring charge states, electrons are transported
through the turnstile one by one. A non-zero voltage, which yields a preferred direction of
tunneling, can be applied since the idle current is ideally zero in the range
$|eV_\mathrm{b}|<2\Delta$ at any constant gate charge value. If the gate signal is
extended to span $k+1$ charge states, one obtains current plateaus with $k$ electrons
pumped per cycle. However, the first plateau around the symmetric (degeneracy of two neighbouring charge states) dc position of the gate is optimal for metrology. Note that if a
non-zero bias voltage is applied across a normal-state SET, a gate span between
different charge states always passes a region where none of the states is stable
and where the current can freely flow through the device (white region in
Fig.~\ref{fig:timantit}(a)). Hence the normal-state SET cannot act as a turnstile even in principle, except for an experimentally unfeasible gate-sequence where $n_g$ jumps abruptly between its extreme values in the Coulomb blockaded parts of the stability diagram.

%\begin{figure}[h]
%    \begin{center}
%    \includegraphics[width=1.0\textwidth]{timantit}
 %   \end{center}
  % \caption{\label{fig:timantit} Schematic picture of pumping (a) with a normal SET, (b) with a
%   hybrid SET with $E_\mathrm{c}=\Delta$, and (c) with a hybrid SET with $E_\mathrm{c}=2\Delta$.
%   The shaded areas are the stability regions of the charge states $n=0$ and $n=1$. The edges of the
%   normal SET stability regions are drawn to all figures with dashed black lines. The long coloured
%   lines represent the transition thresholds from states $n=0$ and $n=1$ by tunneling through the
%   left (L) or the right (R) junction in the wanted forward (F, solid line) or unwanted backward
%   (B, dashed line) direction. We define the bias voltage to be positive in the left electrode.
%   Hence, the forward tunneling direction of electrons is from right to
%   left. The thick black line corresponds to pumping with constant bias voltage $eV_\mathrm{b}/\Delta=1$ and
%   a varying gate voltage [III].}
%\end{figure}

Figure \ref{fig:SINIS1} presents data obtained by a basic turnstile operated under various conditions~\cite{Pekola2008}. Several wide current plateaus with increasing gate amplitude $A_{\rm g}$ can be seen. The gate drive is expressed here as
$n_\mathrm{g}(t)=n_\mathrm{g0}+A_\mathrm{g}w(t)$ where $n_\mathrm{g0}$ and $A_\mathrm{g}$ are
the gate offset and drive amplitude, respectively. The gate waveform of unit amplitude,
is denoted by $w(t)$. The optimal gate drive is symmetric with respect to the two charge
states: therefore in later sections we assume that $n_\mathrm{g0}=1/2$. In these first experiments, accuracy of synchronized charge transport as $I=Nef$, with $N$ the integer index of a plateau, could be verified within about $1$\%.

A rough estimate for the optimal bias voltage $V_{\rm b}$ is obtained by considering the dominant thermal errors~\cite{Pekola2008}. The probability of an electron tunneling against bias, i.e., "in the wrong direction"
is given approximately by $\sim \exp (-eV_\mathrm{b}/k_\mathrm{B}T)$. This error would lead to no net charge transferred during a pumping cycle, but it can be suppressed
by increasing $V_{\rm b}$. On the other hand, increasing $V_{\rm b}$ increases the probability of transporting an extra electron in the forward direction. The magnitude of this kind of an error can be estimated as
$\sim \exp (-(2\Delta-eV_\mathrm{b})/k_\mathrm{B}T)$, since there is an energy cost given by the voltage distance from the conduction threshold at $2\Delta/e$. Combining these conditions, we obtain a trade-off $eV_\mathrm{b}\approx\Delta$
as the optimum bias voltage, where the thermal error probability is approximately $\sim\exp (-\Delta/k_\mathrm{B}T)$. The
combined thermal error probability is $\sim 10^{-9}$ at realistic temperatures of about 100~mK
and with the BCS gap of aluminum, $\Delta/k_\mathrm{B}\approx 2.5$~K. The exact optimum
of the bias close to the value given here depends on many other processes to be discussed below. Experimentally, however, the choice $eV_\mathrm{b}=\Delta$ is a good starting point.

 The optimal gate drive amplitude $A_g$
lies somewhere between the threshold amplitudes for forward and backward tunneling which are
$A_\mathrm{g,ft}=\Delta/4E_\mathrm{c}$ and $A_\mathrm{g,bt}=3\Delta/4E_\mathrm{c}$
for the optimum bias voltage, respectively. The sub-gap
leakage is maximized at the degeneracy point $n_\mathrm{g0}=1/2$. In this respect, a square-wave signal is optimal. On the other
hand, passing the threshold for forward tunneling too quickly tends to heat the island,
whereas a sine signal can also cool it. Hence, the optimal waveform is
of some intermediate form.

The SINIS turnstile presents a choice of a single-electron source which is easy to manufacture and operate, and whose characteristics can be analyzed theoretically into great detail. It promises high accuracy, as will be discussed in the next section. Its operation in a parallel configuration is straightforward thanks to the simple element of a single turnstile, and therefore it can yield higher currents than the other fixed-barrier single-electron sources presented in the previous subsection. Thus it can be considered as a very promising candidate in providing a realization of ampere.

%\subsection{Reaching the metrological level of accuracy and current magnitude in hybrid structures}
%The obtainable accuracy of a single-electron source is of prime interest especially for metrological applications. In this section we consider two main effects which limit the accuracy, namely the higher order tunneling processes and the relaxation of quasiparticle excitations. When higher order tunneling takes place, more than one electron tunnel simultaneously instead of the wanted single electon. The extra electrons leaking through cause excessively high current. The other leakage process arises as transfered electrons create excitations. These excitations lying at high energy states can then easily tunnel again and cause errors unless the relaxation of the electrons is made strong enough. It turns out to be crucial to take care of the relaxation as seen in the following sections.

%In addition to the accuracy of the pumped current, the output signal should be large enough to make measurements in reasonable averaging time. We will address the obtainable output current level and possibility of parallelization in the last part of the section.

\subsubsection{Higher-oder processes}
\label{sec:sinis_higherorder}
As for the fully normal metal pumps, the idealized picture of electron transport based on single electron tunneling is disturbed by simultaneous tunneling of several electrons. Owing to the  gap in the quasiparticle excitation spectrum of a BCS superconductor, elastic co-tunneling takes place only when the bias voltage over the device exceeds $2\Delta/e$. The turnstile operation is achieved with voltages well below this threshold and hence co-tunneling is suppressed, in contrast to purely normal metal devices. As a general rule, any process that leaves behind an unpaired electron on a superconducting electrode incurs an energy penalty equal to $\Delta$.
%This reflects the fact that one needs to provide energy greater or equal to $\Delta$ for each tunneled electron which breaks a Cooper pair.
%For fully normal metallic pumps, the most important higher-order process is co-tunneling where multiple electrons tunnel through Coulomb blockaded islands without suffering any energy cost related to charging. This leads to leakage current proportional to $R_T^{-N}$, where $N$ is the number of tunneling events needed to avoid charge ending up to a Coulomb blockaded islands and $R_T$ is the tunneling resistance of one junction~\cite{Jensen1992}. To obtain the accuracy requirements for metrological applications, one needs to make a chain of islands leading to slower operation speed and more complex control~\cite{Martinis1994,Keller1996,Keller1999}.

\begin{figure}[t]
	\centering
	\includegraphics[width=0.5\textwidth]{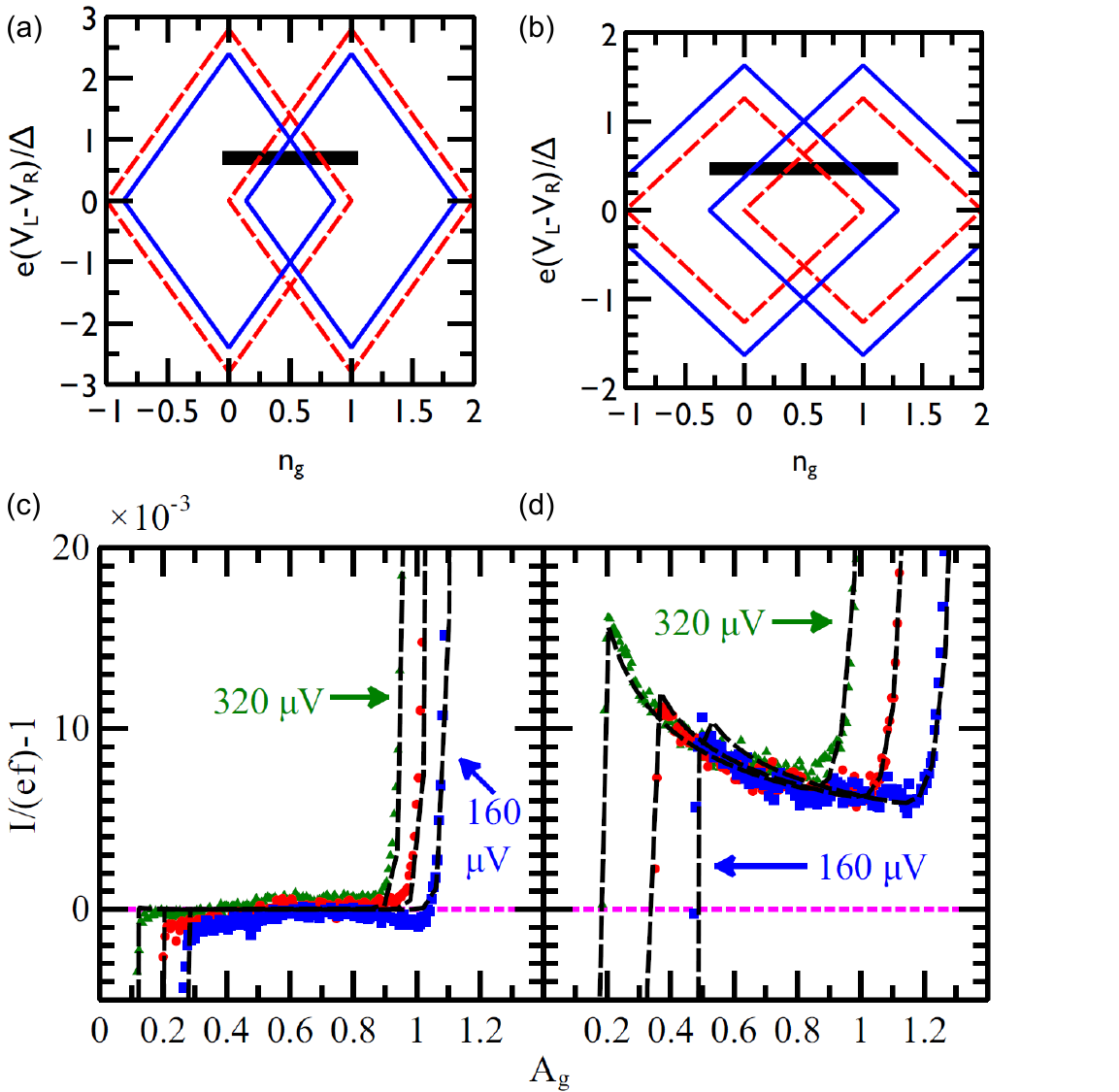}
	\caption{\label{fig:andreevPumping} (a) Stability diamonds for single-electron tunneling (blue solid lines) and Andreev tunneling (red dotted lines) for sample with $E_C > \Delta$. (b) Stability diamonds for $E_C < \Delta$. (c) First pumping plateau of the high $E_C$ device as a function of the gate voltage amplitude $A_g$. The solid symbols show pumped current with $f= 10\ \mathrm{MHz}$ and three different bias voltages. Dotted lines are the simulated traces with the corresponding biasing. (d) The same data as in (c) but now for the low $E_C$ device showing excess current due to Andreev tunneling.}
\end{figure}

For hybrid structures, the lowest order tunneling process where the energy cost of breaking a Cooper pair can be avoided is Andreev tunneling~\cite{Andreev1964}, \ie, a complete Cooper-pair tunneling through a junction. Andreev tunneling has been studied thoroughly with single NIS junctions~\cite{Blonder1982,Eiles1993,Lafarge1993,Hekking1994,Pothier1994,Rajauria2008,Greibe2011,Maisi2011} as well as in so-called Cooper pair splitters where the electrons of a Cooper pair tunnel to different normal metal regions~\cite{Hofstetter2009,Wei2010,Herrmann2010}. In the case of a SINIS turnstile, Andreev tunneling manifests itself as two electrons being added to or removed from the island. Consecutively, increasing the charging energy of a device makes Andreev tunneling energetically unfavorable, suppressing it~\cite{Averin2008,Maisi2011}. The impact of Andreev tunneling on the accuracy of a turnstile has been directly observed on the pumped current~\cite{Aref2011}. In Fig.~\ref{fig:andreevPumping}(a,b), stability diamonds for single electron and Andreev tunneling are shown for a high $E_C$ and low $E_C$ device, respectively. The pumping plateau of the high $E_C$ device, shown in panel (c), is free of Andreev tunneling whereas the low $E_C$ sample exhibits it as seen in panel (d).

For high-charging energy devices where Andreev tunneling is supressed, the process limiting the accuracy of the SINIS turnstile is co-tunneling of a Cooper pair and a single electron~\cite{Averin2008}. In this process, the island will be charged or discharged by a single electron while another electron effectively passes through the device. The net energy change is that of the corresponding single-electron process, plus the energy gained in transporting the Cooper pair from one electrode to another, which equals $2 e V_\txt{b}$ in the forward direction. Hence, the process cannot be made energetically unfavorable in a working turnstile. However, it can be supressed relative to the first-order processes by making the junctions opaque enough. Ideally, to obtain an accuracy of $10^{-7}$, one needs to limit the speed of an aluminum based turnstile to a few tens of $\mathrm{pA}$~\cite{Averin2008}. This theoretically predicted maximum operation speed is expected to slow down by an additional factor of $3$ due to nonuniformity of the tunnel barriers~\cite{Maisi2011,Aref2011}. Thus 10 pA is expected to be the optimum yield per aluminum based turnstile. In addition to the Cooper-pair electron co-tunneling, the co-tunneling of two Cooper pairs through the device increases the leakage current~\cite{Zaikin1994}. In optimized devices discussed above, the Cooper-pair electron co-tunneling is nevertheless the dominant one limiting the accuracy since its threshold is exceeded in the turnstile operation and it is of lower order than the Cooper pair co-tunneling.

\subsubsection{Quasiparticle thermalization}
\label{sec:sinis_qp}
Single-electron tunneling to or from a superconductor will generate quasiparticle excitations. Once created, the excitations carry an energy of $\Delta$, which enables them to cross the tunnel barrier to the normal metal if the electrostatic energy cost is lower than $\Delta$. Hence, they constitute a potential source of pumping errors for the hybrid turnstile. Typically the excitations are injected close to the gap edge. Also, the quasiparticles relax quickly internally compared to the weak recombination rate, so that at low temperatures we can assume them to lie close to the gap edges and have a temperature $T_{\rm qp}$ which is higher than the phonon bath temperature of the system. With this assumption, we can calculate the density of the quasiparticle excitations to be
\begin{equation}
\label{eq:nqp}
\begin{array}{rcl}
n_{qp} & = & 2 D(E_F) \displaystyle\int_\Delta^\infty dE \nu_S(E) e^{-\beta E}\\
            & = &  \displaystyle\sqrt{2 \pi}D(E_F)\Delta\frac{e^{-\beta\Delta}}{\sqrt{\beta\Delta}},
\end{array}
\end{equation}
where $\beta = 1/(\kB T_{\rm qp})$, and we have assumed $e^{-\beta\Delta} \ll 1$ and a negligible branch imbalance~\cite{Clarke1972}. The tunneling rate caused by the excitations can be calculated from the orthodox theory expressions (see Sec.~\ref{sec:seqtunneling}). It depends linearly on the density, and is independent of the biasing at low energies: $\Gamma_{qp} = n_{qp}/[2 e^2 R_T D(E_F)]$. It should be compared to the rate at which we pump electrons. As discussed above, we can obtain roughly $10\ \mathrm{pA}$ from a turnstile free of higher order tunneling errors at accuracy of $10^{-7}$. The tunneling resistance of such a device is approximately $R_T = 1\ \mathrm{M\Omega}$. To ensure that the quasiparticle excitations do not cause errors on this level, we require the tunneling rate to satisfy $\Gamma_{qp} < 10^{-7} \cdot 10\ \mathrm{pA}/e$.  With parameter values $D(E_F) = 1.45 \cdot 10^{47}\ \mathrm{J^{-1}m^{-3}}$ and $\Delta = 200\ \mathrm{\mu eV}$, we need $n_{qp} < 0.04\ \mathrm{\mu m^{-3}}$. Such a level is demonstrated in an experiment without active driving of the system~\cite{Saira2012}, and is sensitive to the filtering and shielding of the sample. Also, the trapping of quasiparticles was shown to be important in this experiment.

%The pumping speed of a turnstile is limited approximately to $f = \Delta/(10 e^2 R_T)$, where we have assumed that one needs $5$ time constants (???) for each tunneling of a single electron. To obtain a relative accuracy of $\delta I/I$, we must have $\Gamma_{qp} < (\delta I/I) f$ leading to condition $n_{qp} < \frac{1}{5} D(E_F) \Delta \times \delta I /I$. With parameter values $D(E_F) = 1.45 \cdot 10^{47}\ \mathrm{J^{-1}m^{-3}}$ and $\Delta = 200\ \mathrm{\mu eV}$, typical for aluminum, and with the asserted accuracy level $\delta I/I = 10^{-7}$, we need $n_{qp} < 0.1\ \mathrm{\mu m^{-3}}$. Such a low level is demonstrated in an experiment without active driving of the system~\cite{Saira2012}, and is sensitive to the filtering and shielding of the sample. Also, the trapping of quasiparticles was shown to be crucial in this experiment.

\begin{figure}[t]
	\centering
	\includegraphics[width=0.5\textwidth]{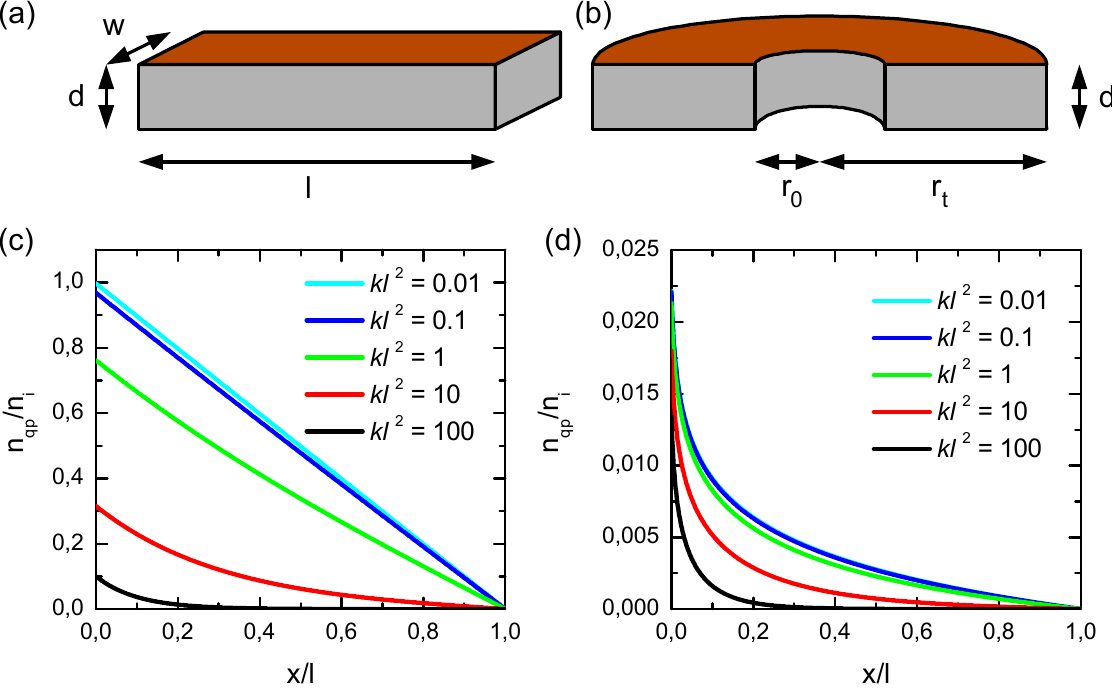}
	\caption{\label{fig:relaxation} Two typical geometries for a superconducting bias lead: (a) A lead having a constant cross-section determined by the thickness $d$ and width $w$. The length of the line is $l$. (b) A sector shaped lead characterized by an opening angle $\theta$, initial radius $r_0$, and final radius $r_t$. For the picture $\theta$ is set to $180^\circ$. The brown parts on top denote a quasiparticle trap connected via an oxide barrier. Panel (c) shows quasiparticle density $n_{qp}$ along a constant cross section line with various oxide trap transparencies $k$, and panel (d) along an opening line. In the plots, $n_{qp}$ is scaled by $n_i = Dl P_{inj}/A_i$, where the injenction area $A_i$ equals $wd$ for panel (c) and $\theta r_0 d$ for panel (d). For the leads in (b) and (d), we also use notation $x = (r-r_0)/l$ with $l = r_t - r_0$, and have used values $r_0 = 20\ \mathrm{nm}$ and $r_t = 5\ \mathrm{\mu m}$.}
\end{figure}

Next, we consider the relaxation of the quasiparticles. In turnstile operation, injection of hot quasiparticles through the tunnel junction drives the quasiparticle system of the superconductor actively out of equilibrium. We model the quasiparticle relaxation in the superconductor in terms of heat flow and obtain a diffusion equation for $n_{qp}$. Such an approach has been used to model several experiments~\cite{Rajauria2009-2,Ullom1998,Oneil2011,Peltonen2011,Knowles2012}. The heat flow of quasiparticles $\vec{J}$ follows the equation $\nabla \cdot \vec{J} = -p$, where $p$ is the power per unit volume removed from the quasiparticles. We use Fourier's law of heat conduction: $\vec{J} = - \kappa_S \nabla T_{\rm qp}$, where $\kappa_S = \frac{6}{\pi^2} \frac{L_0 T_{\rm qp}}{\rho_n} \left(\beta \Delta \right)^2 e^{-\beta \Delta}$ is the heat conductivity of a superconductor~\cite{Bardeen1959}. Here $L_0$ is Lorenz number and $\rho_n$ the resistivity in normal state. By taking the derivatives only over strong exponential dependencies and using Eq.~(\ref{eq:nqp}), we obtain a diffusion equation
\begin{equation}
\label{eq:diff}
D \nabla^2 n_{qp} = p,
\end{equation}
where the coefficient $D = \frac{\sqrt{2}(k_BT_{\rm qp}\Delta)^{1/2}}{\sqrt{\pi}e^2\rho_n D(E_F)}$ is assumed to be constant. To write down the source term on the right side of Eq.~(\ref{eq:diff}), we consider the available mechanisms of heat conduction. Electron-phonon coupling is an inherent relaxation mechanism for quasiparticles inside a superconductor. However, it is so weak that the resulting decay length of $n_{qp}$ is typically on the millimeter scale~\cite{Martinis2009,Peltonen2011}. Typically, to enhance the relaxation, one uses so-called quasiparticle traps~\cite{Pekola2000,Rajauria2009-2,Oneil2011}, which are normal metallic regions connected to the superconductor either directly or via an oxide layer. Once the hot quasiparticles enter the trap, the stronger electron-phonon relaxation in a normal metal removes their excess energy. A perfect quasiparticle trap forces the quasiparticle temperature at the interface to equal the electronic temperature of the normal metal. In the context of Eq.~(\ref{eq:diff}), this can be implemented as a boundary condition for $n_{qp}$. The boundary condition at the junction is obtained by setting the heat flow equal to the power injected by quasiparticle current.

In the case the trap is connected via an oxide barrier, the heat is carried by quasiparticle tunneling. The orthodox theory result for the source term in such a configuration is
\begin{equation}
\label{eq:relaxTrap}
\begin{array}{rcl}
p & = & \displaystyle\frac{2 \sigma_T}{e^2 d}\int_\Delta^\infty dE E n_S(E)(e^{-\beta E}-e^{-\beta_0 E}) \vspace{5pt} \\
   & = & \displaystyle\frac{\sigma_T}{e^2 D(E_F) d}(n_{qp}-n_{qp0}),
\end{array}
\end{equation}
which is obtained by setting the chemical potential difference of the trap and the superconductor to zero and assuming the diffusion to take place in two dimensions which is well justified for the thin films typically used in the samples. We also assumed $k_BT_{\rm qp} \ll \Delta$. Here, $\sigma_T$ is the electrical conductance per unit area of the trap, $d$ thickness of the superconducting film, $\beta = 1/(k_B T_{\rm qp})$, and $\beta_0 = 1/(k_B T_0)$, where $T_0$ is the temperature of the normal metal electrons. We denote by $n_{qp0}$ the quasiparticle density of a fully thermalized superconductor, \ie, one where $T_{\rm qp} = T_0$.

Let us consider some typical geometries of superconducting leads used in devices. First, take a lead with a constant cross-section, as shown in Fig.~\ref{fig:relaxation}(a). We assume that a heat flow $P_{inj}$ is injected at one end of the line, and that the other end is thermally anchored by a direct trap. For the lead itself, we assume a trap connected via an oxide barrier to be located on top. We can solve Eqs.~(\ref{eq:diff})--(\ref{eq:relaxTrap}) analytically in one dimension to obtain $n_{qp}(x) = \frac{1}{D\sqrt{k}} (e^{\sqrt{k}(2l-x)}-e^{\sqrt{k}x}) (e^{2 \sqrt{k}l}+1)^{-1}\frac{P_{inj}}{wd}+n_{qp0}$. Here, $k = \frac{\sqrt{\pi}\rho_n \sigma_T \Delta}{\sqrt{2}(k_BT_{\rm qp}\Delta)^{1/2} d}$, and $x$ is the coordinate along the wire starting at the injection side ($x=0$) and ending at the direct trap ($x =l$). In Fig.~\ref{fig:relaxation} (c), we show the quasiparticle density for various values of $\sigma_T$. The lowest $\sigma_T$ corresponds to the case where the quasiparticles only diffuse through the wire and then relax at the direct contact. At higher transparencies, the oxide trap starts to help for the relaxation as well. If we use parameter values, $d = 50\ \mathrm{nm}$, $w = 100\ \mathrm{nm}$, $l = 1\ \mathrm{\mu m}$ and $T_{\rm qp} = 130\ \mathrm{mK}$ which are typical for fabricated samples, we see that a typical injection power of $P_{inj}= 2\ \mathrm{fW}$ will yield $n_{qp}-n_{qp0} = 10\ \mathrm{\mu m^{-3}}$ without the oxide trap and even with the highest transparency $\sigma_T = (100\ \mathrm{\Omega \mu m^2})^{-1}$, which is possible to fabricate without pinholes~\cite{Brenning2004}, we get only an order of magnitude improvement.

To decrease the quasiparticle density to the acceptable level discussed above, one needs to optimize the lead geometry as well. Therefore, let us consider a lead that widens as shown in  Fig.~\ref{fig:relaxation}(b). In this case, we can solve a one-dimensional diffusion equation in polar coordinates. The junction is assumed to be located at radius $r = r_0$, and the direct contact trap to begin at radius $r = r_t$. Thickness of the lead and the overlaid trap are as in the previous example. The solution of Eqs.~(\ref{eq:diff})-(\ref{eq:relaxTrap}) can be expressed with modified Bessel functions $I_\alpha$ and $K_\alpha$ as $n_{qp}(r) =n_{qp0} + \frac{1}{D\sqrt{k}} \frac{P_{inj}}{\theta r_0 d} \{ (K_1(\sqrt{k}r_0)+\frac{K_0(\sqrt{k}r_t)}{I_0(\sqrt{k}r_t)}I_1(\sqrt{k}r_0))^{-1}K_0(\sqrt{k}r) + (I_1(\sqrt{k}r_0)+\frac{I_0(\sqrt{k}r_t)}{K_0(\sqrt{k}r_t)}K_1(\sqrt{k}r_0))^{-1} I_0(\sqrt{k}r) \}$. In Fig.~\ref{fig:relaxation}(d), we show $n_{qp}(r)$ for various transparencies of the oxide trap. The lowest transparencies, again, correspond to pure diffusion limit. Note that the quasiparticle density at the junction depends only weakly on the transparency of the trap: Due to the logarithmic dependence, changing the transparency by several orders of magnitude makes less than an order of magnitude difference to $n_{qp}(r_0)$. In a widening lead, heat sinking is made efficient by spreading the heat to a larger volume, and the area of the trap contact is also increased. By using realistic parameter values,  $d = 50\ \mathrm{nm}$, $\theta = \pi/2$, $r_0 = 50\ \mathrm{nm}$, $r_t = 5\ \mathrm{\mu m}$, $T_{\rm qp} = 130\ \mathrm{mK}$, $\rho_n = 10\ \mathrm{n\Omega m}$ and $P_{inj}= 2\ \mathrm{fW}$, we see that it is possible to reach $n_{qp} < 1\ \mathrm{\mu m^{-3}}$ at the junction even without an oxide trap. By increasing the thickness of the electrode by a factor of ten would then start to be sufficient for the metrological accuracy requirements.

Several experiments~\cite{Rajauria2009-2,Ullom1998,Oneil2011,Knowles2012} show that the above diffusion model is valid for quasiparticle densities of the order of $n_{qp} \sim 10\ \mathrm{\mu m^{-3}}$. A smaller quasiparticle density required for metrological applications implies that the absolute number of quasiparticles in the conductors becomes very small. With a typical volume of a lead, $100\ \mathrm{nm} \times 100\ \mathrm{\mu m^2}$, the quasiparticle number is $N < 1$ with $n_{qp} < 0.1\ \mathrm{\mu m^{-3}}$. It is not currently obvious if such a situation can be treated with the diffusion model or whether a more elaborated theory is required. Pumping experiments on metrological accuracy can provide a way to shed light on such a situation.

%Now, let us consider a lead with thickness $d$, widht $w$ and lenght $l$. In the case of no trap ($p = 0$), we obtain $n_{qp}$

%by solving Eq.~(\ref{eq:diff}). $P_{inj}$ is the power injected to the line: $P_{inj} = d w (-\kappa_S \nabla T_{qp})$.

\subsection{Quantum-dot-based single-electron pumps and turnstiles}
\label{sec:dotpumps}
In this section, we introduce semiconducting quantum dots and review their applications as single-electron current sources.

\subsubsection{Indroduction to quantum dots}
\label{sec:dotpumps_intro}
In contrast to conventional three-dimensional bulk conductors or two-dimensional more exotic conductors such as quantum Hall systems or graphene, semiconducting quantum dots can be regarded as zero-dimensional conductors, for which the electrons are tightly confined in all three spatial dimensions. Thus quantum dots show truly discrete excitation spectra that are reminiscent to those of natural atoms. One of the early key experiments on these \emph{artificial atoms}~\cite{Kastner1993} was the observation of discrete quantum levels~\cite{Reed1988,Su1992,Johnson1992} and the shell structure in the filling of the electron states~\cite{Tarucha1996}.

The conceptual difference between small metallic islands studied in the previous sections and quantum dots is that the Fermi level and hence the conduction electron density in the metallic islands is high, making the energy spacing between the spatially excited electron states extremely small. The metallic system can be typically described by a constant density of states as opposed to the strongly peaked density of states in quantum dots. Furthermore, quantum dots can contain a low number of electrons in the conduction band ranging from zero~\cite{Ashoori1993,Elzerman2003,Lim2009} to more than hundreds similar to natural atoms, whereas the corresponding number is orders of magnitude higher for metallic systems. In fact, the sharp potential created by a single donor atom in silicon can also be considered to be an ultra-small quantum dot. By connecting such natural atoms to electron reservoirs, for example, single-electron transistors (SETs) have been fabricated~\cite{Lansbergen2008,Tan2010}.

\begin{figure}[t]
\includegraphics[width=8.3cm]{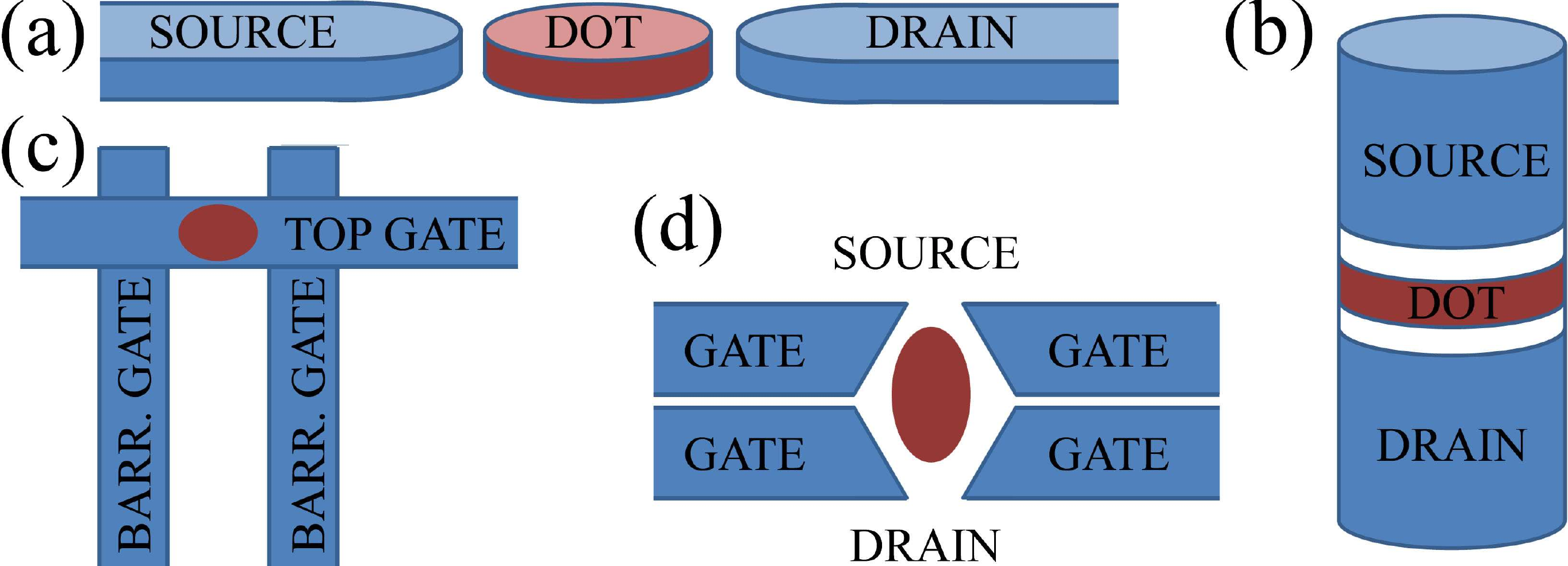}
\caption{\label{fig:3_3_1} (a) Lateral and (b) vertical quantum dot arrangements. All quantum dot pumps and turnstiles discussed further in this paper are in the lateral arrangement. The electrons tunnel between the dot and the source and drain reservoirs. The tunnel barriers between the dot and the reservoirs are created either by (a) the electrostatic potentials of nearby gate electrodes or (b) by materials choise such as AlGaAs. The gate arrangement for (c) the accumulation and (d) depletion mode quantum dots in the lateral arrangement. }
\end{figure}

Figure~\ref{fig:3_3_1} shows different types of quantum dot architectures.
The most conventional quantum dots are based on two-dimensional degenerate electron gas (2DEG) that forms either naturally to the interface between AlGaAs and GaAs~\cite{Chang1974} or that is induced to the interface between silicon and silicon oxide by an external gate~\cite{Ando1982}. Alternatively, quantum dots can be fabricated from epitaxially grown nanowires~\cite{Ohlsson2002,Fasth2007,Nadj2010}. In the conventional dots, the confinement is very strong in the direction perpendicular to the interface. Etching techniques, local anodic oxidation~\cite{Held1997}, or metallic electrodes [see Fig.~\ref{fig:3_3_1}(c) and (d)] can be employed to provide the electrostatic potential defining the well for the electrons in the plane of the interface. The in-plane diameter of this type of dot can vary from tens of nanometers to several micrometers. Thus there are plenty of atoms and electrons in the region of the dot but most of them lie in the valence band and require an energy of the order of one electronvolt to be excited. Since the relevant energy scales for the spatial excitations and the single-electron charging effects are orders of magnitude lower, the occupation of the valence states can be taken as fixed.

In the effective mass approximation~\cite{Ando1982}, the details of the electrostatic potential and the effects of the valence electrons in the solid are coarse grained such that only the electrons in the conduction band are taken into account, and these electrons are treated as particles in the smooth potential defining the dot. This description has proved to reproduce several important experimental findings both qualitatively and quantitatively~\cite{Ando1982}, and it provides insight into the single-electron phenomena in quantum dots. Especially, the potential barriers arising from the gates defining the dot can be visualized just for the small number of electrons in the conduction band.

\subsubsection{Pioneering experiments}
\label{sec:dotpumps_pioexp}
Although quantum dots hold a much smaller number of electrons than metallic islands, probably their greatest benefit is that the tunnel barriers can be formed by electrostatic potentials and controlled externally by gate voltages. Thus the height of the potential barrier, through which the electrons tunnel to the source and drain reservoirs, can be controlled in situ. This property provides fruitful grounds for electron pumping since the dependence of the tunneling rate on the barrier height and hence on the voltage of the gate electrode is typically exponential.

\begin{figure}[t]
\includegraphics[width=8.3cm]{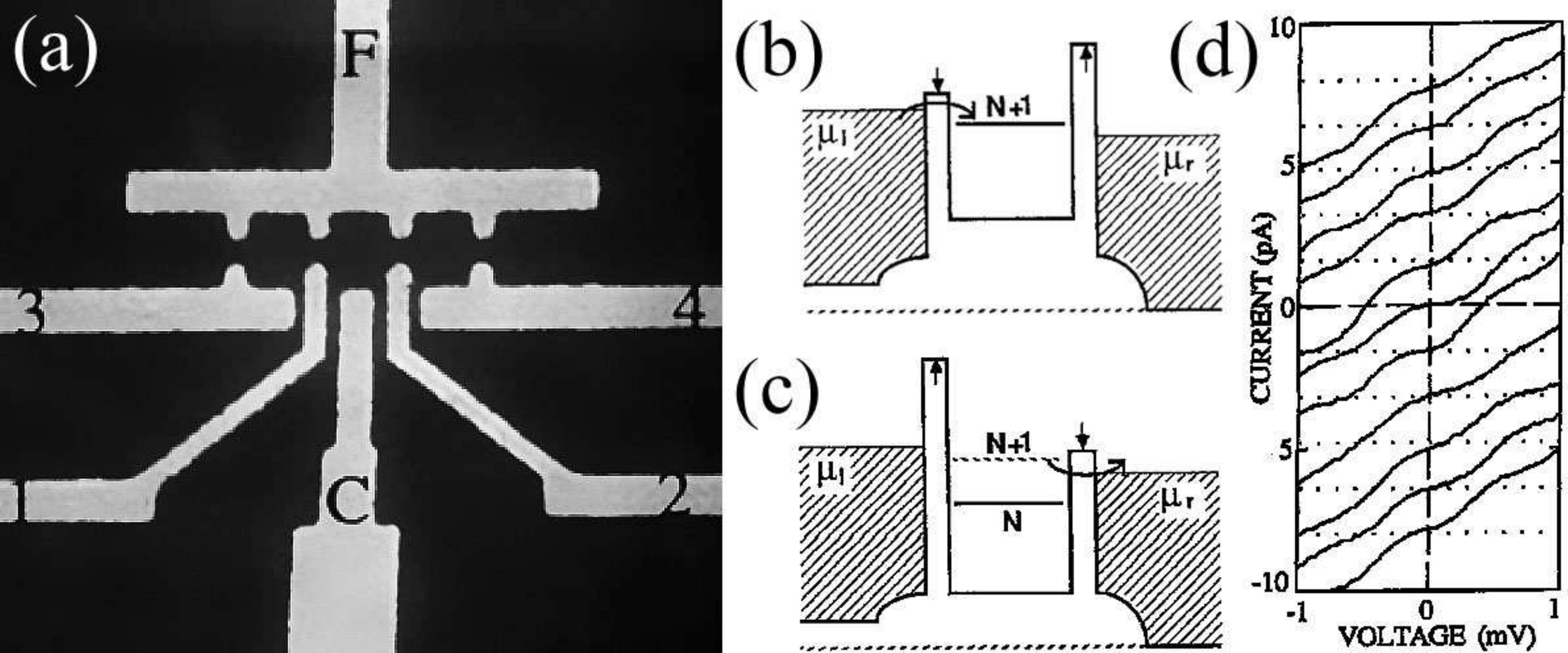}
\caption{\label{fig:3_3_2} The first single-electron current source based on quantum dots by \textcite{Kouwenhoven1991,Kouwenhoven1991_2}. (a) SEM image of the device from the top, (b),(c) operation principle, and (d) measured $IV$ curves reported. The gate configuration corresponds to the case in Fig.~\ref{fig:3_3_1}(d). The different $IV$ curves are measured while driving the turnstile with different center gate [gate C in panel (a)] voltages, rf amplitudes, and phase differences. The curves are not offset and the dashed lines show the current levels $nef$ with $n=-5,\dots,5$.}
\end{figure}

The first experiments employing quantum dots for frequency-locked single-electron transport were reported by %Kouwenhoven et al. in 1991~
\textcite{Kouwenhoven1991,Kouwenhoven1991_2} [see also \textcite{Kouwenhoven1992}]. Here, they used surface-gated GaAs dots as shown in Fig.~\ref{fig:3_3_2}(a). The negative voltages on Gates C, F, 1, and 2 deplete the 2DEG that is located 100 nm below the surface, thus defining the quantum dot in the center with a radius of about 300 nm and charging energy $2E_C=e^2/2C_\Sigma=0.34\textrm{ meV}$. (Gates 3 and 4 are grounded and do not deplete the 2DEG.)

In addition to dc voltages defining the dot, 180-degree-phase-shifted sinusoidal rf drive is superimposed to Gates 1 and 2, lowering one barrier at a time. This rf drive induces a turnstile operation as shown in Figs.~\ref{fig:3_3_2}(b) and \ref{fig:3_3_2}(c) for negative bias voltage on the left side of the dot: when the voltage at Gate 1 is high (low tunnel barrier) and low at Gate 2 (high tunnel barrier), an excess electron enters the dot through the left barrier [Fig.~\ref{fig:3_3_2}(b)], and when the voltage at Gate 1 is low and high at Gate 2, the electron escapes through the right barrier [Fig.~\ref{fig:3_3_2}(c)]. Thus the average dc current through the device in the ideal case is given by $I_p=ef$, where $f$ is the operation frequency. For bias voltages greater than the charging energy, $|eV|\ge E_C$, more than a single electron can be transported in a cycle yielding ideally $I_p=nef$, where $n$ is an integer. Signatures of this type of current quantization were observed in the experiments~\cite{Kouwenhoven1991,Kouwenhoven1991_2,Kouwenhoven1992} and are illustrated in Fig.~\ref{fig:3_3_2}(d). The current through the device as a function of the bias voltage tends clearly to form a staircase-like pattern with the step height $ef$. This was the first experimental demonstration of current quantization in quantum dot structures. Note that in addition to turnstile operation, Fig.~\ref{fig:3_3_2}(d) also shows pumping of electrons against the bias voltage for certain phase differences of the driving signals. The error in the pumped current is a few percent, falling somewhat behind of the first experiments on metallic structures reported by \textcite{Geerligs1990}.

\begin{figure}[t]
\includegraphics[width=8.3cm]{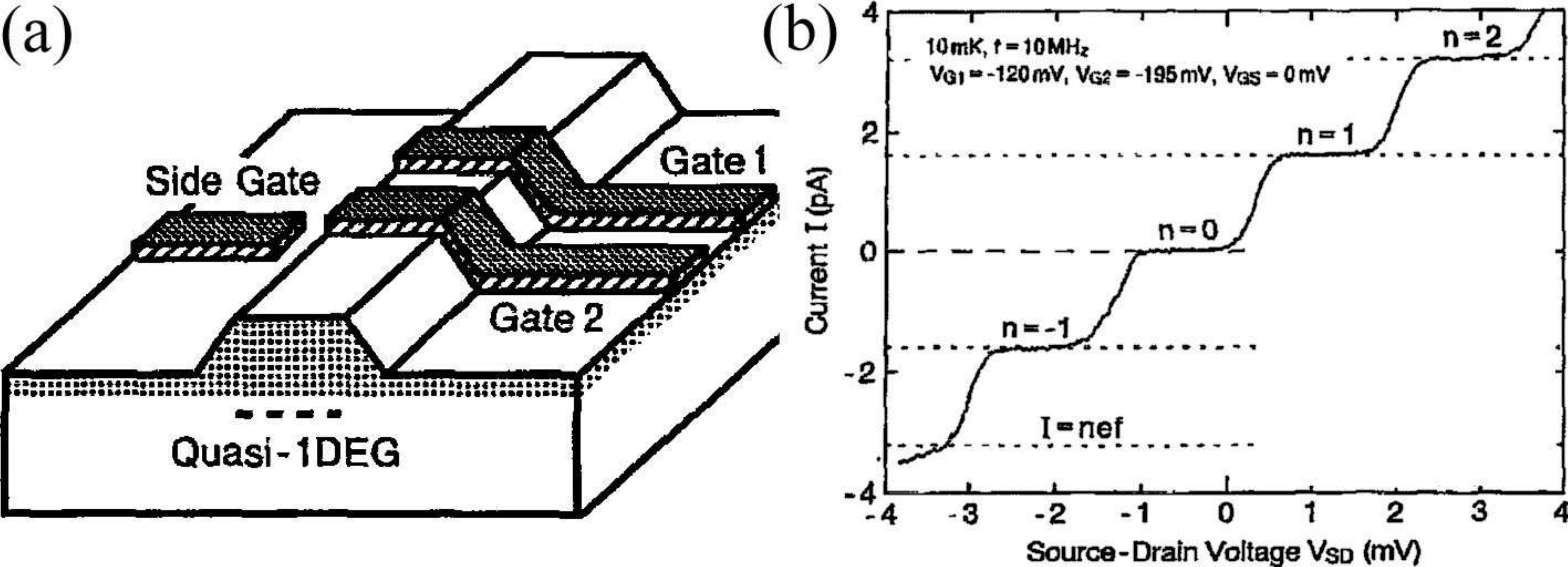}
\caption{\label{fig:3_3_3} (a) Schematic illustration of the device and (b) observed current plateaus during the turnstile operation by \textcite{Nagamune1994}.}
\end{figure}

The second set of experiments on single-electron turnstiles based on quantum dots were published by \textcite{Nagamune1994}. Here, the quantum dot forms into a gallium arsenide 2DEG that is wet etched into a shape of 460-nm-wide wire as illustrated in Fig.~\ref{fig:3_3_3}(a). Two 230-nm-wide metallic gates are deposited perpendicular to the wire with a distance of 330~nm. This different barrier gate configuration and the higher charging energy of $2E_C=1.7\textrm{ meV}$ resulted in a clear improvement of the staircase structure as shown in Fig.~\ref{fig:3_3_3}(b). However, the authors report that a parallel channel forms due to the rf operation and the effect of this channel is substracted from Fig.~\ref{fig:3_3_3}(b). The authors estimate the accuracy of their device to be about 0.4\% if the correction from the parallel channel is taken into account.

\begin{figure}
\includegraphics[width=8.3cm]{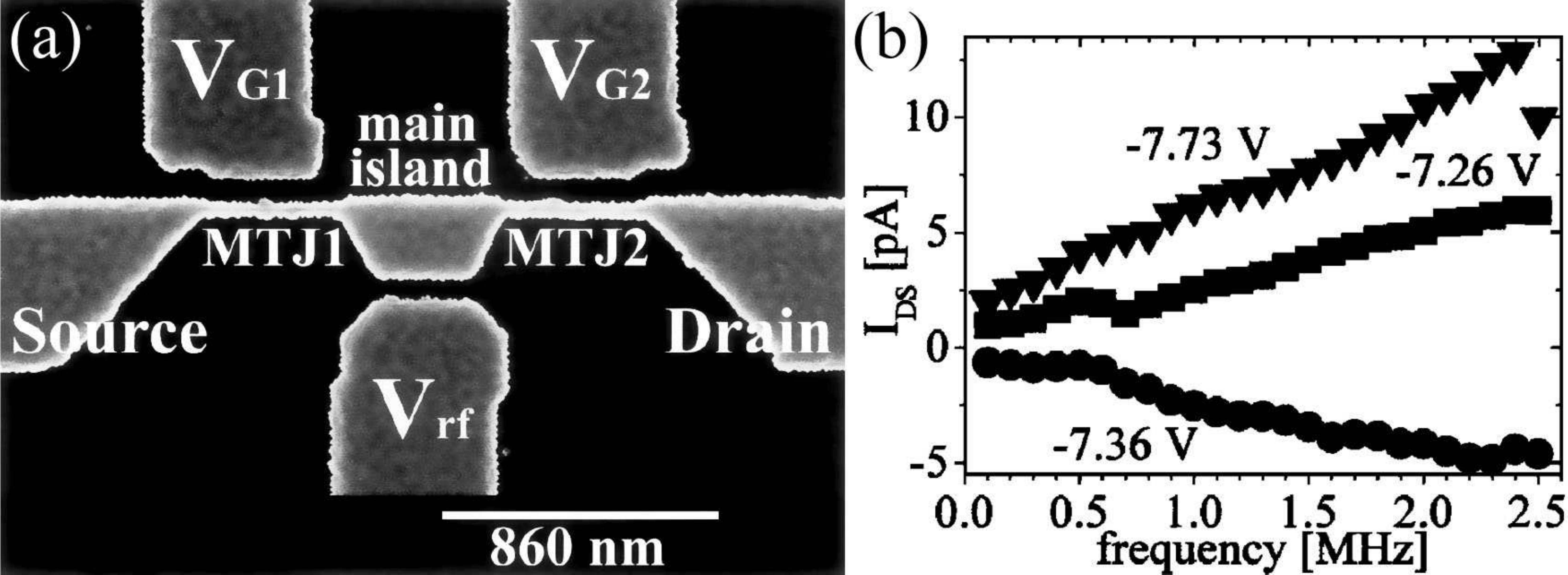}
\caption{\label{fig:3_3_4} (a) SEM image of the device and (b) pumped current through it in the experiments by \textcite{Altebaeumer2001_2}. Different values of the current correspond to different dc voltages $V_{G1}$ [see panel (a)].}
\end{figure}

In 1997--2001, a series of experiments was carried on so-called multiple-tunnel junction devices as electron pumps~\cite{Tsukagoshi1997,Tsukagoshi1998,Altebaeumer2001,Altebaeumer2001_2}. Here, the most common device was either $\delta$-doped GaAs or phosphorus doped silicon that was etched such that a central region is connected to source and drain reservoirs by narrow strips as shown in Fig.\ref{fig:3_3_4}(a). The side gates near the strips are set to a constant potential and an rf drive on the central side gate induces a current that depends linearly on frequency as shown in Fig.~\ref{fig:3_3_4}(b). The explanation of this type of operation is that the dopants and disorder in the strips function as Coulomb blockade devices them selves rather than as single tunnel junctions, which gives rise to the term multiple-tunnel junction. Since these experiments were more motivated by applications in information processing with only few electrons rather than finding a metrological current source, the accuracy of the device was not studied in detail.

\subsubsection{Experiments on silicon quantum dots}
\label{sec:dotpumps_siexp}
\begin{figure}[t]
\includegraphics[width=8.3cm]{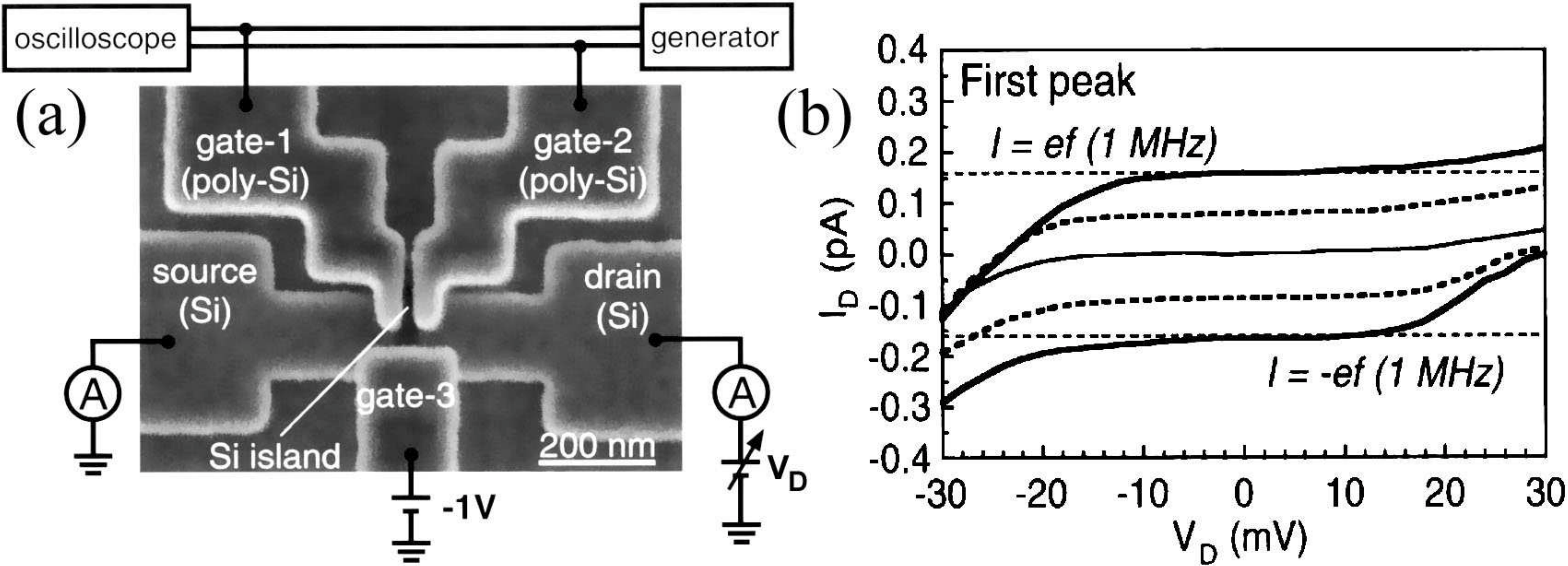}
\caption{\label{fig:3_3_5} (a) SEM image of the device and (b) observed current plateaus up to 1~MHz pumping frequency by \textcite{Ono2003} on a silicon quantum dot.}
\end{figure}

The first step toward single-electron pumping in silicon was taken by~\textcite{Fujiwara2001} as they presented an ultra small charge-coupled device (CCD) and demonstrated that it could be used to trap and move individual holes controllably at the temperature of 25~K. This device was fabricated with silicon-on-insulator techniques \cite{Takahashi1995} and had two adjacent polysilicon gates acting as metal--oxide--semiconductor field effect transistors (MOSFETs). Subsequently, a rather similar device with charging energy $2E_C=30$~meV shown in Fig.~\ref{fig:3_3_5}(a) was utilized for electron pumping by \textcite{Ono2003} at the temperature of 25~K. They obtained an accuracy of the order of $10^{-2}$ up to 1~MHz pumping that was also the limitation set by the calibration of their measurement equipment. Here, the electron pumping was based on two sinusoidal driving signals that are offset less than 180 degrees, which renders the chemical potential of the dot to move during the cycle.

%One could add a schematic picture showing current as a function of the barrier gate voltages and mark the different options for turnstile and pumping operations.

\begin{figure}[t]
\includegraphics[width=8.3cm]{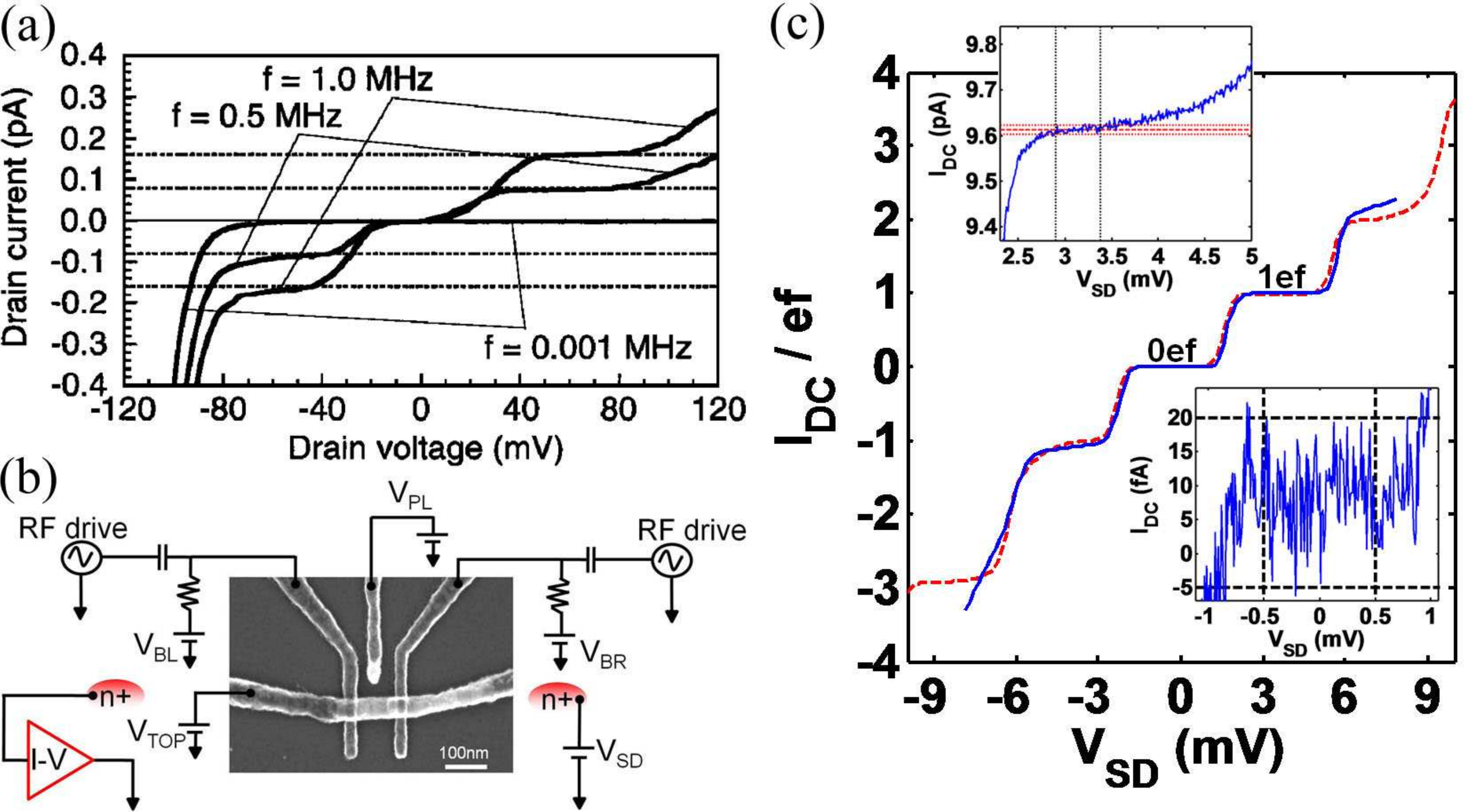}
\caption{\label{fig:3_3_6} (a) Measured current plateaus for different frequencies of the turnstile operation with the device shown in Fig.~\ref{fig:3_3_5}(a) by \textcite{Ono2003_2}. (b) SEM image of the silicon quantum dot device and a schematic measurement setup employed in the experiments by \textcite{Chan2011}. (c) Measured current plateaus (solid line) and the corresponding theoretical curve (dashed line) by \textcite{Chan2011}. The insets show zooms at the $n=0$ (bottom) and $n=1$ (top) plateaus. The red dashed lines show $\pm 10^{-3}$ relative deviation from the ideal $ef$ level.}
\end{figure}

In addition to pumping, \textcite{Ono2003_2} utilized the device shown in Fig.~\ref{fig:3_3_5}(a) as a single-electron turnstile. The operational principle is the same as in the pioneering experiments with GaAs quantum dots described in Fig.~\ref{fig:3_3_2}(b) and (c). \textcite{Ono2003_2} observed current steps of $ef$ up to $f=1$~MHz operation frequencies [see Fig.~\ref{fig:3_3_6}(a)]. The flatness of the plateaus was of the order $10^{-2}$ measured at 25~K. \textcite{Chan2011} used metallic aluminum gates to define a silicon quantum dot in the electron accumulation layer of the device as shown in Fig.~\ref{fig:3_3_6}(b). Although the relative variation of the current at the plateau they measured was below $10^{-3}$ for a broad range of source--drain voltages [see Fig.~\ref{fig:3_3_6}(c)], they could not strictly claim better than 2\% relative accuracy in the current due to the inaccurate calibration of the gain of the transimpedance amplifier employed. These experiments were carried out at 300~mK phonon temperature but the sequential tunneling model used to fit the data by \textcite{Chan2011} suggested that the electron temperature of the dot rose up to 1.5~K. It is to be studied whether 1.5~K was due to power dissipated at the surface mount resistors in the vicinity of the sample or due to the direct heating of the 2DEG from the electrostatic coupling to the driven gate potentials.

\textcite{Fujiwara2008} introduced a single-electron ratchet based on a silicon nanowire quantum dot with two polysilicon gates working as MOSFETs [see also Ref.~\cite{Fujiwara2004}]. In general, ratchets generate directional flow from a non-directional drive due to the asymmetry of the device. Here, an oscillating voltage is applied to one of the gates such that an electron captured through it near the maximum voltage, i.e., minimum barrier height, and ejected through the other barrier near the minimum voltage. In fact, the number of electrons pumped per cycle depends on the applied dc voltages and current plateaus up to $5ef$ were reported. Furthermore, nanoampere pumped current was observed at the $3ef$ plateau with the pumping frequency $f=2.3$~GHz. The error in the current was estimated to be of the order of $10^{-2}$ for the experiment carried out at 20~K temperature.
% COULD ADD A FIGURE ON THE RATCHET IF THERE IS SPACE

The first error counting experiments in silicon were carried out by \textcite{Yamahata2011} [see also Ref.~\cite{Nishiguchi2006}]. In contrast to the pioneering error counting experiments by \textcite{Keller1996}, only a single silicon nanowire quantum dot was used as the current source and the electrons were steered into and out of a quantum dot coupled to a charge sensor. By opening the MOSFET separating the node from the drain reservoir, it was possible to use the same device as a dc current source. The observed pumping error was of the order of $10^{-2}$ and was reported to be dominated by thermal errors at the 17~K temperature of the experiments.
%% COULD ADD A FIGURE ON THIS IS THERE IS SPACE

Very recently, \textcite{Jehl2012} reported on frequency-locked single-electron pumping with a small quantum dot formed in metallic NiSi nanowire interrupted by two MOSFETs controlled by barrier gates. With rf drives on the barrier gates, they were able to pump currents beyond 1~nA but the accuracy of the pump was not studied in detail. The MOSFET channels in this device had very sharp turn-on characteristics requiring only about 4.2~mV of gate voltage to change the conductivity of the channel by a decade, which can be important in reducing unwanted effects from the gate voltage drive such as heating.

\subsubsection{Experiments on gallium arsenide quantum dots}
\label{sec:dotpumps_gaasexp}
After the pioneering experiments discussed in Sec.~\ref{sec:dotpumps_pioexp}, the focus in single-electron sources based on gallium arsenide moved toward the idea of using surface acoustic waves (SAWs) to drive the single electrons in a one-dimensional channel---a topic to be discussed in Sec.~\ref{sec:sawpumps}. In this section, we focus on gate-controlled GaAs pumps.
%To this end, one of the early experiments was put forward by \textcite{Switkes1999} but they did not focus on the metrological aspect of charge pumping, and hence the accuracy of pump was not reported.

\begin{figure}[t]
\includegraphics[width=8.3cm]{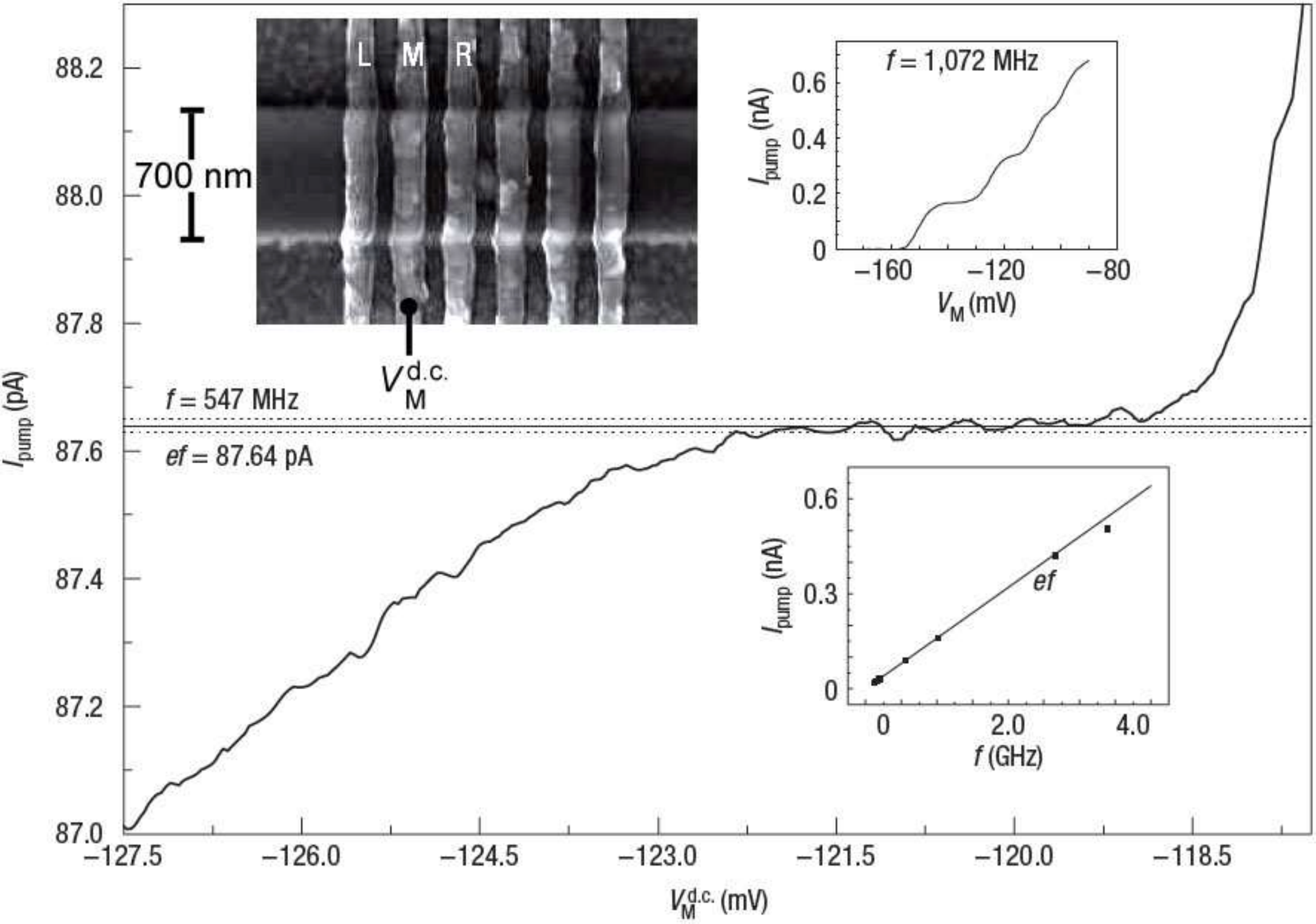}
\caption{\label{fig:3_3_7} Current plateau in the electron pumping experiments by \textcite{Blumenthal2007} as a function of the middle gate voltage at 547~MHz operation frequency. The dashed lines show $\sigma=\pm 10$~fA uncertainty in the electrometer calibration. The top left inset shows the device used as the electron pump. The top right inset shows current plateaus at 1~GHz pumping frequency and the bottom inset shows the pumped current as a function of the operation frequency.}
\end{figure}

The seminal work by \textcite{Blumenthal2007} took gate-controlled GaAs quantum dots a leap closer to a metrological current source, namely, they reported 547~MHz (87.64~pA) single-electron pumping with one-standard-deviation ($1\sigma$) relative uncertainty of $10^{-4}$ (see Fig.~\ref{fig:3_3_7}). However, the authors did not report the full dependence of the pumping errors as functions of all control parameters. As the device, they employed a chemically etched AlGaAs--GaAs wire with overlapping metallic gates as shown in Fig.~\ref{fig:3_3_7}. Only the three leftmost gates L, M, and R were used such that 180-degree-phase-shifted sinusoidal driving signals were applied to the gates L and R in addition to dc voltages applied to all of the three gates. The amplitudes of the rf signals were chosen asymmetric such that the device can work as a pump rather than a turnstile. The charging energy of the device was estimated to be $2E_C=1$~meV %is the charging energy defined as e^2/C_\Sigma in the paper???
and the experiments were carried out at the bath temperature of 300~mK.

With a similar device architecture as shown in Fig.~\ref{fig:3_3_7} but using only two gates instead of three, \textcite{Kaestner2008b} demonstrated that frequency-locked single-electron pumping can be carried out with a single sinusoidal driving voltage, thus decreasing the complexity of the scheme. This type of \emph{single-parameter pumping} with two gates is employed in the rest of the works discussed in this section. \textcite{Maire2008} studied the current noise of a similar single-parameter pump at $f=400$~MHz and estimated based on the noise level that the relative pumping error was below 4 \%. \textcite{Kaestner2008} studied the robustness of the current plateaus as functions of all control parameters of the pump except the source--drain bias. They showed that single-parameter pumping is robust in the sense that wide current plateaus appear in the parameter space but their measurement uncertainty was limited to about $10^{-2}$, and hence a detailed study of the behavior of the accuracy as a function of these parameters was not available.

\begin{figure}[t]
\includegraphics[width=8.3cm]{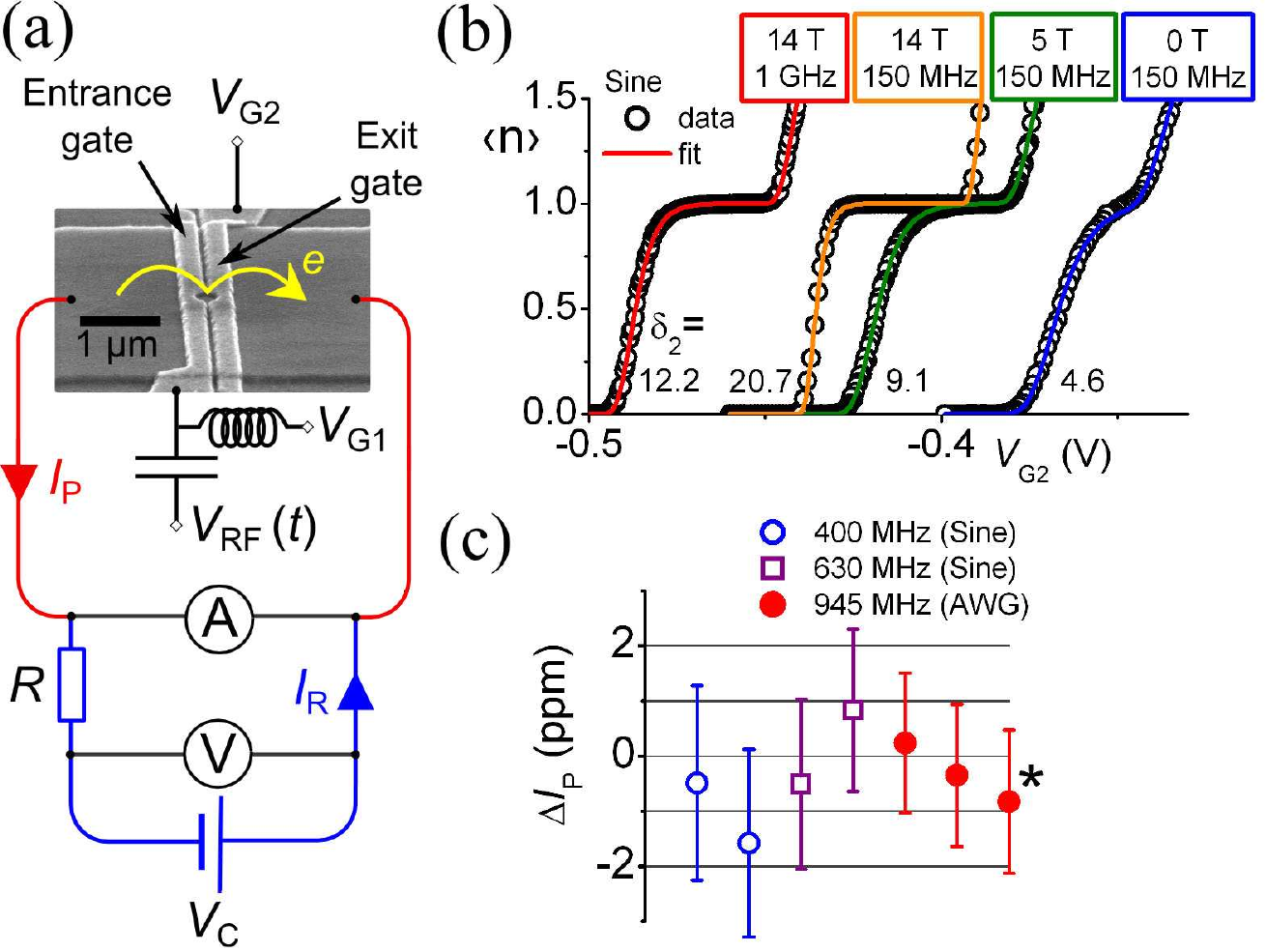}
\caption{\label{fig:3_3_8} (a) SEM image of the device with a schematic measurement setup employed by \textcite{Giblin2012}. (b) Current plateaus obtained by usings a sine wave drive at different frequencies and magnetic fields. (c) Relative difference of the pumped current from $ef$ using a sine waveform and a tailored arbitrary waveform at different frequencies. The rightmost data point denoted by an asterisk shows the result with the potential of the entrance gate shifted by 10~meV from the optimal operation point.}
\end{figure}

\textcite{Wright2008} made an important empirical observation that the accuracy of the single-parameter pump can be improved by an application of perpendicular-to-plane magnetic field [see also Ref.~\cite{Wright2009} and Fig.~\ref{fig:3_3_8}(b)]. They applied fields up to 2.5~T and demonstrated that the $n=1$ plateau as a function of the dc voltage on the non-driven gate widens noticeably with increasing magnetic field. %However, the behavior of the error was not studied in detail.
In the further studies by \textcite{Kaestner2009} and \textcite{Leicht2011} up to magnetic fields of 30~T, a great widening on the plateau was observed, but it essentially stopped at 5~T. On the contrary, high-resolution measurements on the pumped current up to 14~T by \textcite{Fletcher2011} showed a continuous improvement on the pumping accuracy with increasing field [see also Fig.~\ref{fig:3_3_8}(b)]. This discrepancy is possibly explained by the different samples used in the different sets of experiments. %Temperature 30 mK, charging energy=???

\textcite{Giblin2010} employed a magnetic field of 5~T and reported 54~pA of pumped current with $1\sigma=15$~ppm relative uncertainty with a single parameter sinusoidal drive. They were able to measure at such a low uncertainty with a room-temperature current amplifier since they subtracted a reference current from the pumped current and passed less than 100 fA through the amplifier. Thus the uncertainty in the gain of the amplifier did not play a role. The reference current was created by charging a low-loss capacitor and was traceable to primary standards of capacitance.

To date, the most impressive results on single-electron pumping with quantum dots have been reported by \textcite{Giblin2012}. Compared with the previous results in Ref.~\cite{Giblin2010}, they made several changes to improve the results. They used higher magnetic field of 14~T and an advanced generation of samples with a lithographically defined place for the quantum dot in both directions in the plane [see Fig.~\ref{fig:3_3_8}(a)]. Instead of using sinusoidal waveforms, they also tailored the drive voltage so that the cycle time was distributed more evenly for the different parts of the cycle. To make traceable measurements, a reference current was created by an accurate temperature-controlled 1~G$\Omega$ resistor, a voltage source and a high-precision voltmeter. The voltmeter and the resistor were calibrated through intermediate steps against the Josephson voltage standard and the quantum Hall resistance standard, respectively. In this work, \textcite{Giblin2012} reported 150~pA pumped current with relative $1\sigma$ uncertainty of 1.2~ppm [see Fig.~\ref{fig:3_3_8}(c)]. Most of the uncertainty, 0.8~ppm, arose from the calibration of the 1~G$\Omega$ resistor. Thus it is possible that the the electron pumping was actually even more accurate, as suggested by fitting the results to a so-called decay cascade model~\cite{Kashcheyevs2010}. However, there can be processes that are neglected by the model and since there is no experimental evidence on lower than 1.2~ppm uncertainty, it remains the lowest demonstrated upper bound for relative pumping errors for quantum dot single-electron pumps. Error counting, as demonstrated in silicon by \textcite{Yamahata2011} and in aluminum by \textcite{Keller1996}, is a way to measure the pumping errors to a very high precision independent of the other electrical standards and remains to be carried out in future for the GaAs quantum dot pumps.

%NOT CITING TO Nevou2011 (Current quantization in an optically driven electron pump based on self-assembled quantum dots) ANYWHERE... this should be put to Sec. III.E.

\subsection{Surface-acoustic-wave-based charge pumping}
\label{sec:sawpumps}
%could move this before the previous section since taking place before it chronologically
After the pioneering experiments on single-electron sources based on GaAs discussed in Sec.~\ref{sec:dotpumps_pioexp}, the focus in this field moved toward the idea of using surface acoustic waves (SAWs) to drive the single electrons in a one-dimensional channel~\cite{Shilton1996a}. Here, the sinusoidal potential created for the electrons in the piezoelectric GaAs by SAW forms a moving well that can trap an integer number of electrons and transport them in a one-dimensional channel.

The first experiments on this kind of SAW electron pumps were carried out by \cite{Shilton1996}. They employed a SAW frequency of 2.7~GHz and observed a corresponding $n=1$ current current plateau at 433~pA with the uncertainty of the order of $10^{-2}$ at 1~K temperature. \textcite{Talyanskii1997} carried out more detailed experiments on similar samples at two different SAW frequencies and the results were in agreement with $ef$ scaling law. Furthermore, several current plateaus were observed as a function of the gate voltage corresponding to different integer values of pumped electrons per cycle. However, the experimental uncertainty at the plateau was again of the order of $10^{-2}$ and sharp current peaks were observed at various gate voltage values.

After these first experiments, \textcite{Janssen2000,Janssen2000a} studied the accuracy of the SAW pump and reported 431~pA current at the center of the plateau with 200~ppm relative deviation from the ideal value. \textcite{Ebbecke2000} demonstrated SAW pumping up to 4.7~GHz frequencies and with two parallel channels to increase the current, but the measurement accuracy was rather limited here. To improve the quality of the plateau \textcite{Janssen2001} decreased the width of the one-dimensional channel, which helps in general. However, they observed that the required rf power to drive the elcetrons increases with decreasing channel width causing sever rf heating of the sample. This heating caused the quality of the plateau to drop and the conclusion was that materials with lower losses due to rf are needed [see also \cite{Utko2006}]. In fact, \textcite{Ebbecke2003, Flensberg1999} reported that the accuracy of the SAW current is fundamentally limited in one-dimensional channels because of tunneling of electrons out from a moving dot.

To overcome the limitation pointed out in Ref.~\cite{Ebbecke2003}, the charging energy was increased in the system by defining a quantum dot with surface gates rather than utilizing an open one-dimensional channel~\cite{Ebbecke2004}. Thus the applied SAWs modulate both the tunnel barriers between the dot and the reservoirs and the electrochemical potential at the dot. With this technique, current plateaus were observed at a SAW frequency of 3~GHz and the reported relative deviation from the ideal value was of the order of $10^{-3}$. Although the results by \textcite{Janssen2000,Janssen2000a} remain the most accurate ones reported to date with SAW electron pumps, and hence are not valuable for a metrological current source, single-electron transfer with SAWs can be useful in other applications. For example, \textcite{McNeil2011} have shown that an electron taken by SAWs from a quantum dot can be captured by another dot at distance. This kind of electron transport can potentially be used to transport single spins working as quantum bits in a spin-based quantum computer~\cite{Hanson2007,Morello2010}.

\subsection{Superconducting charge pumps}
\label{sec:superpumps}
The envisioned advantage in pumping Cooper pairs instead of electrons is that the supercurrent produced by the Cooper pair pumps is inherently dissipationless and the BCS gap protects the system from microscopic excitations. Thus the operation frequency of the pump can possibly be high with the system still remaining at very low temperature. Another advantage of the supercurrent is that it can sustain its coherence, and hence be virtually noiseless, in contrast to single-electron current that is based on probabilistic tunneling. Furthermore, since the charge of a single Cooper pair is $2e$, single-Cooper-pair pumps yield twice the current compared with single-electron pumps operated at the same frequency. Despite these advantages, the lowest uncertainties in the achieved Cooper pair current is at the percent level~\cite{Vartiainen2007,Gasparinetti2012}. One reason for this is the low impedance of the device rendering it susceptible to current noise.

\begin{figure}[t]
\includegraphics[width=8.3cm]{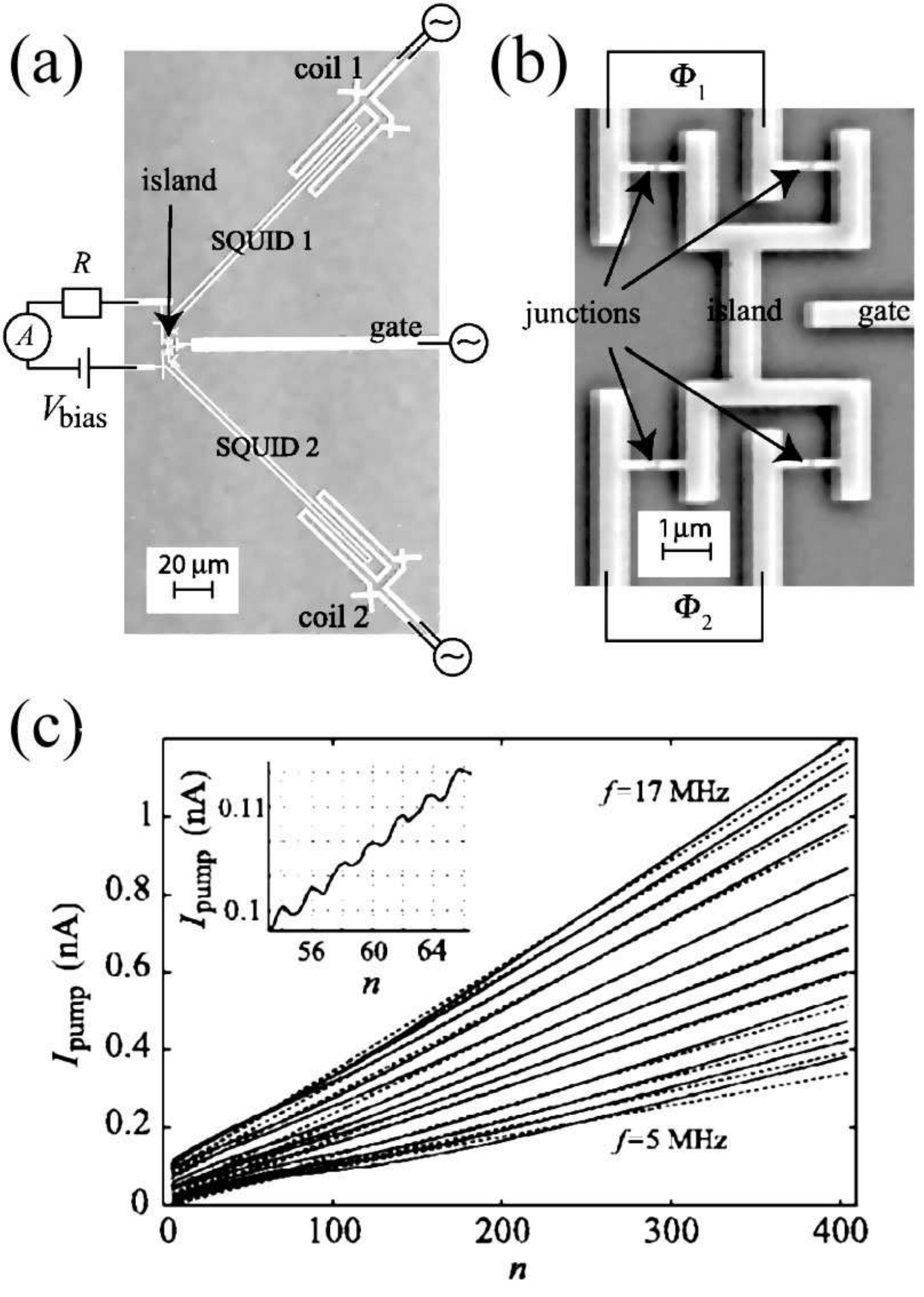}
\caption{\label{fig:3_5_1} (a) Scanning electron micrograph of the sluice used in the experiments by \textcite{Vartiainen2007} with a simplified measurement setup. (b) Magnified view of the island of the device shown in panel (a) with four Josephson junctions. (c) Measured pumped current with the sluice (solid lines) as a function of the magnitude of the gate voltage ramp such that $n$ corresponds to the ideal number of elementary charges $e$ pumped per cycle. The inset shows the step-like behavior observed in the pumped current.}
\end{figure}

Two types of Cooper pair pumps exist in the literature: arrays of superconducting islands \cite{Geerligs1991} with source and drain leads, all separated by single Josephson junctions with fixed tunnel couplings, and a so-called sluice \cite{Niskanen2003,Niskanen2005} that composes of a single island connected to the leads by two SQUIDs that function as tunable Josephson junctions, see Fig.~\ref{fig:3_5_1}. As in the case of single-electron pumps, the device operation is based on Coulomb blockade effects allowing the controlled transfer of individual Cooper pairs, which means in the case of array pumps that the fixed Josephson energies of the junctions must be much lower than the Cooper pair charging energy of the corresponding islands. In the sluice, it is sufficient that the minimum obtainable Josephson energy is much lower than the charging energy. For the arrays, the thermal energy, $k_BT$, must be much lower than the Josephson energy that defines the energy gap between the ground state and the excited state of the quantum system at charge degeneracy. For the sluice, the maximum Josephson energy of the SQUIDs yields the minimum energy gap of the system thus relaxing the constraint on temperature.

The first experiment demonstrating Cooper pair pumping was performed by \textcite{Geerligs1991}. The device is a linear array of three Josephson tunnel junctions. The two
superconducting islands separated by the junctions are capacitively coupled to individual gate electrodes. Except in the vicinity of the charge degeneracy points in the gate voltage space, the
number of Cooper pairs on these islands is rather well defined by the gate voltages because the Coulomb blockade regime is employed.
By biasing the device and
applying sinusoidal ac voltages with appropriate amplitudes to the gates, one obtains a continuously repeated cycle,
during which a Cooper pair is transferred through the device, i.e., Cooper pairs are pumped one by
one. Ideally, this yields a dc current $I=2ef$ that is proportional to the pumping frequency $f$. The driving voltage at each
gate should have the same frequency and a phase difference of $\pi$/2. The pumping direction can be reversed if
the difference is changed by $\pi$. Thus the pumping principle is the same as for a normal pump discussed in Section~\ref{sec:metalpumps}.
The height of the measured current plateau follows rather well the
predicted relation $I=2ef$ at low pumping frequencies, but deviates strongly at higher frequencies. This is
explained by several mechanisms. The uncertainty of the device was not assessed in detail but it seems to lie at least on the percent level with picoampere currents. One of the error mechanisms is the Landau-Zener tunneling when the system is excited
to the higher energy state without transferring a Cooper pair. This was the dominant mechanism at the high-end of the studied pumping frequencies in the experiment by \textcite{Geerligs1991} thus imposing an upper limit on the operation frequency of the device.
Another error source in the device is the
tunneling of non-equilibrium quasiparticles, photon-excited tunneling and relaxation of the excited states produced by Landau-Zener tunneling. In addition, co-tunneling of Cooper pairs through the two junctions produces a step-like feature to
the current plateaus, thus reducing the pumping accuracy. Later, a similar three-junction Cooper pair pump
was studied by \textcite{Toppari2004} and essentially the same conclusions on the pumping accuracy were made. In both
experiments, no 2$e$-periodicity was observed in the dc measurements, which suggests a substantial presence of
nonequilibrium quasiparticles in the system.

The effect of quasiparticles on Cooper pair pumping was also observed in the seven-junction Cooper pair pump~\cite{Aumentado2003}. The device is basically the same as the one used for pumping single electrons in
the earlier experiments in the normal state~\cite{Keller1996}. The pump consists of six
micrometer-scale aluminum islands linked by aluminum-oxide tunnel barriers. Investigation of this circuit in the hold and
pumping modes revealed that besides 2$e$ tunneling events, there is a significant number of 1$e$ events associated
with the quasiparticle tunneling. All these experiments show that in order to obtain accurate Cooper pair
pumping, one must suppress unwanted quasiparticle tunneling. \textcite{Leone2008,Leone2008_2} propose topological protection in pumping Cooper pairs. The charge is expected to be strictly quantized determined by a Chern index. The idea has not been tested experimentally as far as we know.

In order to increase the output dc current and accuracy of a single pump, the sluice pump was introduced by \textcite{Niskanen2003,Niskanen2005}. In the pumping cycle, the two SQUIDs separating the single island work in analogy with valves of a classical pump and the gate voltage controlling the island potential is analogous to a piston. At each moment of time, at least one SQUID is closed (minimum critical current). The gate voltage is used to move Cooper pairs through open SQUIDs (maximum critical current). If the pairs are taken into the island through the left SQUID and out of the island through the right SQUID, the resulting dc current is ideally $I=N2ef$, where the number of pairs transported per cycle $N$ is determined by the span of the gate voltage ramp. In practise, the critical current of the SQUIDs is controlled by flux pulses generated by superconducting on-chip coils. Since each operation cycle can transfer up to several hundreds of Cooper pairs, \textcite{Vartiainen2007} managed to pump roughly 1~nA current with uncertainty less than 2\% and pumping frequency of 10~MHz, see Fig.~\ref{fig:3_5_1}. The investigated high-current Cooper pair pump demonstrated step-like behavior of the pumped current on the gate voltage, however, its accuracy was affected by the residual leakage in the tunnel
junctions and the fact that the SQUIDs did not close completely due to unequal Josephson junctions in the structure.

\begin{figure}[t]
\includegraphics[width=8.3cm]{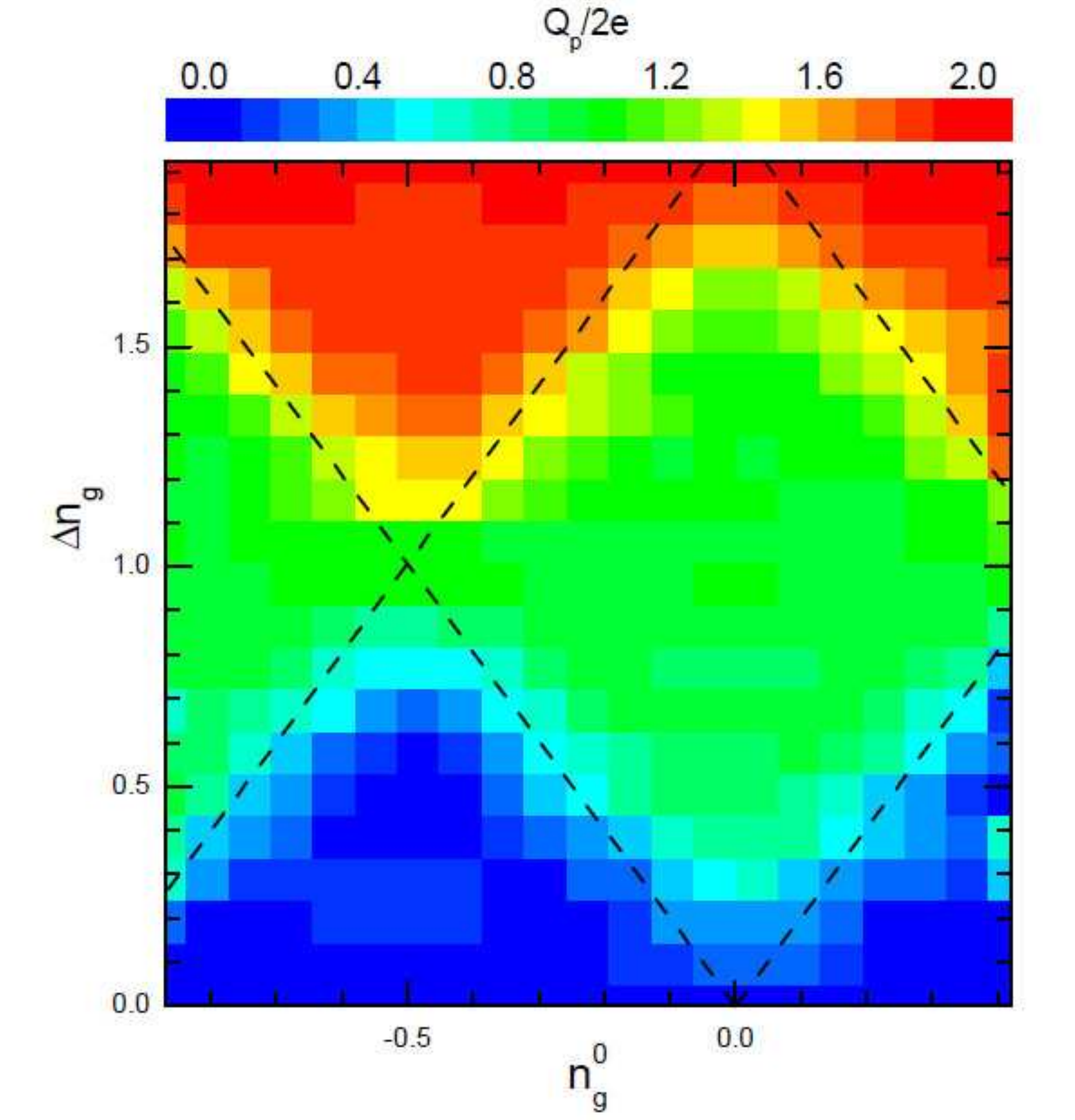}
\caption{\label{fig:3_5_2} Pumped average charge by \textcite{Gasparinetti2012} for a single pumping cycle of a sluice pump near vanishing voltage bias as a function of the gate charge offset, $n_g$, and span during the pumping cycle, $\Delta n_g$.}
\end{figure}

The leakage current in the sluice can be suppressed by working with a phase bias instead of voltage bias which was applied in ~\cite{Niskanen2005,Vartiainen2007}. The only experiment reported for a phase biased pump was carried out by \textcite{Mottonen2008}. They connected a sluice in a superconducting loop with another Josephson junction. By measuring the switching behavior of this junction from the superconducting state to the normal branch with forward and backward pumping, they were able to extract the pumped current of the sluice. However, this type of current detection did not turn out to be as sensitive as the direct measurement with a transimpedance amplifier used in the case of voltage bias. A potential way to improve the sensitivity is, instead of the switching junction, to use a cryogenic current comparator coupled inductively to the superconducting loop. This type of an experiment has not been carried out to date. Instead, \textcite{Gasparinetti2012} have measured a sluice in the vicinity of vanishing voltage bias, where they demonstrated single-Cooper-pair pumping plateaus in both the bias voltage and dc level of the gate voltage, see Fig.~\ref{fig:3_5_2}. The quasiparticle poisoning was reported to be suppressed compared with the previous experiments, and hence they observed clear $2ef$ spacing of the current plateaus.

In addition to the above-mentioned pumping schemes, \textcite{Nguyen2007} have studied how a superconducting quantum bit referred to as quantronium can be used to detect the gate charge ramp arising from a current bias on the gate electrode of the island of the device. The accuracy of this technique on converting the bias current into frequency remains to be studied in detail. \textcite{Hoehne2012} studied another type of a quantum bit, a charge qubit, for pumping Cooper pairs non-adiabatically. The aim here was to increase the pumping speed compared to adiabatic schemes but due to the accumulation of the errors from a pumping cycle to another, a waiting period between the cycles needed to be added. Furthermore, \textcite{Giazotto2011} have showed experimentally how phase oscillations arising from the ac Josephson effect can drive Cooper pairs in a system with no tunnel junctions. However, this kind of pumping was found to be very inaccurate in this proof-of-the-concept experiment.

\subsection{Quantum phase slip pump}
\label{sec:QPS}
There is a proposal to build a source of quantized current based on the effect of quantum phase slip (QPS) in
nanowires made of disordered superconductors \cite{Mooij2006}. Phase slip events occur in thin
superconducting wires where thermodynamic fluctuations of the order parameter become significant
\cite{Arutyunov2008}. During the phase slip, the superconducting order parameter vanishes at a certain
instance and position in the wire, and the phase difference between the wire ends changes by 2$\pi$. This
gives rise to a voltage pulse, in accordance with the Josephson relation. If the phase slips happen
frequently, they produce a finite dc voltage, or a finite resistance.

Phase slips caused by thermal activation broaden the temperature range of superconducting phase transition
and produce a resistive tail below the critical temperature of a superconductor \cite{Tinkham}. At
sufficiently low temperatures quantum fluctuations take over, and the residual temperature independent
resistivity of a nanowire can be attributed to the quantum phase slips (see \cite{Arutyunov2008} and
references therein). Thermally activated phase slips are inherently incoherent. Quantum phase slips may be
coherent provided dissipation associated with every switching event is suppressed. This can be achieved in
superconductors with strong disorder, in which Cooper pairs localize before the superconducting transition
takes place \cite{Feigel'man2007}. Such a localization behavior has been observed by scanning tunneling
microscopy in amorphous TiN and InO$_{\rm x}$ films \cite{Sacepe2010,Sacepe2011}, which are believed to be
the most promising materials for the observation of QPS.

The key parameter describing a nanowire in the quantum phase slip regime is the QPS energy $E_{\rm QPS}=\hbar
\Gamma_{\rm QPS}$, where $\Gamma_{\rm QPS}$ is the QPS rate. Consider a superconducting nanowire of length
$L$, sheet resistance $R_\square$ and made of a superconductor with the superconducting transition
temperature $T_{\rm c}$, coherence length $\xi = (\xi_0\ell)^{1/2}$, where $\xi_0$ is the BCS coherence
length and $\ell$ the electron mean free path ($\ell\ll\xi_0$). Although there is no a commonly accepted
expression for $\Gamma_{\rm QPS}$, it is agreed \cite{Mooij2005,Arutyunov2008} that: $\Gamma_{\rm QPS}
\propto \exp(-0.3A\frac{R_{\rm Q}}{2 R_\square \xi})$, where $A$ is a constant of order unity and $R_{\rm Q} = \pi \hbar/e^2$. Clearly, the
exponential dependence of $\Gamma_{\rm QPS}$ on the wire resistance on the scale of $\xi$, $R_\xi = R_\square
\xi$, requires extremely good control of the film resistivity as well as the wire cross-sectional dimensions.
For the nanowires to be in the quantum phase slip regime rather than in the thermally activated regime,
$E_{\rm QPS}$ should exceed the energy of thermal fluctuations $k_{\rm B}T$. For a typical measurement
temperature of 50 mK, $\Gamma_{\rm QPS}$/2$\pi$ should be higher than 1 GHz. Although the exact estimation of
$R_\xi$ is rather difficult, especially in the case of strongly disordered films, the experimental data
presented by \citetext{Astafiev2012} for InO$_{\rm x}$ films agrees with the following values: $\Gamma_{\rm
QPS}$/2$\pi \approx$ 5 GHz,  $R_\xi$ = 1 k$\Omega$ and $\xi$ = 10 nm.

The first experiment reporting the indirect observation of coherent QPS in nanowires was performed by
\citetext{Hongisto2012}. They studied a transistorlike circuit consisting of two superconducting nanowires
connected in series and separated by a wider gated segment. The circuit was made of amorphous NbSi and
embedded in a network of on-chip 30 nm thick Cr microresistors ensuring a high external electromagnetic
impedance. The NbSi film had a superconducting transition temperature of $\approx$ 1 K and normal-state sheet
resistance about 550 $\Omega$ per square. Provided the nanowires are in the regime of QPS, the circuit is
dual to the dc SQUID. The samples demonstrated appreciable Coulomb blockade voltage (analog of a critical
current of the dc SQUID) and periodic modulation of the blockade by the gate voltage. Such a behavior was
attributed to the quantum interference of voltages in two nanowires that were in the QPS regime. This is
completely analogous to the quantum interference of currents in a dc SQUID.

An unambiguous experimental evidence of coherent QPS was provided in the work by \citetext{Astafiev2012}.
Coherent properties of quantum phase slips were proven by a spectroscopy measurement of a QPS qubit,
which was proposed earlier by \citetext{Mooij2005}. The qubit was a loop that had a 40-nm-wide and about
1-$\mu$m-long constriction. The loop was made of a 35-nm-thick superconducting disordered InO$_{\rm x}$ film
with $T_{\rm c} = 2.7~{\rm K}$ and a sheet resistance of 1.7~k$\Omega$ per square slightly above $T_{\rm c}$. The
qubit was coupled inductively to a step-impedance coplanar waveguide resonator, which was formed due to the
impedance mismatch between an indium oxide strip and Au leads to which it was galvanically connected. The
ground planes on both sides of the strip were made of Au. At the qubit degeneracy point at a flux bias
$(m+1/2)\Phi_{\rm 0}$, where $m$ is an integer, there is an anticrossing in the qubit energy spectrum with a
gap $E_{\rm {QPS}}=\hbar\Gamma_{\rm {QPS}}$. At this flux bias, the two quantum states $|m\rangle$ and
$|m+1\rangle$ corresponding to the loop persistent currents circulating in the opposite directions are
coupled coherently, which gives a gap $E_{\rm {QPS}}$ between the lowest energy bands of the qubit. With the
flux offset $\delta\Phi$ from degeneracy, the gap evolves as $\Delta E=((2I_{\rm p} \delta\Phi)^2 + E_{\rm
QPS}^2)^{1/2}$, where $I_{\rm p}$ is the persistent current in the loop. This gap was revealed in the spectroscopy
measurements by monitoring the resonator transmission as a function of external magnetic field and microwave
frequency. When the microwave frequency matched the qubit energy gap, a dip in the transmission was observed.
The width of the dip $\sim$ 260 MHz close to the degeneracy point indicated rather strong decoherence whose
origin is still to be understood.

Based on the exact duality of QPS and the Josephson effects, it is argued that it should be possible to build
a QPS electric current standard, which is dual to the existing Josephson voltage standard \cite{Mooij2006}.
When biased resistively and irradiated by a high frequency signal, QPS junctions should exhibit current
plateaus, which could provide the basis for the fundamental standard of the electric current. When an ac
signal of frequency $f$ is applied to a Josephson junction, Shapiro voltage steps $V_{\rm n} = n
\frac{hf}{2e}$, where $n$ is an integer, are observed. Similarly, when an ac signal is applied to a QPS
junction, an equivalent of Shapiro steps will occur in form of plateaus at constant current levels
$I_{\rm n} = n 2ef$. One should note, however, that error mechanisms have not yet been analyzed for this type
of quantized current source: thus it is not clear at the moment how accurate this source will be.

From the practical point of view, realization of a QPS current source looks rather challenging because it
requires fabrication of nanowires with an effective diameter $\sim 10$ nm as well as precise control of the
sheet resistance $R_\square$ of the nanowire, which is in the exponent of the expression for $E_{\rm QPS}$.
Various approaches to the nanowire fabrication including step decoration technique, sputtering of a
superconductor on a suspended carbon nanotube, trimming of a nanowire by argon milling, etc. are described in
Ref. \cite{Arutyunov2008}. Another issue is the overheating of the nanowire electron system. Assuming the
phase slip region becomes normal (which is true, for example, for Ti nanowires), for the estimation of the
electron temperature, one can use the expression~(\ref{jh5}) for the power transfer from electrons to phonons. A
nanowire with the cross-sectional dimensions 20 nm $\times$ 20 nm, sheet resistance 1 k$\Omega$ per square
and carrying a dc current of 100 pA will have the effective electron temperature of the order of 250 mK,
which is high enough to smear the current plateaus.

One of the first attempts to observe current plateaus on the current-voltage characteristics of
superconducting nanowires under the rf radiation was reported by \cite{Arutyunov2012}. The nanowires were
made of Ti and had length up to 20 $\mu$m and effective diameter from 40 nm down to about 15 nm. The nanowire
sheet resistance varied from about 20 $\Omega$ up to 1.9 k$\Omega$ per square. They were biased through
high-Ohmic Ti or Bi leads having total resistance of 15 k$\Omega$ and 20 M$\Omega$, respectively. The low-Ohmic
samples biased through 15 k$\Omega$ exhibited weak Coulomb blockade. The estimated $E_{\rm QPS}$ was $\simeq
0.1~\mu$eV only. More resistive nanowires ($R_\square = 180~\Omega$, effective diameter $\approx$ 24 nm)
biased through 20 M$\Omega$ leads had pronounced Coulomb blockade with a critical voltage of up to 0.4 mV.
The thinnest nanowires ($R_\square = 1.9$ k$\Omega$, effective diameter $\leq$ 18 nm) exhibited a Coulomb gap
of a few hundred mV with the largest gap exceeding 600 mV. These gaps did not vanish above $T_{\rm c}$ of Ti,
from which the authors concluded that some weak links were unintentionally formed in the thinnest nanowires.
Despite the fact that nanowires had large variations of parameters, all their current-voltage or ${\rm d}V/{\rm d}I$
characteristics exhibited some quasiregular features under the external rf radiation. Those features were
interpreted as being current steps formed due to the phase-locking of intrinsic oscillations by the external
signal.

It is interesting to note that physics of QPS in superconducting nanowires resembles physics of QPS in
Josephson junction arrays \cite{Fisher1986}. A nanowire can be modeled as a 1D array of small superconducting
islands connected by Josephson junctions. Formation of isolated superconducting regions within a nominally
uniform disordered film was confirmed experimentally \cite{Sacepe2010,Sacepe2011}. Such a weakly connected
array of superconducting islands is characterized by the junction Josephson energy $E_{\rm J}$ and the island
charging energy $E_{\rm c}$. The phase and charge dynamics of the 1D array depends on the ratio $E_{\rm
J}/E_{\rm c}$. In the experiment by \cite{Pop2010} $E_{\rm J}/E_{\rm c}$ in a SQUID array was tuned in
situ by applying a uniform magnetic flux through all SQUIDs. The state of the array was detected by an extra
shunt Josephson junction. The authors deduced the effect of the quantum phase slips on the ground state of the array by measuring the switching current distribution of the entire Josephson circuit as
a function of the external magnetic flux for different values of $E_{\rm J}/E_{\rm c}$,
\subsection{Other realizations and proposals}
\label{sec:otherpumps}
In this subsection we cover various ideas that have been brought up fr experimental demonstration. Although their metrological relevance is still to be proven, we wish to present them for the sake of their complementarity, potential and for completeness.

\subsubsection{Ac-current sources}
The current pumps described in sections \ref{sec:metalpumps}, \ref{sec:hybridpumps} and \ref{sec:dotpumps} can be
considered as single-electron injectors generating dc current. Coulomb blockade ensures a good control of the
electron number on each island during the charge transfer. However, there is no time control of the
electronic injection in metallic and hybrid single-electron pumps because of the stochastic nature of
tunneling. Time-controlled injection can be realized in quantum dots, however, the energy of emitted
electrons spreads randomly in a wide range exceeding $\hbar\Gamma$, where $\Gamma$ is the electron tunneling
rate.

\begin{figure}[t]
\includegraphics[width=8cm]{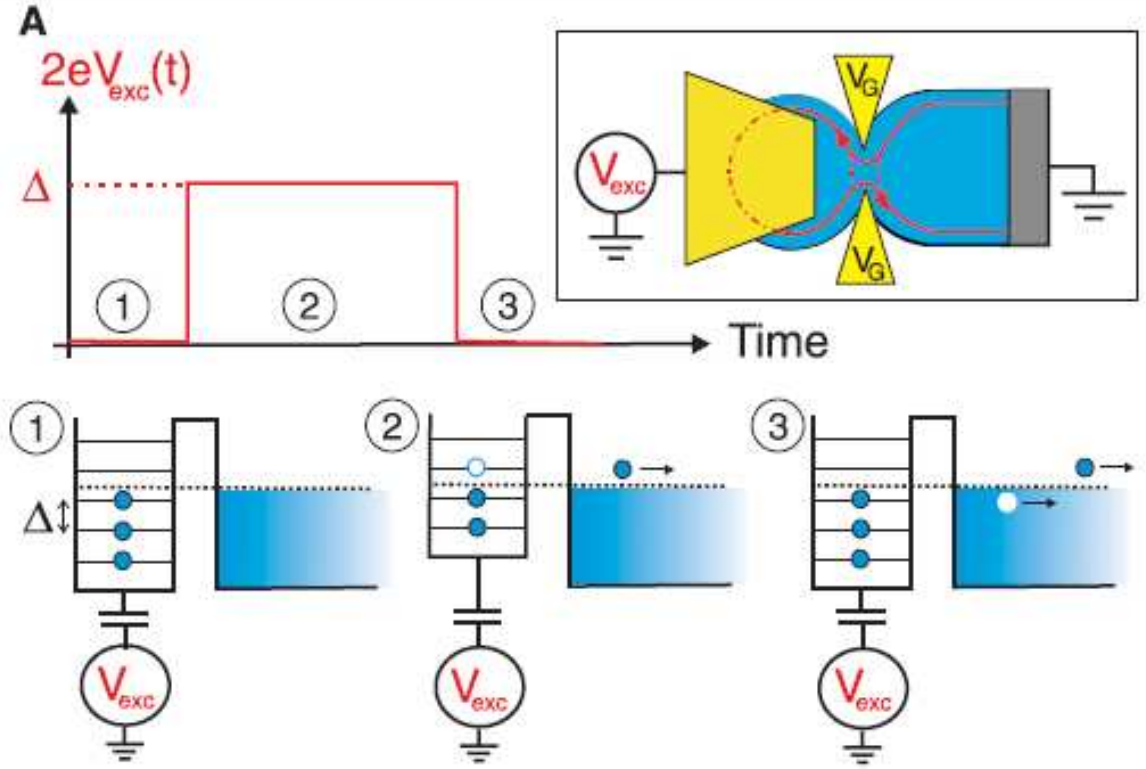}
\caption{\label{fig:ACsource_schematic} Schematic of the single-charge injector and its operation principle.
Starting from step 1 where the Fermi energy level of the conductor lies in between two energy levels of the
dot, the dot potential is increased by $\Delta$ moving one occupied dot level above the Fermi energy (step
2). One electron then escapes from the dot. After that the dot potential is brought back to the initial value
(step 3), where one electron can enter the dot leaving a hole in the conductor. One-edge channel of the
quantum $RC$ circuit is transmitted into the dot, with transmission $D$ tuned by the QPC gate voltage
$V_{\rm G}$. The dot potential is controlled by a radio-frequency excitation $V_{\rm exc}$ applied on a
macroscopic gate located on top of the dot \cite{Feve2007}.}
\end{figure}

A time-controlled single-electron source generating ac current was reported by \textcite{Feve2007}. The
source was made of a quantum dot, realized in a GaAlAs/GaAs heterojunction of nominal density $n_s = 1.7
\times 10^{15}~\rm{m}^{-2}$ and mobility $\mu = 260~\rm{V}^{-1} \rm{m}^2 s^{-1}$ and tunnel-coupled to a
large conductor through a quantum point contact (Fig.~\ref{fig:ACsource_schematic}). The discrete energy
levels of the quantum dot were controlled by the voltage applied to the capacitively coupled metallic gate
located 100~nm above the 2DEG. When the applied voltage is high enough, the electron occupying the highest
energy level tunnels to the conductor. This process is called electron emission. If the applied voltage
is changed abruptly at the subnanosecond time scale (a sudden jump), the emission process is coherent and the
final state of the electron is a coherent wave packet propagating away in the conductor. The energy width of the wave packet is
given by the inverse tunneling time, as required for an on-demand single-particle source. It is also independent
of temperature. Its mean energy is adjusted by the amplitude of the voltage step, $V_{\rm exc}$. For all
measurements, the electronic temperature was about 200~mK. A magnetic field $B \approx 1.3~$T was applied to
the sample so as to work in the quantum Hall regime with no spin degeneracy. The QPC dc gate voltage
$V_{\mathrm G}$ controlled the transmission $D$ of a single edge state as well as the dc dot potential. As
reported by \textcite{Gabelli2006}, this circuit constitutes the effective quantum-coherent $RC$ circuit,
where coherence strongly affects the charge relaxation dynamics. The effective quantum resistance $R$ and
capacitance $C$ are defined as $R=h/2e^2$ and $C=e^2(dN/d\varepsilon)$, where $dN/d\varepsilon$ is the
local density of states of the mode propagating in the dot, taken at the Fermi energy~\cite{Pretre1996}. Interestingly, the QPC
transmission probability $D$ affects the quantum capacitance but not the quantum resistance. In the
experiment, the addition energy $\Delta + e^2/C \approx 2.5$~K was extracted. The large top-gate capacitance
makes the Coulomb energy $e^2/2C$ unusually small, and the total addition energy reduces to the energy-level
spacing $\Delta$.

The single-charge injection was achieved by the application of a high-amplitude excitation voltage $V_{\mathrm{exc}} \sim
\Delta/e$ to the top gate. When an electron level is brought suddenly above the Fermi energy of the
lead, the electron escapes the dot at a typical tunnel rate $\tau^{-1} = D\Delta/h$, where $\Delta/h$ is
the attempt frequency. Typically, the tunnel rates are in the nanosecond time scales, and this makes the single-shot charge detection a
challenging task. In the experiment, a statistical average over many
individual events was used by repeating cycles of single-electron emission with period $T$ followed by
single-electron absorption (or hole emission) as shown in Fig.~\ref{fig:ACsource_schematic}. This was done by
applying a periodic square wave voltage of amplitude $\approx \Delta/e$ to the top gate. Thus, the signal-to-noise
ratio was increased.

\begin{figure}[t]
\includegraphics[width=8 cm]{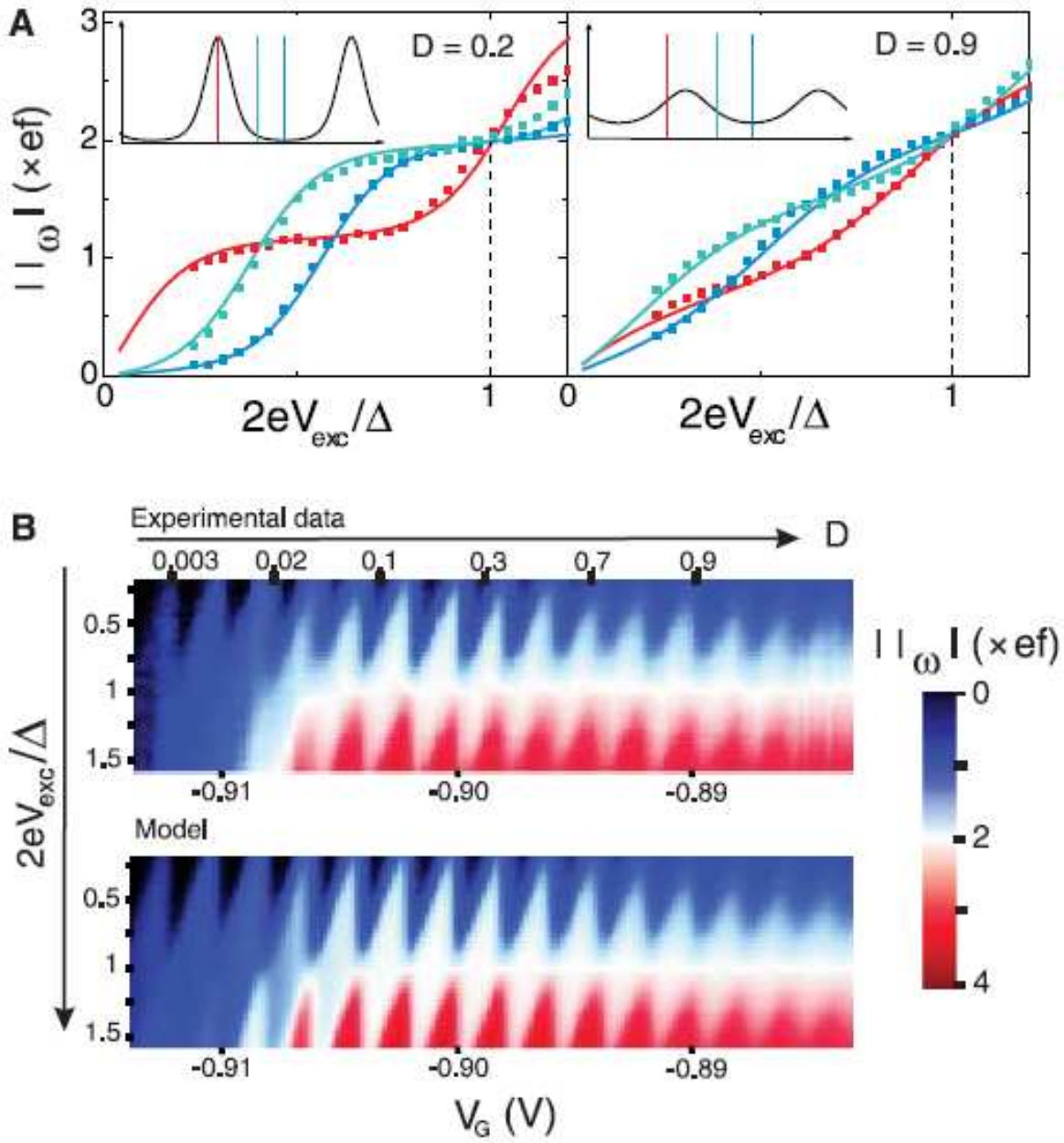}
\caption{\label{fig:ACsource_pumping} AC current quantization. (A) $| I_w |$ as a function of
$2eV_{\mathrm{exc}}/\Delta$ for different dot potentials at $D\approx$ 0.2 (left) and $D\approx$ 0.9 (right).
Points are measured values and lines are theoretical predictions. Insets show schematically the dot density
of states $N(\varepsilon)$. The color vertical lines indicate the dot potential for the corresponding experimental
data. (B) Measured (top) and modeled (bottom) $| I_w |$ as a function of $2eV_{\mathrm{exc}} / \Delta$ and $V_{\rm G}$ shown
as color plots \cite{Feve2007}.}
\end{figure}

Under certain conditions single-electron repeatable injection leads to a good quantization of the ac current. 
$| I_\omega |$ as a function of $V_{\mathrm{exc}}$ for two
values of the dc dot potential at $D \approx$ 0.2 and $D \approx$ 0.9 is shown in Fig.
\ref{fig:ACsource_pumping}. Apparently, a good current plateau is expected when the charge on the dot is well
defined. The latter depends on the transmission and the working point set by the gate voltage. For example, transmission $D
\approx$ 0.2 is low enough and the electronic states in the dot are well resolved, as shown in the inset of Fig.~\ref{fig:ACsource_pumping}
(left). On the other hand, the transmission is large enough for the escape time to be shorter than $T$/2. When the Fermi energy lies
exactly in the middle of the density-of-states valley (blue vertical line in the left inset)), a well-pronounced $| I_\omega |$ = $2ef$ current
plateau is observed centered at $2eV_{\mathrm{exc}}/\Delta = 1$. At this working point, the plateau is rather flat. It is claimed that
the current uncertainty at the plateau is 5\% due to the systematic calibration error.
In contrast, if, with the same transmission, the Fermi energy lies on the peak (light blue vertical line in the left inset), 
there is still a current
plateau, but it is not as flat and very sensitive to parameter variations. When
the energy level is resonant with the Fermi energy, the dot mean occupation at equilibrium is 1/2 and the
height of the plateau is $1/2 \times 2ef = ef$ (Fig.~\ref{fig:ACsource_pumping} A, left), which makes this
working point unsuitable for single-electron pumping. When transmission is increased, the charge fluctuations become stronger and 
the plateau width gets smaller
and finally nearly vanishes at $D \approx$ 0.9 even for the optimal working point, as seen from Fig.~\ref{fig:ACsource_pumping} A, right).
The pumping curves measured at different
transmissions cross at $2eV_{\mathrm{exc}}/\Delta = 1$, which corresponds to the pumped current $| I_w | = 2ef$. Finally, 
Fig.~\ref{fig:ACsource_pumping} B (top) shows domains of good charge quantization, which are presented as
2D color plots, as a function of the excitation voltage $V_{exc}$ and the
gate voltage $V_G$. The white diamonds are the domains of constant current $| I_\omega | =
2ef$ suitable for AC current pumping. At high transmissions, the diamonds become smeared by the dot charge
fluctuations as discussed previously. At small transmissions, even though the dot charge quantization is good,
current quantization is lost because of the long escape time $\omega t \gg 1$, and the current vanishes.
At 180~MHz, the optimal pumping is observed at $D \approx 0.2$. In Fig.~\ref{fig:ACsource_pumping}, the experimental results 
are compared with the theoretical model (solid lines in
Fig.~\ref{fig:ACsource_pumping} A and lower plot in Fig.~\ref{fig:ACsource_pumping} B, respectively) (1D modeling of the
circuit (density of states, transmission, and dot-gate coupling) described in \cite{Gabelli2006} was used),
showing excellent agreement between the two.

The device described above is the electron analog of the single-photon gun. It is not a source of quantized
dc current as the dot emitting the electron can be recharged only though the reverse process of the electron
absorption. However, periodic sequences of single-electron emission and absorption generate a quantized
alternating current $| I_\omega | \approx 2ef$.

\subsubsection{Self-assembled quantum dots in charge pumping}
The idea of using self-assembled quantum dots for charge pumping is based on conversion of optical excitation
into deterministic electric current, see \textcite{Nevou2011}. Unlike in the quantum dots formed in 2DEG that
were described in \ref{sec:dotpumps}, in self-assembled QDs the electrons are confined not due to the Coulomb
blockade effect, but because of the energy level quantization in the ultrasmall island. For the well-studied
InAs/GaAs quantum dots, the characteristic energy scale, i.e., the energy difference between the ground state
and the excited state, is approximately 55~meV and can be tuned to even larger values \cite{Liu2003}. As a
result, an increase in the electron confinement energy from the values $\sim$ 1~meV typical for Coulomb
blockade devices to 55~meV should improve the potential accuracy for a given operation condition.

\begin{figure}[t]
    \begin{center}
        \includegraphics[width=0.49\textwidth]{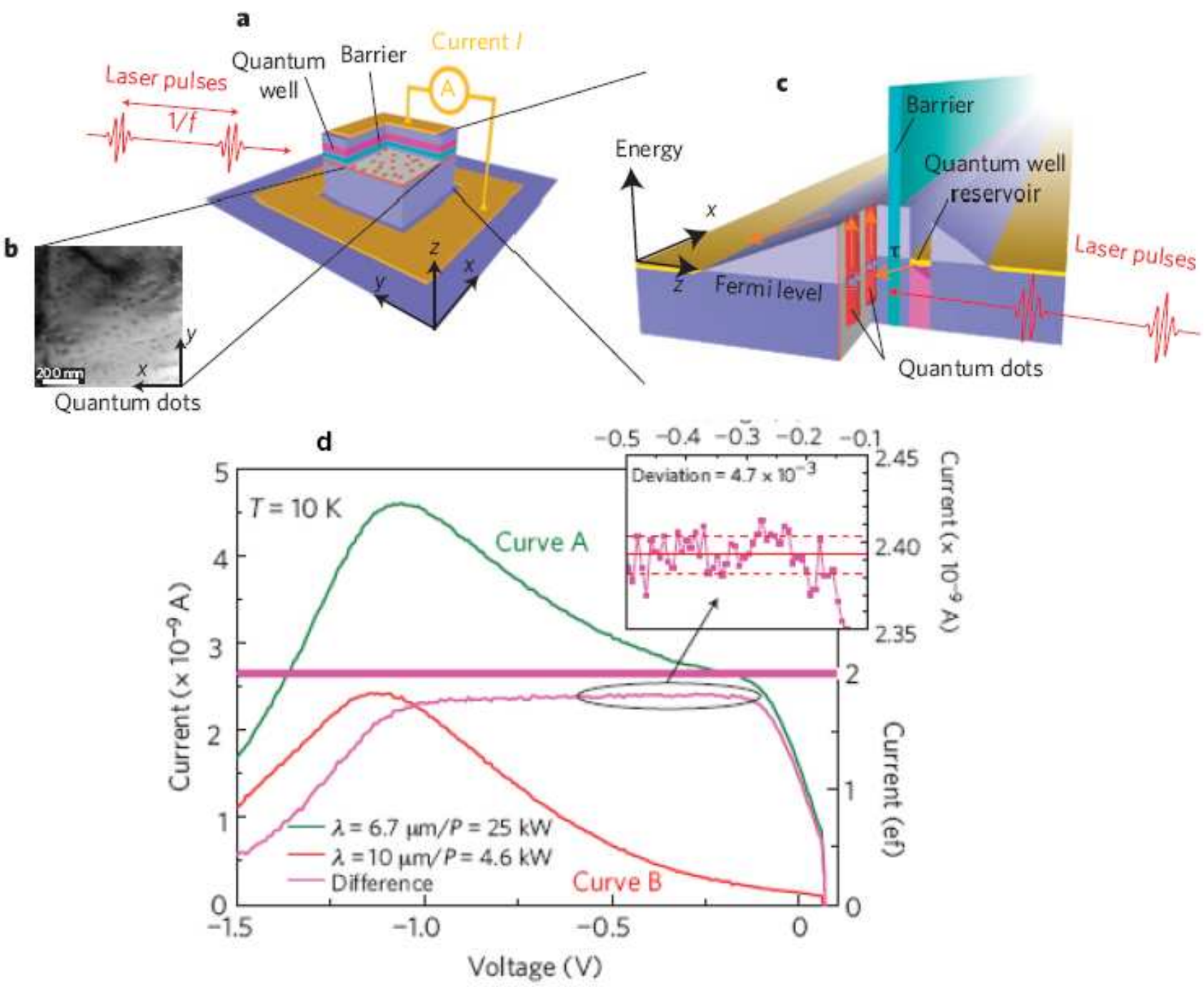}
    \end{center}
   \caption{\label{fig:SAQD_pump} (a) Schematic layout of the self-assembled-quantum-dot electron pump;
   (b) transmission electron microscopy image of the quantum dots; (c) 3D sketch of the conduction band profile
   of the structure under zero bias; (d) Saturation current for two different pump wavelengths ($\lambda = 6.7 \mu$m:
   curve A and $\lambda = 10 \mu$m: curve B. The difference provides a current plateau that should be 2$ef$ (thick
   horizontal line). Inset: Variations of the measured current with respect to the average value \cite{Nevou2010}.}
\end{figure}

In the experiment of \textcite{Nevou2011} a plane of self-assembled InAs quantum dots is coupled to an InGaAs
quantum well reservoir through an Al$_{0.33}$Ga$_{0.67}$As barrier (see Fig.~\ref{fig:SAQD_pump}(a)--(c)).
The structure is sandwiched between two n-doped GaAs regions. The device basically works as a strongly
asymmetric quantum dot infrared photodetector \cite{Nevou2010}. In the absence of any optical excitation,
electrical conduction is inhibited by the AlGaAs barrier. When a laser pulse ionizes the quantum dots,
electrons are excited out of the dot and then swept away by the applied bias voltage, giving rise to a
photocurrent. Electrons are then collected by the contact and thermalized, contributing to the current. Once
all of the electrons are ionized, they will be refilled from the electron reservoir by tunneling through the
AlGaAs barrier. The refilling time $\tau$ can be varied over many orders of magnitude by changing the AlGaAs
barrier width or the Al concentration. As a result, when the pulse energy is high enough to completely ionize
the dot and its duration $t_p$ is much smaller than $\tau$, a fixed number of electrons will be
photo-excited. If the process is repeated at a frequency $f$ , the current will be given by equation $I =
nef$, where $n$ is determined by the number of dots and the number of electrons per dot.

In order to suppress possible sources of error affecting the accuracy of the generated photocurrent, the
probability that some dots are doubly ionized during one optical pulse and the probability that not all dots
are filled during the refilling phase, one needs to satisfy two conditions: $t_p/\tau < \varepsilon$ and
exp$(-1/\tau f) < \varepsilon$, where $\varepsilon=10^{-8}$ is the desired accuracy. Both conditions are
easily satisfied in the experiment. It follows from the two conditions above that $f$ should not exceed
kilohertz range. Although this seems a relatively low frequency, one must keep in mind that in the proposed
device millions of dots are running in parallel, so even this rate could lead to the generation of currents
in the nanoampere range. For example, in the performed pumping experiment, a 210 $\times$ 210 $\mu$m$^2$ mesa was
fabricated by photolithography, and with a dot density of the order of $10^9$ cm$^{-2}$ it contained about 44
million quantum dots. The final source of error, independent of the device design, arises from the
uncertainty in the number of quantum dots contained in one device, as well as the variability in the quantum
dot transition energy. It is claimed that this is more of a technological problem than a fundamental limit
and a variety of solutions can be found to solve this problem.

In the pumping experiment by \textcite{Nevou2011}, refilling time $\tau$ varying from 3 to 5.5 $\mu$s was
measured. For the excitation of the dots and generation of the current, a laser-pumped optical parametric system
with a wavelength $\lambda$ continuously tunable between 3 and 16 $\mu$m was used. It delivered 100-ps-long
pulses with a repetition rate of 1 kHz. The laser source power was sufficient to saturate the quantum dots
during one pulse. The generated current was measured continuously as a function of the bias voltage. The
saturation current depended on the bias voltage, power and pulse wavelength, as shown in
Fig.~\ref{fig:SAQD_pump}(d), which is not suitable for the quantized current source. By subtracting the curve
measured for $\lambda = 10 \mu$m from the one measured for $\lambda = 6.7 \mu$m, a current plateau is
obtained in the voltage range -- 150 mV to -- 1.07 V. By rescaling the current in electron flux per dot
(taking into account the mesa size, the measured dot density and the repetition frequency of the laser
pulses), one obtains the current plateau at the absolute level of 1.8 electrons per dot instead of the
expected value of 2 electrons per dot, giving a pumping accuracy of 10$\%$. This is explained by the
uncertainty in the dot density of 13.6 $\%$ measured by the transmission electron microscopy. Using the same
subtraction method, the dependence of the current plateau height was measured as a function of the pulse
repetition frequency in the range 10 Hz to 100 kHz. As expected, a linear dependence was observed.
The procedure of subtracting the pumped currents measured at different wavelengths was used to compensate for
the strong inhomogeneous broadening of the quantum dot sizes and must be avoided in a real metrological
device. For this, a more uniform set of dots must be fabricated.

\subsubsection{Mechanical single-electron shuttles}
Besides charge pumps with entirely electronic control, there is a group of devices in which a
mechanical degree of freedom is involved. They are called mechanical charge shuttles, because they transfer
either single charges (electrons or Cooper pairs) or portions of charges between the two electrodes due to
the mechanical back and forth motion of a small island between the electrodes. This results in current
flow, either incoherent or coherent. The concept of the mechanical electron shuttle was introduced in 1998 by \cite{Gorelik1998,Isacsson1998}.

\begin{figure}
    \begin{center}
        \includegraphics[width=0.35\textwidth]{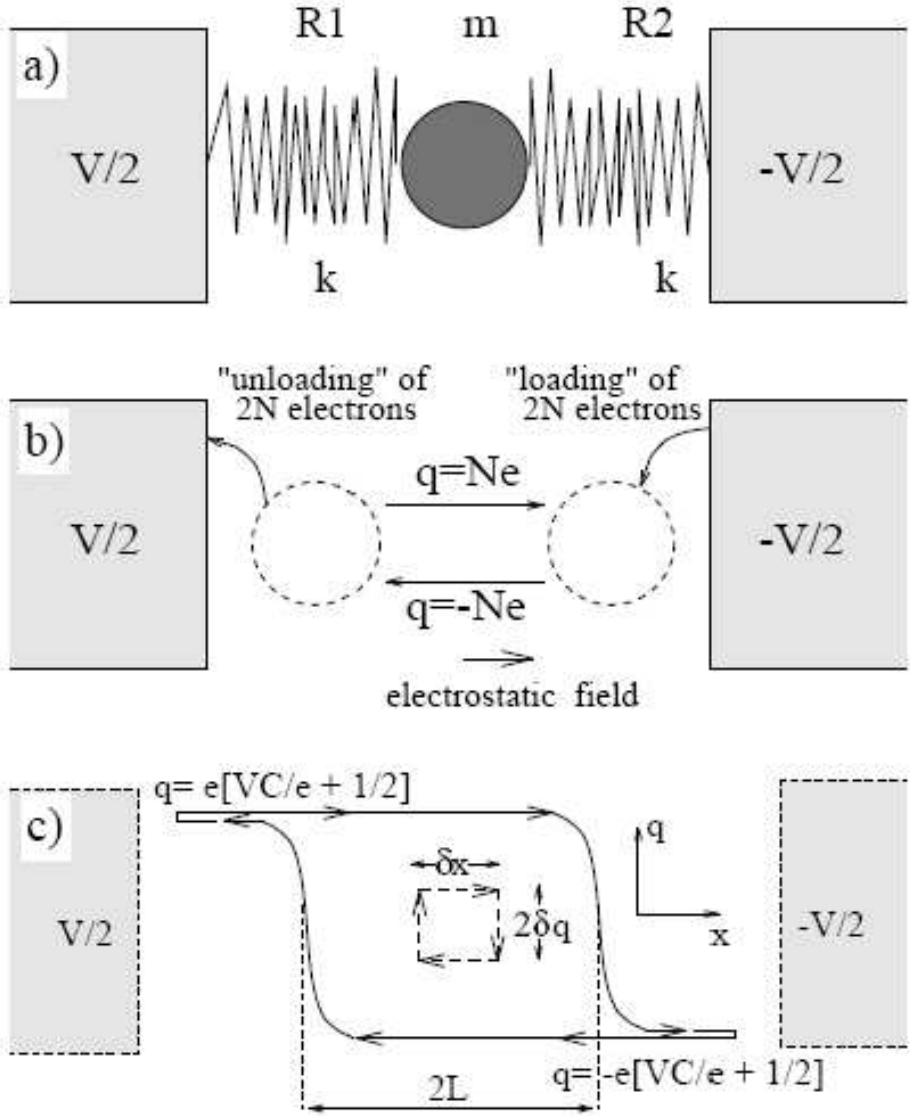}
    \end{center}
   \caption{\label{fig:shuttle} (a) Model of the shuttle device proposed by \cite{Gorelik1998}, where a small
   metallic grain is linked to two electrodes by elastically deformable organic molecular links; (b) Dynamic instabilities
   occur since in the presence of a sufficiently large bias voltage $V$ the grain is accelerated by the electrostatic
   force towards the first electrode, then towards the other one. A cyclic change in direction is caused by the repeated loading
   of electrons near the negatively biased electrode and the subsequent unloading of the same charge at the positively biased
   electrode. As a result, the sign of the net grain charge alternates leading to an oscillatory grain motion and charge
   transport; (c) Charge variations on a cyclically moving metallic island. The force exerted on the island by the electrostatic field along the island displacement between source and drain creates instability
   making the island move. The dashed lines in the middle describe a simplified trajectory in the charge-position plane, when the island motion by
   $\delta x$ and discharge by $2\delta q$ occur instantaneously. The solid trajectory describes the island motion at
   large oscillation amplitudes. Periodic exchange of the charge $2q=2CV+1$ between the island and the leads results
   in charge pumping. From \cite{Gorelik1998}}.
\end{figure}

The proposed device has a small conducting island, which is mechanically attached to electrical leads
with a help of an elastic insulator, see Fig.~\ref{fig:shuttle}(a). The dc voltage applied between the leads
and elastic properties of the insulator together with charging and discharging of the island create
instability and make the island oscillate (see Fig.~\ref{fig:shuttle}(b)). For the proper operation of the
shuttle, two assumptions were made: the amplitude of the mechanical oscillations is much larger than the
electron tunneling distance and the number of electrons on the island is limited. With these assumptions
the island motion and charge fluctuations become strongly coupled. As the shuttle comes into contact
(direct or tunneling) with an electrode, it will pick up charge $q$ and accelerate in the electric field toward
the opposite electrode. Upon contact with this second electrode, the shuttle transfers charge and acquires an
equal but opposite charge $-q$. The shuttle then accelerates back towards the first electrode to begin the
cycle anew, as shown in Fig.~\ref{fig:shuttle}(c). In this manner, a net shuttle current of $I= 2qf$ is
established, where $f$ is the shuttle frequency and $q$ the charge transferred in one cycle. The dc voltage
applied between the source and drain electrodes is sufficient to drive the island. There is a Coulomb force
pushing the island periodically to one lead or the other. Depending on the shuttle details, two regimes can
be distinguished: classical \cite{Gorelik1998,Isacsson1998,Weiss1999} and quantum mechanical
\cite{Armour2002,Fedorets2004,Johansson2008,Cohen2009}. The shuttle, when made superconducting, can transfer
not only electrons but also Cooper pairs \cite{Gorelik2001, Shekhter2003}.

The first experimental realization of the mechanical charge shuttle that operated due to a shuttle
instability was reported by \textcite{Tuominen1999}. This was a rather bulky device even though it was scaled
down considerably in size and operating voltage in comparison to the earlier electrostatic bell versions. The
measured device contained a resonating beam of dimensions $40~\rm{mm} \times 22~\rm{mm} \times 1.6~\rm{mm}$
and a shuttle of mass 0.157~g and radius of about 2~mm. The effective mass of the bending beam was 30~g, its
fundamental vibrational frequency 210~Hz and quality factor 37. The shuttle attached to the beam was placed
in between two voltage biased massive electrodes having a gap adjustable by a micrometer. Besides measuring
the current with a sensitive electrometer, the authors also measured the shuttle frequency using a diode laser
and phototransistor sensor in a photogate arrangement. Both current and frequency were measured as a function
of voltage.  The observed current jumps as a function of the bias voltage as well as hysteresis in the
transport characteristics were the main indications of the shuttling regime of the device.

\begin{figure}
    \begin{center}
        \includegraphics[width=0.3\textwidth]{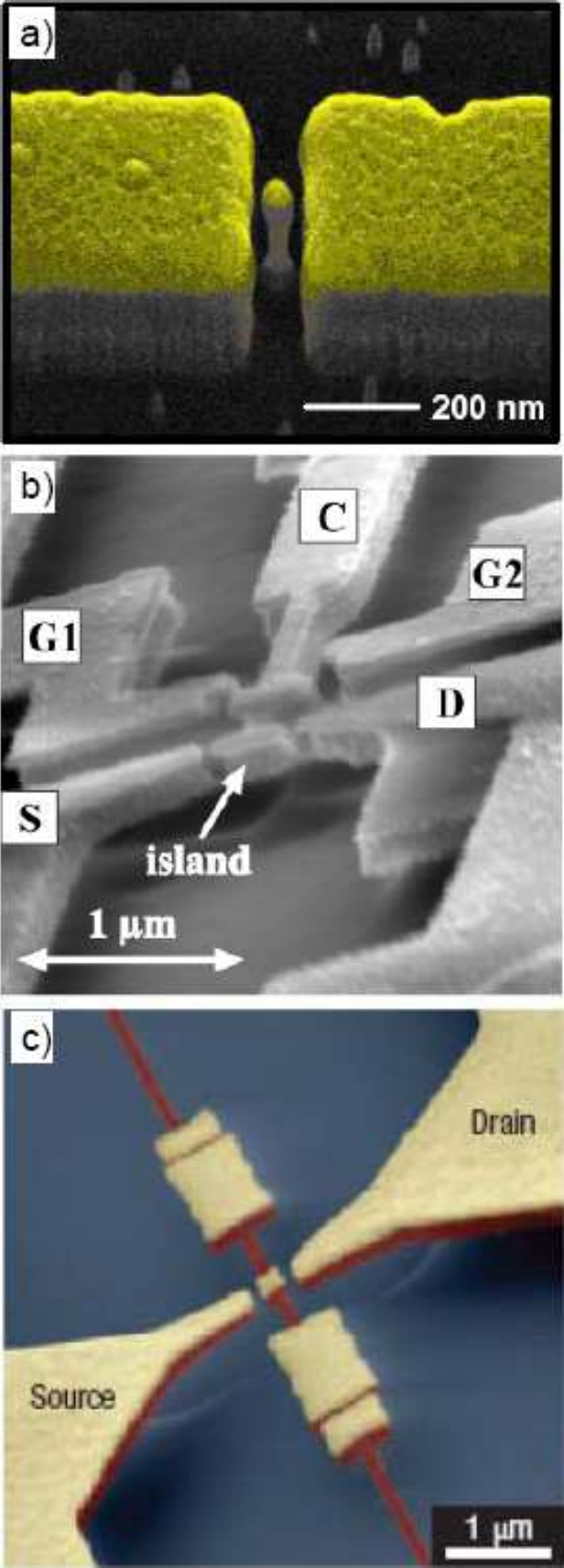}[t]
    \end{center}
   \caption{\label{fig:shuttle_SEM} (a) Scanning electron micrograph of a nanopillar between two electrodes \cite{Kim2010};
(b) Electron micrograph of the quantum bell: The Si beam (clapper) is clamped on the upper side of the
structure. It can be actuated by ac voltage applied to the gates on the left- and right-hand sides (G1 and
G2) of the clapper (C). Electron transport is then observed from source (S) to drain (D) through the island
on top of the clapper. The island is electrically isolated from the rest of the clapper which is grounded
\cite{Erbe2001}; (c) A false-color SEM image taken at an angle to reveal the three-dimensional character of
the nanomechanical SET. A gold (yellow) island is located at the center of a doubly clamped freely suspended
silicon nitride (red) string. The gold island can shuttle electrons between the source and drain electrodes
when excited by ultrasonic waves \cite{Koenig2008}.}
\end{figure}

A nanoscale version of the instability-based electron shuttle was implemented by \textcite{Kim2010}, see
Fig.~\ref{fig:shuttle_SEM}(a)). The device consisted of a 250-nm-tall Si pillar with diameter of 60 nm,
covered on top with a 45 nm thick gold layer. The pillar was placed in the middle of a 110 nm gap between two
electrodes, source and drain, of the central line of a coplanar waveguide. For reference, the authors also
fabricated and measured a similar device without a pillar in the gap. The samples were measured at room
temperature in a probe station evacuated to a pressure of 10$^{-4}$ mbar. The pilar was actuated by applying
a small rf signal together with a dc bias voltage across the source and drain electrodes. The amplitude of the
rf signal alone was not sufficient to make the pillar resonate, however, it was sufficient for driving in
combination with the dc bias. A clear frequency dependence was observed for the sample with a pillar in the
gap, with the resonance frequency of 10.5 MHz and quality factor of about 2.5. It was estimated that the
device shuttles on average 100 electrons per cycle. The resonance was also detected in the mixing experiment.

Another realization of the nanoelectronic shuttling device was reported by
\textcite{Moskalenko2009,Moskalenko2009N}. It had a configuration of a single-electron transistor, whose
island was a 20-nm large gold nanoparticle placed in between the Au source and drain electrodes by means of
an atomic force microscope (AFM). First, planar 30-nm-thick gold electrodes were made by electron-beam
lithography and separated by a gap of 10-20 nm. Then they were covered by flexible organic molecules,
1,8-octanedithiol, with a length of about 1.2~nm. The nanoparticle was linked to the electrodes through
flexible organic molecules, 1,8-octanedithiol, with length of about 1.2~nm. Then nominally 20-nm-diameter
gold nanoparticles were adsorbed by immersion of the electrode assembly into aqueous gold solution. After
that, an AFM image of the sample was taken and one nanoparticle in the vicinity of the gap was identified. It
was pushed to the gap by the AFM tip. Current-voltage characteristics of the fabricated devices were measured
at room temperature in a shielded dry box to protect the structure from moisture and electromagnetic noise. The
authors observed characteristic current jumps in the current-voltage curves and attributed them to the
shuttling effect. They also compared the transport characteristics of their devices with and without the
nanoparticle. They also compared characteristics of a working shuttle device with the same device but from
which the nanoparticle was removed. After this procedure, current through the device dropped below the noise level.

The effect of the mechanical vibrational modes on charge transport in a nanoelectronic device was observed in
the C$_{60}$ single-electron transistor (\cite{Park2000}). In this device, a single C$_{60}$ molecule was
placed in the narrow gap between the two gold electrodes. It was found that the current flowing through the
device increases sharply whenever the applied voltage was sufficient to excite vibrations of the molecule.
Although mechanical vibrations were observed in this structure, they are not related to the shuttling of
electrons.

Externally driven resonant shuttles may be easier to implement in comparison to the instability-based
shuttles, because a much larger displacement amplitude can be achieved. However, the drawback of using a
mechanical resonator is the discrete set of eigenfrequencies, which are determined by geometry and
materials. Therefore, only a limited number of frequencies are available for electron transfer. In the
experiment performed by \textcite{Erbe1998} a Si nanomechanical resonator was placed in between two contacts.
The whole device, being a scaled-down version of the classical bell, was functioning as a mechanically
flexible tunneling contact operating at radio frequencies. The sample was machined out of a single-crystal
silicon-on-insulator substrate by a combined dry- and wet-etch process. It consisted of a metallized
singly-clamped beam (the clapper), whose free end is aligned between the source and the drain. The beam was 1
$\mu$m long, 150 nm wide and 190 nm thick. The beam was driven by an oscillating voltage applied on two gates
at frequencies up to 100 MHz. There was a $\pi$ phase shift between the two gate voltages. The clapper was
biased with a dc voltage and the source was grounded. The resulting clapper-drain dc current as well as rf
response were measured at 300 K and 4.2 K. The current-frequency dependence contained strong peaks, which
were interpreted as being due to the mechanical resonances of the beam, indicating that shuttling was
occurring. The peaks had low quality factors, ranging from 100 to 15, only. The number of electrons $N$
shuttled per cycle was estimated from the current peak height $I$ using the relation $N = I/ef$. Below 20
MHz, 10$^3$-10$^4$ electrons were shuttled in each cycle. On the 73 MHz peak the number was decreased to about 130 electrons
per cycle.

One issue in the device presented by \textcite{Erbe1998} is that there was no island in it, which makes this
system different from the one envisaged by \textcite{Gorelik1998}. This issue was addressed in the later work
by the same group \cite{Erbe2001}. Using a similar fabrication procedure, the authors fabricated a
singly-clamped beam (clapper) with a metal island on its end (see Fig.~\ref{fig:shuttle_SEM}(b)). First,
transport characteristics of the device were measured at 300 K. It was found that there was no detectable
current through the device unless a driving ac voltage ($\pm$ 3 V) was applied to the driving gates. Under
the external drive, the current exhibited several peaks, similar to those in the earlier device
\cite{Erbe1998}, which was attributed to the beam motion. The background current was explained by the thermal
motion of the beam. At a temperature of 4.2 K all the current peaks were suppressed except one at about 120
MHz with much smaller height (only 2.3 pA). This corresponds to shuttling of an average of 0.11 electrons per
cycle.

\textcite{Koenig2008} implemented electromechanical single-electron transistors (MSETs) with a metallic
island placed on a doubly clamped SiN beam (Fig.~\ref{fig:shuttle_SEM}(c)). An array of 44 mechanical SETs
was measured in one run. The sample was placed in a vacuum tube filled with He exchange gas at a pressure of
7.5 $\times$ 10$^{-4}$ mbar. The measurements were performed at a temperature of 20 K. The beams were actuated
using ultrasonic waves produced by a piezo unit. The actuation frequency ranged from 3.5 up to 6 MHz. To
shield the sample from the electrical signal required for the piezo actuation, the sample chip was placed
inside a Faraday cage with solid 3-mm titanium walls. Measurements revealed several resonance features in the
SET dc current, which were attributed to the mechanical resonances. It was argued that the mechanical motion 
of the resonator was strongly nonlinear. This was
imposed by the side electrodes constituting the impacting boundary conditions. The nonlinear nature of the
system resulted in the shape of the resonance curves different from Lorentzian. Although the expected
step-like dependence of the SET current on the source-drain voltage was not observed (because of the high
measurement temperature as compared to the charging energy), the authors made an optimistic conclusion, that
the device may be useful for quantum metrology.

\subsubsection{Electron pumping with graphene mechanical resonators}
An electron pump based on a graphene mechanical resonator in the fundamental flexural mode was introduced by
\textcite{Low2012}. The resonator is actuated electrostatically by a gate electrode. Time varying
deformation of graphene modifies its electronic energy spectrum and in-plain strain. Cyclic variation of
these two properties constitutes the scheme for quantum pumping. To have a nonzero pumping current, spatial asymmetry
must be introduced. It is assumed that the contacts between the graphene layer and the left and right
electrodes are not equivalent, which is modeled by different densities of states. This can be achieved in the
experiment by using different materials for the two electrodes. It is emphasized that Coulomb blockade
effects will favor the transfer of an integer number of electrons per cycle, so that the relation between
current and frequency will be quantized. This is just a proposal and the applicability of this approach for
quantum metrology is still to be verified.

\subsubsection{Magnetic field driven single-electron pump}
Another proposal, not implemented though, is based on using for electron pumping a ferromagnetic
three-tunnel-junction device \cite{Shimada2001}. Its islands and leads are made of ferromagnetic metals with
different coercive forces. Such a device can be operated as a single-electron pump if
controlled not by the gate voltages, but by ac magnetic fields. In addition to the charging effects, it makes
use of the magnetic-field-induced shift of the chemical potential and magnetization reversal in the
ferromagnetic electrode.

The proposed device has intrinsic limitations of the pumping speed, which are determined by the physical time
constraints of the ferromagnet. The pump operation frequency must be much lower than the characteristic
relaxation times. The prospects of this type of an electron pump for quantum metrology are still to be
understood.

\begin{figure}
    \begin{center}
        \includegraphics[width=0.5\textwidth]{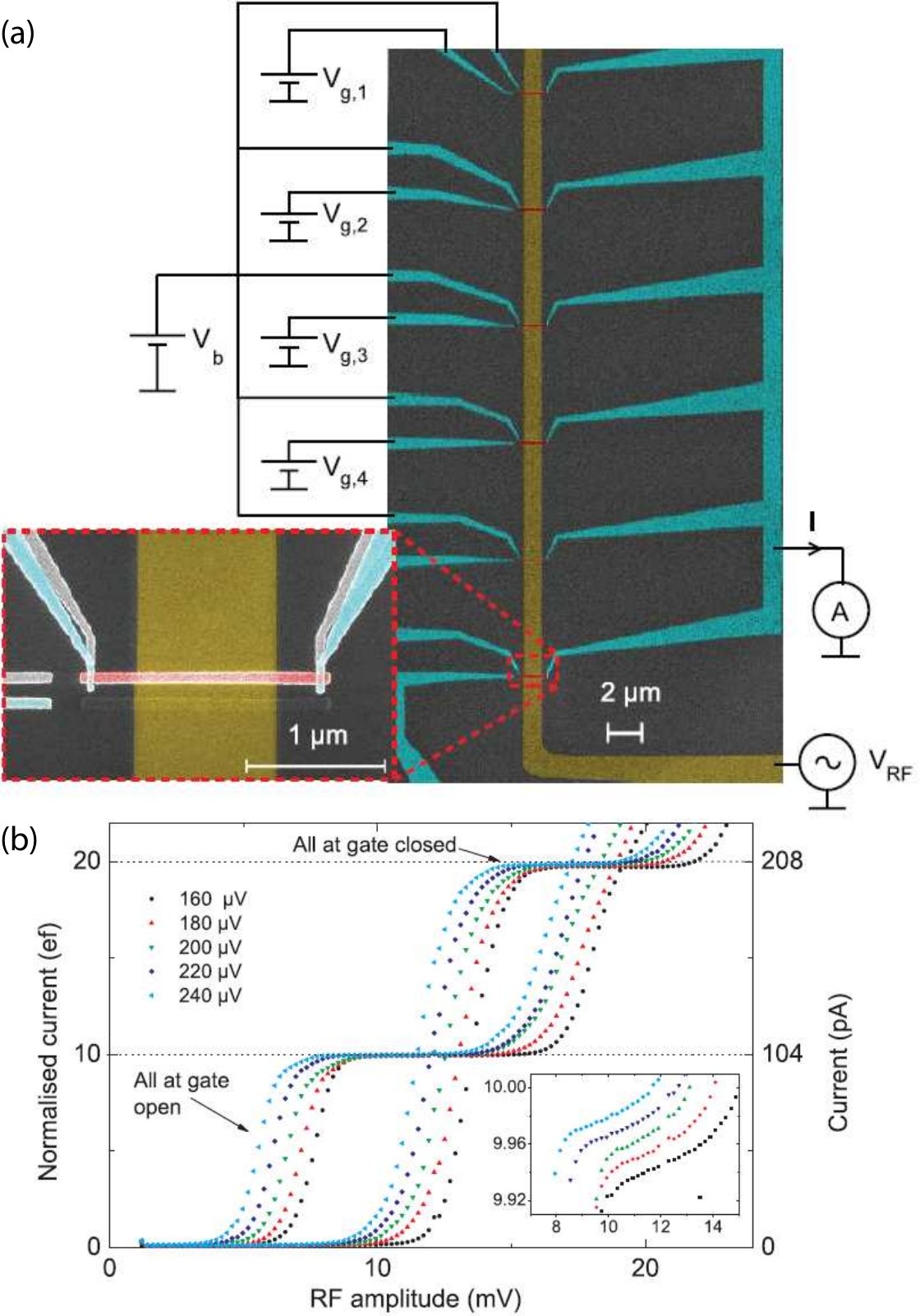}
    \end{center}
   \caption{\label{fig:paralSINIS} (a) Scanning electron micrograph of parallel turnstiles~\cite{Maisi2009}. The tunstiles are biased with a common bias $V_b$ and driven with a common rf gate voltage $V_{\rm rf}$. Gate offset charges are compensated by individual gate voltages $V_{\rm g, i}$. (b) Output current $I$ for ten parallel devices tuned to the same operating point producing current plateaus at $I = 10 Nef$. The curves are taken at different $V_b$ shown in the top left part of the panel.}
\end{figure}

\subsubsection{Device parallelization}
\label{sec:parallel}

As discussed in Sect.~\ref{sec:metalpumps}, it is possible to reach precise electron pumping with a normal metal single-electron pump consisting of a sufficiently long array of islands. With six islands and seven junctions, the accuracy of the pumped current is at $10^{-7}$ level. However, the maximum current is limited to a few $\mathrm{pA}$. To get the current scaled up to $100\ \mathrm{pA}$ level, a requirement for practical metrological applications, see Sect.~\ref{sec:directqmt}, approximately hundred pumps should be operated in parallel. The main reason why parallelization is impractical for normal-metal devices is the tuning of the offset charges~\cite{Keller1996,Camarota2012}. Each island has an individual offset charge that has to be compensated separately. Therefore, a metrological current source implemented as parallellized normal metal pumps would require of the order of 1,000 dc lines.

Compared to normal metal pumps, quantum dot-based pumps allow for higher pumping speeds using fewer control lines thanks to the tunability, see Sect.\ref{sec:dotpumps}. Accuracy of $1.2\ \mathrm{ppm}$ at output current of $150\ \mathrm{pA}$ has been already demonstrated with a single quantum dot~\cite{Giblin2012}. Therefore, parallelization of such pumps may not be even required if the accuracy can be improved without a loss in speed. Nevertheless, parallelization of semiconducting pumps has been considered in the literature. With two pumps, invariance with respect to gate variations has been shown to be below $20\ \mathrm{ppm}$ level~\cite{Wright2009} with output current exceeding $100\ \mathrm{pA}$. In this case, all signals were individually tuned for each device requiring two dc and one rf signal per device. However, it is possible to use common signals for rf drive and for the barrier voltages~\cite{Mirovsky2010}. In this case, only one dc voltage per device is required for tuning the other barrier and possible offset charges. The obtainable accuracy, depending on device uniformity, is still an open question for this approach.

For the hybrid NIS turnstiles, the maximum current per device is limited to few tens of picoamperes, as discussed in Sect.~\ref{sec:sinis_higherorder}. Hence, at least ten devices are to be run in parallel, which has been shown to be experimentally feasible~\cite{Maisi2009}. In Fig.~\ref{fig:paralSINIS} we show a scanning electron micrograph of a sample used in that work and the main experimental findings. The turnstiles in these experiments suffered from photon-assisted tunneling due to insufficient electromagnetic protection, see Section~\ref{sec:environment induced tunneling}, and hence the quantization accuracy was only on $10^{-3}$ level. Improved accuracy is expected for a new generation of turnstile devices~\cite{Pekola2010}. For parallel turnstiles, a common bias voltage can be used as it is determined by the superconducting gap $\Delta$, which is a material constant and varies only very little across a film deposited. Also, the rf drive can be common if the devices have roughly equal $R_T$, $E_C$, and coupling from the rf line to the island. As the error processes that set the ultimate limit on a single turnstile accuracy are not yet determined, the exact requirements on device uniformity cannot be fully resolved. 

\subsection{Single-electron readout and error correction schemes}
\label{sec:readout}

\subsubsection{Techniques for electrometry}
\label{sec:readout_electrometry}

\begin{figure}
    \begin{center}
        \includegraphics[width=0.35\textwidth]{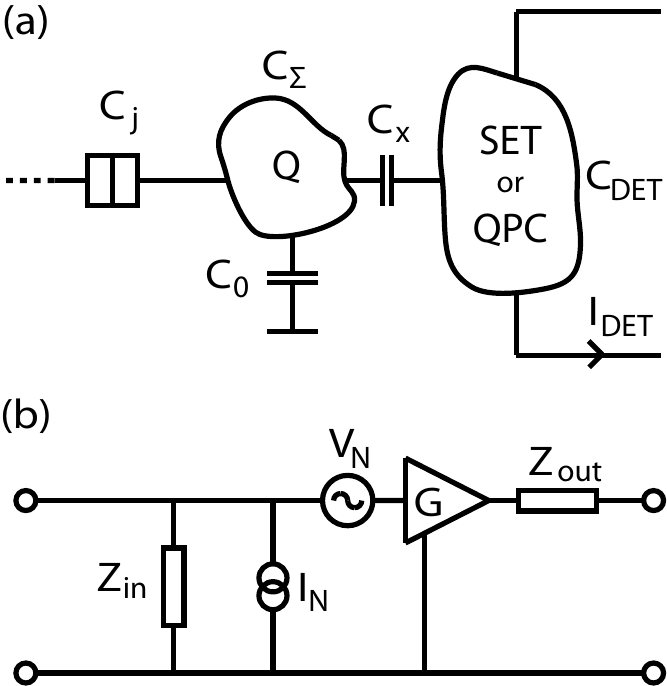}
    \end{center}
   \caption{\label{fig:ops_electrometer} (a) Circuit diagram of a charge counting device. Electric charge $Q$ on the island on the left is monitored. The island is coupled to an electrometer island via capacitor $C_x$, and also tunnel coupled to an external conductor. The single-electron box configuration illustrated here requires only one tunnel junction with capacitance $C_j$. In addition, there is capacitance $C_0$ to ground, which accounts also for gate electrodes and any parasitic capacitances.  The probing current $I_\txt{det}$ through the detector is sensitive to the charge on the coupling capacitor, which is a fraction $C_x/C_\Sigma$ of the total charge $Q$, where $C_\Sigma = C_x + C_0$. The detector is a single-electron tunneling transistor based on Coulomb blockade, and hence the total capacitance of the detector island $C_\txt{det}$ is of the order of 1~fF or less. (b) Circuit diagram of a general noisy electrical amplifier that can also be adapted to describe the electrometers of single-electron experiments. This figure is based on \cite{Devoret2000}. For the configuration shown in panel (a), one has for input impedance $Z_\txt{in}(\omega) = 1/(j \omega C_\txt{in})$, where $C_\txt{in}^{-1} = C_x^{-1} + C_\txt{det}^{-1}.$ The input voltage is related to the island charge $Q$ through $V_\txt{in} = Q / C_{\Sigma}$. Noise source $I_N$ represents back-action and $V_N$ the noise added by the electrometer at the output referred to the input. Gain of the amplifier is given by $G$. Output impedance $Z_\txt{out}$ equals the differential resistance at the amplifier operation point.}
\end{figure}

\begin{figure}
    \begin{center}
        \includegraphics[width=0.35\textwidth]{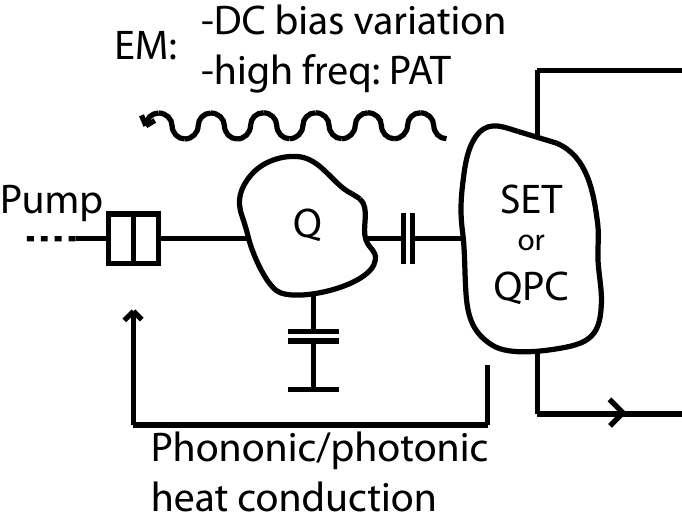}
    \end{center}
   \caption{\label{fig:ops_backaction} Detector back-action mechanisms. The back-action can originate by direct electromagnetic (EM) coupling either by variations in pump biasing or by high frequency photon assisted tunneling (PAT). Another source of back-action is via heat conduction. The detector located in proximity of the device typically heats up. The heat can be then conducted to the device by either phononic or photonic coupling.}
\end{figure}

The electrometer used to detect the presence or absence of individual charge quanta is a central component in schemes for assessing pumping errors and error correction. Figure \ref{fig:ops_electrometer}~(a) introduces the essential components of an electron counting setup. In order to observe proper charge quantization, the counting island is connected to other conductors only via low-transparency tunnel contacts. The electrometer is capacitively coupled to the counting island and biased in such a manner that the small voltage drop of the counting island due to change of its charge state by one electron induces a measurable change in the electrical transport through the detector. The readout performance can be characterized in terms of response time (bandwidth), charge sensitivity, and back-action to the system under measurement. In the present context of electron counting, we define back-action to include all mechanisms by which the presence of the detector changes the charge transport in the measured system.

The two basic electrometer realizations providing sufficient charge sensitivity for electron counting applications are the single-electron transistor (SET)~\cite{Fulton1987,Kuzmin1989} and the quantum point contact (QPC)~\cite{Thornton1986,Berggren1986,Field1993}. From a sample fabrication point of view, it is convenient when the electrometer and the charge pump can be defined in the same process; hence, QPC is the natural charge detector for quantum dots in semiconductor 2DEGs, whereas metallic single-electron devices are typically probed with SETs. There exist also studies where a metallic superconducting SET has been used as the electrometer for a semiconductor QD~\cite{Lu2003,Fujisawa2004,Yuan2011}, and the SET can be realized in the 2DEG as well~\cite{Morello2010}.

The charge sensitivity $\delta q$ is determined by the noise of the system as a whole~\cite{Korotkov1994} and is conveniently expressed in units of $e/\sqrt{\mathrm{Hz}}$ for electrometry applications. For metallic SETs, output voltage fluctuations $\delta V_\txt{out}$ can be related to the charge coupled to the electrometer according to $\delta q = C_g \delta V_\txt{out} / (\partial V_\txt{out} / \partial V_g)$, where $V_g$ is the voltage of the SET gate electrode and $C_g$ is its capacitance to the SET island~\cite{Kuzmin1989}. Here, $C_g$ can be determined reliably in the experiment from the period of Coulomb oscillations. Similar calibration cannot be performed for a QPC and hence the charge sensitivity is expressed in relation to the charge of the neighboring QD~\cite{Cassidy2007}, corresponding to $Q$ in Fig.~\ref{fig:ops_electrometer}(a). Variations of $Q$ and $q$ are related as $\delta q = \kappa\, \delta Q$, where $\kappa = C_x / C_\Sigma$ is the fraction of the island charge that is coupled to the electrometer. For charge counting applications, the relevant parameter is $\delta Q$. The RMS charge noise for a given detection band equals $\Delta q = \sqrt{\int d\omega\,S_{Q_\txt{out}}(\omega)}$, which reduces to $\delta q \times \sqrt{B}$ in the white noise limit, where $B$ is the readout bandwidth. It is possible to pose the charge detection problem in the language of quantum linear amplifiers as shown in Fig.~\ref{fig:ops_electrometer}(b)~\cite{Devoret2000,Averin2003,Clerk2010}. When such a detector is modeled as a linear voltage amplifier, $I_N$ and $V_N$ chracterize the input and output noise, respectively, and the quantum theory limit for the spectral density of fluctuations at signal frequency $\omega$ reads $\sqrt{S_V(\omega) S_I(\omega)} \geq \hbar \omega / 2$. Information about the electronic back-action is contained in the correlator $\left< \delta V_\txt{in} (t)\, \delta V_\txt{in} (t') \right>$ of the induced voltage fluctuations on the counting island. Denoting the total capacitance of the counting island by $C_\Sigma$, the fluctuations in the output charge signal equal $\delta Q_\txt{out}(\omega) = C_\Sigma V_N(\omega)$, and the voltage fluctuations on the counting island are given by $\delta V_\txt{in}(\omega) = I_N(\omega)/(j \omega C_\Sigma)$. One thus finds $\sqrt{S_{Q_\txt{out}}(\omega) S_{V_\txt{in}} (\omega)} \geq \hbar/2$ as the quantum limit.

In theory, quantum-limited operation can be achieved with normal-state SETs operated in the co-tunneling regime \cite{Averin2001}, superconducting SETs~\cite{Zorin1996,Zorin2001}, and QPCs~\cite{Korotkov1999a,Clerk2003,Averin2005}. In practical devices, however, the noise spectrum up to 1-100~kHz depending on the setup is dominated by $1/f$-like charge noise that is intrinsic to the sample but whose microscopic physical origin is still debated~\cite{Starmark1999,Buehler2005,Vandersypen2004}. Above 1~kHz, the charge noise level is usually set by the preamplifier noise, but studies exist where the intrinsic shot noise of the electrometer was comparable to the noise of the readout electronics~\cite{Brenning2006,Kafanov2009a}.
For the normal-state SET, sensitivities of the order of $10^{-7}\ e/\sqrt{\txt{Hz}}$ are attainable in theory with present-day fabrication technology, where the intrinsic noise is due to stochastic character of the tunneling processes and includes both shot and thermal noise \cite{Korotkov1994,Korotkov1999}. The best charge sensitivities reported to date for a single-electron transistor in \cite{Brenning2006} were almost identical in normal and superconducting states, namely $1.0$ and $0.9\times10^{-6}e/\sqrt{\txt{Hz}}$, respectively, at a signal frequency of 1.5~MHz. In \cite{Xue2009}, also the back-action of a superconducting SET was measured and the product of noise and back-action was found to be 3.6 times the quantum limit. For QD charge detection with QPCs, charge sensitivity of $2\times10^{-4}~e/\sqrt{\txt{Hz}}$ referred to the QD charge has been demonstrated~\cite{Cassidy2007}. It appears to be easier to realize large charge coupling fraction $\kappa$ with metallic SETs than with QPCs~\cite{Yuan2011}.

Let us now discuss the back-action mechanisms in more detail. Despite the above quantum theory result connecting back-action and noise, the electronic back-action of electron counter can be addressed in principle independently of its charge noise, as the readout bandwidth (at most 100~MHz, see below) is much below the microwave frequencies that can promote charge transfer errors:  Overcoming even a modest 100~$\mu$V energy barrier requires photon frequencies above 24 GHz if multi-photon processes are neglected. Nevertheless, voltage fluctuations induced by the shot noise of the detector usually have a non-negligible spectral density at microwave frequencies. For SET detectors carrying quasiparticle current, the voltage fluctuations of the island have a spectral density $S_{V_\txt{det}}(\omega) = 2 e I_\txt{det}/(\omega^2 C_\txt{det}^2)$ in the limit $\omega \gg 2 \pi I_\txt{det}/e$, of which fraction $\kappa^2$ is transferred to the counting island via capacitive voltage division.  This can easily be stronger than the equilibrium thermal noise from resistive components at the sample stage; cf.~\cite{Martinis1993}. A full quantum calculation has been presented in~\cite{Johansson2002}. The high-frequency back-action can be attenuated by replacing the capacitive coupling by a lossy wire that acts as a low-pass filter for microwaves but does not affect the charge signal~\cite{Saira2012}. For QPCs, Coulombic back-action can be divided into shot noise, which can be in principle eliminated by circuit design, and fundamental charge noise \cite{Aguado2000,Young2010}.

The low-frequency part of detector back-action manifests itself as variation of the dc bias of the pump or turnstile device. The case of an SET electrometer coupled to a single-electron box was studied in \cite{Turek2005}. In the limit of small coupling capacitance $C_x$, the voltage swing on the counting island due to loading and unloading the detector island equals $\Delta V_\txt{in} = \kappa e / C_\txt{det}$. We note that this is just a fraction $C_x/C_\txt{det} < 1$ of the voltage swing from loading or unloading the actual counting island with an electron. Hence, dc backaction of the detector does not necessarily place an additional constraint on the design of the counting circuit, as the voltage swing $e / C_\Sigma$ needs to be kept small for the operation of the charge pump.

In addition to the electronic back-action described above, one needs to consider the phononic heat conduction from the detector to the charge pump. For reaching the ultimate accuracy, the charge pumps typically require temperatures of the order of 100~mK or lower, where small on-chip dissipation can raise the local temperature significantly due to vanishing heat conductivity in the low temperature limit \cite{Kautz1993,Giazotto2006}, see also Sec.~\ref{sec:heating of single-electron devices}. The average power dissipated by the detector equals $P = \left<I_\txt{det} V_\txt{det} \right>$, and it needs to be transported away by the substrate phonons or electronically via the leads. Requirement for a sufficiently large charge coupling coefficient $\kappa$ limits the distance by which the detector and charge pump can be separated. The temperature increase by dissipated power has been studied on a silicon substate in \cite{Savin2006} and they give the formula
\begin{equation}
T = \left(T_0^4 + \frac{2 f P}{\pi r^2 \nu v} \right)^{1/4},\label{eq:ops_Si_heating}
\end{equation}
where $T$ is the substrate temperature at distance $r$ from a point source of heating power $P$, $T_0$ is the bath temperature, $\nu v = 3600\unit{W m^{-2} K^{-4}}$ is material parameter and $f = 0.72$ is a fitting parameter for their experimental observations. For an illustrative example, we estimate that the dissipated power at the electrometer in the original RF-SET paper \cite{Schoelkopf1998} was 120~fW based on the published numbers. According to Eq.~(\ref{eq:ops_Si_heating}), this will heat the substrate underneath nearby junctions ($r$ = 200~nm) to 140~mK, which is high enough to deteriorate the performance of many single-electron devices below the metrological requirements. In~\cite{Sillanpaa2004}, the readout was coupled to the Josephson inductance of a superconducting SET instead of conductance, reducing the dissipation by two orders of magnitude. Usually, it is possible to assess the severity of detector back-action effects in the experiment by measuring the tunneling rates using different values of $I_\txt{det}$, see, \eg, \cite{Lotkhov2011,Saira2012,Kemppinen2011}, so that any variation of the observed rates can be attributed to back-action. The picture is somewhat different in 2DEG systems due to significantly weakened electron-phonon coupling. Experimental study of phononic back-action in 2DEGs is presented in \cite{Schinner2009,Harbusch2010}.

Finally, let us consider the obtainable bandwidth in single-electron detection. Digital low-pass filtering is applied to the digitized output of the last amplifier stage to reduce the noise level so that some form of threshold detection can be applied for identifying the individual tunneling events. The optimal filter, also called the matched receiver [see, \eg, \cite{Kay1998}], in the case of a known transfer function can be represented in Fourier space as $F(\omega) = A^*(\omega) / \left[|A(\omega)|^2 + S_N(\omega)^2\right / \sigma^2]$, where $A(\omega)$ is the (complex) gain of the readout circuit, $S_N(\omega)$ is the spectral density of noise at the filter input, and $\sigma$ is r.m.s. signal power. The matched receiver will minimize the expected squared error $\left<[Q_\txt{out}(t) - Q(t)]^2\right>$, but in practice any low pass filter with an appropriate cutoff will work. The cutoff may have to be set lower than the intrinsic bandwidth in order to suppress charge noise level to a sufficiently small fraction of $e$. The bandwidth of the readout, $B$, is commonly defined as the corner frequency of the gain from gate charge to output voltage \cite{Visscher1996}. The performance requirements for the charge readout depend on the particular charge counting scheme, but in general the bandwidth $B$ places a limit on the fastest processes that can be detected and hence constrains the magnitude of the electric current that can be reliably monitored.

Both the QPC and SET electrometers have an impedance of the order of $R_K = h/e^2 \approx 25.8\unit{k\Omega}$. For the SET, $R \gtrsim R_K$ is required to realize Coulomb blockade according to the orthodox theory of single-electron tunneling \cite{Averin1991,Ingold1992}. For QPC, the most charge senstive operation point is around a bias point where $\partial V/\partial I = R_K$, midway between the first conductance plateau and pinch-off~\cite{Cassidy2007}. The readout bandwidth is set by the $RC$ constant of the electrometer's differential resistance and the capacitive loading on its outputs. As the barrier capacitance is of the order of 1~fF or less for the devices, the intrinsic bandwidth is in the GHz range. In practice, the capacitance of the biasing leads and the input capacitance of the preamplifier dominate. When the preamplifier is located at room temperature as in the pioneering experiments~\cite{Fulton1987,Kuzmin1989}, the wiring necessarily contributes a capacitance of the order of 0.1--1~nF and henceforth limits the readout bandwidth to the kHz range~\cite{Visscher1996,Pettersson1996}. Readout by a current amplifier from a voltage-biased SET avoids the \emph{RC} cutoff on the gain, but the usable bandwidth is not substantially altered as current noise increases at high frequencies where the cabling capacitance shorts the current amplifier input~\cite{Starmark1999}.

In order to increase the effective readout bandwidth, the SET impedance has to be transformed down towards the cable impedance that is of the order of 50~$\Omega$. Bandwidths up to 700~kHz have been achieved by utilizing a HEMT amplifier with a low impedance output at the sample stage~\cite{Pettersson1996,Visscher1996}. The dissipated power at the HEMT in these studies was 1--10~$\mu$W depending on the biasing, which can easily result in overheating of the electrometer and/or the coupled single-electron device. The best readout configuration to date is the rf reflectometry technique, applicable to both SETs~\cite{Schoelkopf1998} and QPCs~\cite{Qin2006}, where the electrometer is embedded in a radio frequency resonant circuit and the readout is achieved by measuring the damping of the resonator. A read-out bandwidth of 100~MHz was achieved in the original demonstration~\cite{Schoelkopf1998}. The authors also note that their charge sensitivity of $1\times10^{-5}~e/\sqrt{\txt{Hz}}$ yields $\Delta q = 0.1 e$ for the full detection bandwidth, \ie, electron counting at 100~MHz would have been possible in a scenario where the charge coupling fraction $\kappa$ was close to unity.

%\cite{Reilly2007} RF-QPC $4\times 10^$

\subsubsection{Electron-counting schemes}
\label{subsec:ecs}
Realization of a current standard based on electron counting has been one of the key motivators for development of ultrasensitive electrometry~\cite{Schoelkopf1998,Keller2009,Gustavsson2008}. First, let us see why direct current measurement of uncorrelated tunneling events, like those produced by a voltage-biased tunnel junction, cannot be used for a high-precision current standard: Assume a noise-free charge detector that yields the charge state of the counting island with time resolution $\tau = 1/B$, and that Markovian (uncorrelated) tunneling events occur at rate $\Gamma \ll B$. With probability $\Gamma \tau$, a single tunneling event occurs during the time $\tau$ and is correctly counted by the detector. With probability $(\Gamma \tau)^2/2$, two tunneling events occur within $\tau$ and constitute a counting error. Hence, to achieve a relative error rate $p$, one needs $\Gamma < 2 p B$. Even with a noiselees 100~MHz rf-SET, one could not measure a direct current greater than $2 e/\txt{s}$ at metrological accuracy $p = 10^{-8}$ in this manner. Would it be practical to account statistically for the missed events in a manner similar to \cite{Naaman2006a} assuming truly Poissonian tunneling statistics and a well-characterized detector? The answer is unfortunately negative: If $N$ tunneling events are observed, the number of missed events $M$ is a Poissonian variable with a mean of $N \Gamma \tau/2$ and standard deviation $\delta M = \sqrt{N \Gamma \tau/2}$. Requiring $\delta M < p N$ gives $N > \Gamma \tau / (2 p^2)$. For $\Gamma = 1\unit{MHz}$ and $\tau$, $p$ as above, one has to average over $N > 5 \times 10^{13}$ events, which is impractical. A more detailed calculation based on Bayesian inference presented in~\cite{Gustavsson2009} results in the same $N$ dependence.

Charge transport through a 1D-array of tunnel junctions can take place in a form of solitons depending on device parameters~\cite{Likharev1988,Likharev1989}. Propagation of the solitons promotes time correlation in the electron tunneling events, allowing the accuracy limitations of counting uncorrelated electrons presented above to be lifted. A proof-of-concept experimental realization has been presented in~\cite{Bylander2005}. The array is terminated at the middle island of an SET, allowing for unity charge coupling, and a signal centered around frequency $f_c = I/e$ is expected. The authors claim a possible accuracy of $10^{-6}$ based on the charge sensitivity of their electrometer only. However, the spectral peaks in the experimental data appear too wide for an accurate determination of the center frequency. Factors not included in the accuracy estimate are instability of the bias current and SET background charge fluctuations.

\begin{figure}[t]
    \begin{center}
        \includegraphics[width=0.49\textwidth]{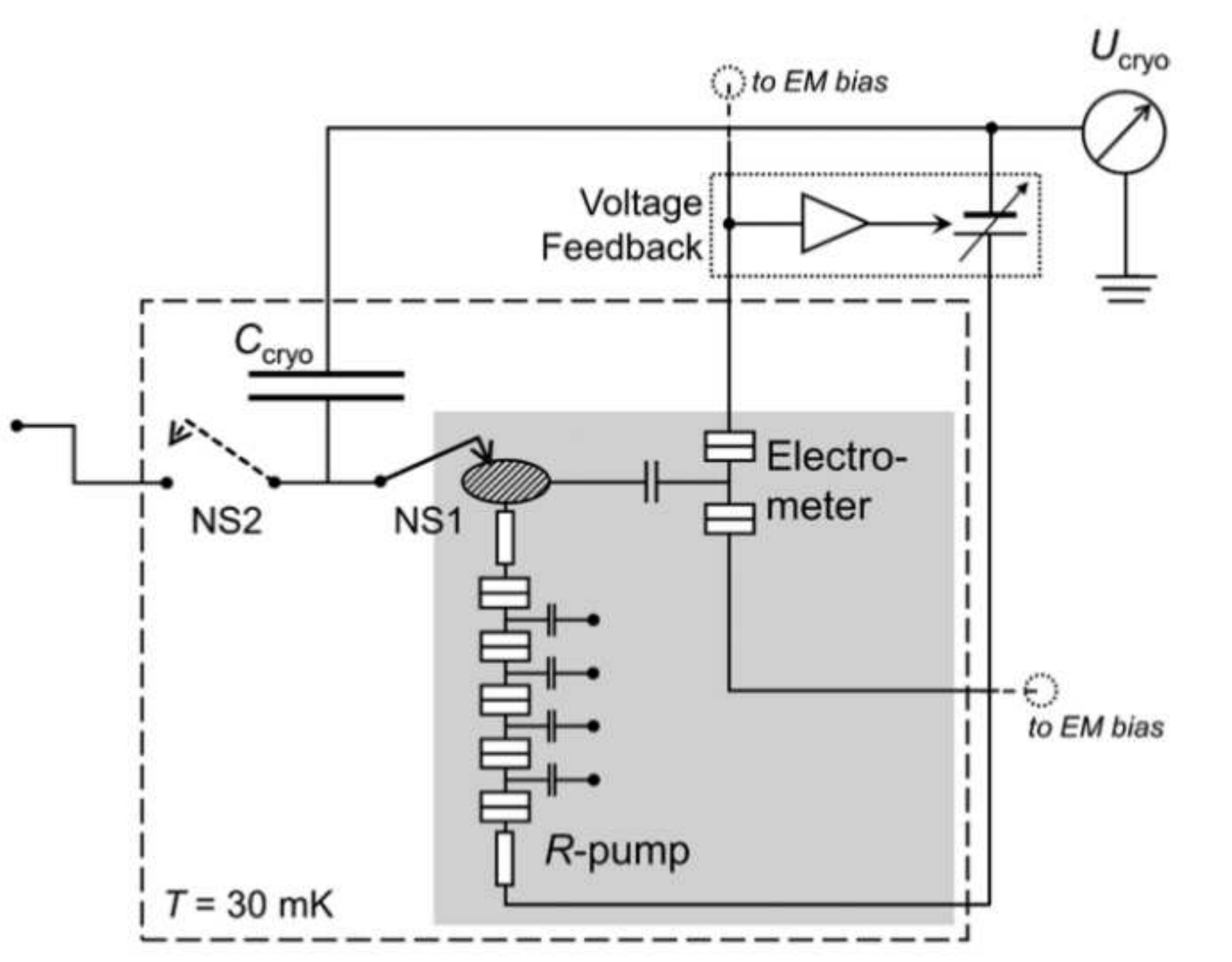}
    \end{center}
   \caption{\label{fig:ops_eccs} Circuit diagram of a practical implementation of the electron counting capacitance standard. Switches NS1 and NS2 are cryogenic needle switches. Figure from Ref.~\cite{Camarota2012}.}
\end{figure}

Single-electron electrometry can be used to count the much rarer pumping errors instead of the total pumping current. Such an approach has been used to study the accuracy of metallic multi-junction pumps that are used in  the electron counting capacitance standard (ECCS)~\cite{Keller1999,Keller2007}. A circuit diagram of an ECCS experiment is shown in Fig.~\ref{fig:ops_eccs}. Two cryogenic needle switches are required to operate the device in different modes: determining coupling capacitances and tuning the pump drive signal (NS1 and NS2 closed), operating the pump to charge $C_\txt{cryo}$ (NS1 closed, NS2 open), and comparing $C_\txt{cryo}$ with an external traceable capacitor (NS1 open, NS2 closed). With NS1 open, the pump can be operated in a shuttle mode: a charge of $e$ is repeatedly pumped back and forth across the pump at the optimal operation frequency (which is above detector bandwidth), and pumping errors appear as discrete jumps in the electrometer output. A relative error rate of $1.5\times 10^{-8}$ at a shuttling frequency of $5.05$~MHz has been demonstrated for a 7-junction pump shown in Fig.~\ref{fig:KellerAPL}, which was used in the NIST ECCS setup~\cite{Keller1996}. At PTB, accuracy of the order of $10^{-7}$ was demonstrated for a 5-junction R-pump operated at a shuttling frequency of $0.5$~MHz~\cite{Camarota2012}. In the actual ECCS operation, the SET electrometer is used as a part of a feedback loop that maintains the voltage of the node at the end of the pump constant, but time-resolved single-electron detection is not needed then.

\cite{Wulf2008} propose various strategies for error accounting utilizing modestly accurate charge pumps and single-electron electrometry. They consider the ECCS, and in addition propose schemes involving several charge pumps connected in series. The advantage of the more complex schemes is that they do not require cryogenic needle switches, and in theory allow one to account for the sign of pumping errors, thus improving the accuracy of the setup beyond the intrinsic accuracy of the pumps. Even though the accuracy requirements on the pumps are relaxed somewhat, the added complexity from operating several pumps and electrometers make the practical realization of these ideas challenging, and none have been presented to date. Measurements of two series-connected semiconductor QD pumps with a QPC charge detector coupled to an island in the middle are presented in \cite{Fricke2011}, although observation of quantized pumping errors was not demonstrated there.

\subsection{Device fabrication}
Fabrication of charge pumps, regardless of their operational principle, requires advanced nanofabrication
methods. These include, for example, electron-beam lithography, various dry etching techniques, and MBE growth of semiconductor
heterostructures. In general, pumping devices can have small feature sizes in multiple layers that
must be aligned with each other accurately. We begin with the description of the fabrication procedure for the metallic single-electron and Cooper pair pumps and turnstiles described in Secs.~\ref{sec:metalpumps}, \ref{sec:hybridpumps}, and \ref{sec:superpumps}. Subsequently, we present the fabrication methods of quantum dot pumps and turnstiles, the operation of which is discussed in Sec.~\ref{sec:dotpumps}.

\subsubsection{Metallic devices}

Metallic single-electron and Cooper pair
pumps and turnstiles are typically made by
the angle deposition technique, which was first introduced by G.\ J.\ Dolan~[\textcite{Dolan1977}] for the
photolithography process and then later adapted by Dolan and Dunsmuir for the electron-beam lithography
process~[\textcite{Dolan1988}]. We note that there is a myriad of different ways fabricating these devices. Below, we describe only a certain fabrication process for these devices in great detail instead of giving a thorough study of all possible variations.

\begin{figure}[t]
    \includegraphics[width=6.5cm]{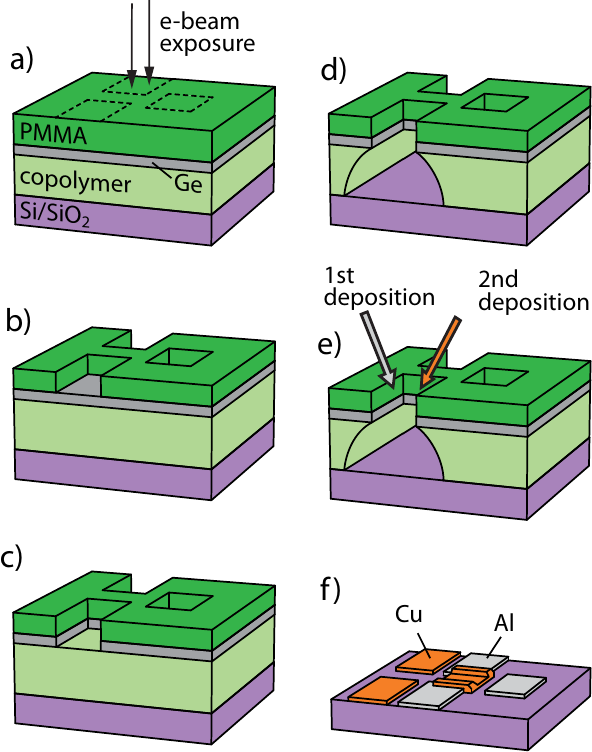}
    \caption{Fabrication of metallic devices. (a) Build-up of a tri-layer resist structure and exposure in the electron-beam writer; (b) Development of the top PMMA layer; (c) Transfer of the pattern formed in the resist into the Ge layer by reactive ion etching; (d) Creation of undercut in the bottom resist and removal of the top resist by oxygen plasma; (e) angle deposition of metals with an oxidation in between; (f) the resulting structure after the liftoff process.}
    \label{fig:fabrication}
\end{figure}

The process starts with the deposition of Au layer on an Si wafer covered by a native silicon oxide. The Au
pattern is formed by a standard photolithography and liftoff process using photoresist S1813 and contains
contact pads and on-chip wiring as well as alignment markers for the deposition of the subsequent layers.
Next, a tri-layer resist structure is built (from bottom to top): copolymer/Ge/poly-methyl-methacrylate
(PMMA) with the thicknesses 200~nm, 20~nm and 50~nm, respectively (See Fig. \ref{fig:fabrication}(a). The
polymer layers are spin-coated on the wafer and baked in a nitrogen oven, and the Ge layer is deposited in an
electron gun evaporator. The wafer is then cleaved into smaller pieces which are exposed and processed
separately. After the exposure of the top PMMA layer on one of the pieces in the electron beam writer, e. g.,
JEOL JBX-5FE, the piece is developed at room temperature in isopropyl alcohol (IPA) mixed with methyl
isobutyl ketone (MIBK) at a ratio of 3:1. Thus, a desired pattern is formed in the PMMA layer (Fig.
\ref{fig:fabrication}(b)). The pattern is transferred into the Ge layer by reactive ion etching in CF$_4$
(Fig. \ref{fig:fabrication}(c)). The sample is then placed in an electron cyclotron resonance (ECR) etcher,
in which an undercut is formed by oxygen plasma. The undercut depth is controlled by the tilt of the sample
stage in the ECR machine. At the same time, the top PMMA layer is etched away. At this stage, each chip has a
Ge mask supported by the copolymer layer (Fig. \ref{fig:fabrication}(d)). Some parts of the mask are
suspended forming the Dolan bridges. Although we described above a method with three layers, in many cases a
bi-layer mask composed of copolymer and PMMA resists is sufficient.

The chips with masks are placed in an electron gun evaporator equipped with a tilting stage. Two consecutive
depositions of metal through the same mask are carried out at different angles to create a partial overlap
between the metal layers (Fig. \ref{fig:fabrication}(e)). If the surface of the bottom layer (typically Al)
is oxidized by introducing oxygen into the evaporation chamber, after the deposition of the top electrode,
the sandwich structure composing of the overlapping metal layers with a thin oxide in between form small
tunnel junctions (Fig. \ref{fig:fabrication}(f)).

The normal-metal or superconducting charge pumps are made entirely of Al, which can be turned normal at low
temperatures by an external magnetic field~[see
\textcite{Geerligs1990,Geerligs1991,Pothier1992,Keller1996,Vartiainen2007}]. In the case of hybrid structures
described in Refs. \textcite{Pekola2008,Kemppinen2008,Kemppinen2009,Maisi2009}, the bottom electrode was Al
and the top one was either Cu or AuPd.

\subsubsection{Quantum dots}
The gate structure of the charge pumps based on quantum dots is also fabricated using electron beam
lithography. The main differences in the fabrication compared with metallic devices are the following: Ohmic contacts have to be made between metallic bonding pads on the surface of the chip and the 2DEG located typically $\sim$100~nm below the surface. Furthermore, the 2DEG has to be either depleted with negative gate voltage from the unwanted positions in the case of GaAs devices [see Fig.~\ref{fig:3_3_1}~(d)] or accumulated with positive gate voltage in the case of metal--oxide--semiconductor (MOS) silicon devices [see Fig.~\ref{fig:3_3_1}~(c)]. For GaAs, also etching techniques have been employed to dispose some parts of the 2DEG leading to a smaller number of required gates [see Fig.~\ref{fig:3_3_3}~(a)]. In GaAs devices, typically a single deposition of metal through a mono-layer PMMA resist is sufficient to create the gate structure. For MOS silicon dots, several aligned layers of gate material are often used. However, only a single layer is typically deposited with each mask in contrast to metallic devices employing angle evaporation.

Let us describe in detail a fabrication process for MOS silicon quantum dots. We begin with a high-resistivity ($\rho > 10\textrm{ k}\Omega$cm at 300~K) near-intrinsic silicon wafer. Phosphorous atoms are deposited on the silicon surface using standard photolithography and they diffuse to a depth of roughly 1.5~$\mu$m during the growth of a 200~nm field silicon oxide on top. All the following process steps involving etching or deposition have to be aligned with the previous ones with the help of alignment markers, a routine we do not discuss separately below. Then a window with size $30\times30$~$\mu$m$^2$ is opened to the field oxide and replaced by 8-nm-thick high-quality SiO$_2$ gate oxide that is grown in ultradry oxidation furnace at 800 $^\circ$C in O$_2$ and dichloroethylene. This thin-oxide window overlaps by a few micrometers with the ends of the metallic phosphorous-rich $n^+$ regions. The field oxide is etched selectively above the other ends of the $n^+$ regions formed in the previous process. The ohmic contacts and the bonding pads are made by depositing metal on these etched regions forming a connection to the $n^+$ silicon. Subsequent annealing is employed to avoid the formation of Schottky barriers.

At this stage, we have bonding pads connected to the metallic $n^+$ regions that extend some hundred micrometers away from the pads to the thin-oxide window with the line width of four micrometers. Electron beam lithography with a 200-nm PMMA resist and metal evaporation with an electron gun evaporator is employed to deposit the first layer of aluminum gates inside the window and their bonding pads outside the window. After the lift-off, the gates are passivated by an Al$_x$O$_y$ layer formed by oxidizing the aluminum gates either by oxygen plasma or thermally on a hot plate (150 $^\circ$C, 5 min). The oxide layer electrically insulates completely the following overlapping layers of aluminum gates that are deposited in the same way with alignment accuracy of $\sim$20~nm.

At least one gate has to overlap with areas where $n^+$ regions extend to the thin-oxide window. By applying positive voltage on these reservoir gates, the electrons from the $n^+$ are attracted to the Si/SiO$_2$ interface below the reservoir gates forming the source and drain reservoirs of the device. For example, the device shown in Fig.~\ref{fig:3_3_6}~(b) is composed of one or two layers of gates: one top gate that induces the source and drain reservoir, two barrier gates below the top gate defining the quantum dot, and a plunger gate in the same layer with the barrier gates. Finally, a forming gas (95\%N$_2$, 5\%H$_2$) anneal is carried out for the sample at 400~$^\circ$C for 15~min to reduce the Si/SiO$_2$ interface trap density to a level $\sim$5$\times 10^{10}$~cm$^{-2}$eV$^{-1}$ near the conduction band edge. Silicon quantum dots can also be fabricated with all-silicon process, in which the aluminum gates are replaced by conducting polysilicon gates shown in Fig.~\ref{fig:3_3_6}~(a).

%
% Chapter 4
%

\section{QUANTUM STANDARDS OF ELECTRIC QUANTITIES AND THE QUANTUM METROLOGY TRIANGLE}
\label{sec:electricstandards}

The ampere is one of the seven base units of \emph{The International System of Units}
(SI) \cite{SI2006} and is defined as follows: "The ampere is that constant current
which, if maintained in two straight parallel conductors of infinite length, of 
negligible
circular cross-section, and placed 1~m apart in vacuum, would produce between these
conductors a force equal to $2\times 10^{–7}$ newton per metre of length."
The present definition is problematic for several reasons: (i) The experiments
required for its realization are beyond the resources of most of the National Metrology
Institutes. (ii) The lowest demonstrated uncertainties are not better than
about $3\times 10^{-7}$, \cite{Clothier1989,Funck1991}. (iii) The definition
involves the unit of newton, $\mathrm{kg}\times\mathrm{m}/\mathrm{s}^2$,
and thus the prototype of the kilogram, which is shown to drift in time
\cite{Quinn1991}. In practice, electric metrologists are working outside the SI
and employing quantum standards of voltage and resistance, based on the Josephson
and quantum Hall effects, respectively.

\subsection{The conventional system of electric units}

According to the ac Josephson effect, $V=(h/2e)\times (\partial\phi/\partial t)$, the
voltage $V$ applied over the Josephson junction induces oscillations of the phase
difference $\phi$ between the macroscopic wavefunctions of the superconducting
electrodes on each side of the junction \cite{Josephson1962}. Phase-locking $\phi$ by a high-frequency ($f_J$)
signal results in quantized voltage plateaus
\begin{equation}\label{eq:shapiro}
V_J=n_J\frac{h}{2e}f_J \equiv \frac{n_Jf_J}{K_J}
\end{equation}
which are often called Shapiro steps~\cite{Shapiro1963}. Here, $K_J$ is called
Josephson constant and $n_J$ is the integer number of cycles of $2\pi$ that $\phi$
evolves during one period of the high-frequency signal.

The Josephson voltage standards (JVS) have been used in electric metrology
since the 1970s, see, e.g., \textcite{Kohlmann2003,Jeanneret2009} for reviews.
The first standards consisted of a single junction and generated voltages
only up to about 10~mV. Arrays of more than 10,000 junctions with the maximum
output of 10~V were developed in the 1980s. They were based
on hysteretic junctions where Shapiro steps with different $n_J$ can exist at the same bias current.
Since 1990s, the research has focused on arrays of nonhysteretic junctions
where $n_J$ can be chosen by the applied current bias. Arrays divided in sections
of $2^m$ junctions ($m=0,1,2\ldots$) are
called \emph{programmable} since one can digitally
select any multiple of $f_J/K_J$ up to the number of junctions as the output
voltage \cite{Hamilton1995,Kohlmann2007}. They are practical for dc voltage
metrology, but are especially developed for generating digitized ac voltage waveforms up to about 1~kHz,
which is an active research topic \cite{Behr2005}. Voltage waveforms at higher frequencies can be
generated by pulse-driven Josephson junction arrays where the desired ac
waveform is synthesized by the delta-sigma modulation of fast voltage
pulses, each having the time integral of one flux quantum $h/(2e)$ \cite{Benz1996}.

The quantum Hall resistance standard (QHR) consists of a two-dimensional
electron gas, which, when placed in a high perpendicular magnetic field,
exhibits plateaus in the Hall voltage $V_H=R_HI$ over the sample in the
direction perpendicular to both the field and the bias current $I$. Here,
\begin{equation}\label{eq:qhe}
R_H=\frac{1}{i_K}\frac{h}{e^2}\equiv \frac{R_K}{i_K}
\end{equation}
is the quantized resistance which is proportional to the von Klitzing constant
$R_K$ and inversely proportional to the integer $i_K$~\cite{Klitzing1980}.
The plateau index $i_K$ can be chosen by tuning the magnetic field.
Usually, the best results are obtained at $i_K=2$.

Quantum Hall standards based on Si MOSFET or
GaAs/AlGaAs heterostructures were taken to routine use
quickly during the 1980s, see, e.g.,
\textcite{Jeckelmann2001,Poirier2009, Weis2011} and Issue 4 of
\emph{C. R. Physique}, vol. 369 (2011) for reviews. Different resistances
can be calibrated against the QHR by using the cryogenic current comparator
(CCC). It is essentially a transformer with an exact transform ratio due to
the Meissner effect of the superconducting loop around the
windings~\cite{Harvey1972,Gallop2006}. Another way to divide or multiply
$R_H$ are parallel and series quantum Hall arrays, respectively, which
are permitted by the technique of multiple connections that suppresses
the contact resistances \cite{Delahaye1993}. One rapidly developing
research topic are ac quantum Hall techniques, which can be
used in impedance standards to expand the traceability to capacitance
and inductance \cite{Schurr2011}. An important recent discovery is that
graphene can be used to realize an accurate and very robust QHR standard
\cite{Zhang2005,Novoselov2007,Tzalenchuk2010,Janssen2011}.

The most precise measurement of $K_J$ within the SI was performed by
a device called liquid-mercury electrometer with the uncertainty
$2.7\times10^{-7}$~\cite{Clothier1989}. The SI value of $R_K$ can
by obtained by comparing the impedance of the QHR and that of the
Thompson-Lampard calculable capacitor~\cite{Thompson1956,Bachmair2009}.
The lowest reported uncertainty of such comparison is
$2.4\times10^{-8}$~\cite{Jeffery1997}. However, both JVS and QHR
are much more reproducible than their uncertainties in the SI, see
Sec.~\ref{sec:universality}. Therefore the consistency of electric
measurements could be improved by defining conventional values for
$R_K$ and $K_J$. Based on the best available data by June 1988, the
member states of the Metre Convention made an agreement of the values
that came into effect in 1990:
\begin{equation}\label{eq:rk90} \begin{array}{rcl}
K_\mathrm{J-90} & = & 483597.9\; \mathrm{GHz/V}\\
R_\mathrm{K-90} & = & 25812.807\;  \Omega\end{array}
\end{equation}
Since then, electric measurements have in practice been performed
using this conventional system which is sometimes emphasized by denoting
the units by $V_{90}$, $\Omega_{90}$, $A_{90}$ etc., and where JVS and
QHR are called \emph{representations} of the units.

\subsection{Universality and exactness of electric quantum standards}
\label{sec:universality}

A theory can never be proven by theory, but, as argued by \textcite{Gallop2005},
theories based on very general principles such as thermodynamics and gauge
invariance are more convincing than microscopic theories such as the
original derivation of the Josephson effect \cite{Josephson1962}. There are rather
strong theoretical arguments for the exactness of JVS: Bloch has shown
that if a Josephson junction is placed in a superconducting ring, the exactness
of $K_J$ can be derived from gauge invariance \cite{Bloch1968, Bloch1970}.
Furthermore, Fulton showed that a dependence of $K_J$ on materials would
violate Faraday's law \cite{Fulton1973}. %, see also \textcite{Langenberg1971,Hartle1971}.
For quantum Hall devices, early
theoretical works argued that the exactness of $R_K$ is a consequence
of gauge invariance \cite{Laughlin1981, Thouless1982}. However, it is very
complicated to model real quantum Hall bars, including dissipation, interactions
etc. and thus the universality and exactness of QHR has sometimes been described
as a "continuing surprise"~\cite{Mohr2005,Keller2008}. Extensive theoretical work,
e.g., on topological Chern numbers, has strengthened the confidence on the
exactness of $R_K$, see \textcite{Avron2003, Doucot2011, Bieri2011} for
introductory reviews. Recent quantum electrodynamics (QED) based theoretical
work predicts that the vacuum polarization can lead to a magnetic field dependence
of both $R_K$ \cite{Penin2009,Penin2010} and $K_J$ \cite{Penin2010a}, but
only at the level of $~10^{-20}$. The case of
single-electron transport has been studied much less and there are no such strong theoretical
arguments for the lack of any corrections for the transported charge
\cite{Gallop2005,Stock2006,Keller2008}.\footnote{A
condensed-matter correction of $\sim10^{-10}e$ for the charge of the electron
was suggested by theory based on QED \cite{Nordtvedt1970}, but it was refuted
in \textcite{Langenberg1971,Hartle1971}.} 

On the experimental side, comparisons between Si and GaAs quantum Hall bars
show no deviations at the experimental uncertainty of $\sim 3\times 10^{-10}$
\cite{Hartland1991}. Recently, an agreement at the uncertainty of
$8.6\times 10^{-11}$ was found between graphene and GaAs devices
\cite{Janssen2011}. This is an extremely important demonstration of the
universality of $R_K$ because the physics of the charge carriers is notably different
in graphene and semiconductors \cite{Goerbig2011}. Comparisons between
JVSs have been summarized recently in \textcite{Wood2009}. The lowest
uncertainties obtained in comparisons between two JVSs are in the range
$10^{-11}$. Even much smaller uncertainties have been obtained in universality
tests of the frequency-to-voltage conversion by applying the same frequency to
two different junctions or junction arrays and detecting the voltage difference by
a SQUID-based null detector. Several accurate experiments have indicated that
the conversion is invariant of, e.g., the superconducting material and the junction
geometry. The lowest demonstrated uncertainty is astonishing:
$3\times10^{-19}$ \cite{Clarke1968,Tsai1983,Jain1987,Kautz1987}.

\begin{figure}[t]
    \begin{center}
    \includegraphics[width=.2\textwidth]{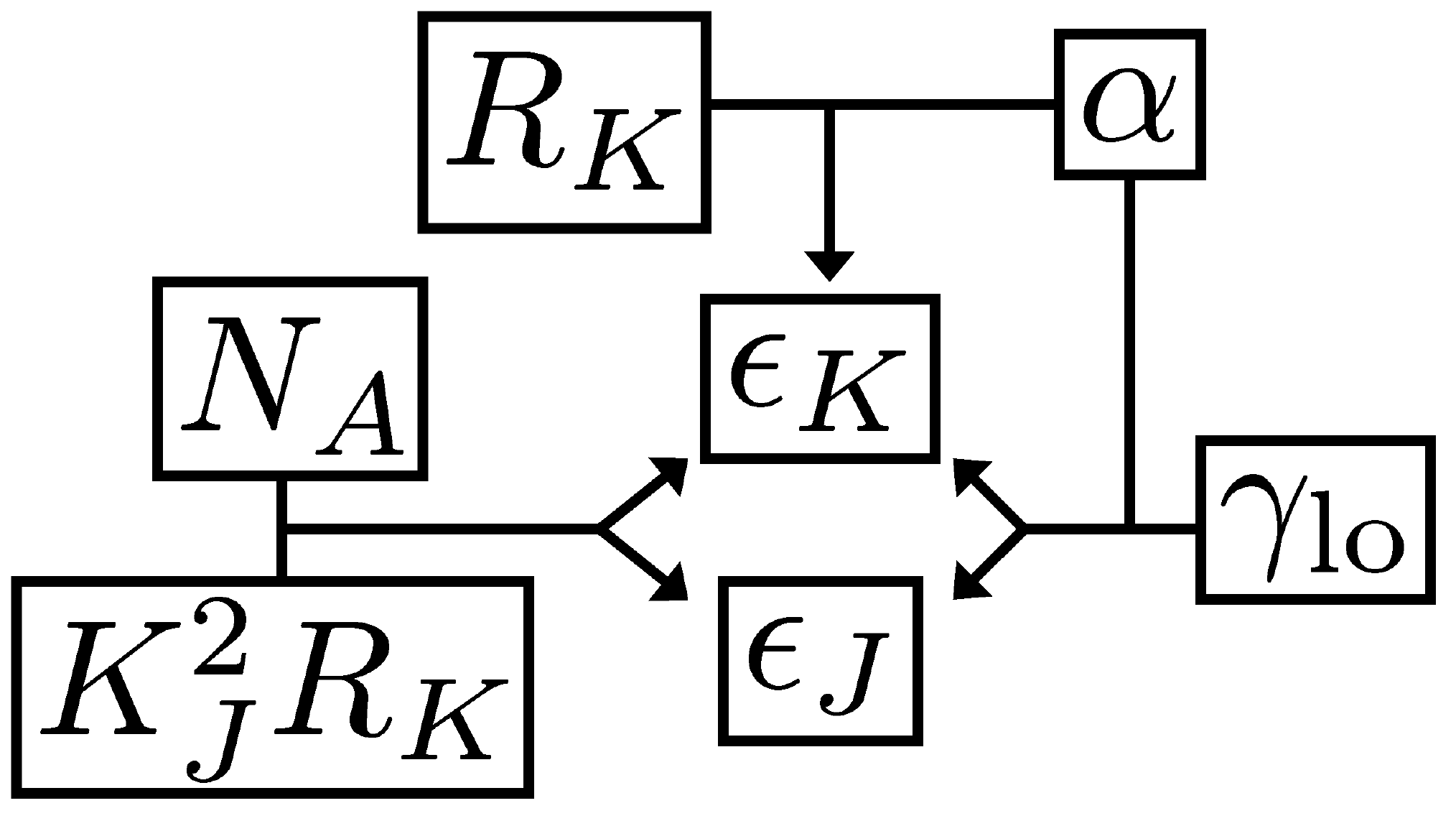}
    \end{center}
   \caption{\label{fig:constants} Simplified sketch of the most accurate
routes to information on $\epsilon_{J,K}$. Direct measurement of $R_K$
together with an independent measurement of the fine structure constant
($\alpha$) yields a value on $\epsilon_K$. Values for the sum of
$\epsilon_J$ and $\epsilon_K$ can be obtained from the combination of
the so called watt balance experiment ($K_J^2R_K$) and a measurement of the
Avogadro constant ($N_A$), or from the combination $\alpha$ and measurements
of low-field gyromagnetic
ratios ($\gamma_\mathrm{lo}$). Less accurate information is provided by
measurements of high-field gyromagnetic ratios, $K_J$, and the Faraday
constant $F=eN_A$, and by the QMT.}
\end{figure}

The reproducibility and universality of the quantum standards are an indication
that Eqs.~(\ref{eq:shapiro}) and (\ref{eq:qhe}) are exact, but a proof
can only be obtained by comparisons to other standards. Any one of the electric
quantities $V$, $I$ or $R$ can be tested by comparison to the other two
via Ohm's law. If all three are quantum standards, such experiment is called
quantum metrology triangle (QMT) \cite{Likharev1985}. It is a major goal in
metrology, but the insufficient performance of single-electron devices has
to date prevented the reaching of low uncertainties, see Sec.~\ref{sec:qmt}.
However, the exactness of Eqs.~(\ref{eq:shapiro}) and (\ref{eq:qhe}) can also
be studied in the
framework of the adjustment of fundamental constants. The most thorough
treatment has been performed by the Committee on Data for Science and
Technology (CODATA). Update papers are nowadays published every four years,
see \textcite{Mohr2000,Mohr2005,Mohr2008,Mohr2012}\footnote{The
adjustments are named after the deadline for the included data, e.g.,
CODATA-10 is based on experimental and theoretical results that were available
by December 31, 2010. The values of the constants and much more information
is available at the Web site physics.nist.gov/constants/.}.
\textcite{Karshenboim2009} provides a useful overview. We review here the
most accurate ($<10^{-7}$) routes to information on the electric quantum standards.
They are also illustrated in Fig.~\ref{fig:constants}.
Most of the equations in this chapter assume that
Eqs.~(\ref{eq:shapiro}) and (\ref{eq:qhe}) are exact, but when referring
to possible deviations, we describe them by symbols $\epsilon_{J,K,S}$:
\begin{equation}\label{eq:epsilon} \begin{array}{rcl}
K_J     & =   &   (1+\epsilon_J)\frac{2e}{h}\\
R_K   &  =   &   (1+\epsilon_K)\frac{h}{e^2}\\
Q_S  &  =  &  (1+\epsilon_S)e. \end{array}
\end{equation}
In this context, the current generated by the single-electron current source
is $I_S=<k_S>Q_Sf$ where $<k_S>$ is the average number of electrons transported
per cycle. Thus any determination of $\epsilon_S$ will require both a measurement
of the macroscopic current and counting the number of transported charges.

There are a number of fundamental constants that are known by much
smaller uncertainties than those related to electric metrology. Some constants,
e.g., the permeability, permittivity, and the speed of light in vacuum, and the
molar mass constant, $\mu_0$, $\epsilon_0$, $c$, and $M_u=1$~g/mol, respectively, are
fixed by the present SI. Examples of constants known with an uncertainty
$\leq 10^{-10}$ are the Rydberg constant $R_\infty$ and several relative atomic
masses, e.g., that of an electron, $A_r(e)$. In the past few years, there has been
tremendous progress in the determination of the fine structure constant $\alpha$.
First, the electron magnetic moment anomaly, $a_e$, was measured with high
accuracy. A separate calculation based on QED gives the function $\alpha(a_e)$.
Together these results yield a value on $\alpha$ with an uncertainty
$0.37\times 10^{-9}$ \cite{Hanneke2008}. Soon after, a measurement of the recoil
velocity of the rubidium atom, when it absorbs a photon, yielded a value for
$\alpha$ with an uncertainty of $0.66\times 10^{-9}$ \cite{Bouchendira2011}. These
two results are in good agreement. Together they give a validity
check for QED since the first result is completely dependent and the latter practically
independent of it.

The fine structure constant is related to $R_K$ by the exact constants
$\mu_0$ and $c$:
\begin{equation}\label{eq:alphark}
\alpha=\frac{\mu_0 c e^2}{2h}=\frac{\mu_0 c}{2R_K}.
\end{equation}
This relationship means that when $R_K$ is measured with a calculable capacitor,
it also yields an estimate for $\alpha$. Thus, a measurement of $R_K$ could also
test QED, but in practice, the atomic recoil measurement is more accurate by about
factor 30. A metrologically more important interpretation of this relation is that
a comparison between $\alpha$ and the weighted mean
of the measurements of $R_K$ yields an estimate of
$\epsilon_K=(26\pm 18)\times10^{-9}$ \cite{Mohr2012}.
There is thus no proof of a nonzero $\epsilon_K$, but several groups are developing
calculable capacitors in order to determine $R_K$ with the uncertainty below
$10^{-8}$ \cite{Poirier2009,Poirier2011}.

The existing data that yields information on $\epsilon_J$ is more discrepant. It
turns out that $\epsilon_J$ is related to measurements of gyromagnetic ratios
(see \textcite{Mohr2000} for a detailed description) and to the realization of
the kilogram \cite{Mohr2008,Mohr2012}.
The gyromagnetic ratio $\gamma$ determines the spin-flip frequency $f$ of a free
particle when it is placed in a magnetic field $B$: $\gamma=2\pi f/B$. The
gyromagnetic ratios of a helium nucleus and a proton are accessible in nuclear and
atomic magnetic resonance experiments. These ratios can be related to the
gyromagnetic ratio of an electron that is linked to $\alpha$ and $h$. There are
two methods to produce the magnetic field: In the low-field method, it is
generated by electric current in a coil and determined from the current and the
geometry. In the high-field method, the field is generated by a permanent
magnet and measured from a current induced in a coil. When the electric current
is determined in terms JVS and QHR, the product
$K_JR_K=\sqrt{2\mu_0c/(h\alpha)}$ will appear either in the numerator or
the denominator of $\gamma$, depending on which method is used. It turns out
that in the low-field method, $h$ cancels out
from the equations of the gyromagnetic ratios, and the experiment yields a
value for $\alpha$. The high-field results also depend on $\alpha$, but since it
is known much more precisely than $h$, they essentially yield a value for $h$.
The high-field results on $h$ are in good agreement with other experiments,
albeit their uncertainty is not better than about $10^{-6}$. However, the low-field
results are discrepant from the CODATA value of $\alpha$. By substituting
Eqs.~(\ref{eq:epsilon}) to the observational equations of the low-field data,
one obtains the estimate $\epsilon_K +\epsilon_J=(-254\pm 93)\times 10^{-9}$
\cite{Cadoret2011,Mohr2012}.
Since the measurements of $R_K$ yield much smaller value for
$\epsilon_K$, the gyromagnetic data seem to imply a significant \emph{negative}
$\epsilon_J$. However, as explained below a \emph{positive} $\epsilon_J$ can be
found from measurements aiming at the redefinition of the kilogram.

There are essentially two candidate methods for the future realization of the
kilogram: the watt balance and silicon sphere methods. The first, suggested
in \textcite{Kibble1975}, relates electric power to the time derivative of
potential energy:
\begin{equation}\label{eq:wattbalance}
mgv=\frac{V^2}{R}\propto \frac{1}{K_J^2R_K}=h/4.
\end{equation}
When the mass $m$, its velocity $v$ and the gravitational acceleration $g$ are
traceable to the SI, the watt balance yields a value on $h$.
Watt balance results have already been published by 
NPL, NIST, METAS, and NRC, and several devices are under development.
See \textcite{Steele2012} for the latest status and
\textcite{Eichenberger2009,Stock2011,Li2012} for reviews.
The silicon sphere approach is so demanding that it is employed only
by the international Avogadro consortium (IAC). The results were published in 2011,
see \textcite{Andreas2011, Andreas2011a} and the whole no.~2 of
\emph{Metrologia}, vol.~48 (2011). This project determines the Avogadro
constant $N_A$ by fabricating spheres of enriched $^{28}$Si whose mass is compared to
the prototype of the kilogram and whose volume is measured by laser interferometry.
The lattice parameter and the relative atomic mass of $^{28}$Si are measured
in different experiments, and the ratio of the relative and absolute mass
densities yields $N_A$. 

Results on $h$ and $N_A$ can be
compared precisely with the help of the molar Planck constant
\begin{equation}\label{eq:nah}
N_Ah=\alpha^2\frac{A_r(e)M_uc}{2R_\infty}.
\end{equation}
Its uncertainty is only $0.7\times10^{-9}$ \cite{Mohr2012} and depends mainly on those of
$A_r(e)$ and $\alpha$. This equation can be derived from the definition of
the Rydberg constant by writing the inaccurate absolute mass of the electron
in terms of its relative mass and $N_A$ which links microscopic and
macroscopic masses. The IAC 2011 result resolved the discrepancy of
$1.2\times 10^{-6}$ between watt balances and the Avogadro constant
determined from a sphere of natural Si that had puzzled metrologists since 1998
\cite{Mohr2000}.  Especially after the new NRC results \cite{Steele2012}
there is no longer any clear discrepancy between the two methods, but the
two most accurate watt balances deviate by factor $~260\times10^{-9}$
which is 3.5 times the uncertainty of their difference
\cite{Steele2012,Steiner2007}. Also the measurements of the isotope ratio
of the silicon sphere deviate more than expected \cite{Yang2012}.
Nevertheless, by combining the Planck constants obtained from watt balance
and silicon sphere experiments, $h_w$ and $h_\mathrm{Avo}$, one can obtain
an estimate of
$\epsilon_J+\epsilon_K/2\simeq (h_\mathrm{Avo}/h_w - 1)/2=(77\pm 18)\times10^{-9}$.
Here, we neglected correlations between experiments. More detailed analysis on
the existence of $\epsilon_{J,K}$ can be found from the CODATA papers
\textcite{Mohr2008,Mohr2012}, see also \textcite{Keller2008}. They executed
the least-squares analysis of fundamental constants several times allowing
either nonzero $\epsilon_K$ or $\epsilon_J$, and including only part of the
data. When they excluded the most accurate but discrepant data, the remaining less precise
but consistent data yielded the conservative estimates $\epsilon_K=(28\pm18)\times10^{-9}$ and
$\epsilon_J=(150\pm490)\times10^{-9}$. Thus the exactness of the
quantum Hall effect is confirmed much better than that of the Josephson effect.

\subsection{The future SI}
\label{sec:si}

Modernizing the SI towards a system based on fundamental constants or
other true invariants of nature has long been a major goal, tracing
back to a proposal by Maxwell in the 19th century, see, e.g.,
\textcite{Flowers2004} and references therein. Atomic clocks and laser
interferometry permitted such a revision of the second and of the meter.
The development of quantum electric standards, watt balance experiments,
the Avogadro project, and measurements of the Boltzmann constant have made
the reform of the ampere, kilogram, mole, and kelvin realistic in the near future.
Specially, suggestions in \textcite{Mills2005} launched an active debate among 
metrologists \cite{Mills2006,Becker2007,Milton2007}.
Soon it was agreed that the SI should not be altered before there are at least
three independent experiments (both from watt balance and Avogadro constant)
with uncertainties $\leq50\times10^{-9}$ that are consistent within the 95\%
confidence intervals, and at least one of them has the uncertainty
$\leq 20\times10^{-9}$ \cite{Glaser2010}. There have also been requests to await
better results from single-electron and QMT experiments
\cite{Borde2005,Milton2007}, and to solve the discrepancy of low-field gyromagnetic
experiments \cite{Cadoret2011}. 

There is already a draft chapter for the SI brochure that would adopt the
new definitions: \textcite{BIPM2010}, see also the whole issue 1953 in
Phil. Trans. Royal Soc. A 369 (2011), especially \textcite{Mills2011}.
In this draft, the whole system of units is scaled by a single sentence that
fixes seven constants. The most substantial changes are that the base units
ampere, kilogram, mole, and kelvin are defined by fixed values of $e$,
$h$, $N_A$, and $k_B$, respectively. The new definition for the ampere would
read: "The ampere, A, is the unit of electric current; its magnitude is set by fixing
the numerical value of the elementary charge to be equal to exactly
$1.60217X\times10^{-19}$ when it is expressed in the unit s A, which is
equal to C." The new definitions will not imply any particular methods for the
realizations of the units. They will be guided by \emph{mises en pratique}, e.g.,
the ampere could be realized with the help of JVS and QHR \cite{ccem2012}.

The new SI would significantly lower the uncertainties of many fundamental
constants, see e.g., \textcite{Mills2011} for evaluations. One should note,
however, that choosing the optimal set of fixed constants is always a tradeoff.
For example, since $\alpha$ is a dimensionless number and thus independent of
the choice of units, one can see from Eq.~(\ref{eq:alphark}) that
fixing $e$ and $h$ would make $\mu_0$ (and $\epsilon_0$) a quantity
that is determined by a measurement of $\alpha$. Presently, $\mu_0$ and
$\epsilon_0$ are fixed by the definition of ampere. However, their uncertainty
would be very low, the same as that of $\alpha$, which is $3.2\times10^{-10}$
\cite{Mohr2012}. One alternative suggestion is to fix $h$ and the Planck charge
$q_p=\sqrt{2\epsilon_0hc}$, which would keep $\mu_0$ and $\epsilon_0$
exact \cite{Stock2006}. It is also worth noting that out of $h$, $N_A$, and the
molar mass of carbon 12, $M(^{12}\mathrm{C})=A_r(^{12}\mathrm{C})M_u$,
only two can be fixed. The suggested SI would release the equality
$M(^{12}\mathrm{C})=0.012$~kg/mol, which has raised criticism. Especially,
there have been claims that the definition of the kilogram based on $h$
would not be understandable for the wider audience, and a definition based
on the mass of a number of elementary particles would be better in this respect
\cite{Becker2007,Milton2007,Leonard2010,Hill2011}. \textcite{Milton2010}
studies two alternatives, fixing either $N_A$ and $h$ or $N_A$ and the atomic
mass constant $m_u=M_u/N_A$, and shows that this choice has little effect on
the uncertainties of fundamental constants, mainly because the ratio $h/m_u$
is known well from the atomic recoil experiments.

\subsection{Quantum metrology triangle}
\label{sec:qmt}

Phase-locked Bloch \cite{Averin1985} and SET \cite{Averin1986} oscillations
in superconducting and normal-state tunnel junctions, respectively, were proposed
as a source of quantized electric current in the mid 1980s, soon after the
discovery of the QHR. Already \textcite{Likharev1985} suggested that the
quantum current standard could provide a consistency check for the existing two
electric quantum standards in an experiment they named as "quantum metrology
triangle". However, the quantized current
turned out to be a much greater challenge than JVS and QHR. Still, after a quarter
of a century, quantum current standards are yet to take their place in metrology.
On the other hand, the progress on the knowledge on $K_J$ and
$R_K$ has also been rather slow: in CODATA-86 the  uncertainties were
$300\times10^{-9}$ and $45\times10^{-9}$, respectively \cite{Cohen1987}.
These uncertainties are essentially on the same level as in CODATA-10 if the
discrepancy of the data is taken into account.

The QMT experiment and its impact has been discussed, e.g., in
\textcite{Piquemal2000,Zimmerman2003,Piquemal2004,Gallop2005,Keller2008,Keller2008a,Feltin2009,Scherer2012,Piquemal2012}.
There are essentially two strategies to close the QMT: by applying the Ohm's law
$V=RI$ or by so-called electron counting capacitance standard (ECCS) which utilizes
the definition of capacitance $C=Q/V$. They are sometimes called direct and
indirect QMT, respectively. Regardless of the variant, QMT compares the macroscopic 
current generated by a single-electron source to the other standards.

\subsubsection{Triangle by Ohm's law}
\label{sec:directqmt}

Applying the Ohm's law is the most obvious way to compare the three quantum
electric standards. It can be realized either as a voltage balance $V_J-R_HI_S$
or as a current balance $V_J/R_H-I_S$. In both cases, substituting
Eqs.~(\ref{eq:epsilon}) to $V_J=R_HI_S$ yields
\begin{equation}\label{eq:triangleresult}
\frac{n_Ji_K}{2k_S}\times\frac{f_J}{f_S}\simeq 1+\epsilon_J+\epsilon_K+\epsilon_S.
\end{equation}

The major difficulty in the realization of the QMT can be outlined as follows. Let
us consider the ideal case where the noise of the experiment is dominated by the
Johnson noise of the resistor. The relative standard deviation of the measurement
result is
\begin{equation}\label{eq:dI}
\frac{\delta I_S}{I_S}=\sqrt{\frac{4k_BT}{tRI_S^2}}.
\end{equation}
By substituting realistic estimates $t=24$~h and $T=100$~mK for the
averaging time and the temperature of the resistor, respectively, and by assuming
that $R=R_K/2$ and $I_S=100$~pA, one obtains the  uncertainty
$\delta I_S/I_S\approx 7\times 10^{-7}$. In practical experiments, $1/f$ noise
and noise of the null detection circuit make the measurement even more demanding,
but this simple model demonstrates that the magnitude of the current should be
at least 100~pA.

\begin{figure}[t]
    \begin{center}
    \includegraphics[width=.5\textwidth]{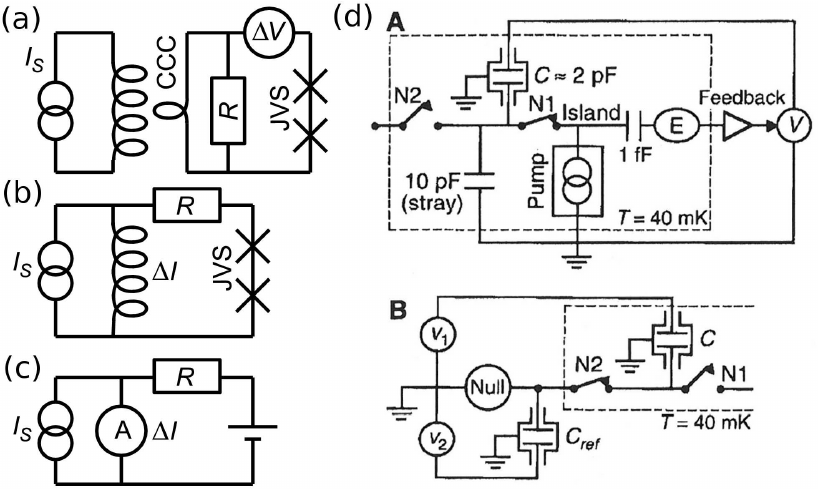}
    \end{center}
   \caption{\label{fig:triangles} (a-c) Variants of Ohm's law triangles where the
quantized current ($I_S$) is compared to resistance ($R$) calibrated against QHR
and to JVS. (a) Quantized current is magnified by a CCC, which allows room temperature
null detection of the voltage difference ($\Delta V$). (b) Triangle with a
high-value cryogenic resistor. The current balance $\Delta I$ can be determined,
e.g., with the help of a CCC. (c) QMT experiment where the null detection is performed 
by a room temperature transimpedance amplifier.
(d) ECCS experiment. In the first phase (A), the electron pump charges the 
cryocapacitor $C\approx 2$~pF. 
An SET electrometer (E) is used to generate a feedback voltage ($V$) that maintains the 
potential of the island at zero. Hence all the charge is accumulated to the
cryocapacitor and not to the stray capacitance. The feedback voltage constitutes
the third part of the indirect QMT $Q=CV$. In the second phase (B), the
cryocapacitor is calibrated against the reference $C_\mathrm{ref}$ which is
traceable to a calculable capacitor. From \textcite{Keller1999}.}
\end{figure}

Another problem is that the product $R_HI_S$ yields a very small voltage,
e.g., $12.9$~k$\Omega\times 100$~pA~$=1.29$~$\mu$V.\footnote{A quantum
voltage standard based on integrating a semiconducting pump with QHR has
been pioneered in \textcite{Hohls2011}.}
Even the voltage of a JVS with only one junction is typically of the order of 
70~GHz$/K_J\approx 140~\mu V$. Such low voltages are also vulnerable to
thermoelectric effects. One way to overcome this problem is to multiply the small
current of the SET by a CCC with a very high winding ratio, $\sim$10,000, as was
suggested in \textcite{Hartland1991,Sese1999,Piquemal2000}, see
Fig.~\ref{fig:triangles}~(a). It allows room-temperature detection, and that JVS,
SET, and QHR can be operated in different refrigerators. This type of efforts have
been described in \textcite{Piquemal2004,Feltin2009,Feltin2011}. Another approach
is to use a high-value cryogenic resistor that is calibrated against
QHR with the help of a CCC \cite{Elmquist2003,Manninen2008}. All parts
of the Ohm's law are in the same cryostat which can reduce thermoelectric effects,
see Fig.~\ref{fig:triangles}(b). Only the difference current $V_J/R_\mathrm{cryo}-I_S$
needs to be multiplied. Despite of persistent efforts, these approaches have not yet
produced any significant results. The best result from the Ohm's law triangle so far
were obtained recently in \textcite{Giblin2012} where a CCC with high winding ratio
was used to calibrate a precision 1~G$\Omega$ room temperature resistor, see 
Fig.~\ref{fig:triangles}(c). This experiment benefited from the relatively large
current of 150~pA that was generated by a semiconducting quantum dot
pump. The uncertainty of the QMT experiment was $1.2\times 10^{-6}$.

\subsubsection{Electron counting capacitance standard}
\label{sec:indirectqmt}

The ECCS experiment was first suggested in \textcite{Williams1992}.
A single-electron current source is used to charge a cryogenic capacitor
$C_\mathrm{cryo}$ by a known number $N_S$ of electrons. The generated voltage
is compared to JVS. The result
\begin{equation}\label{eq:eccs}
C_\mathrm{cryo}=\frac{N_SQ_S}{V}
\end{equation}
thus yields a quantum capacitance standard. The ECCS experiment was pioneered in
\textcite{Keller1999}, see Fig.~\ref{fig:triangles}(d), where an uncertainty of
$0.3\times10^{-6}$ was obtained for the ECCS capacitance. In this approach, ECCS was compared to a calculable capacitor. Then
the observational equation corresponding to  Eq.~(\ref{eq:triangleresult}),
\begin{equation}\label{eq:eccsresult}
\frac{\mu_0cn_Jf_JC}{4\alpha N_S}=1+\epsilon_J+\epsilon_S,
\end{equation}
does not include $\epsilon_K$. However, calculable capacitors have been compared
to QHR with very low uncertainty, and ac QHR techniques
\cite{Schurr2011,Camarota2012} allow
$C_\mathrm{cryo}$ to be compared directly against $R_K$. One should thus obtain
an uncertainty of $\sim{}10^{-8}$ before there is any significant difference
between the implications of the two QMT versions.

A major weakness of the ECCS is that it calibrates $C_\mathrm{cryo}$ at 
$\sim 0.01$~Hz, but commercial capacitance bridges that are used to
compare $C_\mathrm{cryo}$ to the calculable capacitor (and also ac QHR)
operate at $\sim 1000$~Hz. \textcite{Zimmerman2006} presents a model for the
dielectric dispersion of insulating films at the surface of the electrodes of the 
capacitor. They fit this model to measurements of the frequency dependence and its
temperature dependence in the range 100--3000~Hz and 4--300~K. The frequency
dependence decreases at low temperatures. They evaluate that it yields an
uncertainty component of $2\times10^{-7}$ for the QMT. In \textcite{Keller2007}
this estimate was used to finish the uncertainty budget of the NIST ECCS experiment
that closes the QMT at the uncertainty of $0.9\times10^{-6}$. 

Very recently, PTB reached the uncertainty of $1.7\times10^{-6}$ in the ECCS
experiment \cite{Camarota2012}. They used a five-junction R-pump
instead of the seven-junction pump without resistors employed by NIST, but
otherwise the experiments are quite similar. PTB presents their result as "preliminary"
and plan both a more detailed uncertainty budget and several improvements to the
experiment. Besides NIST and PTB, the ECCS has been pursued at
METAS \cite{Rufenacht2010}.

\subsubsection{Metrological implications of single-electron transport and QMT}
\label{sec:setmetrology}

It is clear that when a quantum current standard will be available, it must be
compared against other standards in the QMT. One should, however, be conservative
when interpreting the results. The single-electron sources are proven much more
sensitive to failures than JVS or QHR, and unless the pumping errors are detected 
separately, QMT should only be considered as a way to characterize the current
source. This is the case for \textcite{Giblin2012}. Only the combination of
error counting and QMT allows to separate $\left< k_S \right>$ from $\epsilon_{J,K,S}$. So far,
it has been achieved with a reasonable uncertainty ($\sim 10^{-6}$) only in the ECCS
experiments of NIST \cite{Keller2007} and PTB \cite{Camarota2012}.

As shown in Sec.~\ref{sec:universality}, an uncertainty of $\lesssim 2\times 10^{-8}$
is required to give information on $\epsilon_K$, and an uncertainty of the order of
$10^{-7}$ would strengthen the knowledge on $\epsilon_J$. Thus, the NIST and PTB
results, $\epsilon_S=(-0.10\pm0.92)\times10^{-6}$ and
$\epsilon_S=(-0.3\pm1.7)\times10^{-6}$, respectively, only measure $\epsilon_J$
\cite{Keller2008}.

\textcite{Milton2010} analyze a scenario
where $\epsilon_J$ is an adjusted parameter and $\epsilon_S=\epsilon_K=0$.
They study the effect of QMT on the uncertainties of fundamental constants.
It turns out that when QMT is inaccurate, the uncertainties of $h$, $e$ and
$m_u$ are mainly determined by the Avogadro experiment, but their
uncertainties will be dominated by those of the watt balance and
the direct measurement of $R_K$ when QMT is improved.

One problem of the QMT is that it only gives a value for the sum of the
errors of the quantum standards, and in principle, they could cancel each
other. It is thus useful to have independent test for each standard, and those
for JVS and QHR are discussed in Sec.~\ref{sec:universality}.
A test for the current standard only, i.e., an SI value for $Q_S$, can be obtained
by combining results from three experiments: QMT, measurement of
$R_K$ by a calculable capacitor, and watt balance \cite{Keller2008a}.
Applying Eqs.~(\ref{eq:wattbalance}) and (\ref{eq:eccs}), and
substituting $R$ by $1/(\omega C)$, one obtains
\begin{equation}\label{eq:qssi}
Q_S=\frac{1}{N_S}\sqrt{\frac{mgvC}{\omega}}.
\end{equation}
Also the Ohm's law triangle can be used to yield a similar result, but
in a less direct way. One should note, that JVS and QHR are used here only
as transfer standards. \textcite{Keller2008a} derives a result based on the
NIST ECCS: $Q_S=1.6021763\times 10^{-19}\pm 1.5\times10^{-25}$~C. This could be
compared to the CODATA value on $e$, which, however, depends strongly
on $h$ and on the exactness of $K_J$ and $R_K$. Instead, $Q_S$ can be
compared to another value of $e$ that is also independent of JVS and QHR,
$e=\sqrt{\alpha^3 A_r(e)M_u/(\mu_0 R_\infty N_A)}$ \cite{Feltin2009},
whose uncertainty $\sim 1.5\times 10^{-8}$ is dominated by that of
$N_A$. Using the NIST ECCS result and the IAC value on $N_A$, one
obtains $\epsilon_S=(-0.2\pm 0.9)\times 10^{-6}$.

We note that according to Eq.~(\ref{eq:eccsresult}), QMT also yields a value for
$\alpha$ independently of QHR, which was one of the early motivations for ECCS
\cite{Williams1992}. This fact, however, has little importance before the
uncertainty is competitive with the atomic recoil experiments
($<10^{-9})$. Then QMT would strengthen the verification of QED.

Although single-electron transport would be the conceptually most straightforward 
realization of the ampere in the future SI, it is not likely that it would replace
JVS and QHR as the typical realization in the near future. The exception is naturally
the growing field of metrology for small electric currents, where single electronics
is expected to yield major improvements of accuracy. On the other hand, when the
accuracy of single-electron transport accuracy improves, it can yield vital information 
on other standards and on fundamental constants.

%
% Chapter 5
%

\section{PERSPECTIVES AND OTHER APPLICATIONS}
\label{sec:perspectives}

The quantum dot pump \cite{Kaestner2008,Giblin2012} discussed in Sec.~\ref{sec:dotpumps} has definitely proven its potential to be the basis of the future quantum standard of ampere. The verified uncertainty of the 150~pA output current on the level of 1~ppm and the theoretically predicted 0.01~ppm uncertainty of the present device are truly remarkable figures of merit. On the other hand, a few important questions remain to be answered before one would realize the ampere with the quantum dot pump: The superior device performance depends critically on applying a strong $\gtrsim 10$~T magnetic field on it. This dependence is not fully understood, and the exact magnetic field characteristics seem sample dependent. The reproducibility of the highly accurate pumping results with samples from different fabrication runs remains to be shown. Importantly, error counting experiments on the dot samples have not been carried out, which also prevents
to study possible errors of other quantum standards in the QMT. Future experiments will likely show whether all the relevant error processes have been accounted for in predicting the obtainable accuracy to be on the level of $10^{-8}$. However, even if not in the case of a bare device, the quantum dot pump may perhaps be applicable to the realization of ampere, if the error correction techniques that were described in Sec.~\ref{sec:readout} become feasible experimentally.

Another important development and potential future realization of ampere is the SINIS turnstile introduced in Sec.~\ref{sec:hybridpumps}. Although presently inferior to the quantum dot pump in the level of current output, and consequently with less definite assessment of proven accuracy (present verified uncertainty below $10^{-4}$), this device does not suffer from known obstacles on the way of achieving the required accuracy. Currently, the main error mechanisms have been assessed theoretically and experimentally, including photon assisted tunneling, Andreev current, co-tunneling, residual and generated quasiparticles, and possible residual density of states in a superconductor. Positive conclusions can be drawn from individual experiments with respect to suppressing them in an optimized device. Sample fabrication and reproducibility is currently on a high level, and it has been demonstrated that the requested magnitude of current can be achieved by running many turnstiles in parallel. For the SINIS turnstile, as for the quantum dot pump, the ultimate test would be an error counting experiment and the quantum metrological triangle. At the time of writing this review, such experiments have not been performed for either of these promising new single-electron sources.

Presently several other new proposals are being pushed towards critical tests to study their applicability in current metrology: these include superconducting phase-slip wires, Josephson junction arrays, and mechanical shuttles, just to mention a few less conventional ideas. Although it is not in the horizon at present, it is possible that eventually one of these devices beats the present Coulomb-blockade-based realizations both in current yield and in their robustness against transfer errors.

Developing ever more accurate current sources has constantly been a driving force for understanding the underlying physical phenomena. On the other hand, the studies for the precise control of single electrons and Cooper pairs have created special expertise that is also applicable in variety of other research topics.

In addition to the charge degree of freedom, the electrons hold information in their spin states which have been envisioned \cite{Kane1998,Hollenberg2006} to be utilized~\cite{Morello2010} for quantum information processing. Although the electron transport is typically incoherent in the electron pumps, the spin-encoded information can potentially remain coherent, and hence this information can possibly be transported from the memory cell of the computer to the qubit--qubit interaction cell and back. The transport cycle has to be carried out with high accuracy for fault tolerant computing to be possible, which creates a close connection to the metrological electron pumps.

Geometric phases~\cite{Shapere1989} in quantum mechanics have been studied extensively due to both fundamental scientific curiosity and their applications in geometric quantum computing~\cite{Zanardi1999}. The simplest geometric quantum phase, the Berry, phase has already been measured in the superconducting sluice pump~\cite{Mottonen2006,Mottonen2008} thanks to the development of the sluice for metrology. Some theoretical work on the more complex phases referred to as holonomies has been put forward in the framework of Cooper pair pumps~\cite{Pirkkalainen2010,Solinas2010} but it remains to be seen if these ideas will be implemented experimentally. The main obstacle in practice is perhaps the high level of precision required for the control signals of the pumps, a problem that can possibly be solved with the help of the work on the metrological current source.

Detecting single-electrons and Cooper pairs by single-electron transistors and quantum point contacts has been largely motivated by the need for tests of the charge-transport errors in metrology. During the past decade, these techniques have also been successfully implemented, e.g., in experiments on full counting statistics and noise of charge transport. The experiments on the full counting statistics of current fluctuations in a semiconductor quantum dot by real-time detection of single electron tunneling with a quantum point contact have been successfully performed for instance by Gustavsson et al. \cite{Gustavsson2006,Gustavsson2007}. In these experiments, moments of current up to the fifth one and beyond could be reliably measured. Very recently, single-charge counting experiments have been applied to study energy fluctuation relations~\cite{Evans1993,Jarzynski1997,Crooks1999,Averin2011} in statistical mechanics. The experiments in steady-state non-equilibrium were performed by K\"ung et al. \cite{Kung2012}, and the celebrated Jarzynski and Crooks relations were very recently tested by Saira et al. \cite{Saira2012b}. Single-charge counting experiments allow one to test fundamental statistical mechanics and thermodynamics of classical and quantum systems.

The variety of spin-offs out of the development of single-charge current sources for metrology is certainly expanding. This way the benefits of this research will be obvious not only for the community interested in the system of units and in traceable measurements, but also for other researchers working in basic and applied sciences looking for new tools for measurements that need precise control.

\hspace{10pt}

We thank Antti Manninen, Matthias Meschke, Alexander Savin, Mika Prunnila, Juha Hassel, Heikki Sepp\"a, Panu Helist\"o, Jaw Shen Tsai  and Simone Gasparinetti for useful discussions, and Academy of Finland through its Centre of Excellence Programs, MEXT
Grant-in-Aid "Quantum Cybernetics" and FIRST Project from JSPS and the Finnish National
Graduate School in Nanoscience for financial support during the preparation of this manuscript.

%
% The end
%

\bibliographystyle{apsrmp}
\bibliography{Rmp}

\begin{thebibliography}{387}
\expandafter\ifx\csname natexlab\endcsname\relax\def\natexlab#1{#1}\fi
\expandafter\ifx\csname bibnamefont\endcsname\relax
  \def\bibnamefont#1{#1}\fi
\expandafter\ifx\csname bibfnamefont\endcsname\relax
  \def\bibfnamefont#1{#1}\fi
\expandafter\ifx\csname citenamefont\endcsname\relax
  \def\citenamefont#1{#1}\fi
\expandafter\ifx\csname url\endcsname\relax
  \def\url#1{\texttt{#1}}\fi
\expandafter\ifx\csname urlprefix\endcsname\relax\def\urlprefix{URL }\fi
\providecommand{\bibinfo}[2]{#2}
\providecommand{\eprint}[2][]{\url{#2}}

\bibitem[{\citenamefont{Aguado and Kouwenhoven}(2000)}]{Aguado2000}
\bibinfo{author}{\bibnamefont{Aguado}, \bibfnamefont{R.}}, and
  \bibinfo{author}{\bibfnamefont{L.}~\bibnamefont{Kouwenhoven}},
  \bibinfo{year}{2000}, \bibinfo{journal}{Phys. Rev. Lett.}
  \textbf{\bibinfo{volume}{84}}, \bibinfo{pages}{1986}.

\bibitem[{\citenamefont{Altebaeumer and Ahmed}(2001)}]{Altebaeumer2001_2}
\bibinfo{author}{\bibnamefont{Altebaeumer}, \bibfnamefont{T.}}, and
  \bibinfo{author}{\bibfnamefont{H.}~\bibnamefont{Ahmed}},
  \bibinfo{year}{2001}, \bibinfo{journal}{Jpn. J. Appl. Phys.}
  \textbf{\bibinfo{volume}{40}}, \bibinfo{pages}{80}.

\bibitem[{\citenamefont{Altebaeumer}
  \emph{et~al.}(2001)\citenamefont{Altebaeumer, Amakawa, and
  Ahmed}}]{Altebaeumer2001}
\bibinfo{author}{\bibnamefont{Altebaeumer}, \bibfnamefont{T.}},
  \bibinfo{author}{\bibfnamefont{S.}~\bibnamefont{Amakawa}}, and
  \bibinfo{author}{\bibfnamefont{H.}~\bibnamefont{Ahmed}},
  \bibinfo{year}{2001}, \bibinfo{journal}{Appl. Phys. Lett.}
  \textbf{\bibinfo{volume}{79}}, \bibinfo{pages}{533}.

\bibitem[{\citenamefont{Ambegaokar and Baratoff}(1963)}]{Ambegaokar1963}
\bibinfo{author}{\bibnamefont{Ambegaokar}, \bibfnamefont{V.}}, and
  \bibinfo{author}{\bibfnamefont{A.}~\bibnamefont{Baratoff}},
  \bibinfo{year}{1963}, \bibinfo{journal}{Phys. Rev. Lett.}
  \textbf{\bibinfo{volume}{10}}, \bibinfo{pages}{486}.

\bibitem[{\citenamefont{Ando} \emph{et~al.}(1982)\citenamefont{Ando, Fowler,
  and Stern}}]{Ando1982}
\bibinfo{author}{\bibnamefont{Ando}, \bibfnamefont{T.}},
  \bibinfo{author}{\bibfnamefont{A.~B.} \bibnamefont{Fowler}}, and
  \bibinfo{author}{\bibfnamefont{F.}~\bibnamefont{Stern}},
  \bibinfo{year}{1982}, \bibinfo{journal}{Rev. Mod. Phys.}
  \textbf{\bibinfo{volume}{54}}, \bibinfo{pages}{437}.

\bibitem[{\citenamefont{Andreas}
  \emph{et~al.}(2011{\natexlab{a}})\citenamefont{Andreas, Azuma, Bartl, Becker,
  Bettin, Borys, Busch, Fuchs, Fujii, Fujimoto, Kessler, Krumrey}
  \emph{et~al.}}]{Andreas2011a}
\bibinfo{author}{\bibnamefont{Andreas}, \bibfnamefont{B.}},
  \bibinfo{author}{\bibfnamefont{Y.}~\bibnamefont{Azuma}},
  \bibinfo{author}{\bibfnamefont{G.}~\bibnamefont{Bartl}},
  \bibinfo{author}{\bibfnamefont{P.}~\bibnamefont{Becker}},
  \bibinfo{author}{\bibfnamefont{H.}~\bibnamefont{Bettin}},
  \bibinfo{author}{\bibfnamefont{M.}~\bibnamefont{Borys}},
  \bibinfo{author}{\bibfnamefont{I.}~\bibnamefont{Busch}},
  \bibinfo{author}{\bibfnamefont{P.}~\bibnamefont{Fuchs}},
  \bibinfo{author}{\bibfnamefont{K.}~\bibnamefont{Fujii}},
  \bibinfo{author}{\bibfnamefont{H.}~\bibnamefont{Fujimoto}},
  \bibinfo{author}{\bibfnamefont{E.}~\bibnamefont{Kessler}},
  \bibinfo{author}{\bibfnamefont{M.}~\bibnamefont{Krumrey}}, \emph{et~al.},
  \bibinfo{year}{2011}{\natexlab{a}}, \bibinfo{journal}{Metrologia}
  \textbf{\bibinfo{volume}{48}}, \bibinfo{pages}{S1}.

\bibitem[{\citenamefont{Andreas}
  \emph{et~al.}(2011{\natexlab{b}})\citenamefont{Andreas, Azuma, Bartl, Becker,
  Bettin, Borys, Busch, Gray, Fuchs, Fujii, Fujimoto, Kessler}
  \emph{et~al.}}]{Andreas2011}
\bibinfo{author}{\bibnamefont{Andreas}, \bibfnamefont{B.}},
  \bibinfo{author}{\bibfnamefont{Y.}~\bibnamefont{Azuma}},
  \bibinfo{author}{\bibfnamefont{G.}~\bibnamefont{Bartl}},
  \bibinfo{author}{\bibfnamefont{P.}~\bibnamefont{Becker}},
  \bibinfo{author}{\bibfnamefont{H.}~\bibnamefont{Bettin}},
  \bibinfo{author}{\bibfnamefont{M.}~\bibnamefont{Borys}},
  \bibinfo{author}{\bibfnamefont{I.}~\bibnamefont{Busch}},
  \bibinfo{author}{\bibfnamefont{M.}~\bibnamefont{Gray}},
  \bibinfo{author}{\bibfnamefont{P.}~\bibnamefont{Fuchs}},
  \bibinfo{author}{\bibfnamefont{K.}~\bibnamefont{Fujii}},
  \bibinfo{author}{\bibfnamefont{H.}~\bibnamefont{Fujimoto}},
  \bibinfo{author}{\bibfnamefont{E.}~\bibnamefont{Kessler}}, \emph{et~al.},
  \bibinfo{year}{2011}{\natexlab{b}}, \bibinfo{journal}{Phys. Rev. Lett.}
  \textbf{\bibinfo{volume}{106}}, \bibinfo{pages}{030801}.

\bibitem[{\citenamefont{Andreev}(1964)}]{Andreev1964}
\bibinfo{author}{\bibnamefont{Andreev}, \bibfnamefont{A.~F.}},
  \bibinfo{year}{1964}, \bibinfo{journal}{Sov. Phys. JETP}
  \textbf{\bibinfo{volume}{19}}, \bibinfo{pages}{1228}.

\bibitem[{\citenamefont{Aref} \emph{et~al.}(2011)\citenamefont{Aref, Maisi,
  Gustafsson, Delsing, and Pekola}}]{Aref2011}
\bibinfo{author}{\bibnamefont{Aref}, \bibfnamefont{T.}},
  \bibinfo{author}{\bibfnamefont{V.~F.} \bibnamefont{Maisi}},
  \bibinfo{author}{\bibfnamefont{M.~V.} \bibnamefont{Gustafsson}},
  \bibinfo{author}{\bibfnamefont{P.}~\bibnamefont{Delsing}}, and
  \bibinfo{author}{\bibfnamefont{J.~P.} \bibnamefont{Pekola}},
  \bibinfo{year}{2011}, \bibinfo{journal}{Europhys. Lett.}
  \textbf{\bibinfo{volume}{96}}, \bibinfo{pages}{37008}.

\bibitem[{\citenamefont{Armour and MacKinnon}(2002)}]{Armour2002}
\bibinfo{author}{\bibnamefont{Armour}, \bibfnamefont{A.~D.}}, and
  \bibinfo{author}{\bibfnamefont{A.}~\bibnamefont{MacKinnon}},
  \bibinfo{year}{2002}, \bibinfo{journal}{Phys. Rev. B}
  \textbf{\bibinfo{volume}{66}}, \bibinfo{pages}{035333}.

\bibitem[{\citenamefont{Arutyunov} \emph{et~al.}(2008)\citenamefont{Arutyunov,
  Golubev, and Zaikin}}]{Arutyunov2008}
\bibinfo{author}{\bibnamefont{Arutyunov}, \bibfnamefont{K.~Y.}},
  \bibinfo{author}{\bibfnamefont{D.~S.} \bibnamefont{Golubev}}, and
  \bibinfo{author}{\bibfnamefont{A.~D.} \bibnamefont{Zaikin}},
  \bibinfo{year}{2008}, \bibinfo{journal}{Phys. Rep.}
  \textbf{\bibinfo{volume}{464}}, \bibinfo{pages}{1}.

\bibitem[{\citenamefont{Ashoori} \emph{et~al.}(1993)\citenamefont{Ashoori,
  Stormer, Weiner, Pfeiffer, Baldwin, and West}}]{Ashoori1993}
\bibinfo{author}{\bibnamefont{Ashoori}, \bibfnamefont{R.~C.}},
  \bibinfo{author}{\bibfnamefont{H.~L.} \bibnamefont{Stormer}},
  \bibinfo{author}{\bibfnamefont{J.~S.} \bibnamefont{Weiner}},
  \bibinfo{author}{\bibfnamefont{L.~N.} \bibnamefont{Pfeiffer}},
  \bibinfo{author}{\bibfnamefont{K.~W.} \bibnamefont{Baldwin}}, and
  \bibinfo{author}{\bibfnamefont{K.~W.} \bibnamefont{West}},
  \bibinfo{year}{1993}, \bibinfo{journal}{Phys. Rev. Lett.}
  \textbf{\bibinfo{volume}{71}}, \bibinfo{pages}{613}.

\bibitem[{\citenamefont{Astafiev} \emph{et~al.}(2012)\citenamefont{Astafiev,
  Ioffe, Kafanov, Pashkin, Arutyunov, Shahar, Cohen, and Tsai}}]{Astafiev2012}
\bibinfo{author}{\bibnamefont{Astafiev}, \bibfnamefont{O.~V.}},
  \bibinfo{author}{\bibfnamefont{L.~B.} \bibnamefont{Ioffe}},
  \bibinfo{author}{\bibfnamefont{S.}~\bibnamefont{Kafanov}},
  \bibinfo{author}{\bibfnamefont{Y.~A.} \bibnamefont{Pashkin}},
  \bibinfo{author}{\bibfnamefont{K.~Y.} \bibnamefont{Arutyunov}},
  \bibinfo{author}{\bibfnamefont{D.}~\bibnamefont{Shahar}},
  \bibinfo{author}{\bibfnamefont{O.}~\bibnamefont{Cohen}}, and
  \bibinfo{author}{\bibfnamefont{J.~S.} \bibnamefont{Tsai}},
  \bibinfo{year}{2012}, \bibinfo{journal}{Nature}
  \textbf{\bibinfo{volume}{484}}, \bibinfo{pages}{355}.

\bibitem[{\citenamefont{Aumentado} \emph{et~al.}(2003)\citenamefont{Aumentado,
  Keller, and Martinis}}]{Aumentado2003}
\bibinfo{author}{\bibnamefont{Aumentado}, \bibfnamefont{J.}},
  \bibinfo{author}{\bibfnamefont{M.~W.} \bibnamefont{Keller}}, and
  \bibinfo{author}{\bibfnamefont{J.~M.} \bibnamefont{Martinis}},
  \bibinfo{year}{2003}, \bibinfo{journal}{Physica E}
  \textbf{\bibinfo{volume}{18}}, \bibinfo{pages}{37}.

\bibitem[{\citenamefont{Aunola and Toppari}(2003)}]{Aunola2003}
\bibinfo{author}{\bibnamefont{Aunola}, \bibfnamefont{M.}}, and
  \bibinfo{author}{\bibfnamefont{J.~J.} \bibnamefont{Toppari}},
  \bibinfo{year}{2003}, \bibinfo{journal}{Phys. Rev. B}
  \textbf{\bibinfo{volume}{68}}, \bibinfo{pages}{020502}.

\bibitem[{\citenamefont{Averin and Bardas}(1995)}]{Averin1995}
\bibinfo{author}{\bibnamefont{Averin}, \bibfnamefont{D.}}, and
  \bibinfo{author}{\bibfnamefont{A.}~\bibnamefont{Bardas}},
  \bibinfo{year}{1995}, \bibinfo{journal}{Phys. Rev. B}
  \textbf{\bibinfo{volume}{52}}, \bibinfo{pages}{12873}.

\bibitem[{\citenamefont{Averin}(2000)}]{Averin2000}
\bibinfo{author}{\bibnamefont{Averin}, \bibfnamefont{D.~V.}},
  \bibinfo{year}{2000}, \bibinfo{journal}{Fortschr. Phys.}
  \textbf{\bibinfo{volume}{48}}, \bibinfo{pages}{1055}.

\bibitem[{\citenamefont{Averin}(2001)}]{Averin2001}
\bibinfo{author}{\bibnamefont{Averin}, \bibfnamefont{D.~V.}},
  \bibinfo{year}{2001}, \emph{\bibinfo{title}{Macroscopic quantum coherence and
  quantum computing}} (\bibinfo{publisher}{Kluwer}), p. \bibinfo{pages}{399}.

\bibitem[{\citenamefont{Averin}(2003)}]{Averin2003}
\bibinfo{author}{\bibnamefont{Averin}, \bibfnamefont{D.~V.}},
  \bibinfo{year}{2003}, \emph{\bibinfo{title}{Quantum noise in mesoscopic
  physics}} (\bibinfo{publisher}{Kluwer}), p. \bibinfo{pages}{229}.

\bibitem[{\citenamefont{Averin} \emph{et~al.}(2008)\citenamefont{Averin,
  Bergeman, Hosur, and Bruder}}]{Averin2008b}
\bibinfo{author}{\bibnamefont{Averin}, \bibfnamefont{D.~V.}},
  \bibinfo{author}{\bibfnamefont{T.}~\bibnamefont{Bergeman}},
  \bibinfo{author}{\bibfnamefont{P.~R.} \bibnamefont{Hosur}}, and
  \bibinfo{author}{\bibfnamefont{C.}~\bibnamefont{Bruder}},
  \bibinfo{year}{2008}, \bibinfo{journal}{Phys. Rev. A}
  \textbf{\bibinfo{volume}{78}}, \bibinfo{pages}{031601}.

\bibitem[{\citenamefont{Averin and Likharev}(1986)}]{Averin1986}
\bibinfo{author}{\bibnamefont{Averin}, \bibfnamefont{D.~V.}}, and
  \bibinfo{author}{\bibfnamefont{K.~K.} \bibnamefont{Likharev}},
  \bibinfo{year}{1986}, \bibinfo{journal}{J. Low Temp. Phys.}
  \textbf{\bibinfo{volume}{62}}, \bibinfo{pages}{345}.

\bibitem[{\citenamefont{Averin and Likharev}(1991)}]{Averin1991}
\bibinfo{author}{\bibnamefont{Averin}, \bibfnamefont{D.~V.}}, and
  \bibinfo{author}{\bibfnamefont{K.~K.} \bibnamefont{Likharev}},
  \bibinfo{year}{1991}, \emph{\bibinfo{title}{Mesoscopic Phenomena in Solids}}
  (\bibinfo{publisher}{Nature Publishing Group}, \bibinfo{address}{Amsterdam}),
  p. \bibinfo{pages}{173}.

\bibitem[{\citenamefont{Averin and Nazarov}(1990)}]{Averin1990}
\bibinfo{author}{\bibnamefont{Averin}, \bibfnamefont{D.~V.}}, and
  \bibinfo{author}{\bibfnamefont{Y.~V.} \bibnamefont{Nazarov}},
  \bibinfo{year}{1990}, \bibinfo{journal}{Phys. Rev. Lett.}
  \textbf{\bibinfo{volume}{65}}, \bibinfo{pages}{2446}.

\bibitem[{\citenamefont{Averin and Nazarov}(1992{\natexlab{a}})}]{Averin1992}
\bibinfo{author}{\bibnamefont{Averin}, \bibfnamefont{D.~V.}}, and
  \bibinfo{author}{\bibfnamefont{Y.~V.} \bibnamefont{Nazarov}},
  \bibinfo{year}{1992}{\natexlab{a}}, \emph{\bibinfo{title}{Single charge
  tunneling}} (\bibinfo{publisher}{Plenum Press}, \bibinfo{address}{New York}),
  volume \bibinfo{volume}{294}.

\bibitem[{\citenamefont{Averin and Nazarov}(1992{\natexlab{b}})}]{Averin1992b}
\bibinfo{author}{\bibnamefont{Averin}, \bibfnamefont{D.~V.}}, and
  \bibinfo{author}{\bibfnamefont{Y.~V.} \bibnamefont{Nazarov}},
  \bibinfo{year}{1992}{\natexlab{b}}, \bibinfo{journal}{Phys. Rev. Lett.}
  \textbf{\bibinfo{volume}{69}}, \bibinfo{pages}{1993}.

\bibitem[{\citenamefont{Averin and Odintsov}(1989)}]{Averin1989}
\bibinfo{author}{\bibnamefont{Averin}, \bibfnamefont{D.~V.}}, and
  \bibinfo{author}{\bibfnamefont{A.~A.} \bibnamefont{Odintsov}},
  \bibinfo{year}{1989}, \bibinfo{journal}{Phys. Lett. A}
  \textbf{\bibinfo{volume}{140}}, \bibinfo{pages}{251}.

\bibitem[{\citenamefont{Averin} \emph{et~al.}(1993)\citenamefont{Averin,
  Odintsov, and Vyshenskii}}]{Averin1993}
\bibinfo{author}{\bibnamefont{Averin}, \bibfnamefont{D.~V.}},
  \bibinfo{author}{\bibfnamefont{A.~A.} \bibnamefont{Odintsov}}, and
  \bibinfo{author}{\bibfnamefont{S.~V.} \bibnamefont{Vyshenskii}},
  \bibinfo{year}{1993}, \bibinfo{journal}{J. Appl. Phys.}
  \textbf{\bibinfo{volume}{73}}, \bibinfo{pages}{1297}.

\bibitem[{\citenamefont{Averin and Pekola}(2008)}]{Averin2008}
\bibinfo{author}{\bibnamefont{Averin}, \bibfnamefont{D.~V.}}, and
  \bibinfo{author}{\bibfnamefont{J.~P.} \bibnamefont{Pekola}},
  \bibinfo{year}{2008}, \bibinfo{journal}{Phys. Rev. Lett.}
  \textbf{\bibinfo{volume}{101}}, \bibinfo{pages}{066801}.

\bibitem[{\citenamefont{Averin and Pekola}(2011)}]{Averin2011}
\bibinfo{author}{\bibnamefont{Averin}, \bibfnamefont{D.~V.}}, and
  \bibinfo{author}{\bibfnamefont{J.~P.} \bibnamefont{Pekola}},
  \bibinfo{year}{2011}, \bibinfo{journal}{EPL (Europhysics Letters)}
  \textbf{\bibinfo{volume}{96}}, \bibinfo{pages}{67004}.

\bibitem[{\citenamefont{Averin and Sukhorukov}(2005)}]{Averin2005}
\bibinfo{author}{\bibnamefont{Averin}, \bibfnamefont{D.~V.}}, and
  \bibinfo{author}{\bibfnamefont{E.~V.} \bibnamefont{Sukhorukov}},
  \bibinfo{year}{2005}, \bibinfo{journal}{Phys. Rev. Lett.}
  \textbf{\bibinfo{volume}{95}}, \bibinfo{pages}{126803}.

\bibitem[{\citenamefont{Averin} \emph{et~al.}(1985)\citenamefont{Averin, Zorin,
  and Likharev}}]{Averin1985}
\bibinfo{author}{\bibnamefont{Averin}, \bibfnamefont{D.~V.}},
  \bibinfo{author}{\bibfnamefont{A.~B.} \bibnamefont{Zorin}}, and
  \bibinfo{author}{\bibfnamefont{K.~K.} \bibnamefont{Likharev}},
  \bibinfo{year}{1985}, \bibinfo{journal}{Sov. Phys. JETP}
  \textbf{\bibinfo{volume}{61}}, \bibinfo{pages}{407}.

\bibitem[{\citenamefont{Avron} \emph{et~al.}(2003)\citenamefont{Avron, Osadchy,
  and Seiler}}]{Avron2003}
\bibinfo{author}{\bibnamefont{Avron}, \bibfnamefont{J.~E.}},
  \bibinfo{author}{\bibfnamefont{D.}~\bibnamefont{Osadchy}}, and
  \bibinfo{author}{\bibfnamefont{R.}~\bibnamefont{Seiler}},
  \bibinfo{year}{2003}, \bibinfo{journal}{Phys. Today}
  \textbf{\bibinfo{volume}{56}}, \bibinfo{pages}{38}.

\bibitem[{\citenamefont{Bachmair}(2009)}]{Bachmair2009}
\bibinfo{author}{\bibnamefont{Bachmair}, \bibfnamefont{H.}},
  \bibinfo{year}{2009}, \bibinfo{journal}{Eur. Phys. J. Spec. Top.}
  \textbf{\bibinfo{volume}{172}}, \bibinfo{pages}{257}.

\bibitem[{\citenamefont{Bardeen} \emph{et~al.}(1959)\citenamefont{Bardeen,
  Rickayzen, and Tewordt}}]{Bardeen1959}
\bibinfo{author}{\bibnamefont{Bardeen}, \bibfnamefont{J.}},
  \bibinfo{author}{\bibfnamefont{G.}~\bibnamefont{Rickayzen}}, and
  \bibinfo{author}{\bibfnamefont{L.}~\bibnamefont{Tewordt}},
  \bibinfo{year}{1959}, \bibinfo{journal}{Phys. Rev.}
  \textbf{\bibinfo{volume}{113}}, \bibinfo{pages}{982}.

\bibitem[{\citenamefont{Barends} \emph{et~al.}(2008)\citenamefont{Barends,
  Baselmans, Yates, Gao, Hovenier, and Klapwijk}}]{Barends2008}
\bibinfo{author}{\bibnamefont{Barends}, \bibfnamefont{R.}},
  \bibinfo{author}{\bibfnamefont{J.~J.~A.} \bibnamefont{Baselmans}},
  \bibinfo{author}{\bibfnamefont{S.~J.~C.} \bibnamefont{Yates}},
  \bibinfo{author}{\bibfnamefont{J.~R.} \bibnamefont{Gao}},
  \bibinfo{author}{\bibfnamefont{J.~N.} \bibnamefont{Hovenier}}, and
  \bibinfo{author}{\bibfnamefont{T.~M.} \bibnamefont{Klapwijk}},
  \bibinfo{year}{2008}, \bibinfo{journal}{Phys. Rev. Lett.}
  \textbf{\bibinfo{volume}{100}}, \bibinfo{pages}{257002}.

\bibitem[{\citenamefont{Becker} \emph{et~al.}(2007)\citenamefont{Becker,
  Bievre, Fujii, Glaeser, Inglis, Luebbig, and Mana}}]{Becker2007}
\bibinfo{author}{\bibnamefont{Becker}, \bibfnamefont{P.}},
  \bibinfo{author}{\bibfnamefont{P.~D.} \bibnamefont{Bievre}},
  \bibinfo{author}{\bibfnamefont{K.}~\bibnamefont{Fujii}},
  \bibinfo{author}{\bibfnamefont{M.}~\bibnamefont{Glaeser}},
  \bibinfo{author}{\bibfnamefont{B.}~\bibnamefont{Inglis}},
  \bibinfo{author}{\bibfnamefont{H.}~\bibnamefont{Luebbig}}, and
  \bibinfo{author}{\bibfnamefont{G.}~\bibnamefont{Mana}}, \bibinfo{year}{2007},
  \bibinfo{journal}{Metrologia} \textbf{\bibinfo{volume}{44}},
  \bibinfo{pages}{1}.

\bibitem[{\citenamefont{Behr} \emph{et~al.}(2005)\citenamefont{Behr, Williams,
  Patel, Janssen, Funck, and Klonz}}]{Behr2005}
\bibinfo{author}{\bibnamefont{Behr}, \bibfnamefont{R.}},
  \bibinfo{author}{\bibfnamefont{J.~M.} \bibnamefont{Williams}},
  \bibinfo{author}{\bibfnamefont{P.}~\bibnamefont{Patel}},
  \bibinfo{author}{\bibfnamefont{T.~J. B.~M.} \bibnamefont{Janssen}},
  \bibinfo{author}{\bibfnamefont{T.}~\bibnamefont{Funck}}, and
  \bibinfo{author}{\bibfnamefont{M.}~\bibnamefont{Klonz}},
  \bibinfo{year}{2005}, \bibinfo{journal}{IEEE Trans. Instrum. Meas.}
  \textbf{\bibinfo{volume}{54}}, \bibinfo{pages}{612}.

\bibitem[{\citenamefont{Benz and Hamilton}(1996)}]{Benz1996}
\bibinfo{author}{\bibnamefont{Benz}, \bibfnamefont{S.~P.}}, and
  \bibinfo{author}{\bibfnamefont{C.~A.} \bibnamefont{Hamilton}},
  \bibinfo{year}{1996}, \bibinfo{journal}{Appl. Phys. Lett.}
  \textbf{\bibinfo{volume}{68}}, \bibinfo{pages}{3171}.

\bibitem[{\citenamefont{Berggren} \emph{et~al.}(1986)\citenamefont{Berggren,
  Thornton, Newson, and Pepper}}]{Berggren1986}
\bibinfo{author}{\bibnamefont{Berggren}, \bibfnamefont{K.}},
  \bibinfo{author}{\bibfnamefont{T.}~\bibnamefont{Thornton}},
  \bibinfo{author}{\bibfnamefont{D.}~\bibnamefont{Newson}}, and
  \bibinfo{author}{\bibfnamefont{M.}~\bibnamefont{Pepper}},
  \bibinfo{year}{1986}, \bibinfo{journal}{Phys. Rev. Lett.}
  \textbf{\bibinfo{volume}{57}}, \bibinfo{pages}{1769}.

\bibitem[{\citenamefont{Bieri and Fr\"ohlich}(2011)}]{Bieri2011}
\bibinfo{author}{\bibnamefont{Bieri}, \bibfnamefont{S.}}, and
  \bibinfo{author}{\bibfnamefont{J.}~\bibnamefont{Fr\"ohlich}},
  \bibinfo{year}{2011}, \bibinfo{journal}{Comptes Rendus Physique}
  \textbf{\bibinfo{volume}{12}}, \bibinfo{pages}{332}.

\bibitem[{\citenamefont{BIPM}(2010)}]{BIPM2010}
\bibinfo{author}{\bibnamefont{BIPM}}, \bibinfo{year}{2010}, \bibinfo{title}{{No
  Title}}.

\bibitem[{\citenamefont{Bloch}(1968)}]{Bloch1968}
\bibinfo{author}{\bibnamefont{Bloch}, \bibfnamefont{F.}}, \bibinfo{year}{1968},
  \bibinfo{journal}{Phys. Rev. Lett.} \textbf{\bibinfo{volume}{21}},
  \bibinfo{pages}{1241}.

\bibitem[{\citenamefont{Bloch}(1970)}]{Bloch1970}
\bibinfo{author}{\bibnamefont{Bloch}, \bibfnamefont{F.}}, \bibinfo{year}{1970},
  \bibinfo{journal}{Phys. Rev. B} \textbf{\bibinfo{volume}{2}},
  \bibinfo{pages}{109}.

\bibitem[{\citenamefont{Blonder} \emph{et~al.}(1982)\citenamefont{Blonder,
  Tinkham, and Klapwijk}}]{Blonder1982}
\bibinfo{author}{\bibnamefont{Blonder}, \bibfnamefont{G.~E.}},
  \bibinfo{author}{\bibfnamefont{M.}~\bibnamefont{Tinkham}}, and
  \bibinfo{author}{\bibfnamefont{T.~M.} \bibnamefont{Klapwijk}},
  \bibinfo{year}{1982}, \bibinfo{journal}{Phys. Rev. B}
  \textbf{\bibinfo{volume}{25}}, \bibinfo{pages}{4515}.

\bibitem[{\citenamefont{Blumenthal}
  \emph{et~al.}(2007)\citenamefont{Blumenthal, Kaestner, Li, Giblin, Janssen,
  Pepper, Anderson, Jones, and Ritchie}}]{Blumenthal2007}
\bibinfo{author}{\bibnamefont{Blumenthal}, \bibfnamefont{M.~D.}},
  \bibinfo{author}{\bibfnamefont{B.}~\bibnamefont{Kaestner}},
  \bibinfo{author}{\bibfnamefont{L.}~\bibnamefont{Li}},
  \bibinfo{author}{\bibfnamefont{S.}~\bibnamefont{Giblin}},
  \bibinfo{author}{\bibfnamefont{T.~J. B.~M.} \bibnamefont{Janssen}},
  \bibinfo{author}{\bibfnamefont{M.}~\bibnamefont{Pepper}},
  \bibinfo{author}{\bibfnamefont{D.}~\bibnamefont{Anderson}},
  \bibinfo{author}{\bibfnamefont{G.}~\bibnamefont{Jones}}, and
  \bibinfo{author}{\bibfnamefont{D.~A.} \bibnamefont{Ritchie}},
  \bibinfo{year}{2007}, \bibinfo{journal}{Nature Phys.}
  \textbf{\bibinfo{volume}{3}}, \bibinfo{pages}{343}.

\bibitem[{\citenamefont{Bord\'e}(2005)}]{Borde2005}
\bibinfo{author}{\bibnamefont{Bord\'e}, \bibfnamefont{C.~J.}},
  \bibinfo{year}{2005}, \bibinfo{journal}{Phil. Trans. Royal Soc. A}
  \textbf{\bibinfo{volume}{363}}, \bibinfo{pages}{2177}.

\bibitem[{\citenamefont{Bouchendira}
  \emph{et~al.}(2011)\citenamefont{Bouchendira, Clad\'{e},
  Guellati-Kh\'{e}lifa, Nez, and Biraben}}]{Bouchendira2011}
\bibinfo{author}{\bibnamefont{Bouchendira}, \bibfnamefont{R.}},
  \bibinfo{author}{\bibfnamefont{P.}~\bibnamefont{Clad\'{e}}},
  \bibinfo{author}{\bibfnamefont{S.}~\bibnamefont{Guellati-Kh\'{e}lifa}},
  \bibinfo{author}{\bibfnamefont{F.}~\bibnamefont{Nez}}, and
  \bibinfo{author}{\bibfnamefont{F.}~\bibnamefont{Biraben}},
  \bibinfo{year}{2011}, \bibinfo{journal}{Phys. Rev. Lett.}
  \textbf{\bibinfo{volume}{106}}, \bibinfo{pages}{080801}.

\bibitem[{\citenamefont{Bouchiat} \emph{et~al.}(1998)\citenamefont{Bouchiat,
  Vion, Joyez, Esteve, and H}}]{Bouchiat1998}
\bibinfo{author}{\bibnamefont{Bouchiat}, \bibfnamefont{V.}},
  \bibinfo{author}{\bibfnamefont{D.}~\bibnamefont{Vion}},
  \bibinfo{author}{\bibfnamefont{P.}~\bibnamefont{Joyez}},
  \bibinfo{author}{\bibfnamefont{D.}~\bibnamefont{Esteve}}, and
  \bibinfo{author}{\bibfnamefont{D.~M.} \bibnamefont{H}}, \bibinfo{year}{1998},
  \bibinfo{journal}{Physica Scripta} \textbf{\bibinfo{volume}{T76}},
  \bibinfo{pages}{786}.

\bibitem[{\citenamefont{Brenning} \emph{et~al.}(2006)\citenamefont{Brenning,
  Kafanov, Duty, Kubatkin, and Delsing}}]{Brenning2006}
\bibinfo{author}{\bibnamefont{Brenning}, \bibfnamefont{H.}},
  \bibinfo{author}{\bibfnamefont{S.}~\bibnamefont{Kafanov}},
  \bibinfo{author}{\bibfnamefont{T.}~\bibnamefont{Duty}},
  \bibinfo{author}{\bibfnamefont{S.}~\bibnamefont{Kubatkin}}, and
  \bibinfo{author}{\bibfnamefont{P.}~\bibnamefont{Delsing}},
  \bibinfo{year}{2006}, \bibinfo{journal}{J. Appl. Phys.}
  \textbf{\bibinfo{volume}{100}}, \bibinfo{pages}{114321}.

\bibitem[{\citenamefont{Brenning} \emph{et~al.}(2004)\citenamefont{Brenning,
  S., and P.}}]{Brenning2004}
\bibinfo{author}{\bibnamefont{Brenning}, \bibfnamefont{H.}},
  \bibinfo{author}{\bibfnamefont{K.}~\bibnamefont{S.}}, and
  \bibinfo{author}{\bibfnamefont{D.}~\bibnamefont{P.}}, \bibinfo{year}{2004},
  \bibinfo{journal}{J. Appl. Phys.} \textbf{\bibinfo{volume}{96}},
  \bibinfo{pages}{6822}.

\bibitem[{\citenamefont{Bubanja}(2011)}]{Bubanja2011}
\bibinfo{author}{\bibnamefont{Bubanja}, \bibfnamefont{V.}},
  \bibinfo{year}{2011}, \bibinfo{journal}{Phys. Rev. B}
  \textbf{\bibinfo{volume}{83}}, \bibinfo{pages}{195312}.

\bibitem[{\citenamefont{Buehler} \emph{et~al.}(2005)\citenamefont{Buehler,
  Reilly, Starrett, Greentree, Hamilton, Dzurak, and Clark}}]{Buehler2005}
\bibinfo{author}{\bibnamefont{Buehler}, \bibfnamefont{T.~M.}},
  \bibinfo{author}{\bibfnamefont{D.~J.} \bibnamefont{Reilly}},
  \bibinfo{author}{\bibfnamefont{R.~P.} \bibnamefont{Starrett}},
  \bibinfo{author}{\bibfnamefont{A.~D.} \bibnamefont{Greentree}},
  \bibinfo{author}{\bibfnamefont{A.~R.} \bibnamefont{Hamilton}},
  \bibinfo{author}{\bibfnamefont{A.~S.} \bibnamefont{Dzurak}}, and
  \bibinfo{author}{\bibfnamefont{R.~G.} \bibnamefont{Clark}},
  \bibinfo{year}{2005}, \bibinfo{journal}{Appl. Phys. Lett.}
  \textbf{\bibinfo{volume}{86}}, \bibinfo{pages}{143117}.

\bibitem[{\citenamefont{B\"uttiker}(1987)}]{Buttiker1987}
\bibinfo{author}{\bibnamefont{B\"uttiker}, \bibfnamefont{M.}},
  \bibinfo{year}{1987}, \bibinfo{journal}{Phys. Rev. B.}
  \textbf{\bibinfo{volume}{36}}, \bibinfo{pages}{3548}.

\bibitem[{\citenamefont{Bylander} \emph{et~al.}(2005)\citenamefont{Bylander,
  Duty, and Delsing}}]{Bylander2005}
\bibinfo{author}{\bibnamefont{Bylander}, \bibfnamefont{J.}},
  \bibinfo{author}{\bibfnamefont{T.}~\bibnamefont{Duty}}, and
  \bibinfo{author}{\bibfnamefont{P.}~\bibnamefont{Delsing}},
  \bibinfo{year}{2005}, \bibinfo{journal}{Nature}
  \textbf{\bibinfo{volume}{434}}, \bibinfo{pages}{361}.

\bibitem[{\citenamefont{Cadoret} \emph{et~al.}(2011)\citenamefont{Cadoret,
  de~Mirand\'es, Clad\'e, Guellati-Kh\'elifa, Nez, and Biraben}}]{Cadoret2011}
\bibinfo{author}{\bibnamefont{Cadoret}, \bibfnamefont{M.}},
  \bibinfo{author}{\bibfnamefont{E.}~\bibnamefont{de~Mirand\'es}},
  \bibinfo{author}{\bibfnamefont{P.}~\bibnamefont{Clad\'e}},
  \bibinfo{author}{\bibfnamefont{S.}~\bibnamefont{Guellati-Kh\'elifa}},
  \bibinfo{author}{\bibfnamefont{F.}~\bibnamefont{Nez}}, and
  \bibinfo{author}{\bibfnamefont{F.}~\bibnamefont{Biraben}},
  \bibinfo{year}{2011}, \bibinfo{journal}{Comptes Rendus Physique}
  \textbf{\bibinfo{volume}{12}}, \bibinfo{pages}{379}.

\bibitem[{\citenamefont{Camarota} \emph{et~al.}(2012)\citenamefont{Camarota,
  Scherer, Keller, Lotkhov, Willenberg, and J.}}]{Camarota2012}
\bibinfo{author}{\bibnamefont{Camarota}, \bibfnamefont{B.}},
  \bibinfo{author}{\bibfnamefont{H.}~\bibnamefont{Scherer}},
  \bibinfo{author}{\bibfnamefont{M.~V.} \bibnamefont{Keller}},
  \bibinfo{author}{\bibfnamefont{S.~V.} \bibnamefont{Lotkhov}},
  \bibinfo{author}{\bibfnamefont{G.-D.} \bibnamefont{Willenberg}}, and
  \bibinfo{author}{\bibfnamefont{A.~F.} \bibnamefont{J.}},
  \bibinfo{year}{2012}, \bibinfo{journal}{Metrologia}
  \textbf{\bibinfo{volume}{49}}, \bibinfo{pages}{8}.

\bibitem[{\citenamefont{Cassidy} \emph{et~al.}(2007)\citenamefont{Cassidy,
  Dzurak, Clark, Petersson, Farrer, Ritchie, and Smith}}]{Cassidy2007}
\bibinfo{author}{\bibnamefont{Cassidy}, \bibfnamefont{M.~C.}},
  \bibinfo{author}{\bibfnamefont{A.~S.} \bibnamefont{Dzurak}},
  \bibinfo{author}{\bibfnamefont{R.~G.} \bibnamefont{Clark}},
  \bibinfo{author}{\bibfnamefont{K.~D.} \bibnamefont{Petersson}},
  \bibinfo{author}{\bibfnamefont{I.}~\bibnamefont{Farrer}},
  \bibinfo{author}{\bibfnamefont{D.~A.} \bibnamefont{Ritchie}}, and
  \bibinfo{author}{\bibfnamefont{C.~G.} \bibnamefont{Smith}},
  \bibinfo{year}{2007}, \bibinfo{journal}{Appl. Phys. Lett.}
  \textbf{\bibinfo{volume}{91}}, \bibinfo{pages}{222104}.

\bibitem[{\citenamefont{CCEM}(2012)}]{ccem2012}
\bibinfo{author}{\bibnamefont{CCEM}}, \bibinfo{year}{2012},
  \bibinfo{title}{Draft of mise en pratique for the ampere and other electric
  units in the international system of units}.

\bibitem[{\citenamefont{Chan} \emph{et~al.}(2011)\citenamefont{Chan,
  M\"ott\"onen, Kemppinen, Lai, Tan, Lim, and Dzurak}}]{Chan2011}
\bibinfo{author}{\bibnamefont{Chan}, \bibfnamefont{K.~W.}},
  \bibinfo{author}{\bibfnamefont{M.}~\bibnamefont{M\"ott\"onen}},
  \bibinfo{author}{\bibfnamefont{A.}~\bibnamefont{Kemppinen}},
  \bibinfo{author}{\bibfnamefont{N.~S.} \bibnamefont{Lai}},
  \bibinfo{author}{\bibfnamefont{K.~Y.} \bibnamefont{Tan}},
  \bibinfo{author}{\bibfnamefont{W.~H.} \bibnamefont{Lim}}, and
  \bibinfo{author}{\bibfnamefont{A.~S.} \bibnamefont{Dzurak}},
  \bibinfo{year}{2011}, \bibinfo{journal}{Appl. Phys. Lett.}
  \textbf{\bibinfo{volume}{98}}, \bibinfo{pages}{212103}.

\bibitem[{\citenamefont{Chang} \emph{et~al.}(1974)\citenamefont{Chang, Esaki,
  and Tsu}}]{Chang1974}
\bibinfo{author}{\bibnamefont{Chang}, \bibfnamefont{L.~L.}},
  \bibinfo{author}{\bibfnamefont{L.}~\bibnamefont{Esaki}}, and
  \bibinfo{author}{\bibfnamefont{R.}~\bibnamefont{Tsu}}, \bibinfo{year}{1974},
  \bibinfo{journal}{Appl. Phys. Lett.} \textbf{\bibinfo{volume}{24}},
  \bibinfo{pages}{593}.

\bibitem[{\citenamefont{Cheinet} \emph{et~al.}(2008)\citenamefont{Cheinet,
  Trotzky, Feld, Schnorrberger, Moreno-Cardoner, F\"olling, and
  Bloch}}]{Cheinet2008}
\bibinfo{author}{\bibnamefont{Cheinet}, \bibfnamefont{P.}},
  \bibinfo{author}{\bibfnamefont{S.}~\bibnamefont{Trotzky}},
  \bibinfo{author}{\bibfnamefont{M.}~\bibnamefont{Feld}},
  \bibinfo{author}{\bibfnamefont{U.}~\bibnamefont{Schnorrberger}},
  \bibinfo{author}{\bibfnamefont{M.}~\bibnamefont{Moreno-Cardoner}},
  \bibinfo{author}{\bibfnamefont{S.}~\bibnamefont{F\"olling}}, and
  \bibinfo{author}{\bibfnamefont{I.}~\bibnamefont{Bloch}},
  \bibinfo{year}{2008}, \bibinfo{journal}{Phys. Rev. Lett.}
  \textbf{\bibinfo{volume}{101}}, \bibinfo{pages}{090404}.

\bibitem[{\citenamefont{Clarke}(1968)}]{Clarke1968}
\bibinfo{author}{\bibnamefont{Clarke}, \bibfnamefont{J.}},
  \bibinfo{year}{1968}, \bibinfo{journal}{Phys. Rev. Lett.}
  \textbf{\bibinfo{volume}{21}}, \bibinfo{pages}{1566}.

\bibitem[{\citenamefont{Clarke}(1972)}]{Clarke1972}
\bibinfo{author}{\bibnamefont{Clarke}, \bibfnamefont{J.}},
  \bibinfo{year}{1972}, \bibinfo{journal}{Phys. Rev. Lett.}
  \textbf{\bibinfo{volume}{28}}, \bibinfo{pages}{1363}.

\bibitem[{\citenamefont{Clerk} \emph{et~al.}(2003)\citenamefont{Clerk, Girvin,
  and Stone}}]{Clerk2003}
\bibinfo{author}{\bibnamefont{Clerk}, \bibfnamefont{A.}},
  \bibinfo{author}{\bibfnamefont{S.}~\bibnamefont{Girvin}}, and
  \bibinfo{author}{\bibfnamefont{A.}~\bibnamefont{Stone}},
  \bibinfo{year}{2003}, \bibinfo{journal}{Phys. Rev. B}
  \textbf{\bibinfo{volume}{67}}, \bibinfo{pages}{165324}.

\bibitem[{\citenamefont{Clerk} \emph{et~al.}(2010)\citenamefont{Clerk, Girvin,
  Marquardt, and Schoelkopf}}]{Clerk2010}
\bibinfo{author}{\bibnamefont{Clerk}, \bibfnamefont{A.~A.}},
  \bibinfo{author}{\bibfnamefont{S.~M.} \bibnamefont{Girvin}},
  \bibinfo{author}{\bibfnamefont{F.}~\bibnamefont{Marquardt}}, and
  \bibinfo{author}{\bibfnamefont{R.~J.} \bibnamefont{Schoelkopf}},
  \bibinfo{year}{2010}, \bibinfo{journal}{Rev. Mod. Phys.}
  \textbf{\bibinfo{volume}{82}}, \bibinfo{pages}{1155}.

\bibitem[{\citenamefont{Clothier} \emph{et~al.}(1989)\citenamefont{Clothier,
  Sloggett, Bairnsfather, Currey, and Benjamin}}]{Clothier1989}
\bibinfo{author}{\bibnamefont{Clothier}, \bibfnamefont{W.~K.}},
  \bibinfo{author}{\bibfnamefont{G.~J.} \bibnamefont{Sloggett}},
  \bibinfo{author}{\bibfnamefont{H.}~\bibnamefont{Bairnsfather}},
  \bibinfo{author}{\bibfnamefont{M.~F.} \bibnamefont{Currey}}, and
  \bibinfo{author}{\bibfnamefont{D.~J.} \bibnamefont{Benjamin}},
  \bibinfo{year}{1989}, \bibinfo{journal}{Metrologia}
  \textbf{\bibinfo{volume}{26}}, \bibinfo{pages}{9}.

\bibitem[{\citenamefont{Cohen and Taylor}(1987)}]{Cohen1987}
\bibinfo{author}{\bibnamefont{Cohen}, \bibfnamefont{E.~R.}}, and
  \bibinfo{author}{\bibfnamefont{B.~N.} \bibnamefont{Taylor}},
  \bibinfo{year}{1987}, \bibinfo{journal}{Rev. Mod. Phys.}
  \textbf{\bibinfo{volume}{59}}, \bibinfo{pages}{1121}.

\bibitem[{\citenamefont{Cohen} \emph{et~al.}(2009)\citenamefont{Cohen, Fleurov,
  and Kikoin}}]{Cohen2009}
\bibinfo{author}{\bibnamefont{Cohen}, \bibfnamefont{G.}},
  \bibinfo{author}{\bibfnamefont{V.}~\bibnamefont{Fleurov}}, and
  \bibinfo{author}{\bibfnamefont{K.}~\bibnamefont{Kikoin}},
  \bibinfo{year}{2009}, \bibinfo{journal}{Phys. Rev. B}
  \textbf{\bibinfo{volume}{79}}, \bibinfo{pages}{245307}.

\bibitem[{\citenamefont{Covington} \emph{et~al.}(2000)\citenamefont{Covington,
  Keller, Kautz, and Martinis}}]{Covington2000}
\bibinfo{author}{\bibnamefont{Covington}, \bibfnamefont{M.}},
  \bibinfo{author}{\bibfnamefont{M.~W.} \bibnamefont{Keller}},
  \bibinfo{author}{\bibfnamefont{R.~L.} \bibnamefont{Kautz}}, and
  \bibinfo{author}{\bibfnamefont{J.~M.} \bibnamefont{Martinis}},
  \bibinfo{year}{2000}, \bibinfo{journal}{Phys. Rev. Lett.}
  \textbf{\bibinfo{volume}{84}}, \bibinfo{pages}{5192}.

\bibitem[{\citenamefont{Crooks}(1999)}]{Crooks1999}
\bibinfo{author}{\bibnamefont{Crooks}, \bibfnamefont{G.~E.}},
  \bibinfo{year}{1999}, \bibinfo{journal}{Phys. Rev. E}
  \textbf{\bibinfo{volume}{60}}, \bibinfo{pages}{2721}.

\bibitem[{\citenamefont{{de Visser}} \emph{et~al.}(2011)\citenamefont{{de
  Visser}, Baselmans, Diener, Yates, Endo, and Klapwijk}}]{deVisser2011}
\bibinfo{author}{\bibnamefont{{de Visser}}, \bibfnamefont{P.}},
  \bibinfo{author}{\bibfnamefont{J.}~\bibnamefont{Baselmans}},
  \bibinfo{author}{\bibfnamefont{P.}~\bibnamefont{Diener}},
  \bibinfo{author}{\bibfnamefont{S.}~\bibnamefont{Yates}},
  \bibinfo{author}{\bibfnamefont{A.}~\bibnamefont{Endo}}, and
  \bibinfo{author}{\bibfnamefont{T.}~\bibnamefont{Klapwijk}},
  \bibinfo{year}{2011}, \bibinfo{journal}{Phys. Rev. Lett.}
  \textbf{\bibinfo{volume}{106}}, \bibinfo{pages}{167004}.

\bibitem[{\citenamefont{Delahaye}(1993)}]{Delahaye1993}
\bibinfo{author}{\bibnamefont{Delahaye}, \bibfnamefont{F.}},
  \bibinfo{year}{1993}, \bibinfo{journal}{J. Appl. Phys.}
  \textbf{\bibinfo{volume}{73}}, \bibinfo{pages}{7914}.

\bibitem[{\citenamefont{Devoret} \emph{et~al.}(1990)\citenamefont{Devoret,
  Esteve, Grabert, Ingold, Pothier, and Urbina}}]{Devoret1990}
\bibinfo{author}{\bibnamefont{Devoret}, \bibfnamefont{M.~H.}},
  \bibinfo{author}{\bibfnamefont{D.}~\bibnamefont{Esteve}},
  \bibinfo{author}{\bibfnamefont{H.}~\bibnamefont{Grabert}},
  \bibinfo{author}{\bibfnamefont{G.-L.} \bibnamefont{Ingold}},
  \bibinfo{author}{\bibfnamefont{H.}~\bibnamefont{Pothier}}, and
  \bibinfo{author}{\bibfnamefont{C.}~\bibnamefont{Urbina}},
  \bibinfo{year}{1990}, \bibinfo{journal}{Phys. Rev. Lett.}
  \textbf{\bibinfo{volume}{64}}, \bibinfo{pages}{1824}.

\bibitem[{\citenamefont{Devoret and Schoelkopf}(2000)}]{Devoret2000}
\bibinfo{author}{\bibnamefont{Devoret}, \bibfnamefont{M.~H.}}, and
  \bibinfo{author}{\bibfnamefont{R.~J.} \bibnamefont{Schoelkopf}},
  \bibinfo{year}{2000}, \bibinfo{journal}{Nature}
  \textbf{\bibinfo{volume}{406}}, \bibinfo{pages}{1039}.

\bibitem[{\citenamefont{Dolan}(1977)}]{Dolan1977}
\bibinfo{author}{\bibnamefont{Dolan}, \bibfnamefont{G.~J.}},
  \bibinfo{year}{1977}, \bibinfo{journal}{Appl. Phys. Lett.}
  \textbf{\bibinfo{volume}{31}}, \bibinfo{pages}{337}.

\bibitem[{\citenamefont{Dolan and Dunsmuir}(1988)}]{Dolan1988}
\bibinfo{author}{\bibnamefont{Dolan}, \bibfnamefont{G.~J.}}, and
  \bibinfo{author}{\bibfnamefont{J.~H.} \bibnamefont{Dunsmuir}},
  \bibinfo{year}{1988}, \bibinfo{journal}{Physica B}
  \textbf{\bibinfo{volume}{152}}, \bibinfo{pages}{7}.

\bibitem[{\citenamefont{Doucot}(2011)}]{Doucot2011}
\bibinfo{author}{\bibnamefont{Doucot}, \bibfnamefont{B.}},
  \bibinfo{year}{2011}, \bibinfo{journal}{Comptes Rendus Physique}
  \textbf{\bibinfo{volume}{12}}, \bibinfo{pages}{323}.

\bibitem[{\citenamefont{Durrani}(2009)}]{Durrani2009}
\bibinfo{author}{\bibnamefont{Durrani}, \bibfnamefont{Z.~A.~K.}},
  \bibinfo{year}{2009}, \emph{\bibinfo{title}{Single-electron devices and
  circuits in silicon}} (\bibinfo{publisher}{Imperial College Press}).

\bibitem[{\citenamefont{Ebbecke} \emph{et~al.}(2000)\citenamefont{Ebbecke,
  Bastian, Bl\"{o}cker, Pierz, and Ahlers}}]{Ebbecke2000}
\bibinfo{author}{\bibnamefont{Ebbecke}, \bibfnamefont{J.}},
  \bibinfo{author}{\bibfnamefont{G.}~\bibnamefont{Bastian}},
  \bibinfo{author}{\bibfnamefont{M.}~\bibnamefont{Bl\"{o}cker}},
  \bibinfo{author}{\bibfnamefont{K.}~\bibnamefont{Pierz}}, and
  \bibinfo{author}{\bibfnamefont{F.~J.} \bibnamefont{Ahlers}},
  \bibinfo{year}{2000}, \bibinfo{journal}{Appl. Phys. Lett.}
  \textbf{\bibinfo{volume}{77}}, \bibinfo{pages}{2601}.

\bibitem[{\citenamefont{Ebbecke} \emph{et~al.}(2003)\citenamefont{Ebbecke,
  Fletcher, Ahlers, Hartland, and Janssen}}]{Ebbecke2003}
\bibinfo{author}{\bibnamefont{Ebbecke}, \bibfnamefont{J.}},
  \bibinfo{author}{\bibfnamefont{N.~E.} \bibnamefont{Fletcher}},
  \bibinfo{author}{\bibfnamefont{F.-J.} \bibnamefont{Ahlers}},
  \bibinfo{author}{\bibfnamefont{A.}~\bibnamefont{Hartland}}, and
  \bibinfo{author}{\bibfnamefont{T.~J. B.~M.} \bibnamefont{Janssen}},
  \bibinfo{year}{2003}, \bibinfo{journal}{IEE Trans. Instr. Meas.}
  \textbf{\bibinfo{volume}{52}}, \bibinfo{pages}{594}.

\bibitem[{\citenamefont{Ebbecke} \emph{et~al.}(2004)\citenamefont{Ebbecke,
  Fletcher, Janssen, Ahlers, Pepper, Beere, and Ritchie}}]{Ebbecke2004}
\bibinfo{author}{\bibnamefont{Ebbecke}, \bibfnamefont{J.}},
  \bibinfo{author}{\bibfnamefont{N.~E.} \bibnamefont{Fletcher}},
  \bibinfo{author}{\bibfnamefont{T.~J. B.~M.} \bibnamefont{Janssen}},
  \bibinfo{author}{\bibfnamefont{F.-J.} \bibnamefont{Ahlers}},
  \bibinfo{author}{\bibfnamefont{M.}~\bibnamefont{Pepper}},
  \bibinfo{author}{\bibfnamefont{H.~E.} \bibnamefont{Beere}}, and
  \bibinfo{author}{\bibfnamefont{D.~A.} \bibnamefont{Ritchie}},
  \bibinfo{year}{2004}, \bibinfo{journal}{Appl. Phys. Lett.}
  \textbf{\bibinfo{volume}{84}}, \bibinfo{pages}{4319}.

\bibitem[{\citenamefont{Eichenberger}
  \emph{et~al.}(2009)\citenamefont{Eichenberger, Genev\`{e}s, and
  Gournay}}]{Eichenberger2009}
\bibinfo{author}{\bibnamefont{Eichenberger}, \bibfnamefont{A.}},
  \bibinfo{author}{\bibfnamefont{G.}~\bibnamefont{Genev\`{e}s}}, and
  \bibinfo{author}{\bibfnamefont{P.}~\bibnamefont{Gournay}},
  \bibinfo{year}{2009}, \bibinfo{journal}{Eur. Phys. J. Spec. Top.}
  \textbf{\bibinfo{volume}{172}}, \bibinfo{pages}{363}.

\bibitem[{\citenamefont{Eiles} \emph{et~al.}(1993)\citenamefont{Eiles,
  Martinis, and Devoret}}]{Eiles1993}
\bibinfo{author}{\bibnamefont{Eiles}, \bibfnamefont{T.~M.}},
  \bibinfo{author}{\bibfnamefont{J.~M.} \bibnamefont{Martinis}}, and
  \bibinfo{author}{\bibfnamefont{M.~H.} \bibnamefont{Devoret}},
  \bibinfo{year}{1993}, \bibinfo{journal}{Phys. Rev. Lett.}
  \textbf{\bibinfo{volume}{70}}, \bibinfo{pages}{1862}.

\bibitem[{\citenamefont{Elmquist} \emph{et~al.}(2003)\citenamefont{Elmquist,
  Zimmerman, and Huber}}]{Elmquist2003}
\bibinfo{author}{\bibnamefont{Elmquist}, \bibfnamefont{R.~E.}},
  \bibinfo{author}{\bibfnamefont{N.~M.} \bibnamefont{Zimmerman}}, and
  \bibinfo{author}{\bibfnamefont{W.~H.} \bibnamefont{Huber}},
  \bibinfo{year}{2003}, \bibinfo{journal}{IEEE Trans. Instrum. Meas.}
  \textbf{\bibinfo{volume}{52}}, \bibinfo{pages}{590}.

\bibitem[{\citenamefont{Elzerman} \emph{et~al.}(2003)\citenamefont{Elzerman,
  Hanson, Greidanus, Willems~van Beveren, De~Franceschi, Vandersypen, Tarucha,
  and Kouwenhoven}}]{Elzerman2003}
\bibinfo{author}{\bibnamefont{Elzerman}, \bibfnamefont{J.~M.}},
  \bibinfo{author}{\bibfnamefont{R.}~\bibnamefont{Hanson}},
  \bibinfo{author}{\bibfnamefont{J.~S.} \bibnamefont{Greidanus}},
  \bibinfo{author}{\bibfnamefont{L.~H.} \bibnamefont{Willems~van Beveren}},
  \bibinfo{author}{\bibfnamefont{S.}~\bibnamefont{De~Franceschi}},
  \bibinfo{author}{\bibfnamefont{L.~M.~K.} \bibnamefont{Vandersypen}},
  \bibinfo{author}{\bibfnamefont{S.}~\bibnamefont{Tarucha}}, and
  \bibinfo{author}{\bibfnamefont{L.~P.} \bibnamefont{Kouwenhoven}},
  \bibinfo{year}{2003}, \bibinfo{journal}{Phys. Rev. B}
  \textbf{\bibinfo{volume}{67}}, \bibinfo{pages}{161308}.

\bibitem[{\citenamefont{Erbe} \emph{et~al.}(1998)\citenamefont{Erbe, Blick,
  Tilke, Kriele, and Kotthaus}}]{Erbe1998}
\bibinfo{author}{\bibnamefont{Erbe}, \bibfnamefont{A.}},
  \bibinfo{author}{\bibfnamefont{R.~H.} \bibnamefont{Blick}},
  \bibinfo{author}{\bibfnamefont{A.}~\bibnamefont{Tilke}},
  \bibinfo{author}{\bibfnamefont{A.}~\bibnamefont{Kriele}}, and
  \bibinfo{author}{\bibfnamefont{J.~P.} \bibnamefont{Kotthaus}},
  \bibinfo{year}{1998}, \bibinfo{journal}{Appl. Phys. Lett}
  \textbf{\bibinfo{volume}{73}}, \bibinfo{pages}{3751}.

\bibitem[{\citenamefont{Erbe} \emph{et~al.}(2001)\citenamefont{Erbe, Weiss,
  Zwerger, and Blick}}]{Erbe2001}
\bibinfo{author}{\bibnamefont{Erbe}, \bibfnamefont{A.}},
  \bibinfo{author}{\bibfnamefont{C.}~\bibnamefont{Weiss}},
  \bibinfo{author}{\bibfnamefont{W.}~\bibnamefont{Zwerger}}, and
  \bibinfo{author}{\bibfnamefont{R.~H.} \bibnamefont{Blick}},
  \bibinfo{year}{2001}, \bibinfo{journal}{Phys. Rev. Lett.}
  \textbf{\bibinfo{volume}{87}}, \bibinfo{pages}{096106}.

\bibitem[{\citenamefont{Evans} \emph{et~al.}(1993)\citenamefont{Evans, Cohen,
  and Morriss}}]{Evans1993}
\bibinfo{author}{\bibnamefont{Evans}, \bibfnamefont{D.~J.}},
  \bibinfo{author}{\bibfnamefont{E.~G.~D.} \bibnamefont{Cohen}}, and
  \bibinfo{author}{\bibfnamefont{G.~P.} \bibnamefont{Morriss}},
  \bibinfo{year}{1993}, \bibinfo{journal}{Phys. Rev. Lett.}
  \textbf{\bibinfo{volume}{71}}, \bibinfo{pages}{2401}.

\bibitem[{\citenamefont{Faoro} \emph{et~al.}(2003)\citenamefont{Faoro, Siewert,
  and Fazio}}]{Faoro2003}
\bibinfo{author}{\bibnamefont{Faoro}, \bibfnamefont{L.}},
  \bibinfo{author}{\bibfnamefont{J.}~\bibnamefont{Siewert}}, and
  \bibinfo{author}{\bibfnamefont{R.}~\bibnamefont{Fazio}},
  \bibinfo{year}{2003}, \bibinfo{journal}{Phys. Rev. Lett.}
  \textbf{\bibinfo{volume}{90}}, \bibinfo{pages}{028301}.

\bibitem[{\citenamefont{Fasth} \emph{et~al.}(2007)\citenamefont{Fasth, Fuhrer,
  Samuelson, Golovach, and Loss}}]{Fasth2007}
\bibinfo{author}{\bibnamefont{Fasth}, \bibfnamefont{C.}},
  \bibinfo{author}{\bibfnamefont{A.}~\bibnamefont{Fuhrer}},
  \bibinfo{author}{\bibfnamefont{L.}~\bibnamefont{Samuelson}},
  \bibinfo{author}{\bibfnamefont{V.~N.} \bibnamefont{Golovach}}, and
  \bibinfo{author}{\bibfnamefont{D.}~\bibnamefont{Loss}}, \bibinfo{year}{2007},
  \bibinfo{journal}{Phys. Rev. Lett.} \textbf{\bibinfo{volume}{98}},
  \bibinfo{pages}{266801}.

\bibitem[{\citenamefont{Fedorets} \emph{et~al.}(2004)\citenamefont{Fedorets,
  Gorelik, Shekhter, and Jonson}}]{Fedorets2004}
\bibinfo{author}{\bibnamefont{Fedorets}, \bibfnamefont{D.}},
  \bibinfo{author}{\bibfnamefont{L.~Y.} \bibnamefont{Gorelik}},
  \bibinfo{author}{\bibfnamefont{R.~I.} \bibnamefont{Shekhter}}, and
  \bibinfo{author}{\bibfnamefont{M.}~\bibnamefont{Jonson}},
  \bibinfo{year}{2004}, \bibinfo{journal}{Phys. Rev. Lett.}
  \textbf{\bibinfo{volume}{92}}, \bibinfo{pages}{166801}.

\bibitem[{\citenamefont{Feigel'man}
  \emph{et~al.}(2007)\citenamefont{Feigel'man, Ioffe, Kravtsov, and
  Yuzbashyan}}]{Feigel'man2007}
\bibinfo{author}{\bibnamefont{Feigel'man}, \bibfnamefont{M.~V.}},
  \bibinfo{author}{\bibfnamefont{L.~B.} \bibnamefont{Ioffe}},
  \bibinfo{author}{\bibfnamefont{V.~E.} \bibnamefont{Kravtsov}}, and
  \bibinfo{author}{\bibfnamefont{E.~A.} \bibnamefont{Yuzbashyan}},
  \bibinfo{year}{2007}, \bibinfo{journal}{Phys. Rev. Lett.}
  \textbf{\bibinfo{volume}{98}}, \bibinfo{pages}{027001}.

\bibitem[{\citenamefont{Feltin and Piquemal}(2009)}]{Feltin2009}
\bibinfo{author}{\bibnamefont{Feltin}, \bibfnamefont{N.}}, and
  \bibinfo{author}{\bibfnamefont{F.}~\bibnamefont{Piquemal}},
  \bibinfo{year}{2009}, \bibinfo{journal}{Eur. Phys. J. Spec. Top.}
  \textbf{\bibinfo{volume}{172}}, \bibinfo{pages}{267}.

\bibitem[{\citenamefont{Feltin} \emph{et~al.}(2011)\citenamefont{Feltin, Steck,
  Devoille, Sassine, Chenaud, Poirier, Schopfer, Sprengler, Djordjevic, Seron,
  and Piquemal}}]{Feltin2011}
\bibinfo{author}{\bibnamefont{Feltin}, \bibfnamefont{N.}},
  \bibinfo{author}{\bibfnamefont{B.}~\bibnamefont{Steck}},
  \bibinfo{author}{\bibfnamefont{L.}~\bibnamefont{Devoille}},
  \bibinfo{author}{\bibfnamefont{S.}~\bibnamefont{Sassine}},
  \bibinfo{author}{\bibfnamefont{B.}~\bibnamefont{Chenaud}},
  \bibinfo{author}{\bibfnamefont{W.}~\bibnamefont{Poirier}},
  \bibinfo{author}{\bibfnamefont{F.}~\bibnamefont{Schopfer}},
  \bibinfo{author}{\bibfnamefont{G.}~\bibnamefont{Sprengler}},
  \bibinfo{author}{\bibfnamefont{S.}~\bibnamefont{Djordjevic}},
  \bibinfo{author}{\bibfnamefont{O.}~\bibnamefont{Seron}}, and
  \bibinfo{author}{\bibfnamefont{F.}~\bibnamefont{Piquemal}},
  \bibinfo{year}{2011}, \bibinfo{journal}{Revue Francaise de M\'etrologie}
  \textbf{\bibinfo{volume}{2011-1}}, \bibinfo{pages}{3}.

\bibitem[{\citenamefont{F\`{e}ve} \emph{et~al.}(2007)\citenamefont{F\`{e}ve,
  Mah\'{e}, Berroir, Kontos, Pla\c{c}ais, Glattli, Cavanna, Etienne, and
  Jin}}]{Feve2007}
\bibinfo{author}{\bibnamefont{F\`{e}ve}, \bibfnamefont{G.}},
  \bibinfo{author}{\bibfnamefont{A.}~\bibnamefont{Mah\'{e}}},
  \bibinfo{author}{\bibfnamefont{J.-M.} \bibnamefont{Berroir}},
  \bibinfo{author}{\bibfnamefont{T.}~\bibnamefont{Kontos}},
  \bibinfo{author}{\bibfnamefont{B.}~\bibnamefont{Pla\c{c}ais}},
  \bibinfo{author}{\bibfnamefont{D.~C.} \bibnamefont{Glattli}},
  \bibinfo{author}{\bibfnamefont{A.}~\bibnamefont{Cavanna}},
  \bibinfo{author}{\bibfnamefont{B.}~\bibnamefont{Etienne}}, and
  \bibinfo{author}{\bibfnamefont{Y.}~\bibnamefont{Jin}}, \bibinfo{year}{2007},
  \bibinfo{journal}{Science} \textbf{\bibinfo{volume}{316}},
  \bibinfo{pages}{1169}.

\bibitem[{\citenamefont{Field} \emph{et~al.}(1993)\citenamefont{Field, Smith,
  Pepper, Ritchie, Frost, Jones, and Hasko}}]{Field1993}
\bibinfo{author}{\bibnamefont{Field}, \bibfnamefont{M.}},
  \bibinfo{author}{\bibfnamefont{C.}~\bibnamefont{Smith}},
  \bibinfo{author}{\bibfnamefont{M.}~\bibnamefont{Pepper}},
  \bibinfo{author}{\bibfnamefont{D.}~\bibnamefont{Ritchie}},
  \bibinfo{author}{\bibfnamefont{J.}~\bibnamefont{Frost}},
  \bibinfo{author}{\bibfnamefont{G.}~\bibnamefont{Jones}}, and
  \bibinfo{author}{\bibfnamefont{D.}~\bibnamefont{Hasko}},
  \bibinfo{year}{1993}, \bibinfo{journal}{Phys. Rev. Lett.}
  \textbf{\bibinfo{volume}{70}}, \bibinfo{pages}{1311}.

\bibitem[{\citenamefont{Fisher}(1986)}]{Fisher1986}
\bibinfo{author}{\bibnamefont{Fisher}, \bibfnamefont{M.~P.~A.}},
  \bibinfo{year}{1986}, \bibinfo{journal}{Phys. Rev. Lett.}
  \textbf{\bibinfo{volume}{57}}, \bibinfo{pages}{885}.

\bibitem[{\citenamefont{Flensberg} \emph{et~al.}(1999)\citenamefont{Flensberg,
  Niu, and Pustilnik}}]{Flensberg1999}
\bibinfo{author}{\bibnamefont{Flensberg}, \bibfnamefont{K.}},
  \bibinfo{author}{\bibfnamefont{Q.}~\bibnamefont{Niu}}, and
  \bibinfo{author}{\bibfnamefont{M.}~\bibnamefont{Pustilnik}},
  \bibinfo{year}{1999}, \bibinfo{journal}{Phys. Rev. B}
  \textbf{\bibinfo{volume}{60}}, \bibinfo{pages}{16291(R)}.

\bibitem[{\citenamefont{Fletcher} \emph{et~al.}(2011)\citenamefont{Fletcher,
  Kataoka, Giblin, Park, Sim, See, Janssen, Griffiths, Jones, Beere, and
  Ritchie}}]{Fletcher2011}
\bibinfo{author}{\bibnamefont{Fletcher}, \bibfnamefont{J.~D.}},
  \bibinfo{author}{\bibfnamefont{M.}~\bibnamefont{Kataoka}},
  \bibinfo{author}{\bibfnamefont{S.~P.} \bibnamefont{Giblin}},
  \bibinfo{author}{\bibfnamefont{S.}~\bibnamefont{Park}},
  \bibinfo{author}{\bibfnamefont{H.-S.} \bibnamefont{Sim}},
  \bibinfo{author}{\bibfnamefont{P.}~\bibnamefont{See}},
  \bibinfo{author}{\bibfnamefont{T.~J. B.~M.} \bibnamefont{Janssen}},
  \bibinfo{author}{\bibfnamefont{J.~P.} \bibnamefont{Griffiths}},
  \bibinfo{author}{\bibfnamefont{G.~A.~C.} \bibnamefont{Jones}},
  \bibinfo{author}{\bibfnamefont{H.~E.} \bibnamefont{Beere}}, and
  \bibinfo{author}{\bibfnamefont{D.~A.} \bibnamefont{Ritchie}},
  \bibinfo{year}{2011}, \bibinfo{journal}{{arXiv:1107.4560}} .

\bibitem[{\citenamefont{Flowers}(2004)}]{Flowers2004}
\bibinfo{author}{\bibnamefont{Flowers}, \bibfnamefont{J.}},
  \bibinfo{year}{2004}, \bibinfo{journal}{Science}
  \textbf{\bibinfo{volume}{306}}, \bibinfo{pages}{1324}.

\bibitem[{\citenamefont{F\"olling} \emph{et~al.}(2007)\citenamefont{F\"olling,
  Trotzky, Cheinet, Feld, Saers, Widera, M\"uller, and Bloch}}]{Folling2007}
\bibinfo{author}{\bibnamefont{F\"olling}, \bibfnamefont{S.}},
  \bibinfo{author}{\bibfnamefont{S.}~\bibnamefont{Trotzky}},
  \bibinfo{author}{\bibfnamefont{P.}~\bibnamefont{Cheinet}},
  \bibinfo{author}{\bibfnamefont{M.}~\bibnamefont{Feld}},
  \bibinfo{author}{\bibfnamefont{R.}~\bibnamefont{Saers}},
  \bibinfo{author}{\bibfnamefont{A.}~\bibnamefont{Widera}},
  \bibinfo{author}{\bibfnamefont{T.}~\bibnamefont{M\"uller}}, and
  \bibinfo{author}{\bibfnamefont{I.}~\bibnamefont{Bloch}},
  \bibinfo{year}{2007}, \bibinfo{journal}{Nature}
  \textbf{\bibinfo{volume}{448}}, \bibinfo{pages}{1029}.

\bibitem[{\citenamefont{Fricke} \emph{et~al.}(2011)\citenamefont{Fricke, Hohls,
  Ubbelohde, Kaestner, Kashcheyevs, Leicht, Mirovsky, Pierz, Schumacher, and
  Haug}}]{Fricke2011}
\bibinfo{author}{\bibnamefont{Fricke}, \bibfnamefont{L.}},
  \bibinfo{author}{\bibfnamefont{F.}~\bibnamefont{Hohls}},
  \bibinfo{author}{\bibfnamefont{N.}~\bibnamefont{Ubbelohde}},
  \bibinfo{author}{\bibfnamefont{B.}~\bibnamefont{Kaestner}},
  \bibinfo{author}{\bibfnamefont{V.}~\bibnamefont{Kashcheyevs}},
  \bibinfo{author}{\bibfnamefont{C.}~\bibnamefont{Leicht}},
  \bibinfo{author}{\bibfnamefont{P.}~\bibnamefont{Mirovsky}},
  \bibinfo{author}{\bibfnamefont{K.}~\bibnamefont{Pierz}},
  \bibinfo{author}{\bibfnamefont{H.~W.} \bibnamefont{Schumacher}}, and
  \bibinfo{author}{\bibfnamefont{R.~J.} \bibnamefont{Haug}},
  \bibinfo{year}{2011}, \bibinfo{journal}{Phys. Rev. B}
  \textbf{\bibinfo{volume}{83}}, \bibinfo{pages}{193306}.

\bibitem[{\citenamefont{Fujisawa} \emph{et~al.}(2004)\citenamefont{Fujisawa,
  Hayashi, Hirayama, Cheong, and Jeong}}]{Fujisawa2004}
\bibinfo{author}{\bibnamefont{Fujisawa}, \bibfnamefont{T.}},
  \bibinfo{author}{\bibfnamefont{T.}~\bibnamefont{Hayashi}},
  \bibinfo{author}{\bibfnamefont{Y.}~\bibnamefont{Hirayama}},
  \bibinfo{author}{\bibfnamefont{H.~D.} \bibnamefont{Cheong}}, and
  \bibinfo{author}{\bibfnamefont{Y.~H.} \bibnamefont{Jeong}},
  \bibinfo{year}{2004}, \bibinfo{journal}{Appl. Phys. Lett.}
  \textbf{\bibinfo{volume}{84}}, \bibinfo{pages}{2343}.

\bibitem[{\citenamefont{Fujiwara} \emph{et~al.}(2008)\citenamefont{Fujiwara,
  Nishiguchi, and Ono}}]{Fujiwara2008}
\bibinfo{author}{\bibnamefont{Fujiwara}, \bibfnamefont{A.}},
  \bibinfo{author}{\bibfnamefont{K.}~\bibnamefont{Nishiguchi}}, and
  \bibinfo{author}{\bibfnamefont{Y.}~\bibnamefont{Ono}}, \bibinfo{year}{2008},
  \bibinfo{journal}{Appl. Phys. Lett.} \textbf{\bibinfo{volume}{92}},
  \bibinfo{pages}{042102}.

\bibitem[{\citenamefont{Fujiwara and Takahashi}(2001)}]{Fujiwara2001}
\bibinfo{author}{\bibnamefont{Fujiwara}, \bibfnamefont{A.}}, and
  \bibinfo{author}{\bibfnamefont{Y.}~\bibnamefont{Takahashi}},
  \bibinfo{year}{2001}, \bibinfo{journal}{Nature}
  \textbf{\bibinfo{volume}{410}}, \bibinfo{pages}{560}.

\bibitem[{\citenamefont{Fujiwara} \emph{et~al.}(2004)\citenamefont{Fujiwara,
  Zimmerman, Ono, and Takahashi}}]{Fujiwara2004}
\bibinfo{author}{\bibnamefont{Fujiwara}, \bibfnamefont{A.}},
  \bibinfo{author}{\bibfnamefont{N.~M.} \bibnamefont{Zimmerman}},
  \bibinfo{author}{\bibfnamefont{Y.}~\bibnamefont{Ono}}, and
  \bibinfo{author}{\bibfnamefont{Y.}~\bibnamefont{Takahashi}},
  \bibinfo{year}{2004}, \bibinfo{journal}{Appl. Phys. Lett.}
  \textbf{\bibinfo{volume}{84}}, \bibinfo{pages}{1323}.

\bibitem[{\citenamefont{Fulton}(1973)}]{Fulton1973}
\bibinfo{author}{\bibnamefont{Fulton}, \bibfnamefont{T.~A.}},
  \bibinfo{year}{1973}, \bibinfo{journal}{Phys. Rev. B}
  \textbf{\bibinfo{volume}{7}}, \bibinfo{pages}{981}.

\bibitem[{\citenamefont{Fulton and Dolan}(1987)}]{Fulton1987}
\bibinfo{author}{\bibnamefont{Fulton}, \bibfnamefont{T.~A.}}, and
  \bibinfo{author}{\bibfnamefont{G.}~\bibnamefont{Dolan}},
  \bibinfo{year}{1987}, \bibinfo{journal}{Physical Review Letters}
  \textbf{\bibinfo{volume}{59}}, \bibinfo{pages}{109}.

\bibitem[{\citenamefont{Funck and Sienknecht}(1991)}]{Funck1991}
\bibinfo{author}{\bibnamefont{Funck}, \bibfnamefont{T.}}, and
  \bibinfo{author}{\bibfnamefont{V.}~\bibnamefont{Sienknecht}},
  \bibinfo{year}{1991}, \bibinfo{journal}{IEEE Trans. Instrum. Meas.}
  \textbf{\bibinfo{volume}{40}}, \bibinfo{pages}{158}.

\bibitem[{\citenamefont{Gabelli} \emph{et~al.}(2006)\citenamefont{Gabelli,
  F\`{e}ve, Berroir, Pla\c{c}ais, Cavanna, Etienne, Jin, and
  Glattli}}]{Gabelli2006}
\bibinfo{author}{\bibnamefont{Gabelli}, \bibfnamefont{J.}},
  \bibinfo{author}{\bibfnamefont{G.}~\bibnamefont{F\`{e}ve}},
  \bibinfo{author}{\bibfnamefont{J.-M.} \bibnamefont{Berroir}},
  \bibinfo{author}{\bibfnamefont{B.}~\bibnamefont{Pla\c{c}ais}},
  \bibinfo{author}{\bibfnamefont{A.}~\bibnamefont{Cavanna}},
  \bibinfo{author}{\bibfnamefont{B.}~\bibnamefont{Etienne}},
  \bibinfo{author}{\bibfnamefont{Y.}~\bibnamefont{Jin}}, and
  \bibinfo{author}{\bibfnamefont{D.~C.} \bibnamefont{Glattli}},
  \bibinfo{year}{2006}, \bibinfo{journal}{Science}
  \textbf{\bibinfo{volume}{313}}, \bibinfo{pages}{499}.

\bibitem[{\citenamefont{Gallop}(2005)}]{Gallop2005}
\bibinfo{author}{\bibnamefont{Gallop}, \bibfnamefont{J.~C.}},
  \bibinfo{year}{2005}, \bibinfo{journal}{Phil. Trans. Royal Soc. A}
  \textbf{\bibinfo{volume}{363}}, \bibinfo{pages}{2221}.

\bibitem[{\citenamefont{Gallop and Piquemal}(2006)}]{Gallop2006}
\bibinfo{author}{\bibnamefont{Gallop}, \bibfnamefont{J.~C.}}, and
  \bibinfo{author}{\bibfnamefont{F.}~\bibnamefont{Piquemal}},
  \bibinfo{year}{2006}, \emph{\bibinfo{title}{SQUIDs for Standards and
  Metrology}} (\bibinfo{publisher}{Wiley-VCH Verlag GmbH \& Co. KGaA,
  Weinheim}), chapter~\bibinfo{chapter}{4}, p.~\bibinfo{pages}{95}.

\bibitem[{\citenamefont{Gasparinetti}
  \emph{et~al.}(2012)\citenamefont{Gasparinetti, Solinas, Yoon, and
  Pekola}}]{Gasparinetti2012}
\bibinfo{author}{\bibnamefont{Gasparinetti}, \bibfnamefont{S.}},
  \bibinfo{author}{\bibfnamefont{P.}~\bibnamefont{Solinas}},
  \bibinfo{author}{\bibfnamefont{Y.}~\bibnamefont{Yoon}}, and
  \bibinfo{author}{\bibfnamefont{J.~P.} \bibnamefont{Pekola}},
  \bibinfo{year}{2012}, \bibinfo{journal}{arXiv:1206.0193} .

\bibitem[{\citenamefont{Geerligs}
  \emph{et~al.}(1990{\natexlab{a}})\citenamefont{Geerligs, Averin, and
  Mooij}}]{Geerligs1990c}
\bibinfo{author}{\bibnamefont{Geerligs}, \bibfnamefont{L.}},
  \bibinfo{author}{\bibfnamefont{D.}~\bibnamefont{Averin}}, and
  \bibinfo{author}{\bibfnamefont{J.}~\bibnamefont{Mooij}},
  \bibinfo{year}{1990}{\natexlab{a}}, \bibinfo{journal}{Phys. Rev. Lett.}
  \textbf{\bibinfo{volume}{65}}, \bibinfo{pages}{3037}.

\bibitem[{\citenamefont{Geerligs}
  \emph{et~al.}(1990{\natexlab{b}})\citenamefont{Geerligs, Anderegg, Holweg,
  Mooij, Pothier, Esteve, Urbina, and Devoret}}]{Geerligs1990}
\bibinfo{author}{\bibnamefont{Geerligs}, \bibfnamefont{L.~J.}},
  \bibinfo{author}{\bibfnamefont{V.~F.} \bibnamefont{Anderegg}},
  \bibinfo{author}{\bibfnamefont{P.~A.~M.} \bibnamefont{Holweg}},
  \bibinfo{author}{\bibfnamefont{J.~E.} \bibnamefont{Mooij}},
  \bibinfo{author}{\bibfnamefont{H.}~\bibnamefont{Pothier}},
  \bibinfo{author}{\bibfnamefont{D.}~\bibnamefont{Esteve}},
  \bibinfo{author}{\bibfnamefont{C.}~\bibnamefont{Urbina}}, and
  \bibinfo{author}{\bibfnamefont{M.~H.} \bibnamefont{Devoret}},
  \bibinfo{year}{1990}{\natexlab{b}}, \bibinfo{journal}{Phys. Rev. Lett.}
  \textbf{\bibinfo{volume}{64}}, \bibinfo{pages}{2691}.

\bibitem[{\citenamefont{Geerligs} \emph{et~al.}(1991)\citenamefont{Geerligs,
  Verbrugh, Hadley, Mooij, Pothier, Lafarge, Urbina, Esteve, and
  Devoret}}]{Geerligs1991}
\bibinfo{author}{\bibnamefont{Geerligs}, \bibfnamefont{L.~J.}},
  \bibinfo{author}{\bibfnamefont{S.~M.} \bibnamefont{Verbrugh}},
  \bibinfo{author}{\bibfnamefont{P.}~\bibnamefont{Hadley}},
  \bibinfo{author}{\bibfnamefont{J.~E.} \bibnamefont{Mooij}},
  \bibinfo{author}{\bibfnamefont{H.}~\bibnamefont{Pothier}},
  \bibinfo{author}{\bibfnamefont{P.}~\bibnamefont{Lafarge}},
  \bibinfo{author}{\bibfnamefont{C.}~\bibnamefont{Urbina}},
  \bibinfo{author}{\bibfnamefont{D.}~\bibnamefont{Esteve}}, and
  \bibinfo{author}{\bibfnamefont{M.~H.} \bibnamefont{Devoret}},
  \bibinfo{year}{1991}, \bibinfo{journal}{Z. Phys. B}
  \textbf{\bibinfo{volume}{85}}, \bibinfo{pages}{349}.

\bibitem[{\citenamefont{Giaever}(1960)}]{Giaever1960}
\bibinfo{author}{\bibnamefont{Giaever}, \bibfnamefont{I.}},
  \bibinfo{year}{1960}, \bibinfo{journal}{Phys. Rev. Lett.}
  \textbf{\bibinfo{volume}{5}}, \bibinfo{pages}{147}.

\bibitem[{\citenamefont{Giazotto} \emph{et~al.}(2006)\citenamefont{Giazotto,
  Heikkil\"{a}, Luukanen, Savin, and Pekola}}]{Giazotto2006}
\bibinfo{author}{\bibnamefont{Giazotto}, \bibfnamefont{F.}},
  \bibinfo{author}{\bibfnamefont{T.~T.} \bibnamefont{Heikkil\"{a}}},
  \bibinfo{author}{\bibfnamefont{A.}~\bibnamefont{Luukanen}},
  \bibinfo{author}{\bibfnamefont{A.~M.} \bibnamefont{Savin}}, and
  \bibinfo{author}{\bibfnamefont{J.~P.} \bibnamefont{Pekola}},
  \bibinfo{year}{2006}, \bibinfo{journal}{Rev. Mod. Phys.}
  \textbf{\bibinfo{volume}{78}}, \bibinfo{pages}{217}.

\bibitem[{\citenamefont{Giazotto} \emph{et~al.}(2011)\citenamefont{Giazotto,
  Spathis, Roddaro, Biswas, Taddei, Governale, and Sorba}}]{Giazotto2011}
\bibinfo{author}{\bibnamefont{Giazotto}, \bibfnamefont{F.}},
  \bibinfo{author}{\bibfnamefont{P.}~\bibnamefont{Spathis}},
  \bibinfo{author}{\bibfnamefont{S.}~\bibnamefont{Roddaro}},
  \bibinfo{author}{\bibfnamefont{S.}~\bibnamefont{Biswas}},
  \bibinfo{author}{\bibfnamefont{F.}~\bibnamefont{Taddei}},
  \bibinfo{author}{\bibfnamefont{M.}~\bibnamefont{Governale}}, and
  \bibinfo{author}{\bibfnamefont{L.}~\bibnamefont{Sorba}},
  \bibinfo{year}{2011}, \bibinfo{journal}{Nature Phys.}
  \textbf{\bibinfo{volume}{7}}, \bibinfo{pages}{857}.

\bibitem[{\citenamefont{Giblin} \emph{et~al.}(2012)\citenamefont{Giblin,
  Kataoka, Fletcher, See, Janssen, Griffiths, Jones, Farrer, and
  Ritchie}}]{Giblin2012}
\bibinfo{author}{\bibnamefont{Giblin}, \bibfnamefont{S.~P.}},
  \bibinfo{author}{\bibfnamefont{M.}~\bibnamefont{Kataoka}},
  \bibinfo{author}{\bibfnamefont{J.~D.} \bibnamefont{Fletcher}},
  \bibinfo{author}{\bibfnamefont{P.}~\bibnamefont{See}},
  \bibinfo{author}{\bibfnamefont{T.~J. B.~M.} \bibnamefont{Janssen}},
  \bibinfo{author}{\bibfnamefont{J.~P.} \bibnamefont{Griffiths}},
  \bibinfo{author}{\bibfnamefont{G.~A.~C.} \bibnamefont{Jones}},
  \bibinfo{author}{\bibfnamefont{I.}~\bibnamefont{Farrer}}, and
  \bibinfo{author}{\bibfnamefont{D.~A.} \bibnamefont{Ritchie}},
  \bibinfo{year}{2012}, \bibinfo{journal}{Nature Comm.}
  \textbf{\bibinfo{volume}{3}}, \bibinfo{pages}{930}.

\bibitem[{\citenamefont{Giblin} \emph{et~al.}(2010)\citenamefont{Giblin,
  Wright, Fletcher, Kataoka, Pepper, Janssen, Ritchie, Nicoll, Anderson, and
  Jones}}]{Giblin2010}
\bibinfo{author}{\bibnamefont{Giblin}, \bibfnamefont{S.~P.}},
  \bibinfo{author}{\bibfnamefont{S.~J.} \bibnamefont{Wright}},
  \bibinfo{author}{\bibfnamefont{J.~D.} \bibnamefont{Fletcher}},
  \bibinfo{author}{\bibfnamefont{M.}~\bibnamefont{Kataoka}},
  \bibinfo{author}{\bibfnamefont{M.}~\bibnamefont{Pepper}},
  \bibinfo{author}{\bibfnamefont{T.~J. B.~M.} \bibnamefont{Janssen}},
  \bibinfo{author}{\bibfnamefont{D.~A.} \bibnamefont{Ritchie}},
  \bibinfo{author}{\bibfnamefont{C.~A.} \bibnamefont{Nicoll}},
  \bibinfo{author}{\bibfnamefont{D.}~\bibnamefont{Anderson}}, and
  \bibinfo{author}{\bibfnamefont{G.~A.~C.} \bibnamefont{Jones}},
  \bibinfo{year}{2010}, \bibinfo{journal}{New J. Phys.}
  \textbf{\bibinfo{volume}{12}}, \bibinfo{pages}{073013}.

\bibitem[{\citenamefont{Girvin} \emph{et~al.}(1990)\citenamefont{Girvin,
  Glazman, Jonson, Penn, and Stiles}}]{Girvin1990}
\bibinfo{author}{\bibnamefont{Girvin}, \bibfnamefont{S.~M.}},
  \bibinfo{author}{\bibfnamefont{L.~I.} \bibnamefont{Glazman}},
  \bibinfo{author}{\bibfnamefont{M.}~\bibnamefont{Jonson}},
  \bibinfo{author}{\bibfnamefont{D.~R.} \bibnamefont{Penn}}, and
  \bibinfo{author}{\bibfnamefont{M.~D.} \bibnamefont{Stiles}},
  \bibinfo{year}{1990}, \bibinfo{journal}{Phys. Rev. Lett.}
  \textbf{\bibinfo{volume}{64}}, \bibinfo{pages}{3183}.

\bibitem[{\citenamefont{Gl\"aser} \emph{et~al.}(2010)\citenamefont{Gl\"aser,
  Borys, Ratschko, and Schwartz}}]{Glaser2010}
\bibinfo{author}{\bibnamefont{Gl\"aser}, \bibfnamefont{M.}},
  \bibinfo{author}{\bibfnamefont{M.}~\bibnamefont{Borys}},
  \bibinfo{author}{\bibfnamefont{D.}~\bibnamefont{Ratschko}}, and
  \bibinfo{author}{\bibfnamefont{R.}~\bibnamefont{Schwartz}},
  \bibinfo{year}{2010}, \bibinfo{journal}{Metrologia}
  \textbf{\bibinfo{volume}{47}}, \bibinfo{pages}{419}.

\bibitem[{\citenamefont{Goerbig}(2011)}]{Goerbig2011}
\bibinfo{author}{\bibnamefont{Goerbig}, \bibfnamefont{M.~O.}},
  \bibinfo{year}{2011}, \bibinfo{journal}{Comptes Rendus Physique}
  \textbf{\bibinfo{volume}{12}}, \bibinfo{pages}{369}.

\bibitem[{\citenamefont{Gorelik} \emph{et~al.}(2001)\citenamefont{Gorelik,
  Isacsson, Galperin, Shekhter, and Jonson}}]{Gorelik2001}
\bibinfo{author}{\bibnamefont{Gorelik}, \bibfnamefont{L.~Y.}},
  \bibinfo{author}{\bibfnamefont{A.}~\bibnamefont{Isacsson}},
  \bibinfo{author}{\bibfnamefont{Y.~M.} \bibnamefont{Galperin}},
  \bibinfo{author}{\bibfnamefont{R.~I.} \bibnamefont{Shekhter}}, and
  \bibinfo{author}{\bibfnamefont{M.}~\bibnamefont{Jonson}},
  \bibinfo{year}{2001}, \bibinfo{journal}{Nature}
  \textbf{\bibinfo{volume}{411}}, \bibinfo{pages}{454}.

\bibitem[{\citenamefont{Gorelik} \emph{et~al.}(1998)\citenamefont{Gorelik,
  Isacsson, Voinova, Kasemo, Shekhter, and Jonson}}]{Gorelik1998}
\bibinfo{author}{\bibnamefont{Gorelik}, \bibfnamefont{L.~Y.}},
  \bibinfo{author}{\bibfnamefont{A.}~\bibnamefont{Isacsson}},
  \bibinfo{author}{\bibfnamefont{M.~V.} \bibnamefont{Voinova}},
  \bibinfo{author}{\bibfnamefont{B.}~\bibnamefont{Kasemo}},
  \bibinfo{author}{\bibfnamefont{R.~I.} \bibnamefont{Shekhter}}, and
  \bibinfo{author}{\bibfnamefont{M.}~\bibnamefont{Jonson}},
  \bibinfo{year}{1998}, \bibinfo{journal}{Phys. Rev. Lett.}
  \textbf{\bibinfo{volume}{80}}, \bibinfo{pages}{4526}.

\bibitem[{\citenamefont{Greibe} \emph{et~al.}(2011)\citenamefont{Greibe,
  Stenberg, Wilson, Bauch, Shumeiko, and Delsing}}]{Greibe2011}
\bibinfo{author}{\bibnamefont{Greibe}, \bibfnamefont{T.}},
  \bibinfo{author}{\bibfnamefont{M.~P.~V.} \bibnamefont{Stenberg}},
  \bibinfo{author}{\bibfnamefont{C.~M.} \bibnamefont{Wilson}},
  \bibinfo{author}{\bibfnamefont{T.}~\bibnamefont{Bauch}},
  \bibinfo{author}{\bibfnamefont{V.~S.} \bibnamefont{Shumeiko}}, and
  \bibinfo{author}{\bibfnamefont{P.}~\bibnamefont{Delsing}},
  \bibinfo{year}{2011}, \bibinfo{journal}{Phys. Rev. Lett.}
  \textbf{\bibinfo{volume}{106}}, \bibinfo{pages}{097001}.

\bibitem[{\citenamefont{Gustavsson}
  \emph{et~al.}(2007)\citenamefont{Gustavsson, Leturcq, Ihn, Ensslin, Reinwald,
  and Wegscheider}}]{Gustavsson2007}
\bibinfo{author}{\bibnamefont{Gustavsson}, \bibfnamefont{S.}},
  \bibinfo{author}{\bibfnamefont{R.}~\bibnamefont{Leturcq}},
  \bibinfo{author}{\bibfnamefont{T.}~\bibnamefont{Ihn}},
  \bibinfo{author}{\bibfnamefont{K.}~\bibnamefont{Ensslin}},
  \bibinfo{author}{\bibfnamefont{M.}~\bibnamefont{Reinwald}}, and
  \bibinfo{author}{\bibfnamefont{W.}~\bibnamefont{Wegscheider}},
  \bibinfo{year}{2007}, \bibinfo{journal}{Phys. Rev. B}
  \textbf{\bibinfo{volume}{75}}, \bibinfo{pages}{075314}.

\bibitem[{\citenamefont{Gustavsson}
  \emph{et~al.}(2009)\citenamefont{Gustavsson, Leturcq, Studer, Shorubalko,
  Ihn, Ensslin, Driscoll, and Gossard}}]{Gustavsson2009}
\bibinfo{author}{\bibnamefont{Gustavsson}, \bibfnamefont{S.}},
  \bibinfo{author}{\bibfnamefont{R.}~\bibnamefont{Leturcq}},
  \bibinfo{author}{\bibfnamefont{M.}~\bibnamefont{Studer}},
  \bibinfo{author}{\bibfnamefont{I.}~\bibnamefont{Shorubalko}},
  \bibinfo{author}{\bibfnamefont{T.}~\bibnamefont{Ihn}},
  \bibinfo{author}{\bibfnamefont{K.}~\bibnamefont{Ensslin}},
  \bibinfo{author}{\bibfnamefont{D.~C.} \bibnamefont{Driscoll}}, and
  \bibinfo{author}{\bibfnamefont{A.~C.} \bibnamefont{Gossard}},
  \bibinfo{year}{2009}, \bibinfo{journal}{Surf. Sci. Rep.}
  \textbf{\bibinfo{volume}{64}}, \bibinfo{pages}{191}.

\bibitem[{\citenamefont{Gustavsson}
  \emph{et~al.}(2006)\citenamefont{Gustavsson, Leturcq,
  Simovi\ifmmode~\check{c}\else \v{c}\fi{}, Schleser, Ihn, Studerus, Ensslin,
  Driscoll, and Gossard}}]{Gustavsson2006}
\bibinfo{author}{\bibnamefont{Gustavsson}, \bibfnamefont{S.}},
  \bibinfo{author}{\bibfnamefont{R.}~\bibnamefont{Leturcq}},
  \bibinfo{author}{\bibfnamefont{B.}~\bibnamefont{Simovi\ifmmode~\check{c}\else
  \v{c}\fi{}}}, \bibinfo{author}{\bibfnamefont{R.}~\bibnamefont{Schleser}},
  \bibinfo{author}{\bibfnamefont{T.}~\bibnamefont{Ihn}},
  \bibinfo{author}{\bibfnamefont{P.}~\bibnamefont{Studerus}},
  \bibinfo{author}{\bibfnamefont{K.}~\bibnamefont{Ensslin}},
  \bibinfo{author}{\bibfnamefont{D.~C.} \bibnamefont{Driscoll}}, and
  \bibinfo{author}{\bibfnamefont{A.~C.} \bibnamefont{Gossard}},
  \bibinfo{year}{2006}, \bibinfo{journal}{Phys. Rev. Lett.}
  \textbf{\bibinfo{volume}{96}}, \bibinfo{pages}{076605}.

\bibitem[{\citenamefont{Gustavsson}
  \emph{et~al.}(2008)\citenamefont{Gustavsson, Shorubalko, Leturcq, Schon, and
  Ensslin}}]{Gustavsson2008}
\bibinfo{author}{\bibnamefont{Gustavsson}, \bibfnamefont{S.}},
  \bibinfo{author}{\bibfnamefont{I.}~\bibnamefont{Shorubalko}},
  \bibinfo{author}{\bibfnamefont{R.}~\bibnamefont{Leturcq}},
  \bibinfo{author}{\bibfnamefont{S.}~\bibnamefont{Schon}}, and
  \bibinfo{author}{\bibfnamefont{K.}~\bibnamefont{Ensslin}},
  \bibinfo{year}{2008}, \bibinfo{journal}{Appl. Phys. Lett.}
  \textbf{\bibinfo{volume}{92}}, \bibinfo{pages}{152101}.

\bibitem[{\citenamefont{Hamilton} \emph{et~al.}(1995)\citenamefont{Hamilton,
  Burroughs, and Kautz}}]{Hamilton1995}
\bibinfo{author}{\bibnamefont{Hamilton}, \bibfnamefont{C.}},
  \bibinfo{author}{\bibfnamefont{C.}~\bibnamefont{Burroughs}}, and
  \bibinfo{author}{\bibfnamefont{R.}~\bibnamefont{Kautz}},
  \bibinfo{year}{1995}, \bibinfo{journal}{IEEE Trans. Instrum. Meas.}
  \textbf{\bibinfo{volume}{44}}, \bibinfo{pages}{223}.

\bibitem[{\citenamefont{Hanneke} \emph{et~al.}(2008)\citenamefont{Hanneke,
  Fogwell, and Gabrielse}}]{Hanneke2008}
\bibinfo{author}{\bibnamefont{Hanneke}, \bibfnamefont{D.}},
  \bibinfo{author}{\bibfnamefont{S.}~\bibnamefont{Fogwell}}, and
  \bibinfo{author}{\bibfnamefont{G.}~\bibnamefont{Gabrielse}},
  \bibinfo{year}{2008}, \bibinfo{journal}{Phys. Rev. Lett.}
  \textbf{\bibinfo{volume}{100}}, \bibinfo{pages}{120801}.

\bibitem[{\citenamefont{Hanson} \emph{et~al.}(2007)\citenamefont{Hanson,
  Kouwenhoven, Petta, Tarucha, and Vandersypen}}]{Hanson2007}
\bibinfo{author}{\bibnamefont{Hanson}, \bibfnamefont{R.}},
  \bibinfo{author}{\bibfnamefont{L.~P.} \bibnamefont{Kouwenhoven}},
  \bibinfo{author}{\bibfnamefont{J.~R.} \bibnamefont{Petta}},
  \bibinfo{author}{\bibfnamefont{S.}~\bibnamefont{Tarucha}}, and
  \bibinfo{author}{\bibfnamefont{L.~M.~K.} \bibnamefont{Vandersypen}},
  \bibinfo{year}{2007}, \bibinfo{journal}{Rev. Mod. Phys.}
  \textbf{\bibinfo{volume}{79}}, \bibinfo{pages}{1217}.

\bibitem[{\citenamefont{Harbusch} \emph{et~al.}(2010)\citenamefont{Harbusch,
  Taubert, Tranitz, Wegscheider, and Ludwig}}]{Harbusch2010}
\bibinfo{author}{\bibnamefont{Harbusch}, \bibfnamefont{D.}},
  \bibinfo{author}{\bibfnamefont{D.}~\bibnamefont{Taubert}},
  \bibinfo{author}{\bibfnamefont{H.~P.} \bibnamefont{Tranitz}},
  \bibinfo{author}{\bibfnamefont{W.}~\bibnamefont{Wegscheider}}, and
  \bibinfo{author}{\bibfnamefont{S.}~\bibnamefont{Ludwig}},
  \bibinfo{year}{2010}, \bibinfo{journal}{Phys. Rev. Lett.}
  \textbf{\bibinfo{volume}{104}}, \bibinfo{pages}{196801}.

\bibitem[{\citenamefont{Hartland} \emph{et~al.}(1991)\citenamefont{Hartland,
  Jones, Williams, Gallagher, and Galloway}}]{Hartland1991}
\bibinfo{author}{\bibnamefont{Hartland}, \bibfnamefont{A.}},
  \bibinfo{author}{\bibfnamefont{K.}~\bibnamefont{Jones}},
  \bibinfo{author}{\bibfnamefont{J.~M.} \bibnamefont{Williams}},
  \bibinfo{author}{\bibfnamefont{B.~L.} \bibnamefont{Gallagher}}, and
  \bibinfo{author}{\bibfnamefont{T.}~\bibnamefont{Galloway}},
  \bibinfo{year}{1991}, \bibinfo{journal}{Phys. Rev. Lett.}
  \textbf{\bibinfo{volume}{66}}, \bibinfo{pages}{969}.

\bibitem[{\citenamefont{Hartle} \emph{et~al.}(1971)\citenamefont{Hartle,
  Scalapino, and Sugar}}]{Hartle1971}
\bibinfo{author}{\bibnamefont{Hartle}, \bibfnamefont{J.~B.}},
  \bibinfo{author}{\bibfnamefont{D.~J.} \bibnamefont{Scalapino}}, and
  \bibinfo{author}{\bibfnamefont{R.~L.} \bibnamefont{Sugar}},
  \bibinfo{year}{1971}, \bibinfo{journal}{Phys. Rev. B}
  \textbf{\bibinfo{volume}{3}}, \bibinfo{pages}{1778}.

\bibitem[{\citenamefont{Harvey}(1972)}]{Harvey1972}
\bibinfo{author}{\bibnamefont{Harvey}, \bibfnamefont{I.~K.}},
  \bibinfo{year}{1972}, \bibinfo{journal}{Rev. Sci. Instrum.}
  \textbf{\bibinfo{volume}{43}}, \bibinfo{pages}{1626}.

\bibitem[{\citenamefont{Hekking and Nazarov}(1994)}]{Hekking1994}
\bibinfo{author}{\bibnamefont{Hekking}, \bibfnamefont{F.~W.~J.}}, and
  \bibinfo{author}{\bibfnamefont{Y.~V.} \bibnamefont{Nazarov}},
  \bibinfo{year}{1994}, \bibinfo{journal}{Phys. Rev. B}
  \textbf{\bibinfo{volume}{49}}, \bibinfo{pages}{6847}.

\bibitem[{\citenamefont{Held} \emph{et~al.}(1997)\citenamefont{Held, Heinzel,
  Studerus, Ensslin, and Holland}}]{Held1997}
\bibinfo{author}{\bibnamefont{Held}, \bibfnamefont{R.}},
  \bibinfo{author}{\bibfnamefont{T.}~\bibnamefont{Heinzel}},
  \bibinfo{author}{\bibfnamefont{P.}~\bibnamefont{Studerus}},
  \bibinfo{author}{\bibfnamefont{K.}~\bibnamefont{Ensslin}}, and
  \bibinfo{author}{\bibfnamefont{M.}~\bibnamefont{Holland}},
  \bibinfo{year}{1997}, \bibinfo{journal}{Apl. Phys. Lett.}
  \textbf{\bibinfo{volume}{71}}, \bibinfo{pages}{2689}.

\bibitem[{\citenamefont{Hergenrother}
  \emph{et~al.}(1995)\citenamefont{Hergenrother, Lu, Tuominen, Ralph, and
  Tinkham}}]{Hergenrother1995}
\bibinfo{author}{\bibnamefont{Hergenrother}, \bibfnamefont{J.~M.}},
  \bibinfo{author}{\bibfnamefont{J.~G.} \bibnamefont{Lu}},
  \bibinfo{author}{\bibfnamefont{M.~T.} \bibnamefont{Tuominen}},
  \bibinfo{author}{\bibfnamefont{D.~C.} \bibnamefont{Ralph}}, and
  \bibinfo{author}{\bibfnamefont{M.}~\bibnamefont{Tinkham}},
  \bibinfo{year}{1995}, \bibinfo{journal}{Phys. Rev. B}
  \textbf{\bibinfo{volume}{51}}, \bibinfo{pages}{9407}.

\bibitem[{\citenamefont{Herrmann} \emph{et~al.}(2010)\citenamefont{Herrmann,
  Portier, Roche, Yeyati, Kontos, and Strunk}}]{Herrmann2010}
\bibinfo{author}{\bibnamefont{Herrmann}, \bibfnamefont{L.~G.}},
  \bibinfo{author}{\bibfnamefont{F.}~\bibnamefont{Portier}},
  \bibinfo{author}{\bibfnamefont{P.}~\bibnamefont{Roche}},
  \bibinfo{author}{\bibfnamefont{A.~L.} \bibnamefont{Yeyati}},
  \bibinfo{author}{\bibfnamefont{T.}~\bibnamefont{Kontos}}, and
  \bibinfo{author}{\bibfnamefont{C.}~\bibnamefont{Strunk}},
  \bibinfo{year}{2010}, \bibinfo{journal}{Phys. Rev. Lett.}
  \textbf{\bibinfo{volume}{104}}, \bibinfo{pages}{026801}.

\bibitem[{\citenamefont{Hill} \emph{et~al.}(2011)\citenamefont{Hill, Miller,
  and Censullo}}]{Hill2011}
\bibinfo{author}{\bibnamefont{Hill}, \bibfnamefont{T.~P.}},
  \bibinfo{author}{\bibfnamefont{J.}~\bibnamefont{Miller}}, and
  \bibinfo{author}{\bibfnamefont{A.~C.} \bibnamefont{Censullo}},
  \bibinfo{year}{2011}, \bibinfo{journal}{Metrologia}
  \textbf{\bibinfo{volume}{48}}, \bibinfo{pages}{83}.

\bibitem[{\citenamefont{Hoehne} \emph{et~al.}(2012)\citenamefont{Hoehne,
  Pashkin, Astafiev, M\"ott\"onen, Pekola, and Tsai}}]{Hoehne2012}
\bibinfo{author}{\bibnamefont{Hoehne}, \bibfnamefont{F.}},
  \bibinfo{author}{\bibfnamefont{Y.~A.} \bibnamefont{Pashkin}},
  \bibinfo{author}{\bibfnamefont{O.~V.} \bibnamefont{Astafiev}},
  \bibinfo{author}{\bibfnamefont{M.}~\bibnamefont{M\"ott\"onen}},
  \bibinfo{author}{\bibfnamefont{J.~P.} \bibnamefont{Pekola}}, and
  \bibinfo{author}{\bibfnamefont{J.~S.} \bibnamefont{Tsai}},
  \bibinfo{year}{2012}, \bibinfo{journal}{Phys. Rev. B}
  \textbf{\bibinfo{volume}{85}}, \bibinfo{pages}{140504}.

\bibitem[{\citenamefont{Hofstetter}
  \emph{et~al.}(2009)\citenamefont{Hofstetter, Csonka, and
  Nygård}}]{Hofstetter2009}
\bibinfo{author}{\bibnamefont{Hofstetter}, \bibfnamefont{L.}},
  \bibinfo{author}{\bibfnamefont{S.}~\bibnamefont{Csonka}}, and
  \bibinfo{author}{\bibfnamefont{J.~a. S.~C.} \bibnamefont{Nygård}},
  \bibinfo{year}{2009}, \bibinfo{journal}{Nature}
  \textbf{\bibinfo{volume}{461}}, \bibinfo{pages}{960}.

\bibitem[{\citenamefont{Hohls} \emph{et~al.}(2011)\citenamefont{Hohls, Welker,
  Leicht, Fricke, Kaestner, Mirovsky, M\"uller, Pierz, Siegner, and
  Schumacher}}]{Hohls2011}
\bibinfo{author}{\bibnamefont{Hohls}, \bibfnamefont{F.}},
  \bibinfo{author}{\bibfnamefont{A.~C.} \bibnamefont{Welker}},
  \bibinfo{author}{\bibfnamefont{C.}~\bibnamefont{Leicht}},
  \bibinfo{author}{\bibfnamefont{L.}~\bibnamefont{Fricke}},
  \bibinfo{author}{\bibfnamefont{B.}~\bibnamefont{Kaestner}},
  \bibinfo{author}{\bibfnamefont{P.}~\bibnamefont{Mirovsky}},
  \bibinfo{author}{\bibfnamefont{A.}~\bibnamefont{M\"uller}},
  \bibinfo{author}{\bibfnamefont{K.}~\bibnamefont{Pierz}},
  \bibinfo{author}{\bibfnamefont{U.}~\bibnamefont{Siegner}}, and
  \bibinfo{author}{\bibfnamefont{H.~W.} \bibnamefont{Schumacher}},
  \bibinfo{year}{2011}, \bibinfo{journal}{arXiv:1103.1746} .

\bibitem[{\citenamefont{Hollenberg}
  \emph{et~al.}(2006)\citenamefont{Hollenberg, Greentree, Fowler, and
  Wellard}}]{Hollenberg2006}
\bibinfo{author}{\bibnamefont{Hollenberg}, \bibfnamefont{L.~C.~L.}},
  \bibinfo{author}{\bibfnamefont{A.~D.} \bibnamefont{Greentree}},
  \bibinfo{author}{\bibfnamefont{A.~G.} \bibnamefont{Fowler}}, and
  \bibinfo{author}{\bibfnamefont{C.~J.} \bibnamefont{Wellard}},
  \bibinfo{year}{2006}, \bibinfo{journal}{Phys. Rev. B}
  \textbf{\bibinfo{volume}{74}}, \bibinfo{pages}{045311}.

\bibitem[{\citenamefont{Ingold and Nazarov}(1992)}]{Ingold1992}
\bibinfo{author}{\bibnamefont{Ingold}, \bibfnamefont{G.~L.}}, and
  \bibinfo{author}{\bibfnamefont{Y.~V.} \bibnamefont{Nazarov}},
  \bibinfo{year}{1992}, \emph{\bibinfo{title}{Single charge tunneling}}
  (\bibinfo{publisher}{Plenum Press}, \bibinfo{address}{New York}), volume
  \bibinfo{volume}{294} of \emph{\bibinfo{series}{NATO ASI Series B}},
  p.~\bibinfo{pages}{21}.

\bibitem[{\citenamefont{Isacsson} \emph{et~al.}(1998)\citenamefont{Isacsson,
  Gorelik, Voinova, Kasemo, Shekhter, and Jonson}}]{Isacsson1998}
\bibinfo{author}{\bibnamefont{Isacsson}, \bibfnamefont{A.}},
  \bibinfo{author}{\bibfnamefont{L.~Y.} \bibnamefont{Gorelik}},
  \bibinfo{author}{\bibfnamefont{M.~V.} \bibnamefont{Voinova}},
  \bibinfo{author}{\bibfnamefont{B.}~\bibnamefont{Kasemo}},
  \bibinfo{author}{\bibfnamefont{R.~I.} \bibnamefont{Shekhter}}, and
  \bibinfo{author}{\bibfnamefont{M.}~\bibnamefont{Jonson}},
  \bibinfo{year}{1998}, \bibinfo{journal}{Physica B}
  \textbf{\bibinfo{volume}{255}}, \bibinfo{pages}{150}.

\bibitem[{\citenamefont{Jain} \emph{et~al.}(1987)\citenamefont{Jain, Lukens,
  and Tsai}}]{Jain1987}
\bibinfo{author}{\bibnamefont{Jain}, \bibfnamefont{A.~K.}},
  \bibinfo{author}{\bibfnamefont{J.~E.} \bibnamefont{Lukens}}, and
  \bibinfo{author}{\bibfnamefont{J.~S.} \bibnamefont{Tsai}},
  \bibinfo{year}{1987}, \bibinfo{journal}{Phys. Rev. Lett.}
  \textbf{\bibinfo{volume}{58}}, \bibinfo{pages}{1165}.

\bibitem[{\citenamefont{Janssen} \emph{et~al.}(2011)\citenamefont{Janssen,
  Fletcher, Goebel, Williams, Tzalenchuk, Yakimova, Kubatkin, Lara-Avila, and
  Falko}}]{Janssen2011}
\bibinfo{author}{\bibnamefont{Janssen}, \bibfnamefont{T.~J. B.~M.}},
  \bibinfo{author}{\bibfnamefont{N.~E.} \bibnamefont{Fletcher}},
  \bibinfo{author}{\bibfnamefont{R.}~\bibnamefont{Goebel}},
  \bibinfo{author}{\bibfnamefont{J.~M.} \bibnamefont{Williams}},
  \bibinfo{author}{\bibfnamefont{A.}~\bibnamefont{Tzalenchuk}},
  \bibinfo{author}{\bibfnamefont{R.}~\bibnamefont{Yakimova}},
  \bibinfo{author}{\bibfnamefont{S.}~\bibnamefont{Kubatkin}},
  \bibinfo{author}{\bibfnamefont{S.}~\bibnamefont{Lara-Avila}}, and
  \bibinfo{author}{\bibfnamefont{V.~I.} \bibnamefont{Falko}},
  \bibinfo{year}{2011}, \bibinfo{journal}{New J. Phys.}
  \textbf{\bibinfo{volume}{13}}, \bibinfo{pages}{093026}.

\bibitem[{\citenamefont{Janssen and
  Hartland}(2000{\natexlab{a}})}]{Janssen2000}
\bibinfo{author}{\bibnamefont{Janssen}, \bibfnamefont{T.~J. B.~M.}}, and
  \bibinfo{author}{\bibfnamefont{A.}~\bibnamefont{Hartland}},
  \bibinfo{year}{2000}{\natexlab{a}}, \bibinfo{journal}{Physica B}
  \textbf{\bibinfo{volume}{284}}, \bibinfo{pages}{1790}.

\bibitem[{\citenamefont{Janssen and
  Hartland}(2000{\natexlab{b}})}]{Janssen2000a}
\bibinfo{author}{\bibnamefont{Janssen}, \bibfnamefont{T.~J. B.~M.}}, and
  \bibinfo{author}{\bibfnamefont{A.}~\bibnamefont{Hartland}},
  \bibinfo{year}{2000}{\natexlab{b}}, \bibinfo{journal}{IEE Proc.-Sci. Meas.
  Technol.} \textbf{\bibinfo{volume}{147}}, \bibinfo{pages}{174}.

\bibitem[{\citenamefont{Janssen and Hartland}(2001)}]{Janssen2001}
\bibinfo{author}{\bibnamefont{Janssen}, \bibfnamefont{T.~J. B.~M.}}, and
  \bibinfo{author}{\bibfnamefont{A.}~\bibnamefont{Hartland}},
  \bibinfo{year}{2001}, \bibinfo{journal}{IEEE Trans. Instr. Meas.}
  \textbf{\bibinfo{volume}{50}}, \bibinfo{pages}{227}.

\bibitem[{\citenamefont{Jarzynski}(1997)}]{Jarzynski1997}
\bibinfo{author}{\bibnamefont{Jarzynski}, \bibfnamefont{C.}},
  \bibinfo{year}{1997}, \bibinfo{journal}{Phys. Rev. Lett.}
  \textbf{\bibinfo{volume}{78}}, \bibinfo{pages}{2690}.

\bibitem[{\citenamefont{Jeanneret and Benz}(2009)}]{Jeanneret2009}
\bibinfo{author}{\bibnamefont{Jeanneret}, \bibfnamefont{B.}}, and
  \bibinfo{author}{\bibfnamefont{S.~P.} \bibnamefont{Benz}},
  \bibinfo{year}{2009}, \bibinfo{journal}{Eur. Phys. J. Spec. Top.}
  \textbf{\bibinfo{volume}{172}}, \bibinfo{pages}{181}.

\bibitem[{\citenamefont{Jeckelmann and Jeanneret}(2001)}]{Jeckelmann2001}
\bibinfo{author}{\bibnamefont{Jeckelmann}, \bibfnamefont{B.}}, and
  \bibinfo{author}{\bibfnamefont{B.}~\bibnamefont{Jeanneret}},
  \bibinfo{year}{2001}, \bibinfo{journal}{Rep. Prog. Phys.}
  \textbf{\bibinfo{volume}{64}}, \bibinfo{pages}{1603}.

\bibitem[{\citenamefont{Jeffery} \emph{et~al.}(1997)\citenamefont{Jeffery,
  Elmquist, Lee, Shields, and Dziuba}}]{Jeffery1997}
\bibinfo{author}{\bibnamefont{Jeffery}, \bibfnamefont{A.-M.}},
  \bibinfo{author}{\bibfnamefont{R.}~\bibnamefont{Elmquist}},
  \bibinfo{author}{\bibfnamefont{L.}~\bibnamefont{Lee}},
  \bibinfo{author}{\bibfnamefont{J.}~\bibnamefont{Shields}}, and
  \bibinfo{author}{\bibfnamefont{R.}~\bibnamefont{Dziuba}},
  \bibinfo{year}{1997}, \bibinfo{journal}{IEEE Trans. Instrum. Meas.}
  \textbf{\bibinfo{volume}{46}}, \bibinfo{pages}{264}.

\bibitem[{\citenamefont{Jehl} \emph{et~al.}(2003)\citenamefont{Jehl, Keller,
  Kautz, Aumentado, and Martinis}}]{Jehl2003}
\bibinfo{author}{\bibnamefont{Jehl}, \bibfnamefont{X.}},
  \bibinfo{author}{\bibfnamefont{M.~W.} \bibnamefont{Keller}},
  \bibinfo{author}{\bibfnamefont{R.~L.} \bibnamefont{Kautz}},
  \bibinfo{author}{\bibfnamefont{J.}~\bibnamefont{Aumentado}}, and
  \bibinfo{author}{\bibfnamefont{J.~M.} \bibnamefont{Martinis}},
  \bibinfo{year}{2003}, \bibinfo{journal}{Phys. Rev. B}
  \textbf{\bibinfo{volume}{67}}, \bibinfo{pages}{165331}.

\bibitem[{\citenamefont{Jehl} \emph{et~al.}(2012)\citenamefont{Jehl, Voisin,
  Sanquer, Wacquez, and Vinet}}]{Jehl2012}
\bibinfo{author}{\bibnamefont{Jehl}, \bibfnamefont{X.}},
  \bibinfo{author}{\bibfnamefont{B.}~\bibnamefont{Voisin}},
  \bibinfo{author}{\bibfnamefont{M.}~\bibnamefont{Sanquer}},
  \bibinfo{author}{\bibfnamefont{R.}~\bibnamefont{Wacquez}}, and
  \bibinfo{author}{\bibfnamefont{M.}~\bibnamefont{Vinet}},
  \bibinfo{year}{2012}, \bibinfo{journal}{unpublished} .

\bibitem[{\citenamefont{Jensen and Martinis}(1992)}]{Jensen1992}
\bibinfo{author}{\bibnamefont{Jensen}, \bibfnamefont{H.~D.}}, and
  \bibinfo{author}{\bibfnamefont{J.~M.} \bibnamefont{Martinis}},
  \bibinfo{year}{1992}, \bibinfo{journal}{Phys. Rev. B}
  \textbf{\bibinfo{volume}{46}}, \bibinfo{pages}{13407}.

\bibitem[{\citenamefont{Johansson} \emph{et~al.}(2002)\citenamefont{Johansson,
  K\"ack, and Wendin}}]{Johansson2002}
\bibinfo{author}{\bibnamefont{Johansson}, \bibfnamefont{G.}},
  \bibinfo{author}{\bibfnamefont{A.}~\bibnamefont{K\"ack}}, and
  \bibinfo{author}{\bibfnamefont{G.}~\bibnamefont{Wendin}},
  \bibinfo{year}{2002}, \bibinfo{journal}{Phys. Rev. Lett.}
  \textbf{\bibinfo{volume}{88}}, \bibinfo{pages}{046802}.

\bibitem[{\citenamefont{Johansson} \emph{et~al.}(2008)\citenamefont{Johansson,
  Mourokh, Smirnov, and Nori}}]{Johansson2008}
\bibinfo{author}{\bibnamefont{Johansson}, \bibfnamefont{J.~R.}},
  \bibinfo{author}{\bibfnamefont{L.~G.} \bibnamefont{Mourokh}},
  \bibinfo{author}{\bibfnamefont{A.~Y.} \bibnamefont{Smirnov}}, and
  \bibinfo{author}{\bibfnamefont{F.}~\bibnamefont{Nori}}, \bibinfo{year}{2008},
  \bibinfo{journal}{Phys. Rev. B} \textbf{\bibinfo{volume}{77}},
  \bibinfo{pages}{035428}.

\bibitem[{\citenamefont{Johnson} \emph{et~al.}(1992)\citenamefont{Johnson,
  Kouwenhoven, de~Jong, van~der Vaart, Harmans, and Foxon}}]{Johnson1992}
\bibinfo{author}{\bibnamefont{Johnson}, \bibfnamefont{A.~T.}},
  \bibinfo{author}{\bibfnamefont{L.~P.} \bibnamefont{Kouwenhoven}},
  \bibinfo{author}{\bibfnamefont{W.}~\bibnamefont{de~Jong}},
  \bibinfo{author}{\bibfnamefont{N.~C.} \bibnamefont{van~der Vaart}},
  \bibinfo{author}{\bibfnamefont{C.~J. P.~M.} \bibnamefont{Harmans}}, and
  \bibinfo{author}{\bibfnamefont{C.~T.} \bibnamefont{Foxon}},
  \bibinfo{year}{1992}, \bibinfo{journal}{Phys. Rev. Lett.}
  \textbf{\bibinfo{volume}{69}}, \bibinfo{pages}{1592}.

\bibitem[{\citenamefont{Johnson}(1928)}]{Johnson1928}
\bibinfo{author}{\bibnamefont{Johnson}, \bibfnamefont{J.~B.}},
  \bibinfo{year}{1928}, \bibinfo{journal}{Phys. Rev.}
  \textbf{\bibinfo{volume}{32}}, \bibinfo{pages}{97}.

\bibitem[{\citenamefont{Josephson}(1962)}]{Josephson1962}
\bibinfo{author}{\bibnamefont{Josephson}, \bibfnamefont{B.~D.}},
  \bibinfo{year}{1962}, \bibinfo{journal}{Phys. Lett.}
  \textbf{\bibinfo{volume}{1}}, \bibinfo{pages}{251}.

\bibitem[{\citenamefont{Kaestner}
  \emph{et~al.}(2008{\natexlab{a}})\citenamefont{Kaestner, Kashcheyevs,
  Amakawa, Blumenthal, Li, Janssen, Hein, Pierz, Weimann, Siegner, and
  Schumacher}}]{Kaestner2008b}
\bibinfo{author}{\bibnamefont{Kaestner}, \bibfnamefont{B.}},
  \bibinfo{author}{\bibfnamefont{V.}~\bibnamefont{Kashcheyevs}},
  \bibinfo{author}{\bibfnamefont{S.}~\bibnamefont{Amakawa}},
  \bibinfo{author}{\bibfnamefont{M.~D.} \bibnamefont{Blumenthal}},
  \bibinfo{author}{\bibfnamefont{L.}~\bibnamefont{Li}},
  \bibinfo{author}{\bibfnamefont{T.~J. B.~M.} \bibnamefont{Janssen}},
  \bibinfo{author}{\bibfnamefont{G.}~\bibnamefont{Hein}},
  \bibinfo{author}{\bibfnamefont{K.}~\bibnamefont{Pierz}},
  \bibinfo{author}{\bibfnamefont{T.}~\bibnamefont{Weimann}},
  \bibinfo{author}{\bibfnamefont{U.}~\bibnamefont{Siegner}}, and
  \bibinfo{author}{\bibfnamefont{H.~W.} \bibnamefont{Schumacher}},
  \bibinfo{year}{2008}{\natexlab{a}}, \bibinfo{journal}{Phys. Rev. B}
  \textbf{\bibinfo{volume}{77}}, \bibinfo{pages}{153301}.

\bibitem[{\citenamefont{Kaestner}
  \emph{et~al.}(2008{\natexlab{b}})\citenamefont{Kaestner, Kashcheyevs, Hein,
  Pierz, Siegner, and Schumacher}}]{Kaestner2008}
\bibinfo{author}{\bibnamefont{Kaestner}, \bibfnamefont{B.}},
  \bibinfo{author}{\bibfnamefont{V.}~\bibnamefont{Kashcheyevs}},
  \bibinfo{author}{\bibfnamefont{G.}~\bibnamefont{Hein}},
  \bibinfo{author}{\bibfnamefont{K.}~\bibnamefont{Pierz}},
  \bibinfo{author}{\bibfnamefont{U.}~\bibnamefont{Siegner}}, and
  \bibinfo{author}{\bibfnamefont{H.~W.} \bibnamefont{Schumacher}},
  \bibinfo{year}{2008}{\natexlab{b}}, \bibinfo{journal}{Appl. Phys. Lett.}
  \textbf{\bibinfo{volume}{92}}, \bibinfo{pages}{192106}.

\bibitem[{\citenamefont{Kaestner} \emph{et~al.}(2009)\citenamefont{Kaestner,
  Leicht, Kashcheyevs, Pierz, Siegner, and Schumacher}}]{Kaestner2009}
\bibinfo{author}{\bibnamefont{Kaestner}, \bibfnamefont{B.}},
  \bibinfo{author}{\bibfnamefont{C.}~\bibnamefont{Leicht}},
  \bibinfo{author}{\bibfnamefont{V.}~\bibnamefont{Kashcheyevs}},
  \bibinfo{author}{\bibfnamefont{K.}~\bibnamefont{Pierz}},
  \bibinfo{author}{\bibfnamefont{U.}~\bibnamefont{Siegner}}, and
  \bibinfo{author}{\bibfnamefont{H.~W.} \bibnamefont{Schumacher}},
  \bibinfo{year}{2009}, \bibinfo{journal}{Appl. Phys. Lett.}
  \textbf{\bibinfo{volume}{94}}, \bibinfo{pages}{012106}.

\bibitem[{\citenamefont{Kafanov and Delsing}(2009)}]{Kafanov2009a}
\bibinfo{author}{\bibnamefont{Kafanov}, \bibfnamefont{S.}}, and
  \bibinfo{author}{\bibfnamefont{P.}~\bibnamefont{Delsing}},
  \bibinfo{year}{2009}, \bibinfo{journal}{Phys. Rev. B}
  \textbf{\bibinfo{volume}{80}}, \bibinfo{pages}{155320}.

\bibitem[{\citenamefont{Kafanov} \emph{et~al.}(2009)\citenamefont{Kafanov,
  Kemppinen, Pashkin, Meschke, Tsai, and Pekola}}]{Kafanov2009}
\bibinfo{author}{\bibnamefont{Kafanov}, \bibfnamefont{S.}},
  \bibinfo{author}{\bibfnamefont{A.}~\bibnamefont{Kemppinen}},
  \bibinfo{author}{\bibfnamefont{Y.~A.} \bibnamefont{Pashkin}},
  \bibinfo{author}{\bibfnamefont{M.}~\bibnamefont{Meschke}},
  \bibinfo{author}{\bibfnamefont{J.~S.} \bibnamefont{Tsai}}, and
  \bibinfo{author}{\bibfnamefont{J.~P.} \bibnamefont{Pekola}},
  \bibinfo{year}{2009}, \bibinfo{journal}{Phys. Rev. Lett.}
  \textbf{\bibinfo{volume}{103}}, \bibinfo{pages}{120801}.

\bibitem[{\citenamefont{Kane}(1998)}]{Kane1998}
\bibinfo{author}{\bibnamefont{Kane}, \bibfnamefont{B.~E.}},
  \bibinfo{year}{1998}, \bibinfo{journal}{Nature}
  \textbf{\bibinfo{volume}{393}}, \bibinfo{pages}{133}.

\bibitem[{\citenamefont{Karshenboim}(2009)}]{Karshenboim2009}
\bibinfo{author}{\bibnamefont{Karshenboim}, \bibfnamefont{S.~G.}},
  \bibinfo{year}{2009}, \bibinfo{journal}{Eur. Phys. J. Spec. Top.}
  \textbf{\bibinfo{volume}{172}}, \bibinfo{pages}{385}.

\bibitem[{\citenamefont{Kashcheyevs and Kaestner}(2010)}]{Kashcheyevs2010}
\bibinfo{author}{\bibnamefont{Kashcheyevs}, \bibfnamefont{V.}}, and
  \bibinfo{author}{\bibfnamefont{B.}~\bibnamefont{Kaestner}},
  \bibinfo{year}{2010}, \bibinfo{journal}{Phys. Rev. Lett.}
  \textbf{\bibinfo{volume}{104}}, \bibinfo{pages}{186805}.

\bibitem[{\citenamefont{Kastner}(1993)}]{Kastner1993}
\bibinfo{author}{\bibnamefont{Kastner}, \bibfnamefont{M.~A.}},
  \bibinfo{year}{1993}, \bibinfo{journal}{Physics Today}
  \textbf{\bibinfo{volume}{46}}, \bibinfo{pages}{24}.

\bibitem[{\citenamefont{Kautz} \emph{et~al.}(1999)\citenamefont{Kautz, Keller,
  and Martinis}}]{Kautz1999}
\bibinfo{author}{\bibnamefont{Kautz}, \bibfnamefont{R.~L.}},
  \bibinfo{author}{\bibfnamefont{M.~W.} \bibnamefont{Keller}}, and
  \bibinfo{author}{\bibfnamefont{J.~M.} \bibnamefont{Martinis}},
  \bibinfo{year}{1999}, \bibinfo{journal}{Phys. Rev. B}
  \textbf{\bibinfo{volume}{60}}, \bibinfo{pages}{8199}.

\bibitem[{\citenamefont{Kautz} \emph{et~al.}(2000)\citenamefont{Kautz, Keller,
  and Martinis}}]{Kautz2000}
\bibinfo{author}{\bibnamefont{Kautz}, \bibfnamefont{R.~L.}},
  \bibinfo{author}{\bibfnamefont{M.~W.} \bibnamefont{Keller}}, and
  \bibinfo{author}{\bibfnamefont{J.~M.} \bibnamefont{Martinis}},
  \bibinfo{year}{2000}, \bibinfo{journal}{Phys. Rev. B}
  \textbf{\bibinfo{volume}{62}}, \bibinfo{pages}{15888}.

\bibitem[{\citenamefont{Kautz and Lloyd}(1987)}]{Kautz1987}
\bibinfo{author}{\bibnamefont{Kautz}, \bibfnamefont{R.~L.}}, and
  \bibinfo{author}{\bibfnamefont{F.~L.} \bibnamefont{Lloyd}},
  \bibinfo{year}{1987}, \bibinfo{journal}{Appl. Phys. Lett.}
  \textbf{\bibinfo{volume}{51}}, \bibinfo{pages}{2043}.

\bibitem[{\citenamefont{Kautz} \emph{et~al.}(1993)\citenamefont{Kautz,
  Zimmerli, and Martinis}}]{Kautz1993}
\bibinfo{author}{\bibnamefont{Kautz}, \bibfnamefont{R.~L.}},
  \bibinfo{author}{\bibfnamefont{G.}~\bibnamefont{Zimmerli}}, and
  \bibinfo{author}{\bibfnamefont{J.~M.} \bibnamefont{Martinis}},
  \bibinfo{year}{1993}, \bibinfo{journal}{J. Appl. Phys.}
  \textbf{\bibinfo{volume}{73}}, \bibinfo{pages}{2386}.

\bibitem[{\citenamefont{Kay}(1998)}]{Kay1998}
\bibinfo{author}{\bibnamefont{Kay}, \bibfnamefont{S.~M.}},
  \bibinfo{year}{1998}, \emph{\bibinfo{title}{{Fundamentals of Statistical
  Signal Processing, Volume 2: Detection Theory}}}, volume~\bibinfo{volume}{II}
  of \emph{\bibinfo{series}{Prentice Hall Signal Processing Series}}
  (\bibinfo{publisher}{Prentice Hall}).

\bibitem[{\citenamefont{Keller}(2008)}]{Keller2008}
\bibinfo{author}{\bibnamefont{Keller}, \bibfnamefont{M.~W.}},
  \bibinfo{year}{2008}, \bibinfo{journal}{Metrologia}
  \textbf{\bibinfo{volume}{45}}, \bibinfo{pages}{102}.

\bibitem[{\citenamefont{Keller}(2009)}]{Keller2009}
\bibinfo{author}{\bibnamefont{Keller}, \bibfnamefont{M.~W.}},
  \bibinfo{year}{2009}, \bibinfo{journal}{Eur. Phys. J. Spec. Top.}
  \textbf{\bibinfo{volume}{172}}, \bibinfo{pages}{297}.

\bibitem[{\citenamefont{Keller} \emph{et~al.}(1999)\citenamefont{Keller,
  Eichenberger, Martinis, and Zimmerman}}]{Keller1999}
\bibinfo{author}{\bibnamefont{Keller}, \bibfnamefont{M.~W.}},
  \bibinfo{author}{\bibfnamefont{A.~L.} \bibnamefont{Eichenberger}},
  \bibinfo{author}{\bibfnamefont{J.~M.} \bibnamefont{Martinis}}, and
  \bibinfo{author}{\bibfnamefont{N.~M.} \bibnamefont{Zimmerman}},
  \bibinfo{year}{1999}, \bibinfo{journal}{Science}
  \textbf{\bibinfo{volume}{285}}, \bibinfo{pages}{1706}.

\bibitem[{\citenamefont{Keller} \emph{et~al.}(1996)\citenamefont{Keller,
  Martinis, Zimmerman, and Steinbach}}]{Keller1996}
\bibinfo{author}{\bibnamefont{Keller}, \bibfnamefont{M.~W.}},
  \bibinfo{author}{\bibfnamefont{J.~M.} \bibnamefont{Martinis}},
  \bibinfo{author}{\bibfnamefont{N.~M.} \bibnamefont{Zimmerman}}, and
  \bibinfo{author}{\bibfnamefont{A.~H.} \bibnamefont{Steinbach}},
  \bibinfo{year}{1996}, \bibinfo{journal}{Appl. Phys. Lett.}
  \textbf{\bibinfo{volume}{69}}, \bibinfo{pages}{1804}.

\bibitem[{\citenamefont{Keller} \emph{et~al.}(2008)\citenamefont{Keller,
  Piquemal, Feltin, Steck, and Devoille}}]{Keller2008a}
\bibinfo{author}{\bibnamefont{Keller}, \bibfnamefont{M.~W.}},
  \bibinfo{author}{\bibfnamefont{F.}~\bibnamefont{Piquemal}},
  \bibinfo{author}{\bibfnamefont{N.}~\bibnamefont{Feltin}},
  \bibinfo{author}{\bibfnamefont{B.}~\bibnamefont{Steck}}, and
  \bibinfo{author}{\bibfnamefont{L.}~\bibnamefont{Devoille}},
  \bibinfo{year}{2008}, \bibinfo{journal}{Metrologia}
  \textbf{\bibinfo{volume}{45}}, \bibinfo{pages}{330}.

\bibitem[{\citenamefont{Keller} \emph{et~al.}(2007)\citenamefont{Keller,
  Zimmerman, and Eichenberger}}]{Keller2007}
\bibinfo{author}{\bibnamefont{Keller}, \bibfnamefont{M.~W.}},
  \bibinfo{author}{\bibfnamefont{N.~M.} \bibnamefont{Zimmerman}}, and
  \bibinfo{author}{\bibfnamefont{A.~L.} \bibnamefont{Eichenberger}},
  \bibinfo{year}{2007}, \bibinfo{journal}{Metrologia}
  \textbf{\bibinfo{volume}{44}}, \bibinfo{pages}{505}.

\bibitem[{\citenamefont{Kemppinen}(2009)}]{Kemppinen2009a}
\bibinfo{author}{\bibnamefont{Kemppinen}, \bibfnamefont{A.}},
  \bibinfo{year}{2009}, \emph{\bibinfo{title}{PhD Thesis}}
  (\bibinfo{publisher}{Helsinki University of Technology (unpublished)}).

\bibitem[{\citenamefont{Kemppinen}
  \emph{et~al.}(2009{\natexlab{a}})\citenamefont{Kemppinen, Kafanov, Pashkin,
  Tsai, Averin, and Pekola}}]{Kemppinen2009}
\bibinfo{author}{\bibnamefont{Kemppinen}, \bibfnamefont{A.}},
  \bibinfo{author}{\bibfnamefont{S.}~\bibnamefont{Kafanov}},
  \bibinfo{author}{\bibfnamefont{Y.~A.} \bibnamefont{Pashkin}},
  \bibinfo{author}{\bibfnamefont{J.~S.} \bibnamefont{Tsai}},
  \bibinfo{author}{\bibfnamefont{D.~V.} \bibnamefont{Averin}}, and
  \bibinfo{author}{\bibfnamefont{J.~P.} \bibnamefont{Pekola}},
  \bibinfo{year}{2009}{\natexlab{a}}, \bibinfo{journal}{Appl. Phys. Lett.}
  \textbf{\bibinfo{volume}{94}}, \bibinfo{pages}{172108}.

\bibitem[{\citenamefont{Kemppinen} \emph{et~al.}(2011)\citenamefont{Kemppinen,
  Lotkhov, Saira, Zorin, Pekola, and Manninen}}]{Kemppinen2011}
\bibinfo{author}{\bibnamefont{Kemppinen}, \bibfnamefont{A.}},
  \bibinfo{author}{\bibfnamefont{S.~V.} \bibnamefont{Lotkhov}},
  \bibinfo{author}{\bibfnamefont{O.-P.} \bibnamefont{Saira}},
  \bibinfo{author}{\bibfnamefont{A.~B.} \bibnamefont{Zorin}},
  \bibinfo{author}{\bibfnamefont{J.~P.} \bibnamefont{Pekola}}, and
  \bibinfo{author}{\bibfnamefont{A.~J.} \bibnamefont{Manninen}},
  \bibinfo{year}{2011}, \bibinfo{journal}{Appl. Phys. Lett.}
  \textbf{\bibinfo{volume}{99}}, \bibinfo{pages}{142106}.

\bibitem[{\citenamefont{Kemppinen}
  \emph{et~al.}(2009{\natexlab{b}})\citenamefont{Kemppinen, Meschke,
  M�tt�nen, Averin, and Pekola}}]{Kemppinen2008}
\bibinfo{author}{\bibnamefont{Kemppinen}, \bibfnamefont{A.}},
  \bibinfo{author}{\bibfnamefont{M.}~\bibnamefont{Meschke}},
  \bibinfo{author}{\bibfnamefont{M.}~\bibnamefont{M�tt�nen}},
  \bibinfo{author}{\bibfnamefont{D.~V.} \bibnamefont{Averin}}, and
  \bibinfo{author}{\bibfnamefont{J.~P.} \bibnamefont{Pekola}},
  \bibinfo{year}{2009}{\natexlab{b}}, \bibinfo{journal}{Eur. Phys. J. Spec.
  Top.} \textbf{\bibinfo{volume}{172}}, \bibinfo{pages}{311}.

\bibitem[{\citenamefont{Kibble}(1975)}]{Kibble1975}
\bibinfo{author}{\bibnamefont{Kibble}, \bibfnamefont{B.~P.}},
  \bibinfo{year}{1975}, \emph{\bibinfo{title}{Atomic masses and fundamental
  constants 5}} (\bibinfo{publisher}{Plenum, New York}).

\bibitem[{\citenamefont{Kim} \emph{et~al.}(2010)\citenamefont{Kim, Qin, and
  Blick}}]{Kim2010}
\bibinfo{author}{\bibnamefont{Kim}, \bibfnamefont{H.~S.}},
  \bibinfo{author}{\bibfnamefont{H.}~\bibnamefont{Qin}}, and
  \bibinfo{author}{\bibfnamefont{R.~H.} \bibnamefont{Blick}},
  \bibinfo{year}{2010}, \bibinfo{journal}{New J. Phys.}
  \textbf{\bibinfo{volume}{12}}, \bibinfo{pages}{033008}.

\bibitem[{\citenamefont{von Klitzing} \emph{et~al.}(1980)\citenamefont{von
  Klitzing, Dorda, and Pepper}}]{Klitzing1980}
\bibinfo{author}{\bibnamefont{von Klitzing}, \bibfnamefont{K.}},
  \bibinfo{author}{\bibfnamefont{G.}~\bibnamefont{Dorda}}, and
  \bibinfo{author}{\bibfnamefont{M.}~\bibnamefont{Pepper}},
  \bibinfo{year}{1980}, \bibinfo{journal}{Phys. Rev. Lett.}
  \textbf{\bibinfo{volume}{45}}, \bibinfo{pages}{494}.

\bibitem[{\citenamefont{Knowles} \emph{et~al.}(2012)\citenamefont{Knowles,
  Maisi, and Pekola}}]{Knowles2012}
\bibinfo{author}{\bibnamefont{Knowles}, \bibfnamefont{H.~S.}},
  \bibinfo{author}{\bibfnamefont{V.~F.} \bibnamefont{Maisi}}, and
  \bibinfo{author}{\bibfnamefont{J.~P.} \bibnamefont{Pekola}},
  \bibinfo{year}{2012}, \bibinfo{journal}{Appl. Phys. Lett.}
  \textbf{\bibinfo{volume}{100}}, \bibinfo{pages}{262601}.

\bibitem[{\citenamefont{Koenig} \emph{et~al.}(2008)\citenamefont{Koenig, Weig,
  and Kotthaus}}]{Koenig2008}
\bibinfo{author}{\bibnamefont{Koenig}, \bibfnamefont{D.~R.}},
  \bibinfo{author}{\bibfnamefont{E.~M.} \bibnamefont{Weig}}, and
  \bibinfo{author}{\bibfnamefont{J.~P.} \bibnamefont{Kotthaus}},
  \bibinfo{year}{2008}, \bibinfo{journal}{Nature Nanotech.}
  \textbf{\bibinfo{volume}{3}}, \bibinfo{pages}{482}.

\bibitem[{\citenamefont{Kohlmann} \emph{et~al.}(2003)\citenamefont{Kohlmann,
  Behr, and Funck}}]{Kohlmann2003}
\bibinfo{author}{\bibnamefont{Kohlmann}, \bibfnamefont{J.}},
  \bibinfo{author}{\bibfnamefont{R.}~\bibnamefont{Behr}}, and
  \bibinfo{author}{\bibfnamefont{T.}~\bibnamefont{Funck}},
  \bibinfo{year}{2003}, \bibinfo{journal}{Meas. Sci. and Tech.}
  \textbf{\bibinfo{volume}{14}}, \bibinfo{pages}{1216}.

\bibitem[{\citenamefont{Kohlmann} \emph{et~al.}(2007)\citenamefont{Kohlmann,
  Muller, Kieler, Behr, Palafox, Kahmann, and Niemeyer}}]{Kohlmann2007}
\bibinfo{author}{\bibnamefont{Kohlmann}, \bibfnamefont{J.}},
  \bibinfo{author}{\bibfnamefont{F.}~\bibnamefont{Muller}},
  \bibinfo{author}{\bibfnamefont{O.}~\bibnamefont{Kieler}},
  \bibinfo{author}{\bibfnamefont{R.}~\bibnamefont{Behr}},
  \bibinfo{author}{\bibfnamefont{L.}~\bibnamefont{Palafox}},
  \bibinfo{author}{\bibfnamefont{M.}~\bibnamefont{Kahmann}}, and
  \bibinfo{author}{\bibfnamefont{J.}~\bibnamefont{Niemeyer}},
  \bibinfo{year}{2007}, \bibinfo{journal}{IEEE Trans. Instrum. Meas.}
  \textbf{\bibinfo{volume}{56}}, \bibinfo{pages}{472}.

\bibitem[{\citenamefont{Korotkov}(1999)}]{Korotkov1999a}
\bibinfo{author}{\bibnamefont{Korotkov}, \bibfnamefont{A.}},
  \bibinfo{year}{1999}, \bibinfo{journal}{Phys. Rev. B}
  \textbf{\bibinfo{volume}{60}}, \bibinfo{pages}{5737}.

\bibitem[{\citenamefont{Korotkov}(1994)}]{Korotkov1994}
\bibinfo{author}{\bibnamefont{Korotkov}, \bibfnamefont{A.~N.}},
  \bibinfo{year}{1994}, \bibinfo{journal}{Phys. Rev. B}
  \textbf{\bibinfo{volume}{49}}, \bibinfo{pages}{10381}.

\bibitem[{\citenamefont{Korotkov and Paalanen}(1999)}]{Korotkov1999}
\bibinfo{author}{\bibnamefont{Korotkov}, \bibfnamefont{A.~N.}}, and
  \bibinfo{author}{\bibfnamefont{M.~A.} \bibnamefont{Paalanen}},
  \bibinfo{year}{1999}, \bibinfo{journal}{Appl. Phys. Lett.}
  \textbf{\bibinfo{volume}{74}}, \bibinfo{pages}{4052}.

\bibitem[{\citenamefont{Kouwenhoven}(1992)}]{Kouwenhoven1992}
\bibinfo{author}{\bibnamefont{Kouwenhoven}, \bibfnamefont{L.~P.}},
  \bibinfo{year}{1992}, \bibinfo{journal}{Phys. Scr.}
  \textbf{\bibinfo{volume}{T42}}, \bibinfo{pages}{133}.

\bibitem[{\citenamefont{Kouwenhoven}
  \emph{et~al.}(1991{\natexlab{a}})\citenamefont{Kouwenhoven, Johnson, van~der
  Vaart, van~der Enden, Harmans, and Foxon}}]{Kouwenhoven1991_2}
\bibinfo{author}{\bibnamefont{Kouwenhoven}, \bibfnamefont{L.~P.}},
  \bibinfo{author}{\bibfnamefont{A.~T.} \bibnamefont{Johnson}},
  \bibinfo{author}{\bibfnamefont{N.~C.} \bibnamefont{van~der Vaart}},
  \bibinfo{author}{\bibfnamefont{A.}~\bibnamefont{van~der Enden}},
  \bibinfo{author}{\bibfnamefont{C.~J. P.~M.} \bibnamefont{Harmans}}, and
  \bibinfo{author}{\bibfnamefont{C.~T.} \bibnamefont{Foxon}},
  \bibinfo{year}{1991}{\natexlab{a}}, \bibinfo{journal}{Z. Phys. B}
  \textbf{\bibinfo{volume}{85}}, \bibinfo{pages}{381}.

\bibitem[{\citenamefont{Kouwenhoven}
  \emph{et~al.}(1991{\natexlab{b}})\citenamefont{Kouwenhoven, Johnson, van~der
  Vaart, Harmans, and Foxon}}]{Kouwenhoven1991}
\bibinfo{author}{\bibnamefont{Kouwenhoven}, \bibfnamefont{L.~P.}},
  \bibinfo{author}{\bibfnamefont{A.~T.} \bibnamefont{Johnson}},
  \bibinfo{author}{\bibfnamefont{N.~C.} \bibnamefont{van~der Vaart}},
  \bibinfo{author}{\bibfnamefont{C.~J. P.~M.} \bibnamefont{Harmans}}, and
  \bibinfo{author}{\bibfnamefont{C.~T.} \bibnamefont{Foxon}},
  \bibinfo{year}{1991}{\natexlab{b}}, \bibinfo{journal}{Phys. Rev. Lett.}
  \textbf{\bibinfo{volume}{67}}, \bibinfo{pages}{1626}.

\bibitem[{\citenamefont{Krupenin}(1998)}]{Krupenin1998}
\bibinfo{author}{\bibnamefont{Krupenin}, \bibfnamefont{V.~A.}},
  \bibinfo{year}{1998}, \bibinfo{journal}{J. Appl. Phys.}
  \textbf{\bibinfo{volume}{84}}, \bibinfo{pages}{3212}.

\bibitem[{\citenamefont{K\"ung} \emph{et~al.}(2012)\citenamefont{K\"ung,
  R\"ossler, Beck, Marthaler, Golubev, Utsumi, Ihn, and Ensslin}}]{Kung2012}
\bibinfo{author}{\bibnamefont{K\"ung}, \bibfnamefont{B.}},
  \bibinfo{author}{\bibfnamefont{C.}~\bibnamefont{R\"ossler}},
  \bibinfo{author}{\bibfnamefont{M.}~\bibnamefont{Beck}},
  \bibinfo{author}{\bibfnamefont{M.}~\bibnamefont{Marthaler}},
  \bibinfo{author}{\bibfnamefont{D.~S.} \bibnamefont{Golubev}},
  \bibinfo{author}{\bibfnamefont{Y.}~\bibnamefont{Utsumi}},
  \bibinfo{author}{\bibfnamefont{T.}~\bibnamefont{Ihn}}, and
  \bibinfo{author}{\bibfnamefont{K.}~\bibnamefont{Ensslin}},
  \bibinfo{year}{2012}, \bibinfo{journal}{Phys. Rev. X}
  \textbf{\bibinfo{volume}{2}}, \bibinfo{pages}{011001}.

\bibitem[{\citenamefont{Kuzmin} \emph{et~al.}(1989)\citenamefont{Kuzmin,
  Delsing, Claeson, and Likharev}}]{Kuzmin1989}
\bibinfo{author}{\bibnamefont{Kuzmin}, \bibfnamefont{L.}},
  \bibinfo{author}{\bibfnamefont{P.}~\bibnamefont{Delsing}},
  \bibinfo{author}{\bibfnamefont{T.}~\bibnamefont{Claeson}}, and
  \bibinfo{author}{\bibfnamefont{K.}~\bibnamefont{Likharev}},
  \bibinfo{year}{1989}, \bibinfo{journal}{Phys. Rev. Lett.}
  \textbf{\bibinfo{volume}{62}}, \bibinfo{pages}{2539}.

\bibitem[{\citenamefont{Lafarge} \emph{et~al.}(1993)\citenamefont{Lafarge,
  Joyez, Esteve, C., and H.}}]{Lafarge1993}
\bibinfo{author}{\bibnamefont{Lafarge}, \bibfnamefont{P.}},
  \bibinfo{author}{\bibfnamefont{P.}~\bibnamefont{Joyez}},
  \bibinfo{author}{\bibfnamefont{D.}~\bibnamefont{Esteve}},
  \bibinfo{author}{\bibfnamefont{U.}~\bibnamefont{C.}}, and
  \bibinfo{author}{\bibfnamefont{D.~M.} \bibnamefont{H.}},
  \bibinfo{year}{1993}, \bibinfo{journal}{Nature}
  \textbf{\bibinfo{volume}{365}}, \bibinfo{pages}{422}.

\bibitem[{\citenamefont{Lafarge} \emph{et~al.}(1991)\citenamefont{Lafarge,
  Pothier, Williams, Esteve, Urbina, and Devoret}}]{Lafarge1991}
\bibinfo{author}{\bibnamefont{Lafarge}, \bibfnamefont{P.}},
  \bibinfo{author}{\bibfnamefont{H.}~\bibnamefont{Pothier}},
  \bibinfo{author}{\bibfnamefont{E.}~\bibnamefont{Williams}},
  \bibinfo{author}{\bibfnamefont{D.}~\bibnamefont{Esteve}},
  \bibinfo{author}{\bibfnamefont{C.}~\bibnamefont{Urbina}}, and
  \bibinfo{author}{\bibfnamefont{M.}~\bibnamefont{Devoret}},
  \bibinfo{year}{1991}, \bibinfo{journal}{Z. Phys B}
  \textbf{\bibinfo{volume}{85}}, \bibinfo{pages}{327}.

\bibitem[{\citenamefont{Landau and
  Lifshitz}(1980{\natexlab{a}})}]{Landau1980_Sec2}
\bibinfo{author}{\bibnamefont{Landau}, \bibfnamefont{L.}}, and
  \bibinfo{author}{\bibfnamefont{E.}~\bibnamefont{Lifshitz}},
  \bibinfo{year}{1980}{\natexlab{a}}, \emph{\bibinfo{title}{Electrodynamics of
  continuous media}} (\bibinfo{publisher}{Pergamon}), p. \bibinfo{pages}{Sec.
  2}.

\bibitem[{\citenamefont{Landau and
  Lifshitz}(1980{\natexlab{b}})}]{Landau1980_Sec43}
\bibinfo{author}{\bibnamefont{Landau}, \bibfnamefont{L.}}, and
  \bibinfo{author}{\bibfnamefont{E.}~\bibnamefont{Lifshitz}},
  \bibinfo{year}{1980}{\natexlab{b}}, \emph{\bibinfo{title}{Quantum mchanics}}
  (\bibinfo{publisher}{Pergamon}), p. \bibinfo{pages}{Sec. 43}.

\bibitem[{\citenamefont{Langenberg and Schrieffer}(1971)}]{Langenberg1971}
\bibinfo{author}{\bibnamefont{Langenberg}, \bibfnamefont{D.~N.}}, and
  \bibinfo{author}{\bibfnamefont{J.~R.} \bibnamefont{Schrieffer}},
  \bibinfo{year}{1971}, \bibinfo{journal}{Phys. Rev. B}
  \textbf{\bibinfo{volume}{3}}, \bibinfo{pages}{1776}.

\bibitem[{\citenamefont{Lansbergen}
  \emph{et~al.}(2008)\citenamefont{Lansbergen, Rahman, Wellard, Woo, Caro,
  Collaert, Biesemans, Klimeck, Hollenberg, and Rogge}}]{Lansbergen2008}
\bibinfo{author}{\bibnamefont{Lansbergen}, \bibfnamefont{G.~P.}},
  \bibinfo{author}{\bibfnamefont{R.}~\bibnamefont{Rahman}},
  \bibinfo{author}{\bibfnamefont{C.~J.} \bibnamefont{Wellard}},
  \bibinfo{author}{\bibfnamefont{I.}~\bibnamefont{Woo}},
  \bibinfo{author}{\bibfnamefont{J.}~\bibnamefont{Caro}},
  \bibinfo{author}{\bibfnamefont{N.}~\bibnamefont{Collaert}},
  \bibinfo{author}{\bibfnamefont{S.}~\bibnamefont{Biesemans}},
  \bibinfo{author}{\bibfnamefont{G.}~\bibnamefont{Klimeck}},
  \bibinfo{author}{\bibfnamefont{L.~C.~L.} \bibnamefont{Hollenberg}}, and
  \bibinfo{author}{\bibfnamefont{S.}~\bibnamefont{Rogge}},
  \bibinfo{year}{2008}, \bibinfo{journal}{Nature Phys.}
  \textbf{\bibinfo{volume}{4}}, \bibinfo{pages}{656}.

\bibitem[{\citenamefont{Laughlin}(1981)}]{Laughlin1981}
\bibinfo{author}{\bibnamefont{Laughlin}, \bibfnamefont{R.~B.}},
  \bibinfo{year}{1981}, \bibinfo{journal}{Phys. Rev. B}
  \textbf{\bibinfo{volume}{23}}, \bibinfo{pages}{5632}.

\bibitem[{\citenamefont{Lehtinen} \emph{et~al.}(2012)\citenamefont{Lehtinen,
  Zakharov, and Arutyunov}}]{Arutyunov2012}
\bibinfo{author}{\bibnamefont{Lehtinen}, \bibfnamefont{J.~S.}},
  \bibinfo{author}{\bibfnamefont{K.}~\bibnamefont{Zakharov}}, and
  \bibinfo{author}{\bibfnamefont{K.~Y.} \bibnamefont{Arutyunov}},
  \bibinfo{year}{2012}, \bibinfo{journal}{Phys. Rep.} , \bibinfo{pages}{1}.

\bibitem[{\citenamefont{Leicht} \emph{et~al.}(2011)\citenamefont{Leicht,
  Mirovsky, Kaestner, Hols, Kaschcheyevs, Kurganova, Zeitler, Weinmann, Pierz,
  and Schumacher}}]{Leicht2011}
\bibinfo{author}{\bibnamefont{Leicht}, \bibfnamefont{C.}},
  \bibinfo{author}{\bibfnamefont{P.}~\bibnamefont{Mirovsky}},
  \bibinfo{author}{\bibfnamefont{B.}~\bibnamefont{Kaestner}},
  \bibinfo{author}{\bibfnamefont{F.}~\bibnamefont{Hols}},
  \bibinfo{author}{\bibfnamefont{V.}~\bibnamefont{Kaschcheyevs}},
  \bibinfo{author}{\bibfnamefont{E.~V.} \bibnamefont{Kurganova}},
  \bibinfo{author}{\bibfnamefont{U.}~\bibnamefont{Zeitler}},
  \bibinfo{author}{\bibfnamefont{T.}~\bibnamefont{Weinmann}},
  \bibinfo{author}{\bibfnamefont{K.}~\bibnamefont{Pierz}}, and
  \bibinfo{author}{\bibfnamefont{H.~W.} \bibnamefont{Schumacher}},
  \bibinfo{year}{2011}, \bibinfo{journal}{Semicond. Sci. Technol.}
  \textbf{\bibinfo{volume}{26}}, \bibinfo{pages}{055010}.

\bibitem[{\citenamefont{Leonard}(2010)}]{Leonard2010}
\bibinfo{author}{\bibnamefont{Leonard}, \bibfnamefont{B.~P.}},
  \bibinfo{year}{2010}, \bibinfo{journal}{Metrologia}
  \textbf{\bibinfo{volume}{47}}, \bibinfo{pages}{L5}.

\bibitem[{\citenamefont{Leone and L\'evy}(2008)}]{Leone2008_2}
\bibinfo{author}{\bibnamefont{Leone}, \bibfnamefont{R.}}, and
  \bibinfo{author}{\bibfnamefont{L.}~\bibnamefont{L\'evy}},
  \bibinfo{year}{2008}, \bibinfo{journal}{Phys. Rev. B}
  \textbf{\bibinfo{volume}{77}}, \bibinfo{pages}{064524}.

\bibitem[{\citenamefont{Leone} \emph{et~al.}(2008)\citenamefont{Leone, L\'evy,
  and Lafarge}}]{Leone2008}
\bibinfo{author}{\bibnamefont{Leone}, \bibfnamefont{R.}},
  \bibinfo{author}{\bibfnamefont{L.~P.} \bibnamefont{L\'evy}}, and
  \bibinfo{author}{\bibfnamefont{P.}~\bibnamefont{Lafarge}},
  \bibinfo{year}{2008}, \bibinfo{journal}{Phys. Rev. Lett.}
  \textbf{\bibinfo{volume}{100}}, \bibinfo{pages}{117001}.

\bibitem[{\citenamefont{Li} \emph{et~al.}(2012)\citenamefont{Li, Han, Li, and
  Lan}}]{Li2012}
\bibinfo{author}{\bibnamefont{Li}, \bibfnamefont{S.}},
  \bibinfo{author}{\bibfnamefont{B.}~\bibnamefont{Han}},
  \bibinfo{author}{\bibfnamefont{Z.}~\bibnamefont{Li}}, and
  \bibinfo{author}{\bibfnamefont{J.}~\bibnamefont{Lan}}, \bibinfo{year}{2012},
  \bibinfo{journal}{Measurement} \textbf{\bibinfo{volume}{45}},
  \bibinfo{pages}{1}.

\bibitem[{\citenamefont{Likharev}(1988)}]{Likharev1988}
\bibinfo{author}{\bibnamefont{Likharev}, \bibfnamefont{K.}},
  \bibinfo{year}{1988}, \bibinfo{journal}{IBM J. Res. Dev.}
  \textbf{\bibinfo{volume}{32}}, \bibinfo{pages}{144}.

\bibitem[{\citenamefont{Likharev} \emph{et~al.}(1989)\citenamefont{Likharev,
  Bakhvalov, Kazacha, and Serdyokova}}]{Likharev1989}
\bibinfo{author}{\bibnamefont{Likharev}, \bibfnamefont{K.}},
  \bibinfo{author}{\bibfnamefont{N.}~\bibnamefont{Bakhvalov}},
  \bibinfo{author}{\bibfnamefont{G.}~\bibnamefont{Kazacha}}, and
  \bibinfo{author}{\bibfnamefont{S.}~\bibnamefont{Serdyokova}},
  \bibinfo{year}{1989}, \bibinfo{journal}{IEEE Trans. Magn.}
  \textbf{\bibinfo{volume}{25}}, \bibinfo{pages}{1436}.

\bibitem[{\citenamefont{Likharev}(1987)}]{Likharev1987}
\bibinfo{author}{\bibnamefont{Likharev}, \bibfnamefont{K.~K.}},
  \bibinfo{year}{1987}, \bibinfo{journal}{IEEE Trans. Magn.}
  \textbf{\bibinfo{volume}{23}}, \bibinfo{pages}{1142}.

\bibitem[{\citenamefont{Likharev and Zorin}(1985)}]{Likharev1985}
\bibinfo{author}{\bibnamefont{Likharev}, \bibfnamefont{K.~K.}}, and
  \bibinfo{author}{\bibfnamefont{A.~B.} \bibnamefont{Zorin}},
  \bibinfo{year}{1985}, \bibinfo{journal}{J. Low Temp. Phys.}
  \textbf{\bibinfo{volume}{59}}, \bibinfo{pages}{347}.

\bibitem[{\citenamefont{Lim} \emph{et~al.}(2009)\citenamefont{Lim, Zwanenburg,
  Huebl, M\"ott\"onen, Chan, Morello, and Dzurak}}]{Lim2009}
\bibinfo{author}{\bibnamefont{Lim}, \bibfnamefont{W.~H.}},
  \bibinfo{author}{\bibfnamefont{F.~A.} \bibnamefont{Zwanenburg}},
  \bibinfo{author}{\bibfnamefont{H.}~\bibnamefont{Huebl}},
  \bibinfo{author}{\bibfnamefont{M.}~\bibnamefont{M\"ott\"onen}},
  \bibinfo{author}{\bibfnamefont{K.~W.} \bibnamefont{Chan}},
  \bibinfo{author}{\bibfnamefont{A.}~\bibnamefont{Morello}}, and
  \bibinfo{author}{\bibfnamefont{A.~S.} \bibnamefont{Dzurak}},
  \bibinfo{year}{2009}, \bibinfo{journal}{Appl. Phys. Lett.}
  \textbf{\bibinfo{volume}{95}}, \bibinfo{pages}{242102}.

\bibitem[{\citenamefont{Liu and Hopkinson}(2003)}]{Liu2003}
\bibinfo{author}{\bibnamefont{Liu}, \bibfnamefont{H.~Y.}}, and
  \bibinfo{author}{\bibfnamefont{M.}~\bibnamefont{Hopkinson}},
  \bibinfo{year}{2003}, \bibinfo{journal}{Appl. Phys. Lett}
  \textbf{\bibinfo{volume}{82}}, \bibinfo{pages}{3644}.

\bibitem[{\citenamefont{Lotkhov} \emph{et~al.}(2001)\citenamefont{Lotkhov,
  Bogoslovsky, Zorin, and Niemeyer}}]{Lotkhov2001}
\bibinfo{author}{\bibnamefont{Lotkhov}, \bibfnamefont{S.~V.}},
  \bibinfo{author}{\bibfnamefont{S.~A.} \bibnamefont{Bogoslovsky}},
  \bibinfo{author}{\bibfnamefont{A.~B.} \bibnamefont{Zorin}}, and
  \bibinfo{author}{\bibfnamefont{J.}~\bibnamefont{Niemeyer}},
  \bibinfo{year}{2001}, \bibinfo{journal}{Appl. Phys. Lett.}
  \textbf{\bibinfo{volume}{78}}, \bibinfo{pages}{946}.

\bibitem[{\citenamefont{Lotkhov} \emph{et~al.}(2011)\citenamefont{Lotkhov,
  Saira, Pekola, and Zorin}}]{Lotkhov2011}
\bibinfo{author}{\bibnamefont{Lotkhov}, \bibfnamefont{S.~V.}},
  \bibinfo{author}{\bibfnamefont{O.-P.} \bibnamefont{Saira}},
  \bibinfo{author}{\bibfnamefont{J.~P.} \bibnamefont{Pekola}}, and
  \bibinfo{author}{\bibfnamefont{A.~B.} \bibnamefont{Zorin}},
  \bibinfo{year}{2011}, \bibinfo{journal}{New J. Phys.}
  \textbf{\bibinfo{volume}{13}}, \bibinfo{pages}{013040}.

\bibitem[{\citenamefont{Low} \emph{et~al.}(2012)\citenamefont{Low, Jiang,
  Katsnelson, and Guinea}}]{Low2012}
\bibinfo{author}{\bibnamefont{Low}, \bibfnamefont{T.}},
  \bibinfo{author}{\bibfnamefont{Y.}~\bibnamefont{Jiang}},
  \bibinfo{author}{\bibfnamefont{M.}~\bibnamefont{Katsnelson}}, and
  \bibinfo{author}{\bibfnamefont{F.}~\bibnamefont{Guinea}},
  \bibinfo{year}{2012}, \bibinfo{journal}{Nano Lett.}
  \textbf{\bibinfo{volume}{12}}, \bibinfo{pages}{850}.

\bibitem[{\citenamefont{Lu} \emph{et~al.}(2003)\citenamefont{Lu, Ji, Pfeiffer,
  West, and Rimberg}}]{Lu2003}
\bibinfo{author}{\bibnamefont{Lu}, \bibfnamefont{W.}},
  \bibinfo{author}{\bibfnamefont{Z.}~\bibnamefont{Ji}},
  \bibinfo{author}{\bibfnamefont{L.}~\bibnamefont{Pfeiffer}},
  \bibinfo{author}{\bibfnamefont{K.~W.} \bibnamefont{West}}, and
  \bibinfo{author}{\bibfnamefont{A.~J.} \bibnamefont{Rimberg}},
  \bibinfo{year}{2003}, \bibinfo{journal}{Nature}
  \textbf{\bibinfo{volume}{423}}, \bibinfo{pages}{422}.

\bibitem[{\citenamefont{M. and C.}(1962)}]{Cohen1962}
\bibinfo{author}{\bibnamefont{M.}, \bibfnamefont{C.~M. H. F.~L.}}, and
  \bibinfo{author}{\bibfnamefont{P.~T.} \bibnamefont{C.}},
  \bibinfo{year}{1962}, \bibinfo{journal}{Phys. Rev. Lett.}
  \textbf{\bibinfo{volume}{8}}, \bibinfo{pages}{316}.

\bibitem[{\citenamefont{Maire} \emph{et~al.}(2008)\citenamefont{Maire, Hols,
  Kaestner, Pierz, Schumacher, and Haug}}]{Maire2008}
\bibinfo{author}{\bibnamefont{Maire}, \bibfnamefont{N.}},
  \bibinfo{author}{\bibfnamefont{F.}~\bibnamefont{Hols}},
  \bibinfo{author}{\bibfnamefont{B.}~\bibnamefont{Kaestner}},
  \bibinfo{author}{\bibfnamefont{K.}~\bibnamefont{Pierz}},
  \bibinfo{author}{\bibfnamefont{H.~W.} \bibnamefont{Schumacher}}, and
  \bibinfo{author}{\bibfnamefont{R.~J.} \bibnamefont{Haug}},
  \bibinfo{year}{2008}, \bibinfo{journal}{Appl. Phys. Lett.}
  \textbf{\bibinfo{volume}{92}}, \bibinfo{pages}{082112}.

\bibitem[{\citenamefont{Maisi} \emph{et~al.}(2009)\citenamefont{Maisi, Pashkin,
  Kafanov, Tsai, and Pekola}}]{Maisi2009}
\bibinfo{author}{\bibnamefont{Maisi}, \bibfnamefont{V.~F.}},
  \bibinfo{author}{\bibfnamefont{Y.~A.} \bibnamefont{Pashkin}},
  \bibinfo{author}{\bibfnamefont{S.}~\bibnamefont{Kafanov}},
  \bibinfo{author}{\bibfnamefont{J.~S.} \bibnamefont{Tsai}}, and
  \bibinfo{author}{\bibfnamefont{J.~P.} \bibnamefont{Pekola}},
  \bibinfo{year}{2009}, \bibinfo{journal}{New J. Phys.}
  \textbf{\bibinfo{volume}{11}}, \bibinfo{pages}{113057}.

\bibitem[{\citenamefont{Maisi} \emph{et~al.}(2011)\citenamefont{Maisi, Saira,
  Pashkin, Tsai, Averin, and Pekola}}]{Maisi2011}
\bibinfo{author}{\bibnamefont{Maisi}, \bibfnamefont{V.~F.}},
  \bibinfo{author}{\bibfnamefont{O.-P.} \bibnamefont{Saira}},
  \bibinfo{author}{\bibfnamefont{Y.~A.} \bibnamefont{Pashkin}},
  \bibinfo{author}{\bibfnamefont{J.~S.} \bibnamefont{Tsai}},
  \bibinfo{author}{\bibfnamefont{D.~V.} \bibnamefont{Averin}}, and
  \bibinfo{author}{\bibfnamefont{J.~P.} \bibnamefont{Pekola}},
  \bibinfo{year}{2011}, \bibinfo{journal}{Phys. Rev. Lett.}
  \textbf{\bibinfo{volume}{106}}, \bibinfo{pages}{217003}.

\bibitem[{\citenamefont{Makhlin} \emph{et~al.}(2001)\citenamefont{Makhlin,
  Sch\"on, and Shnirman}}]{Makhlin2001}
\bibinfo{author}{\bibnamefont{Makhlin}, \bibfnamefont{Y.}},
  \bibinfo{author}{\bibfnamefont{G.}~\bibnamefont{Sch\"on}}, and
  \bibinfo{author}{\bibfnamefont{A.}~\bibnamefont{Shnirman}},
  \bibinfo{year}{2001}, \bibinfo{journal}{Rev. Mod. Phys.}
  \textbf{\bibinfo{volume}{73}}, \bibinfo{pages}{357}.

\bibitem[{\citenamefont{Manninen} \emph{et~al.}(2010)\citenamefont{Manninen,
  Hahtela, Hakonen, Hassel, Helist\"o, Kemppinen, M\"ott\"onen, Paalanen,
  Pekola, Satrapinski, and Sepp\"{a}}}]{Manninen2008}
\bibinfo{author}{\bibnamefont{Manninen}, \bibfnamefont{A.}},
  \bibinfo{author}{\bibfnamefont{O.}~\bibnamefont{Hahtela}},
  \bibinfo{author}{\bibfnamefont{P.}~\bibnamefont{Hakonen}},
  \bibinfo{author}{\bibfnamefont{J.}~\bibnamefont{Hassel}},
  \bibinfo{author}{\bibfnamefont{P.}~\bibnamefont{Helist\"o}},
  \bibinfo{author}{\bibfnamefont{A.}~\bibnamefont{Kemppinen}},
  \bibinfo{author}{\bibfnamefont{M.}~\bibnamefont{M\"ott\"onen}},
  \bibinfo{author}{\bibfnamefont{M.}~\bibnamefont{Paalanen}},
  \bibinfo{author}{\bibfnamefont{J.}~\bibnamefont{Pekola}},
  \bibinfo{author}{\bibfnamefont{A.}~\bibnamefont{Satrapinski}}, and
  \bibinfo{author}{\bibfnamefont{H.}~\bibnamefont{Sepp\"{a}}},
  \bibinfo{year}{2010}, in \emph{\bibinfo{booktitle}{CPEM 2008 digest}}
  (\bibinfo{address}{Broomfield, Colorado, USA}), p. \bibinfo{pages}{630}.

\bibitem[{\citenamefont{Martinis} \emph{et~al.}(2009)\citenamefont{Martinis,
  Ansmann, and Aumentado}}]{Martinis2009}
\bibinfo{author}{\bibnamefont{Martinis}, \bibfnamefont{J.~M.}},
  \bibinfo{author}{\bibfnamefont{M.}~\bibnamefont{Ansmann}}, and
  \bibinfo{author}{\bibfnamefont{J.}~\bibnamefont{Aumentado}},
  \bibinfo{year}{2009}, \bibinfo{journal}{Phys. Rev. Lett.}
  \textbf{\bibinfo{volume}{103}}, \bibinfo{pages}{097002}.

\bibitem[{\citenamefont{Martinis and Nahum}(1993)}]{Martinis1993}
\bibinfo{author}{\bibnamefont{Martinis}, \bibfnamefont{J.~M.}}, and
  \bibinfo{author}{\bibfnamefont{M.}~\bibnamefont{Nahum}},
  \bibinfo{year}{1993}, \bibinfo{journal}{Phys. Rev. B}
  \textbf{\bibinfo{volume}{48}}, \bibinfo{pages}{18316}.

\bibitem[{\citenamefont{Martinis} \emph{et~al.}(1994)\citenamefont{Martinis,
  Nahum, and Jensen}}]{Martinis1994}
\bibinfo{author}{\bibnamefont{Martinis}, \bibfnamefont{J.~M.}},
  \bibinfo{author}{\bibfnamefont{M.}~\bibnamefont{Nahum}}, and
  \bibinfo{author}{\bibfnamefont{H.~D.} \bibnamefont{Jensen}},
  \bibinfo{year}{1994}, \bibinfo{journal}{Phys. Rev. Lett.}
  \textbf{\bibinfo{volume}{72}}, \bibinfo{pages}{904}.

\bibitem[{\citenamefont{\mbox{Bureau International des Poids et
  Mesures}}(2006)}]{SI2006}
\bibinfo{author}{\bibnamefont{\mbox{Bureau International des Poids et
  Mesures}}}, \bibinfo{year}{2006}, \emph{\bibinfo{title}{The International
  System of Units (SI)}} (\bibinfo{publisher}{STEDI Media},
  \bibinfo{address}{Paris}), \bibinfo{edition}{8} edition.

\bibitem[{\citenamefont{McNeil} \emph{et~al.}(2011)\citenamefont{McNeil,
  Kataoka, Ford, Barnes, Anderson, Jones, Farrer, and Ritchie}}]{McNeil2011}
\bibinfo{author}{\bibnamefont{McNeil}, \bibfnamefont{R.~P.~G.}},
  \bibinfo{author}{\bibfnamefont{M.}~\bibnamefont{Kataoka}},
  \bibinfo{author}{\bibfnamefont{C.~J.~B.} \bibnamefont{Ford}},
  \bibinfo{author}{\bibfnamefont{C.~H.~W.} \bibnamefont{Barnes}},
  \bibinfo{author}{\bibfnamefont{D.}~\bibnamefont{Anderson}},
  \bibinfo{author}{\bibfnamefont{G.~A.~C.} \bibnamefont{Jones}},
  \bibinfo{author}{\bibfnamefont{I.}~\bibnamefont{Farrer}}, and
  \bibinfo{author}{\bibfnamefont{D.~A.} \bibnamefont{Ritchie}},
  \bibinfo{year}{2011}, \bibinfo{journal}{Nature}
  \textbf{\bibinfo{volume}{477}}, \bibinfo{pages}{439}.

\bibitem[{\citenamefont{Meschke} \emph{et~al.}(2006)\citenamefont{Meschke,
  Guichard, and Pekola}}]{Meschke2006}
\bibinfo{author}{\bibnamefont{Meschke}, \bibfnamefont{M.}},
  \bibinfo{author}{\bibfnamefont{W.}~\bibnamefont{Guichard}}, and
  \bibinfo{author}{\bibfnamefont{J.~P.} \bibnamefont{Pekola}},
  \bibinfo{year}{2006}, \bibinfo{journal}{Nature}
  \textbf{\bibinfo{volume}{444}}, \bibinfo{pages}{187}.

\bibitem[{\citenamefont{Millikan}(1911)}]{Millikan1911}
\bibinfo{author}{\bibnamefont{Millikan}, \bibfnamefont{R.}},
  \bibinfo{year}{1911}, \bibinfo{journal}{Phys. Rev. (Ser. I)}
  \textbf{\bibinfo{volume}{32}}, \bibinfo{pages}{349}.

\bibitem[{\citenamefont{Mills} \emph{et~al.}(2005)\citenamefont{Mills, Mohr,
  Quinn, Taylor, and Williams}}]{Mills2005}
\bibinfo{author}{\bibnamefont{Mills}, \bibfnamefont{I.~M.}},
  \bibinfo{author}{\bibfnamefont{P.~J.} \bibnamefont{Mohr}},
  \bibinfo{author}{\bibfnamefont{T.~J.} \bibnamefont{Quinn}},
  \bibinfo{author}{\bibfnamefont{B.~N.} \bibnamefont{Taylor}}, and
  \bibinfo{author}{\bibfnamefont{E.~R.} \bibnamefont{Williams}},
  \bibinfo{year}{2005}, \bibinfo{journal}{Metrologia}
  \textbf{\bibinfo{volume}{42}}, \bibinfo{pages}{71}.

\bibitem[{\citenamefont{Mills} \emph{et~al.}(2006)\citenamefont{Mills, Mohr,
  Quinn, Taylor, and Williams}}]{Mills2006}
\bibinfo{author}{\bibnamefont{Mills}, \bibfnamefont{I.~M.}},
  \bibinfo{author}{\bibfnamefont{P.~J.} \bibnamefont{Mohr}},
  \bibinfo{author}{\bibfnamefont{T.~J.} \bibnamefont{Quinn}},
  \bibinfo{author}{\bibfnamefont{B.~N.} \bibnamefont{Taylor}}, and
  \bibinfo{author}{\bibfnamefont{E.~R.} \bibnamefont{Williams}},
  \bibinfo{year}{2006}, \bibinfo{journal}{Metrologia}
  \textbf{\bibinfo{volume}{43}}, \bibinfo{pages}{227}.

\bibitem[{\citenamefont{Mills} \emph{et~al.}(2011)\citenamefont{Mills, Mohr,
  Quinn, Taylor, and Williams}}]{Mills2011}
\bibinfo{author}{\bibnamefont{Mills}, \bibfnamefont{I.~M.}},
  \bibinfo{author}{\bibfnamefont{P.~J.} \bibnamefont{Mohr}},
  \bibinfo{author}{\bibfnamefont{T.~J.} \bibnamefont{Quinn}},
  \bibinfo{author}{\bibfnamefont{B.~N.} \bibnamefont{Taylor}}, and
  \bibinfo{author}{\bibfnamefont{E.~R.} \bibnamefont{Williams}},
  \bibinfo{year}{2011}, \bibinfo{journal}{Phil. Trans. Royal Soc. A}
  \textbf{\bibinfo{volume}{369}}, \bibinfo{pages}{3907}.

\bibitem[{\citenamefont{Milton} \emph{et~al.}(2007)\citenamefont{Milton,
  Williams, and Bennett}}]{Milton2007}
\bibinfo{author}{\bibnamefont{Milton}, \bibfnamefont{M.~J.~T.}},
  \bibinfo{author}{\bibfnamefont{J.~M.} \bibnamefont{Williams}}, and
  \bibinfo{author}{\bibfnamefont{S.~J.} \bibnamefont{Bennett}},
  \bibinfo{year}{2007}, \bibinfo{journal}{Metrologia}
  \textbf{\bibinfo{volume}{44}}, \bibinfo{pages}{356}.

\bibitem[{\citenamefont{Milton} \emph{et~al.}(2010)\citenamefont{Milton,
  Williams, and Forbes}}]{Milton2010}
\bibinfo{author}{\bibnamefont{Milton}, \bibfnamefont{M.~J.~T.}},
  \bibinfo{author}{\bibfnamefont{J.~M.} \bibnamefont{Williams}}, and
  \bibinfo{author}{\bibfnamefont{A.~B.} \bibnamefont{Forbes}},
  \bibinfo{year}{2010}, \bibinfo{journal}{Metrologia}
  \textbf{\bibinfo{volume}{47}}, \bibinfo{pages}{279}.

\bibitem[{\citenamefont{Mirovsky} \emph{et~al.}(2010)\citenamefont{Mirovsky,
  Kaestner, Leicht, Welker, Weimann, Pierz, and Schumacher}}]{Mirovsky2010}
\bibinfo{author}{\bibnamefont{Mirovsky}, \bibfnamefont{P.}},
  \bibinfo{author}{\bibfnamefont{B.}~\bibnamefont{Kaestner}},
  \bibinfo{author}{\bibfnamefont{C.}~\bibnamefont{Leicht}},
  \bibinfo{author}{\bibfnamefont{A.~C.} \bibnamefont{Welker}},
  \bibinfo{author}{\bibfnamefont{T.}~\bibnamefont{Weimann}},
  \bibinfo{author}{\bibfnamefont{K.}~\bibnamefont{Pierz}}, and
  \bibinfo{author}{\bibfnamefont{H.~W.} \bibnamefont{Schumacher}},
  \bibinfo{year}{2010}, \bibinfo{journal}{Appl. Phys. Lett.}
  \textbf{\bibinfo{volume}{97}}, \bibinfo{pages}{252104}.

\bibitem[{\citenamefont{Mohr and Taylor}(2000)}]{Mohr2000}
\bibinfo{author}{\bibnamefont{Mohr}, \bibfnamefont{P.~J.}}, and
  \bibinfo{author}{\bibfnamefont{B.~N.} \bibnamefont{Taylor}},
  \bibinfo{year}{2000}, \bibinfo{journal}{Rev. Mod. Phys.}
  \textbf{\bibinfo{volume}{72}}, \bibinfo{pages}{351}.

\bibitem[{\citenamefont{Mohr and Taylor}(2005)}]{Mohr2005}
\bibinfo{author}{\bibnamefont{Mohr}, \bibfnamefont{P.~J.}}, and
  \bibinfo{author}{\bibfnamefont{B.~N.} \bibnamefont{Taylor}},
  \bibinfo{year}{2005}, \bibinfo{journal}{Rev. Mod. Phys.}
  \textbf{\bibinfo{volume}{77}}, \bibinfo{pages}{1}.

\bibitem[{\citenamefont{Mohr} \emph{et~al.}(2008)\citenamefont{Mohr, Taylor,
  and Newell}}]{Mohr2008}
\bibinfo{author}{\bibnamefont{Mohr}, \bibfnamefont{P.~J.}},
  \bibinfo{author}{\bibfnamefont{B.~N.} \bibnamefont{Taylor}}, and
  \bibinfo{author}{\bibfnamefont{D.~B.} \bibnamefont{Newell}},
  \bibinfo{year}{2008}, \bibinfo{journal}{Rev. Mod. Phys.}
  \textbf{\bibinfo{volume}{80}}, \bibinfo{pages}{633}.

\bibitem[{\citenamefont{Mohr} \emph{et~al.}(2012)\citenamefont{Mohr, Taylor,
  and Newell}}]{Mohr2012}
\bibinfo{author}{\bibnamefont{Mohr}, \bibfnamefont{P.~J.}},
  \bibinfo{author}{\bibfnamefont{B.~N.} \bibnamefont{Taylor}}, and
  \bibinfo{author}{\bibfnamefont{D.~B.} \bibnamefont{Newell}},
  \bibinfo{year}{2012}, \bibinfo{journal}{arXiv:1203.5425} .

\bibitem[{\citenamefont{Mooij and Harmans}(2005)}]{Mooij2005}
\bibinfo{author}{\bibnamefont{Mooij}, \bibfnamefont{J.~E.}}, and
  \bibinfo{author}{\bibfnamefont{C.~J. P.~M.} \bibnamefont{Harmans}},
  \bibinfo{year}{2005}, \bibinfo{journal}{New J. Phys.}
  \textbf{\bibinfo{volume}{7}}, \bibinfo{pages}{219}.

\bibitem[{\citenamefont{Mooij and Nazarov}(2006)}]{Mooij2006}
\bibinfo{author}{\bibnamefont{Mooij}, \bibfnamefont{J.~E.}}, and
  \bibinfo{author}{\bibfnamefont{Y.~V.} \bibnamefont{Nazarov}},
  \bibinfo{year}{2006}, \bibinfo{journal}{Nature Phys.}
  \textbf{\bibinfo{volume}{2}}, \bibinfo{pages}{169}.

\bibitem[{\citenamefont{Morello} \emph{et~al.}(2010)\citenamefont{Morello, Pla,
  Zwanenburg, Chan, Huebl, M\"{o}tt\"{o}nen, Nugroho, Yang, van Donkelaar,
  Alves, Jamieson, Escott} \emph{et~al.}}]{Morello2010}
\bibinfo{author}{\bibnamefont{Morello}, \bibfnamefont{A.}},
  \bibinfo{author}{\bibfnamefont{J.~J.} \bibnamefont{Pla}},
  \bibinfo{author}{\bibfnamefont{F.~A.} \bibnamefont{Zwanenburg}},
  \bibinfo{author}{\bibfnamefont{K.~W.} \bibnamefont{Chan}},
  \bibinfo{author}{\bibfnamefont{H.}~\bibnamefont{Huebl}},
  \bibinfo{author}{\bibfnamefont{M.}~\bibnamefont{M\"{o}tt\"{o}nen}},
  \bibinfo{author}{\bibfnamefont{C.~D.} \bibnamefont{Nugroho}},
  \bibinfo{author}{\bibfnamefont{C.}~\bibnamefont{Yang}},
  \bibinfo{author}{\bibfnamefont{J.~A.} \bibnamefont{van Donkelaar}},
  \bibinfo{author}{\bibfnamefont{A.}~\bibnamefont{Alves}},
  \bibinfo{author}{\bibfnamefont{D.~N.} \bibnamefont{Jamieson}},
  \bibinfo{author}{\bibfnamefont{C.~C.} \bibnamefont{Escott}}, \emph{et~al.},
  \bibinfo{year}{2010}, \bibinfo{journal}{Nature}
  \textbf{\bibinfo{volume}{467}}, \bibinfo{pages}{687}.

\bibitem[{\citenamefont{Moskalenko}
  \emph{et~al.}(2009{\natexlab{a}})\citenamefont{Moskalenko, Gordeev,
  Koentjoro, Raithby, French, and Savel'ev}}]{Moskalenko2009N}
\bibinfo{author}{\bibnamefont{Moskalenko}, \bibfnamefont{A.~V.}},
  \bibinfo{author}{\bibfnamefont{S.~N.} \bibnamefont{Gordeev}},
  \bibinfo{author}{\bibfnamefont{O.~F.} \bibnamefont{Koentjoro}},
  \bibinfo{author}{\bibfnamefont{P.~R.} \bibnamefont{Raithby}},
  \bibinfo{author}{\bibfnamefont{F.}~\bibnamefont{French},
  \bibfnamefont{R.~W.~Marken}}, and \bibinfo{author}{\bibfnamefont{S.~E.}
  \bibnamefont{Savel'ev}}, \bibinfo{year}{2009}{\natexlab{a}},
  \bibinfo{journal}{Nanotechnology} \textbf{\bibinfo{volume}{20}},
  \bibinfo{pages}{485202}.

\bibitem[{\citenamefont{Moskalenko}
  \emph{et~al.}(2009{\natexlab{b}})\citenamefont{Moskalenko, Gordeev,
  Koentjoro, Raithby, French, and Savel'ev}}]{Moskalenko2009}
\bibinfo{author}{\bibnamefont{Moskalenko}, \bibfnamefont{A.~V.}},
  \bibinfo{author}{\bibfnamefont{S.~N.} \bibnamefont{Gordeev}},
  \bibinfo{author}{\bibfnamefont{O.~F.} \bibnamefont{Koentjoro}},
  \bibinfo{author}{\bibfnamefont{P.~R.} \bibnamefont{Raithby}},
  \bibinfo{author}{\bibfnamefont{F.}~\bibnamefont{French},
  \bibfnamefont{R.~W.~Marken}}, and \bibinfo{author}{\bibfnamefont{S.~E.}
  \bibnamefont{Savel'ev}}, \bibinfo{year}{2009}{\natexlab{b}},
  \bibinfo{journal}{Phys. Rev. B} \textbf{\bibinfo{volume}{79}},
  \bibinfo{pages}{241403}.

\bibitem[{\citenamefont{M\"ott\"onen}
  \emph{et~al.}(2006)\citenamefont{M\"ott\"onen, Pekola, Vartiainen, Brosco,
  and Hekking}}]{Mottonen2006}
\bibinfo{author}{\bibnamefont{M\"ott\"onen}, \bibfnamefont{M.}},
  \bibinfo{author}{\bibfnamefont{J.~P.} \bibnamefont{Pekola}},
  \bibinfo{author}{\bibfnamefont{J.~J.} \bibnamefont{Vartiainen}},
  \bibinfo{author}{\bibfnamefont{V.}~\bibnamefont{Brosco}}, and
  \bibinfo{author}{\bibfnamefont{F.~W.~J.} \bibnamefont{Hekking}},
  \bibinfo{year}{2006}, \bibinfo{journal}{Phys. Rev. B}
  \textbf{\bibinfo{volume}{73}}, \bibinfo{pages}{214523}.

\bibitem[{\citenamefont{M\"{o}tt\"{o}nen}
  \emph{et~al.}(2008)\citenamefont{M\"{o}tt\"{o}nen, Vartiainen, and
  Pekola}}]{Mottonen2008}
\bibinfo{author}{\bibnamefont{M\"{o}tt\"{o}nen}, \bibfnamefont{M.}},
  \bibinfo{author}{\bibfnamefont{J.~J.} \bibnamefont{Vartiainen}}, and
  \bibinfo{author}{\bibfnamefont{J.~P.} \bibnamefont{Pekola}},
  \bibinfo{year}{2008}, \bibinfo{journal}{Phys. Rev. Lett.}
  \textbf{\bibinfo{volume}{100}}, \bibinfo{pages}{177201}.

\bibitem[{\citenamefont{Naaman and Aumentado}(2006)}]{Naaman2006a}
\bibinfo{author}{\bibnamefont{Naaman}, \bibfnamefont{O.}}, and
  \bibinfo{author}{\bibfnamefont{J.}~\bibnamefont{Aumentado}},
  \bibinfo{year}{2006}, \bibinfo{journal}{Physical Review Letters}
  \textbf{\bibinfo{volume}{96}}, \bibinfo{pages}{172504}.

\bibitem[{\citenamefont{Nadj-Perge}
  \emph{et~al.}(2010)\citenamefont{Nadj-Perge, Frolov, Bakkers, and
  Kouwenhoven}}]{Nadj2010}
\bibinfo{author}{\bibnamefont{Nadj-Perge}, \bibfnamefont{S.}},
  \bibinfo{author}{\bibfnamefont{S.~M.} \bibnamefont{Frolov}},
  \bibinfo{author}{\bibfnamefont{E.~P. A.~M.} \bibnamefont{Bakkers}}, and
  \bibinfo{author}{\bibfnamefont{L.~P.} \bibnamefont{Kouwenhoven}},
  \bibinfo{year}{2010}, \bibinfo{journal}{Nature}
  \textbf{\bibinfo{volume}{468}}, \bibinfo{pages}{1084}.

\bibitem[{\citenamefont{Nagamune} \emph{et~al.}(1994)\citenamefont{Nagamune,
  Sakaki, Kouwenhoven, Mur, Harmans, Motohisa, and Noge}}]{Nagamune1994}
\bibinfo{author}{\bibnamefont{Nagamune}, \bibfnamefont{Y.}},
  \bibinfo{author}{\bibfnamefont{H.}~\bibnamefont{Sakaki}},
  \bibinfo{author}{\bibfnamefont{L.~P.} \bibnamefont{Kouwenhoven}},
  \bibinfo{author}{\bibfnamefont{L.~C.} \bibnamefont{Mur}},
  \bibinfo{author}{\bibfnamefont{C.~J. P.~M.} \bibnamefont{Harmans}},
  \bibinfo{author}{\bibfnamefont{J.}~\bibnamefont{Motohisa}}, and
  \bibinfo{author}{\bibfnamefont{H.}~\bibnamefont{Noge}}, \bibinfo{year}{1994},
  \bibinfo{journal}{Appl. Phys. Lett.} \textbf{\bibinfo{volume}{64}},
  \bibinfo{pages}{2379}.

\bibitem[{\citenamefont{Nakamura} \emph{et~al.}(1999)\citenamefont{Nakamura,
  Pashkin, and Tsai}}]{Nakamura1999}
\bibinfo{author}{\bibnamefont{Nakamura}, \bibfnamefont{Y.}},
  \bibinfo{author}{\bibfnamefont{Y.~A.} \bibnamefont{Pashkin}}, and
  \bibinfo{author}{\bibfnamefont{J.~S.} \bibnamefont{Tsai}},
  \bibinfo{year}{1999}, \bibinfo{journal}{Nature}
  \textbf{\bibinfo{volume}{398}}, \bibinfo{pages}{786}.

\bibitem[{\citenamefont{Nevou} \emph{et~al.}(2010)\citenamefont{Nevou,
  Liverini, Castellano, Bismuto, and J.}}]{Nevou2010}
\bibinfo{author}{\bibnamefont{Nevou}, \bibfnamefont{L.}},
  \bibinfo{author}{\bibfnamefont{V.}~\bibnamefont{Liverini}},
  \bibinfo{author}{\bibfnamefont{F.}~\bibnamefont{Castellano}},
  \bibinfo{author}{\bibfnamefont{A.}~\bibnamefont{Bismuto}}, and
  \bibinfo{author}{\bibfnamefont{F.}~\bibnamefont{J.}}, \bibinfo{year}{2010},
  \bibinfo{journal}{Appl. Phys. Lett} \textbf{\bibinfo{volume}{97}},
  \bibinfo{pages}{023505}.

\bibitem[{\citenamefont{Nevou} \emph{et~al.}(2011)\citenamefont{Nevou,
  Liverini, Friedli, Castellano, Bismuto, Sigg, Gramm, M?ller, and
  Faist}}]{Nevou2011}
\bibinfo{author}{\bibnamefont{Nevou}, \bibfnamefont{L.}},
  \bibinfo{author}{\bibfnamefont{V.}~\bibnamefont{Liverini}},
  \bibinfo{author}{\bibfnamefont{P.}~\bibnamefont{Friedli}},
  \bibinfo{author}{\bibfnamefont{F.}~\bibnamefont{Castellano}},
  \bibinfo{author}{\bibfnamefont{A.}~\bibnamefont{Bismuto}},
  \bibinfo{author}{\bibfnamefont{H.}~\bibnamefont{Sigg}},
  \bibinfo{author}{\bibfnamefont{F.}~\bibnamefont{Gramm}},
  \bibinfo{author}{\bibfnamefont{E.}~\bibnamefont{M?ller}}, and
  \bibinfo{author}{\bibfnamefont{J.}~\bibnamefont{Faist}},
  \bibinfo{year}{2011}, \bibinfo{journal}{Nature Phys.}
  \textbf{\bibinfo{volume}{7}}, \bibinfo{pages}{423}.

\bibitem[{\citenamefont{Nguyen} \emph{et~al.}(2007)\citenamefont{Nguyen,
  Boulant, Ithier, Bertet, Pothier, Vion, and Esteve}}]{Nguyen2007}
\bibinfo{author}{\bibnamefont{Nguyen}, \bibfnamefont{F.}},
  \bibinfo{author}{\bibfnamefont{N.}~\bibnamefont{Boulant}},
  \bibinfo{author}{\bibfnamefont{G.}~\bibnamefont{Ithier}},
  \bibinfo{author}{\bibfnamefont{P.}~\bibnamefont{Bertet}},
  \bibinfo{author}{\bibfnamefont{H.}~\bibnamefont{Pothier}},
  \bibinfo{author}{\bibfnamefont{D.}~\bibnamefont{Vion}}, and
  \bibinfo{author}{\bibfnamefont{D.}~\bibnamefont{Esteve}},
  \bibinfo{year}{2007}, \bibinfo{journal}{Phys. Rev. Lett.}
  \textbf{\bibinfo{volume}{99}}, \bibinfo{pages}{187005}.

\bibitem[{\citenamefont{Nishiguchi}
  \emph{et~al.}(2006)\citenamefont{Nishiguchi, Fujiwara, Ono, Inokawa, and
  Takahashi}}]{Nishiguchi2006}
\bibinfo{author}{\bibnamefont{Nishiguchi}, \bibfnamefont{K.}},
  \bibinfo{author}{\bibfnamefont{A.}~\bibnamefont{Fujiwara}},
  \bibinfo{author}{\bibfnamefont{Y.}~\bibnamefont{Ono}},
  \bibinfo{author}{\bibfnamefont{H.}~\bibnamefont{Inokawa}}, and
  \bibinfo{author}{\bibfnamefont{Y.}~\bibnamefont{Takahashi}},
  \bibinfo{year}{2006}, \bibinfo{journal}{Appl. Phys. Lett.}
  \textbf{\bibinfo{volume}{88}}, \bibinfo{pages}{183101}.

\bibitem[{\citenamefont{Niskanen} \emph{et~al.}(2005)\citenamefont{Niskanen,
  Kivioja, Sepp\"a, and Pekola}}]{Niskanen2005}
\bibinfo{author}{\bibnamefont{Niskanen}, \bibfnamefont{A.~O.}},
  \bibinfo{author}{\bibfnamefont{J.~M.} \bibnamefont{Kivioja}},
  \bibinfo{author}{\bibfnamefont{H.}~\bibnamefont{Sepp\"a}}, and
  \bibinfo{author}{\bibfnamefont{J.~P.} \bibnamefont{Pekola}},
  \bibinfo{year}{2005}, \bibinfo{journal}{Phys. Rev. B}
  \textbf{\bibinfo{volume}{71}}, \bibinfo{pages}{012513}.

\bibitem[{\citenamefont{Niskanen} \emph{et~al.}(2003)\citenamefont{Niskanen,
  Pekola, and Sepp\"{a}}}]{Niskanen2003}
\bibinfo{author}{\bibnamefont{Niskanen}, \bibfnamefont{A.~O.}},
  \bibinfo{author}{\bibfnamefont{J.~P.} \bibnamefont{Pekola}}, and
  \bibinfo{author}{\bibfnamefont{H.}~\bibnamefont{Sepp\"{a}}},
  \bibinfo{year}{2003}, \bibinfo{journal}{Phys. Rev. Lett.}
  \textbf{\bibinfo{volume}{91}}, \bibinfo{pages}{177003}.

\bibitem[{\citenamefont{Nordtvedt}(1970)}]{Nordtvedt1970}
\bibinfo{author}{\bibnamefont{Nordtvedt}, \bibfnamefont{K.}},
  \bibinfo{year}{1970}, \bibinfo{journal}{Phys. Rev. B}
  \textbf{\bibinfo{volume}{1}}, \bibinfo{pages}{81}.

\bibitem[{\citenamefont{Novoselov} \emph{et~al.}(2007)\citenamefont{Novoselov,
  Jiang, Zhang, Morozov, Stormer, Zeitler, Maan, Boebinger, Kim, and
  Geim}}]{Novoselov2007}
\bibinfo{author}{\bibnamefont{Novoselov}, \bibfnamefont{K.~S.}},
  \bibinfo{author}{\bibfnamefont{Z.}~\bibnamefont{Jiang}},
  \bibinfo{author}{\bibfnamefont{Y.}~\bibnamefont{Zhang}},
  \bibinfo{author}{\bibfnamefont{S.~V.} \bibnamefont{Morozov}},
  \bibinfo{author}{\bibfnamefont{H.~L.} \bibnamefont{Stormer}},
  \bibinfo{author}{\bibfnamefont{U.}~\bibnamefont{Zeitler}},
  \bibinfo{author}{\bibfnamefont{J.~C.} \bibnamefont{Maan}},
  \bibinfo{author}{\bibfnamefont{G.~S.} \bibnamefont{Boebinger}},
  \bibinfo{author}{\bibfnamefont{P.}~\bibnamefont{Kim}}, and
  \bibinfo{author}{\bibfnamefont{A.~K.} \bibnamefont{Geim}},
  \bibinfo{year}{2007}, \bibinfo{journal}{Science}
  \textbf{\bibinfo{volume}{315}}, \bibinfo{pages}{1379}.

\bibitem[{\citenamefont{Nyquist}(1928)}]{Nyquist1928}
\bibinfo{author}{\bibnamefont{Nyquist}, \bibfnamefont{H.}},
  \bibinfo{year}{1928}, \bibinfo{journal}{Phys. Rev.}
  \textbf{\bibinfo{volume}{32}}, \bibinfo{pages}{110}.

\bibitem[{\citenamefont{Ohlsson} \emph{et~al.}(2002)\citenamefont{Ohlsson,
  Bj\"ork, Persson, Thelander, Wallenberg, Magnusson, Deppert, and
  Samuelson}}]{Ohlsson2002}
\bibinfo{author}{\bibnamefont{Ohlsson}, \bibfnamefont{B.~J.}},
  \bibinfo{author}{\bibfnamefont{M.~T.} \bibnamefont{Bj\"ork}},
  \bibinfo{author}{\bibfnamefont{A.~I.} \bibnamefont{Persson}},
  \bibinfo{author}{\bibfnamefont{C.}~\bibnamefont{Thelander}},
  \bibinfo{author}{\bibfnamefont{L.~R.} \bibnamefont{Wallenberg}},
  \bibinfo{author}{\bibfnamefont{M.~H.} \bibnamefont{Magnusson}},
  \bibinfo{author}{\bibfnamefont{K.}~\bibnamefont{Deppert}}, and
  \bibinfo{author}{\bibfnamefont{L.}~\bibnamefont{Samuelson}},
  \bibinfo{year}{2002}, \bibinfo{journal}{Physica (Amsterdam)}
  \textbf{\bibinfo{volume}{13E}}, \bibinfo{pages}{1126}.

\bibitem[{\citenamefont{O'Neil} \emph{et~al.}(2011)\citenamefont{O'Neil,
  Lowell, Underwood, and Ullom}}]{Oneil2011}
\bibinfo{author}{\bibnamefont{O'Neil}, \bibfnamefont{G.~C.}},
  \bibinfo{author}{\bibfnamefont{P.~J.} \bibnamefont{Lowell}},
  \bibinfo{author}{\bibfnamefont{J.~M.} \bibnamefont{Underwood}}, and
  \bibinfo{author}{\bibfnamefont{J.~N.} \bibnamefont{Ullom}},
  \bibinfo{year}{2011}, \bibinfo{journal}{arXiv:1109.1273v1} .

\bibitem[{\citenamefont{Ono and Takahashi}(2003)}]{Ono2003}
\bibinfo{author}{\bibnamefont{Ono}, \bibfnamefont{Y.}}, and
  \bibinfo{author}{\bibfnamefont{Y.}~\bibnamefont{Takahashi}},
  \bibinfo{year}{2003}, \bibinfo{journal}{Appl. Phys. Lett.}
  \textbf{\bibinfo{volume}{82}}, \bibinfo{pages}{1221}.

\bibitem[{\citenamefont{Ono} \emph{et~al.}(2003)\citenamefont{Ono, Zimmerman,
  Yamazaki, and Takahashi}}]{Ono2003_2}
\bibinfo{author}{\bibnamefont{Ono}, \bibfnamefont{Y.}},
  \bibinfo{author}{\bibfnamefont{N.~M.} \bibnamefont{Zimmerman}},
  \bibinfo{author}{\bibfnamefont{K.}~\bibnamefont{Yamazaki}}, and
  \bibinfo{author}{\bibfnamefont{Y.}~\bibnamefont{Takahashi}},
  \bibinfo{year}{2003}, \bibinfo{journal}{Jpn. J. Appl. Phys., Part 2}
  \textbf{\bibinfo{volume}{42}}, \bibinfo{pages}{L1109}.

\bibitem[{\citenamefont{Park} \emph{et~al.}(2000)\citenamefont{Park, Park, Lim,
  Anderson, Alivisatos, and McEuen}}]{Park2000}
\bibinfo{author}{\bibnamefont{Park}, \bibfnamefont{H.}},
  \bibinfo{author}{\bibfnamefont{J.}~\bibnamefont{Park}},
  \bibinfo{author}{\bibfnamefont{A.~K.~L.} \bibnamefont{Lim}},
  \bibinfo{author}{\bibfnamefont{E.~H.} \bibnamefont{Anderson}},
  \bibinfo{author}{\bibfnamefont{A.~P.} \bibnamefont{Alivisatos}}, and
  \bibinfo{author}{\bibfnamefont{P.~L.} \bibnamefont{McEuen}},
  \bibinfo{year}{2000}, \bibinfo{journal}{Nature}
  \textbf{\bibinfo{volume}{407}}, \bibinfo{pages}{57}.

\bibitem[{\citenamefont{Pekola} \emph{et~al.}(2000)\citenamefont{Pekola,
  Anghel, Suppula, Suoknuuti, Manninen, and Manninen}}]{Pekola2000}
\bibinfo{author}{\bibnamefont{Pekola}, \bibfnamefont{J.~P.}},
  \bibinfo{author}{\bibfnamefont{D.~V.} \bibnamefont{Anghel}},
  \bibinfo{author}{\bibfnamefont{T.~I.} \bibnamefont{Suppula}},
  \bibinfo{author}{\bibfnamefont{J.~K.} \bibnamefont{Suoknuuti}},
  \bibinfo{author}{\bibfnamefont{A.~J.} \bibnamefont{Manninen}}, and
  \bibinfo{author}{\bibfnamefont{M.}~\bibnamefont{Manninen}},
  \bibinfo{year}{2000}, \bibinfo{journal}{Appl. Phys. Lett.}
  \textbf{\bibinfo{volume}{76}}, \bibinfo{pages}{2782}.

\bibitem[{\citenamefont{Pekola and Hekking}(2007)}]{Pekola2007a}
\bibinfo{author}{\bibnamefont{Pekola}, \bibfnamefont{J.~P.}}, and
  \bibinfo{author}{\bibfnamefont{F.~W.~J.} \bibnamefont{Hekking}},
  \bibinfo{year}{2007}, \bibinfo{journal}{Phys. Rev. Lett.}
  \textbf{\bibinfo{volume}{98}}, \bibinfo{pages}{210604}.

\bibitem[{\citenamefont{Pekola} \emph{et~al.}(2010)\citenamefont{Pekola, Maisi,
  Kafanov, Chekurov, Kemppinen, Pashkin, Saira, M\"ott\"onen, and
  Tsai}}]{Pekola2010}
\bibinfo{author}{\bibnamefont{Pekola}, \bibfnamefont{J.~P.}},
  \bibinfo{author}{\bibfnamefont{V.~F.} \bibnamefont{Maisi}},
  \bibinfo{author}{\bibfnamefont{S.}~\bibnamefont{Kafanov}},
  \bibinfo{author}{\bibfnamefont{N.}~\bibnamefont{Chekurov}},
  \bibinfo{author}{\bibfnamefont{A.}~\bibnamefont{Kemppinen}},
  \bibinfo{author}{\bibfnamefont{Y.~A.} \bibnamefont{Pashkin}},
  \bibinfo{author}{\bibfnamefont{O.-P.} \bibnamefont{Saira}},
  \bibinfo{author}{\bibfnamefont{M.}~\bibnamefont{M\"ott\"onen}}, and
  \bibinfo{author}{\bibfnamefont{J.~S.} \bibnamefont{Tsai}},
  \bibinfo{year}{2010}, \bibinfo{journal}{Phys. Rev. Lett.}
  \textbf{\bibinfo{volume}{105}}, \bibinfo{pages}{026803}.

\bibitem[{\citenamefont{Pekola} \emph{et~al.}(1999)\citenamefont{Pekola,
  Toppari, Aunola, Savolainen, and Averin}}]{Pekola1999}
\bibinfo{author}{\bibnamefont{Pekola}, \bibfnamefont{J.~P.}},
  \bibinfo{author}{\bibfnamefont{J.~J.} \bibnamefont{Toppari}},
  \bibinfo{author}{\bibfnamefont{M.}~\bibnamefont{Aunola}},
  \bibinfo{author}{\bibfnamefont{M.~T.} \bibnamefont{Savolainen}}, and
  \bibinfo{author}{\bibfnamefont{D.~V.} \bibnamefont{Averin}},
  \bibinfo{year}{1999}, \bibinfo{journal}{Phys. Rev. B}
  \textbf{\bibinfo{volume}{60}}, \bibinfo{pages}{R9931}.

\bibitem[{\citenamefont{Pekola} \emph{et~al.}(2008)\citenamefont{Pekola,
  Vartiainen, M\"{o}tt\"{o}nen, Saira, Meschke, and Averin}}]{Pekola2008}
\bibinfo{author}{\bibnamefont{Pekola}, \bibfnamefont{J.~P.}},
  \bibinfo{author}{\bibfnamefont{J.~J.} \bibnamefont{Vartiainen}},
  \bibinfo{author}{\bibfnamefont{M.}~\bibnamefont{M\"{o}tt\"{o}nen}},
  \bibinfo{author}{\bibfnamefont{O.-P.} \bibnamefont{Saira}},
  \bibinfo{author}{\bibfnamefont{M.}~\bibnamefont{Meschke}}, and
  \bibinfo{author}{\bibfnamefont{D.~V.} \bibnamefont{Averin}},
  \bibinfo{year}{2008}, \bibinfo{journal}{Nature Phys.}
  \textbf{\bibinfo{volume}{4}}, \bibinfo{pages}{120}.

\bibitem[{\citenamefont{Peltonen} \emph{et~al.}(2011)\citenamefont{Peltonen,
  Muhonen, Meschke, Kopnin, and Pekola}}]{Peltonen2011}
\bibinfo{author}{\bibnamefont{Peltonen}, \bibfnamefont{J.~T.}},
  \bibinfo{author}{\bibfnamefont{J.~T.} \bibnamefont{Muhonen}},
  \bibinfo{author}{\bibfnamefont{M.}~\bibnamefont{Meschke}},
  \bibinfo{author}{\bibfnamefont{N.~B.} \bibnamefont{Kopnin}}, and
  \bibinfo{author}{\bibfnamefont{J.~P.} \bibnamefont{Pekola}},
  \bibinfo{year}{2011}, \bibinfo{journal}{Phys. Rev. B}
  \textbf{\bibinfo{volume}{84}}, \bibinfo{pages}{220502}.

\bibitem[{\citenamefont{Pendry}(1983)}]{Pendry1983}
\bibinfo{author}{\bibnamefont{Pendry}, \bibfnamefont{J.~B.}},
  \bibinfo{year}{1983}, \bibinfo{journal}{J. Phys. A}
  \textbf{\bibinfo{volume}{16}}, \bibinfo{pages}{2161}.

\bibitem[{\citenamefont{Penin}(2009)}]{Penin2009}
\bibinfo{author}{\bibnamefont{Penin}, \bibfnamefont{A.~A.}},
  \bibinfo{year}{2009}, \bibinfo{journal}{Phys. Rev. B}
  \textbf{\bibinfo{volume}{79}}, \bibinfo{pages}{113303}.

\bibitem[{\citenamefont{Penin}(2010{\natexlab{a}})}]{Penin2010}
\bibinfo{author}{\bibnamefont{Penin}, \bibfnamefont{A.~A.}},
  \bibinfo{year}{2010}{\natexlab{a}}, \bibinfo{journal}{Phys. Rev. B}
  \textbf{\bibinfo{volume}{81}}, \bibinfo{pages}{089902}.

\bibitem[{\citenamefont{Penin}(2010{\natexlab{b}})}]{Penin2010a}
\bibinfo{author}{\bibnamefont{Penin}, \bibfnamefont{A.~A.}},
  \bibinfo{year}{2010}{\natexlab{b}}, \bibinfo{journal}{Phys. Rev. Lett.}
  \textbf{\bibinfo{volume}{104}}, \bibinfo{pages}{097003}.

\bibitem[{\citenamefont{Pettersson}
  \emph{et~al.}(1996)\citenamefont{Pettersson, Wahlgren, Delsing, Haviland,
  Claeson, Rorsman, and Zirath}}]{Pettersson1996}
\bibinfo{author}{\bibnamefont{Pettersson}, \bibfnamefont{J.}},
  \bibinfo{author}{\bibfnamefont{P.}~\bibnamefont{Wahlgren}},
  \bibinfo{author}{\bibfnamefont{P.}~\bibnamefont{Delsing}},
  \bibinfo{author}{\bibfnamefont{D.}~\bibnamefont{Haviland}},
  \bibinfo{author}{\bibfnamefont{T.}~\bibnamefont{Claeson}},
  \bibinfo{author}{\bibfnamefont{N.}~\bibnamefont{Rorsman}}, and
  \bibinfo{author}{\bibfnamefont{H.}~\bibnamefont{Zirath}},
  \bibinfo{year}{1996}, \bibinfo{journal}{Physical Review B}
  \textbf{\bibinfo{volume}{53}}, \bibinfo{pages}{R13272}.

\bibitem[{\citenamefont{Piquemal}(2004)}]{Piquemal2004}
\bibinfo{author}{\bibnamefont{Piquemal}, \bibfnamefont{F.}},
  \bibinfo{year}{2004}, \bibinfo{journal}{C.R. Physique}
  \textbf{\bibinfo{volume}{5}}, \bibinfo{pages}{857}.

\bibitem[{\citenamefont{Piquemal}(2012)}]{Piquemal2012}
\bibinfo{author}{\bibnamefont{Piquemal}, \bibfnamefont{F.}},
  \bibinfo{year}{2012}, \bibinfo{journal}{Meas. Sci. Technol. (in preparation)}
  .

\bibitem[{\citenamefont{Piquemal and Geneves}(2000)}]{Piquemal2000}
\bibinfo{author}{\bibnamefont{Piquemal}, \bibfnamefont{F.}}, and
  \bibinfo{author}{\bibfnamefont{G.}~\bibnamefont{Geneves}},
  \bibinfo{year}{2000}, \bibinfo{journal}{Metrologia}
  \textbf{\bibinfo{volume}{37}}, \bibinfo{pages}{207}.

\bibitem[{\citenamefont{Pirkkalainen}
  \emph{et~al.}(2010)\citenamefont{Pirkkalainen, Solinas, Pekola, and
  M\"ott\"onen}}]{Pirkkalainen2010}
\bibinfo{author}{\bibnamefont{Pirkkalainen}, \bibfnamefont{J.-M.}},
  \bibinfo{author}{\bibfnamefont{P.}~\bibnamefont{Solinas}},
  \bibinfo{author}{\bibfnamefont{J.~P.} \bibnamefont{Pekola}}, and
  \bibinfo{author}{\bibfnamefont{M.}~\bibnamefont{M\"ott\"onen}},
  \bibinfo{year}{2010}, \bibinfo{journal}{Phys. Rev. B}
  \textbf{\bibinfo{volume}{81}}, \bibinfo{pages}{174506}.

\bibitem[{\citenamefont{Poirier and Schopfer}(2009)}]{Poirier2009}
\bibinfo{author}{\bibnamefont{Poirier}, \bibfnamefont{W.}}, and
  \bibinfo{author}{\bibfnamefont{F.}~\bibnamefont{Schopfer}},
  \bibinfo{year}{2009}, \bibinfo{journal}{Eur. Phys. J. Spec. Top.}
  \textbf{\bibinfo{volume}{172}}, \bibinfo{pages}{207}.

\bibitem[{\citenamefont{Poirier} \emph{et~al.}(2011)\citenamefont{Poirier,
  Schopfer, Guignard, Th\'evenot, and Gournay}}]{Poirier2011}
\bibinfo{author}{\bibnamefont{Poirier}, \bibfnamefont{W.}},
  \bibinfo{author}{\bibfnamefont{F.}~\bibnamefont{Schopfer}},
  \bibinfo{author}{\bibfnamefont{J.}~\bibnamefont{Guignard}},
  \bibinfo{author}{\bibfnamefont{O.}~\bibnamefont{Th\'evenot}}, and
  \bibinfo{author}{\bibfnamefont{P.}~\bibnamefont{Gournay}},
  \bibinfo{year}{2011}, \bibinfo{journal}{Comptes Rendus Physique}
  \textbf{\bibinfo{volume}{12}}, \bibinfo{pages}{347}.

\bibitem[{\citenamefont{Pop} \emph{et~al.}(2010)\citenamefont{Pop, Protopopov,
  Lecocq, Peng, Pannetier, Buisson, and Guichard}}]{Pop2010}
\bibinfo{author}{\bibnamefont{Pop}, \bibfnamefont{I.~M.}},
  \bibinfo{author}{\bibfnamefont{I.}~\bibnamefont{Protopopov}},
  \bibinfo{author}{\bibfnamefont{F.}~\bibnamefont{Lecocq}},
  \bibinfo{author}{\bibfnamefont{Z.}~\bibnamefont{Peng}},
  \bibinfo{author}{\bibfnamefont{B.}~\bibnamefont{Pannetier}},
  \bibinfo{author}{\bibfnamefont{O.}~\bibnamefont{Buisson}}, and
  \bibinfo{author}{\bibfnamefont{W.}~\bibnamefont{Guichard}},
  \bibinfo{year}{2010}, \bibinfo{journal}{Nature Phys.}
  \textbf{\bibinfo{volume}{6}}, \bibinfo{pages}{589}.

\bibitem[{\citenamefont{Pothier}(1991)}]{Pothier1991a}
\bibinfo{author}{\bibnamefont{Pothier}, \bibfnamefont{H.}},
  \bibinfo{year}{1991}, \emph{\bibinfo{title}{PhD Thesis}}
  (\bibinfo{publisher}{University of Paris 6 (unpublished)}).

\bibitem[{\citenamefont{Pothier} \emph{et~al.}(1994)\citenamefont{Pothier,
  Gu\'eron, Esteve, and Devoret}}]{Pothier1994}
\bibinfo{author}{\bibnamefont{Pothier}, \bibfnamefont{H.}},
  \bibinfo{author}{\bibfnamefont{S.}~\bibnamefont{Gu\'eron}},
  \bibinfo{author}{\bibfnamefont{D.}~\bibnamefont{Esteve}}, and
  \bibinfo{author}{\bibfnamefont{M.~H.} \bibnamefont{Devoret}},
  \bibinfo{year}{1994}, \bibinfo{journal}{Phys. Rev. Lett.}
  \textbf{\bibinfo{volume}{73}}, \bibinfo{pages}{2488}.

\bibitem[{\citenamefont{Pothier} \emph{et~al.}(1991)\citenamefont{Pothier,
  Lafarge, Orfila, Urbina, Esteve, and Devoret}}]{Pothier1991}
\bibinfo{author}{\bibnamefont{Pothier}, \bibfnamefont{H.}},
  \bibinfo{author}{\bibfnamefont{P.}~\bibnamefont{Lafarge}},
  \bibinfo{author}{\bibfnamefont{P.~F.} \bibnamefont{Orfila}},
  \bibinfo{author}{\bibfnamefont{C.}~\bibnamefont{Urbina}},
  \bibinfo{author}{\bibfnamefont{D.}~\bibnamefont{Esteve}}, and
  \bibinfo{author}{\bibfnamefont{M.~H.} \bibnamefont{Devoret}},
  \bibinfo{year}{1991}, \bibinfo{journal}{Physica B: Condensed Matter}
  \textbf{\bibinfo{volume}{169}}, \bibinfo{pages}{573}.

\bibitem[{\citenamefont{Pothier} \emph{et~al.}(1992)\citenamefont{Pothier,
  Lafarge, Urbina, Esteve, and Devoret}}]{Pothier1992}
\bibinfo{author}{\bibnamefont{Pothier}, \bibfnamefont{H.}},
  \bibinfo{author}{\bibfnamefont{P.}~\bibnamefont{Lafarge}},
  \bibinfo{author}{\bibfnamefont{C.}~\bibnamefont{Urbina}},
  \bibinfo{author}{\bibfnamefont{D.}~\bibnamefont{Esteve}}, and
  \bibinfo{author}{\bibfnamefont{M.~H.} \bibnamefont{Devoret}},
  \bibinfo{year}{1992}, \bibinfo{journal}{Europhys. Lett.}
  \textbf{\bibinfo{volume}{17}}, \bibinfo{pages}{249}.

\bibitem[{\citenamefont{Pr\^etre} \emph{et~al.}(1996)\citenamefont{Pr\^etre,
  Thomas, and B\"uttiker}}]{Pretre1996}
\bibinfo{author}{\bibnamefont{Pr\^etre}, \bibfnamefont{A.}},
  \bibinfo{author}{\bibfnamefont{H.}~\bibnamefont{Thomas}}, and
  \bibinfo{author}{\bibfnamefont{M.}~\bibnamefont{B\"uttiker}},
  \bibinfo{year}{1996}, \bibinfo{journal}{Phys. Rev. B}
  \textbf{\bibinfo{volume}{54}}, \bibinfo{pages}{8130}.

\bibitem[{\citenamefont{Prunnila} \emph{et~al.}(2010)\citenamefont{Prunnila,
  Meschke, Gunnarsson, Enouz-Vedrenne, Kivioja, and Pekola}}]{Prunnila2010}
\bibinfo{author}{\bibnamefont{Prunnila}, \bibfnamefont{M.}},
  \bibinfo{author}{\bibfnamefont{M.}~\bibnamefont{Meschke}},
  \bibinfo{author}{\bibfnamefont{D.}~\bibnamefont{Gunnarsson}},
  \bibinfo{author}{\bibfnamefont{S.}~\bibnamefont{Enouz-Vedrenne}},
  \bibinfo{author}{\bibfnamefont{J.~M.} \bibnamefont{Kivioja}}, and
  \bibinfo{author}{\bibfnamefont{J.~P.} \bibnamefont{Pekola}},
  \bibinfo{year}{2010}, \bibinfo{journal}{Journal of Vacuum Science and
  Technology B: Microelectronics and Nanometer Structures}
  \textbf{\bibinfo{volume}{28}}(\bibinfo{number}{5}), \bibinfo{pages}{1026}.

\bibitem[{\citenamefont{Qin and Williams}(2006)}]{Qin2006}
\bibinfo{author}{\bibnamefont{Qin}, \bibfnamefont{H.}}, and
  \bibinfo{author}{\bibfnamefont{D.~A.} \bibnamefont{Williams}},
  \bibinfo{year}{2006}, \bibinfo{journal}{Applied Physics Letters}
  \textbf{\bibinfo{volume}{88}}, \bibinfo{pages}{203506}.

\bibitem[{\citenamefont{Quinn}(1991)}]{Quinn1991}
\bibinfo{author}{\bibnamefont{Quinn}, \bibfnamefont{T.~J.}},
  \bibinfo{year}{1991}, \bibinfo{journal}{IEEE Trans. Instrum. Meas.}
  \textbf{\bibinfo{volume}{40}}, \bibinfo{pages}{81}.

\bibitem[{\citenamefont{Rajauria} \emph{et~al.}(2009)\citenamefont{Rajauria,
  Courtois, and Pannetier}}]{Rajauria2009-2}
\bibinfo{author}{\bibnamefont{Rajauria}, \bibfnamefont{S.}},
  \bibinfo{author}{\bibfnamefont{H.}~\bibnamefont{Courtois}}, and
  \bibinfo{author}{\bibfnamefont{B.}~\bibnamefont{Pannetier}},
  \bibinfo{year}{2009}, \bibinfo{journal}{Phys. Rev. B}
  \textbf{\bibinfo{volume}{80}}, \bibinfo{pages}{214521}.

\bibitem[{\citenamefont{Rajauria} \emph{et~al.}(2008)\citenamefont{Rajauria,
  Gandit, Fournier, Hekking, Pannetier, and Courtois}}]{Rajauria2008}
\bibinfo{author}{\bibnamefont{Rajauria}, \bibfnamefont{S.}},
  \bibinfo{author}{\bibfnamefont{P.}~\bibnamefont{Gandit}},
  \bibinfo{author}{\bibfnamefont{T.}~\bibnamefont{Fournier}},
  \bibinfo{author}{\bibfnamefont{F.~W.~J.} \bibnamefont{Hekking}},
  \bibinfo{author}{\bibfnamefont{B.}~\bibnamefont{Pannetier}}, and
  \bibinfo{author}{\bibfnamefont{H.}~\bibnamefont{Courtois}},
  \bibinfo{year}{2008}, \bibinfo{journal}{Phys. Rev. Lett.}
  \textbf{\bibinfo{volume}{100}}, \bibinfo{pages}{207002}.

\bibitem[{\citenamefont{Reed} \emph{et~al.}(1988)\citenamefont{Reed, Randall,
  Aggarwal, Matyi, Moore, and Wetsel}}]{Reed1988}
\bibinfo{author}{\bibnamefont{Reed}, \bibfnamefont{M.~A.}},
  \bibinfo{author}{\bibfnamefont{J.~N.} \bibnamefont{Randall}},
  \bibinfo{author}{\bibfnamefont{R.~J.} \bibnamefont{Aggarwal}},
  \bibinfo{author}{\bibfnamefont{R.~J.} \bibnamefont{Matyi}},
  \bibinfo{author}{\bibfnamefont{T.~M.} \bibnamefont{Moore}}, and
  \bibinfo{author}{\bibfnamefont{A.~E.} \bibnamefont{Wetsel}},
  \bibinfo{year}{1988}, \bibinfo{journal}{Phys. Rev. Lett.}
  \textbf{\bibinfo{volume}{60}}, \bibinfo{pages}{535}.

\bibitem[{\citenamefont{Roschier} \emph{et~al.}(2001)\citenamefont{Roschier,
  Tarkiainen, Ahlskog, Paalanen, and Hakonen}}]{Roschier2001}
\bibinfo{author}{\bibnamefont{Roschier}, \bibfnamefont{L.}},
  \bibinfo{author}{\bibfnamefont{R.}~\bibnamefont{Tarkiainen}},
  \bibinfo{author}{\bibfnamefont{M.}~\bibnamefont{Ahlskog}},
  \bibinfo{author}{\bibfnamefont{M.}~\bibnamefont{Paalanen}}, and
  \bibinfo{author}{\bibfnamefont{P.}~\bibnamefont{Hakonen}},
  \bibinfo{year}{2001}, \bibinfo{journal}{Appl. Phys. Lett.}
  \textbf{\bibinfo{volume}{78}}, \bibinfo{pages}{3295}.

\bibitem[{\citenamefont{Rothwarf and Taylor}(1967)}]{Rothwarf1967}
\bibinfo{author}{\bibnamefont{Rothwarf}, \bibfnamefont{A.}}, and
  \bibinfo{author}{\bibfnamefont{B.~N.} \bibnamefont{Taylor}},
  \bibinfo{year}{1967}, \bibinfo{journal}{Phys. Rev. Lett.}
  \textbf{\bibinfo{volume}{19}}, \bibinfo{pages}{27}.

\bibitem[{\citenamefont{R\"ufenacht}
  \emph{et~al.}(2010)\citenamefont{R\"ufenacht, Jeanneret, and
  Lotkhov}}]{Rufenacht2010}
\bibinfo{author}{\bibnamefont{R\"ufenacht}, \bibfnamefont{A.}},
  \bibinfo{author}{\bibfnamefont{B.}~\bibnamefont{Jeanneret}}, and
  \bibinfo{author}{\bibfnamefont{S.~V.} \bibnamefont{Lotkhov}},
  \bibinfo{year}{2010}, in \emph{\bibinfo{booktitle}{CPEM 2010 digest}}, p.
  \bibinfo{pages}{498}.

\bibitem[{\citenamefont{Sac\'{e}p\'{e}}
  \emph{et~al.}(2010)\citenamefont{Sac\'{e}p\'{e}, Chapelier, Baturina,
  Vinokur, R., and Sanquer}}]{Sacepe2010}
\bibinfo{author}{\bibnamefont{Sac\'{e}p\'{e}}, \bibfnamefont{B.}},
  \bibinfo{author}{\bibfnamefont{C.}~\bibnamefont{Chapelier}},
  \bibinfo{author}{\bibfnamefont{T.~I.} \bibnamefont{Baturina}},
  \bibinfo{author}{\bibfnamefont{V.~M.} \bibnamefont{Vinokur}},
  \bibinfo{author}{\bibfnamefont{B.~M.} \bibnamefont{R.}}, and
  \bibinfo{author}{\bibfnamefont{M.}~\bibnamefont{Sanquer}},
  \bibinfo{year}{2010}, \bibinfo{journal}{Nature Commun.}
  \textbf{\bibinfo{volume}{1}}, \bibinfo{pages}{140}.

\bibitem[{\citenamefont{Sac\'{e}p\'{e}}
  \emph{et~al.}(2011)\citenamefont{Sac\'{e}p\'{e}, Dubouchet, Chapelier,
  Sanquer, Ovadia, Shahar, Feigel'man, and Ioffe}}]{Sacepe2011}
\bibinfo{author}{\bibnamefont{Sac\'{e}p\'{e}}, \bibfnamefont{B.}},
  \bibinfo{author}{\bibfnamefont{T.}~\bibnamefont{Dubouchet}},
  \bibinfo{author}{\bibfnamefont{C.}~\bibnamefont{Chapelier}},
  \bibinfo{author}{\bibfnamefont{M.}~\bibnamefont{Sanquer}},
  \bibinfo{author}{\bibfnamefont{M.}~\bibnamefont{Ovadia}},
  \bibinfo{author}{\bibfnamefont{D.}~\bibnamefont{Shahar}},
  \bibinfo{author}{\bibfnamefont{M.}~\bibnamefont{Feigel'man}}, and
  \bibinfo{author}{\bibfnamefont{L.}~\bibnamefont{Ioffe}},
  \bibinfo{year}{2011}, \bibinfo{journal}{Nature Phys.}
  \textbf{\bibinfo{volume}{7}}, \bibinfo{pages}{239}.

\bibitem[{\citenamefont{Saira}
  \emph{et~al.}(2012{\natexlab{a}})\citenamefont{Saira, Kemppinen, Maisi, and
  Pekola}}]{Saira2012}
\bibinfo{author}{\bibnamefont{Saira}, \bibfnamefont{O.-P.}},
  \bibinfo{author}{\bibfnamefont{A.}~\bibnamefont{Kemppinen}},
  \bibinfo{author}{\bibfnamefont{V.~F.} \bibnamefont{Maisi}}, and
  \bibinfo{author}{\bibfnamefont{J.~P.} \bibnamefont{Pekola}},
  \bibinfo{year}{2012}{\natexlab{a}}, \bibinfo{journal}{Phys. Rev. B}
  \textbf{\bibinfo{volume}{85}}, \bibinfo{pages}{012504}.

\bibitem[{\citenamefont{Saira} \emph{et~al.}(2010)\citenamefont{Saira,
  M\"ott\"onen, Maisi, and Pekola}}]{Saira2010}
\bibinfo{author}{\bibnamefont{Saira}, \bibfnamefont{O.-P.}},
  \bibinfo{author}{\bibfnamefont{M.}~\bibnamefont{M\"ott\"onen}},
  \bibinfo{author}{\bibfnamefont{V.~F.} \bibnamefont{Maisi}}, and
  \bibinfo{author}{\bibfnamefont{J.~P.} \bibnamefont{Pekola}},
  \bibinfo{year}{2010}, \bibinfo{journal}{Phys. Rev. B}
  \textbf{\bibinfo{volume}{82}}, \bibinfo{pages}{155443}.

\bibitem[{\citenamefont{Saira}
  \emph{et~al.}(2012{\natexlab{b}})\citenamefont{Saira, Yoon, Tanttu,
  M\"ott\"onen, Averin, and Pekola}}]{Saira2012b}
\bibinfo{author}{\bibnamefont{Saira}, \bibfnamefont{O.-P.}},
  \bibinfo{author}{\bibfnamefont{Y.}~\bibnamefont{Yoon}},
  \bibinfo{author}{\bibfnamefont{T.}~\bibnamefont{Tanttu}},
  \bibinfo{author}{\bibfnamefont{M.}~\bibnamefont{M\"ott\"onen}},
  \bibinfo{author}{\bibfnamefont{D.~V.} \bibnamefont{Averin}}, and
  \bibinfo{author}{\bibfnamefont{J.~P.} \bibnamefont{Pekola}},
  \bibinfo{year}{2012}{\natexlab{b}}, \bibinfo{journal}{arXiv:1206.7049} .

\bibitem[{\citenamefont{Savin} \emph{et~al.}(2006)\citenamefont{Savin, Pekola,
  Averin, and Semenov}}]{Savin2006}
\bibinfo{author}{\bibnamefont{Savin}, \bibfnamefont{A.~M.}},
  \bibinfo{author}{\bibfnamefont{J.~P.} \bibnamefont{Pekola}},
  \bibinfo{author}{\bibfnamefont{D.~V.} \bibnamefont{Averin}}, and
  \bibinfo{author}{\bibfnamefont{V.~K.} \bibnamefont{Semenov}},
  \bibinfo{year}{2006}, \bibinfo{journal}{Journal of Applied Physics}
  \textbf{\bibinfo{volume}{99}}, \bibinfo{pages}{084501}.

\bibitem[{\citenamefont{Scherer and Camarota}(2012)}]{Scherer2012}
\bibinfo{author}{\bibnamefont{Scherer}, \bibfnamefont{H.}}, and
  \bibinfo{author}{\bibfnamefont{B.}~\bibnamefont{Camarota}},
  \bibinfo{year}{2012}, \bibinfo{journal}{Meas. Sci. Technol. (in preparation)}
  .

\bibitem[{\citenamefont{Schinner} \emph{et~al.}(2009)\citenamefont{Schinner,
  Tranitz, Wegscheider, Kotthaus, and Ludwig}}]{Schinner2009}
\bibinfo{author}{\bibnamefont{Schinner}, \bibfnamefont{G.~J.}},
  \bibinfo{author}{\bibfnamefont{H.~P.} \bibnamefont{Tranitz}},
  \bibinfo{author}{\bibfnamefont{W.}~\bibnamefont{Wegscheider}},
  \bibinfo{author}{\bibfnamefont{J.~P.} \bibnamefont{Kotthaus}}, and
  \bibinfo{author}{\bibfnamefont{S.}~\bibnamefont{Ludwig}},
  \bibinfo{year}{2009}, \bibinfo{journal}{Phys. Rev. Lett.}
  \textbf{\bibinfo{volume}{102}}, \bibinfo{pages}{186801}.

\bibitem[{\citenamefont{Schmidt} \emph{et~al.}(2004)\citenamefont{Schmidt,
  Schoelkopf, and Cleland}}]{Schmidt2004}
\bibinfo{author}{\bibnamefont{Schmidt}, \bibfnamefont{D.~R.}},
  \bibinfo{author}{\bibfnamefont{R.~J.} \bibnamefont{Schoelkopf}}, and
  \bibinfo{author}{\bibfnamefont{A.~N.} \bibnamefont{Cleland}},
  \bibinfo{year}{2004}, \bibinfo{journal}{Phys. Rev. Lett.}
  \textbf{\bibinfo{volume}{93}}, \bibinfo{pages}{045901}.

\bibitem[{\citenamefont{Schoelkopf}
  \emph{et~al.}(1998)\citenamefont{Schoelkopf, P., A., P., and
  E.}}]{Schoelkopf1998}
\bibinfo{author}{\bibnamefont{Schoelkopf}, \bibfnamefont{R.~J.}},
  \bibinfo{author}{\bibfnamefont{W.}~\bibnamefont{P.}},
  \bibinfo{author}{\bibfnamefont{K.~A.} \bibnamefont{A.}},
  \bibinfo{author}{\bibfnamefont{D.}~\bibnamefont{P.}}, and
  \bibinfo{author}{\bibfnamefont{P.~D.} \bibnamefont{E.}},
  \bibinfo{year}{1998}, \bibinfo{journal}{Science}
  \textbf{\bibinfo{volume}{280}}, \bibinfo{pages}{1238}.

\bibitem[{\citenamefont{Sch\"{o}n and Zaikin}(1994)}]{Schon1994}
\bibinfo{author}{\bibnamefont{Sch\"{o}n}, \bibfnamefont{G.}}, and
  \bibinfo{author}{\bibfnamefont{A.}~\bibnamefont{Zaikin}},
  \bibinfo{year}{1994}, \bibinfo{journal}{Europhys. Lett.}
  \textbf{\bibinfo{volume}{26}}, \bibinfo{pages}{695}.

\bibitem[{\citenamefont{Schurr} \emph{et~al.}(2011)\citenamefont{Schurr,
  Ku\v{c}era, Pierz, and Kibble}}]{Schurr2011}
\bibinfo{author}{\bibnamefont{Schurr}, \bibfnamefont{J.}},
  \bibinfo{author}{\bibfnamefont{J.}~\bibnamefont{Ku\v{c}era}},
  \bibinfo{author}{\bibfnamefont{K.}~\bibnamefont{Pierz}}, and
  \bibinfo{author}{\bibfnamefont{B.~P.} \bibnamefont{Kibble}},
  \bibinfo{year}{2011}, \bibinfo{journal}{Metrologia}
  \textbf{\bibinfo{volume}{48}}, \bibinfo{pages}{47}.

\bibitem[{\citenamefont{Sese} \emph{et~al.}(1999)\citenamefont{Sese, Rietveld,
  Camon, Rillo, Vargas, Christian, Brons, Hiddink, Flokstra, Rogalla, Jaszczuk,
  and Altenburg}}]{Sese1999}
\bibinfo{author}{\bibnamefont{Sese}, \bibfnamefont{J.}},
  \bibinfo{author}{\bibfnamefont{G.}~\bibnamefont{Rietveld}},
  \bibinfo{author}{\bibfnamefont{A.}~\bibnamefont{Camon}},
  \bibinfo{author}{\bibfnamefont{C.}~\bibnamefont{Rillo}},
  \bibinfo{author}{\bibfnamefont{L.}~\bibnamefont{Vargas}},
  \bibinfo{author}{\bibfnamefont{G.}~\bibnamefont{Christian}},
  \bibinfo{author}{\bibfnamefont{S.}~\bibnamefont{Brons}},
  \bibinfo{author}{\bibfnamefont{M.}~\bibnamefont{Hiddink}},
  \bibinfo{author}{\bibfnamefont{J.}~\bibnamefont{Flokstra}},
  \bibinfo{author}{\bibfnamefont{H.}~\bibnamefont{Rogalla}},
  \bibinfo{author}{\bibfnamefont{W.}~\bibnamefont{Jaszczuk}}, and
  \bibinfo{author}{\bibfnamefont{H.}~\bibnamefont{Altenburg}},
  \bibinfo{year}{1999}, \bibinfo{journal}{Instrumentation and Measurement, IEEE
  Transactions on DOI - 10.1109/19.769603} \textbf{\bibinfo{volume}{48}},
  \bibinfo{pages}{370}.

\bibitem[{\citenamefont{Shapere and Wilczek~(eds)}(1989)}]{Shapere1989}
\bibinfo{author}{\bibnamefont{Shapere}, \bibfnamefont{A.}}, and
  \bibinfo{author}{\bibfnamefont{F.}~\bibnamefont{Wilczek~(eds)}},
  \bibinfo{year}{1989}, \emph{\bibinfo{title}{Geometric Phases in Physics}}
  (\bibinfo{publisher}{World Scientific, Singapore}).

\bibitem[{\citenamefont{Shapiro}(1963)}]{Shapiro1963}
\bibinfo{author}{\bibnamefont{Shapiro}, \bibfnamefont{S.}},
  \bibinfo{year}{1963}, \bibinfo{journal}{Phys. Rev. Lett.}
  \textbf{\bibinfo{volume}{11}}, \bibinfo{pages}{80}.

\bibitem[{\citenamefont{Shekhter} \emph{et~al.}(2003)\citenamefont{Shekhter,
  Galperin, Gorelik, Isacsson, and Jonson}}]{Shekhter2003}
\bibinfo{author}{\bibnamefont{Shekhter}, \bibfnamefont{R.~I.}},
  \bibinfo{author}{\bibfnamefont{Y.}~\bibnamefont{Galperin}},
  \bibinfo{author}{\bibfnamefont{L.~Y.} \bibnamefont{Gorelik}},
  \bibinfo{author}{\bibfnamefont{A.}~\bibnamefont{Isacsson}}, and
  \bibinfo{author}{\bibfnamefont{M.}~\bibnamefont{Jonson}},
  \bibinfo{year}{2003}, \bibinfo{journal}{J. Phys. Condens. Matter}
  \textbf{\bibinfo{volume}{15}}, \bibinfo{pages}{R441}.

\bibitem[{\citenamefont{Shilton}
  \emph{et~al.}(1996{\natexlab{a}})\citenamefont{Shilton, Mace, Talyanskii,
  Galperin, Simmons, Pepper, and Ritchie}}]{Shilton1996a}
\bibinfo{author}{\bibnamefont{Shilton}, \bibfnamefont{J.~M.}},
  \bibinfo{author}{\bibfnamefont{D.~R.} \bibnamefont{Mace}},
  \bibinfo{author}{\bibfnamefont{V.~I.} \bibnamefont{Talyanskii}},
  \bibinfo{author}{\bibfnamefont{Y.}~\bibnamefont{Galperin}},
  \bibinfo{author}{\bibfnamefont{M.~Y.} \bibnamefont{Simmons}},
  \bibinfo{author}{\bibfnamefont{M.}~\bibnamefont{Pepper}}, and
  \bibinfo{author}{\bibfnamefont{D.~A.} \bibnamefont{Ritchie}},
  \bibinfo{year}{1996}{\natexlab{a}}, \bibinfo{journal}{J. Phys. Condens.
  Matter} \textbf{\bibinfo{volume}{8}}, \bibinfo{pages}{L337}.

\bibitem[{\citenamefont{Shilton}
  \emph{et~al.}(1996{\natexlab{b}})\citenamefont{Shilton, Talyanskii, Pepper,
  Ritchie, Frost, Ford, Smith, and Jones}}]{Shilton1996}
\bibinfo{author}{\bibnamefont{Shilton}, \bibfnamefont{J.~M.}},
  \bibinfo{author}{\bibfnamefont{V.~I.} \bibnamefont{Talyanskii}},
  \bibinfo{author}{\bibfnamefont{M.}~\bibnamefont{Pepper}},
  \bibinfo{author}{\bibfnamefont{D.~A.} \bibnamefont{Ritchie}},
  \bibinfo{author}{\bibfnamefont{J.~E.~F.} \bibnamefont{Frost}},
  \bibinfo{author}{\bibfnamefont{C.~J.~B.} \bibnamefont{Ford}},
  \bibinfo{author}{\bibfnamefont{C.~G.} \bibnamefont{Smith}}, and
  \bibinfo{author}{\bibfnamefont{G.~A.~C.} \bibnamefont{Jones}},
  \bibinfo{year}{1996}{\natexlab{b}}, \bibinfo{journal}{J. Phys. Condens.
  Matter} \textbf{\bibinfo{volume}{8}}, \bibinfo{pages}{L531}.

\bibitem[{\citenamefont{Shimada and Ootuka}(2001)}]{Shimada2001}
\bibinfo{author}{\bibnamefont{Shimada}, \bibfnamefont{H.}}, and
  \bibinfo{author}{\bibfnamefont{Y.}~\bibnamefont{Ootuka}},
  \bibinfo{year}{2001}, \bibinfo{journal}{Phys. Rev. B}
  \textbf{\bibinfo{volume}{64}}, \bibinfo{pages}{235418}.

\bibitem[{\citenamefont{Sillanp\"{a}\"{a}}
  \emph{et~al.}(2004)\citenamefont{Sillanp\"{a}\"{a}, Roschier, and
  Hakonen}}]{Sillanpaa2004}
\bibinfo{author}{\bibnamefont{Sillanp\"{a}\"{a}}, \bibfnamefont{M.~A.}},
  \bibinfo{author}{\bibfnamefont{L.}~\bibnamefont{Roschier}}, and
  \bibinfo{author}{\bibfnamefont{P.~J.} \bibnamefont{Hakonen}},
  \bibinfo{year}{2004}, \bibinfo{journal}{Physical Review Letters}
  \textbf{\bibinfo{volume}{93}}, \bibinfo{pages}{066805}.

\bibitem[{\citenamefont{Sohn} \emph{et~al.}(1997)\citenamefont{Sohn,
  Kouverhowven, and Sch\"{o}n}}]{Sohn1997}
\bibinfo{editor}{\bibnamefont{Sohn}, \bibfnamefont{L.~L.}},
  \bibinfo{editor}{\bibfnamefont{L.~P.} \bibnamefont{Kouverhowven}}, and
  \bibinfo{editor}{\bibfnamefont{G.}~\bibnamefont{Sch\"{o}n}} (eds.),
  \bibinfo{year}{1997} (\bibinfo{publisher}{Kluwer}).

\bibitem[{\citenamefont{Solinas} \emph{et~al.}(2010)\citenamefont{Solinas,
  Pirkkalainen, and M\"ott\"onen}}]{Solinas2010}
\bibinfo{author}{\bibnamefont{Solinas}, \bibfnamefont{P.}},
  \bibinfo{author}{\bibfnamefont{J.-M.} \bibnamefont{Pirkkalainen}}, and
  \bibinfo{author}{\bibfnamefont{M.}~\bibnamefont{M\"ott\"onen}},
  \bibinfo{year}{2010}, \bibinfo{journal}{Phys. Rev. A}
  \textbf{\bibinfo{volume}{82}}, \bibinfo{pages}{052304}.

\bibitem[{\citenamefont{Starmark} \emph{et~al.}(1999)\citenamefont{Starmark,
  Henning, Claeson, Delsing, and Korotkov}}]{Starmark1999}
\bibinfo{author}{\bibnamefont{Starmark}, \bibfnamefont{B.}},
  \bibinfo{author}{\bibfnamefont{T.}~\bibnamefont{Henning}},
  \bibinfo{author}{\bibfnamefont{T.}~\bibnamefont{Claeson}},
  \bibinfo{author}{\bibfnamefont{P.}~\bibnamefont{Delsing}}, and
  \bibinfo{author}{\bibfnamefont{A.~N.} \bibnamefont{Korotkov}},
  \bibinfo{year}{1999}, \bibinfo{journal}{Journal of Applied Physics}
  \textbf{\bibinfo{volume}{86}}, \bibinfo{pages}{2132}.

\bibitem[{\citenamefont{Steele} \emph{et~al.}(2012)\citenamefont{Steele, Meija,
  Sanchez, Yang, Wood, Sturgeon, Mester, and Inglis}}]{Steele2012}
\bibinfo{author}{\bibnamefont{Steele}, \bibfnamefont{A.~G.}},
  \bibinfo{author}{\bibfnamefont{J.}~\bibnamefont{Meija}},
  \bibinfo{author}{\bibfnamefont{C.~A.} \bibnamefont{Sanchez}},
  \bibinfo{author}{\bibfnamefont{L.}~\bibnamefont{Yang}},
  \bibinfo{author}{\bibfnamefont{B.~M.} \bibnamefont{Wood}},
  \bibinfo{author}{\bibfnamefont{R.~E.} \bibnamefont{Sturgeon}},
  \bibinfo{author}{\bibfnamefont{Z.}~\bibnamefont{Mester}}, and
  \bibinfo{author}{\bibfnamefont{A.~D.} \bibnamefont{Inglis}},
  \bibinfo{year}{2012}, \bibinfo{journal}{Metrologia}
  \textbf{\bibinfo{volume}{49}}, \bibinfo{pages}{L8}.

\bibitem[{\citenamefont{Steiner} \emph{et~al.}(2007)\citenamefont{Steiner,
  Williams, Newell, and Lui}}]{Steiner2007}
\bibinfo{author}{\bibnamefont{Steiner}, \bibfnamefont{R.}},
  \bibinfo{author}{\bibfnamefont{E.~R.} \bibnamefont{Williams}},
  \bibinfo{author}{\bibfnamefont{D.~B.} \bibnamefont{Newell}}, and
  \bibinfo{author}{\bibfnamefont{R.}~\bibnamefont{Lui}}, \bibinfo{year}{2007},
  \bibinfo{journal}{IEEE Trans. Instrum. Meas.} \textbf{\bibinfo{volume}{56}},
  \bibinfo{pages}{592}.

\bibitem[{\citenamefont{Stock}(2011)}]{Stock2011}
\bibinfo{author}{\bibnamefont{Stock}, \bibfnamefont{M.}}, \bibinfo{year}{2011},
  \bibinfo{journal}{Phil. Trans. Royal Soc. A} \textbf{\bibinfo{volume}{369}},
  \bibinfo{pages}{3936}.

\bibitem[{\citenamefont{Stock and Witt}(2006)}]{Stock2006}
\bibinfo{author}{\bibnamefont{Stock}, \bibfnamefont{M.}}, and
  \bibinfo{author}{\bibfnamefont{T.~J.} \bibnamefont{Witt}},
  \bibinfo{year}{2006}, \bibinfo{journal}{Metrologia}
  \textbf{\bibinfo{volume}{43}}, \bibinfo{pages}{583}.

\bibitem[{\citenamefont{Su} \emph{et~al.}(1992)\citenamefont{Su, Goldman, and
  Cunningham}}]{Su1992}
\bibinfo{author}{\bibnamefont{Su}, \bibfnamefont{B.}},
  \bibinfo{author}{\bibfnamefont{V.~J.} \bibnamefont{Goldman}}, and
  \bibinfo{author}{\bibfnamefont{J.~E.} \bibnamefont{Cunningham}},
  \bibinfo{year}{1992}, \bibinfo{journal}{Science}
  \textbf{\bibinfo{volume}{255}}, \bibinfo{pages}{313}.

\bibitem[{\citenamefont{Takahashi} \emph{et~al.}(1995)\citenamefont{Takahashi,
  Nagase, Namatsu, Kurihara, Iwadate, Nakajima, Horiguchi, Murase, and
  Tabe}}]{Takahashi1995}
\bibinfo{author}{\bibnamefont{Takahashi}, \bibfnamefont{Y.}},
  \bibinfo{author}{\bibfnamefont{M.}~\bibnamefont{Nagase}},
  \bibinfo{author}{\bibfnamefont{H.}~\bibnamefont{Namatsu}},
  \bibinfo{author}{\bibfnamefont{K.}~\bibnamefont{Kurihara}},
  \bibinfo{author}{\bibfnamefont{K.}~\bibnamefont{Iwadate}},
  \bibinfo{author}{\bibfnamefont{Y.}~\bibnamefont{Nakajima}},
  \bibinfo{author}{\bibfnamefont{S.}~\bibnamefont{Horiguchi}},
  \bibinfo{author}{\bibfnamefont{K.}~\bibnamefont{Murase}}, and
  \bibinfo{author}{\bibfnamefont{M.}~\bibnamefont{Tabe}}, \bibinfo{year}{1995},
  \bibinfo{journal}{Electron. Lett.} \textbf{\bibinfo{volume}{31}},
  \bibinfo{pages}{136}.

\bibitem[{\citenamefont{Talyanskii}
  \emph{et~al.}(1997)\citenamefont{Talyanskii, Shilton, Pepper, Smith, Ford,
  Linfield, Ritchie, and Jones}}]{Talyanskii1997}
\bibinfo{author}{\bibnamefont{Talyanskii}, \bibfnamefont{V.~I.}},
  \bibinfo{author}{\bibfnamefont{J.~M.} \bibnamefont{Shilton}},
  \bibinfo{author}{\bibfnamefont{M.}~\bibnamefont{Pepper}},
  \bibinfo{author}{\bibfnamefont{C.~G.} \bibnamefont{Smith}},
  \bibinfo{author}{\bibfnamefont{C.~J.~B.} \bibnamefont{Ford}},
  \bibinfo{author}{\bibfnamefont{E.~H.} \bibnamefont{Linfield}},
  \bibinfo{author}{\bibfnamefont{D.~A.} \bibnamefont{Ritchie}}, and
  \bibinfo{author}{\bibfnamefont{G.~A.~C.} \bibnamefont{Jones}},
  \bibinfo{year}{1997}, \bibinfo{journal}{Phys. Rev. B}
  \textbf{\bibinfo{volume}{56}}, \bibinfo{pages}{15180}.

\bibitem[{\citenamefont{Tan} \emph{et~al.}(2010)\citenamefont{Tan, Chan,
  M\"{o}tt\"{o}nen, Morello, Yang, van Donkelaar, Alves, Pirkkalainen,
  Jamieson, Clark, and Dzurak}}]{Tan2010}
\bibinfo{author}{\bibnamefont{Tan}, \bibfnamefont{K.~Y.}},
  \bibinfo{author}{\bibfnamefont{K.~W.} \bibnamefont{Chan}},
  \bibinfo{author}{\bibfnamefont{M.}~\bibnamefont{M\"{o}tt\"{o}nen}},
  \bibinfo{author}{\bibfnamefont{A.}~\bibnamefont{Morello}},
  \bibinfo{author}{\bibfnamefont{C.}~\bibnamefont{Yang}},
  \bibinfo{author}{\bibfnamefont{J.}~\bibnamefont{van Donkelaar}},
  \bibinfo{author}{\bibfnamefont{A.}~\bibnamefont{Alves}},
  \bibinfo{author}{\bibfnamefont{J.-M.} \bibnamefont{Pirkkalainen}},
  \bibinfo{author}{\bibfnamefont{D.~N.} \bibnamefont{Jamieson}},
  \bibinfo{author}{\bibfnamefont{R.~G.} \bibnamefont{Clark}}, and
  \bibinfo{author}{\bibfnamefont{A.~S.} \bibnamefont{Dzurak}},
  \bibinfo{year}{2010}, \bibinfo{journal}{Nano Lett.}
  \textbf{\bibinfo{volume}{10}}, \bibinfo{pages}{11}.

\bibitem[{\citenamefont{Tan} \emph{et~al.}(2008)\citenamefont{Tan, Patel, Liu,
  Lukens, Likharev, and Zhu}}]{Tan2008}
\bibinfo{author}{\bibnamefont{Tan}, \bibfnamefont{Z.}},
  \bibinfo{author}{\bibfnamefont{V.}~\bibnamefont{Patel}},
  \bibinfo{author}{\bibfnamefont{X.}~\bibnamefont{Liu}},
  \bibinfo{author}{\bibfnamefont{J.}~\bibnamefont{Lukens}},
  \bibinfo{author}{\bibfnamefont{K.}~\bibnamefont{Likharev}}, and
  \bibinfo{author}{\bibfnamefont{Y.}~\bibnamefont{Zhu}}, \bibinfo{year}{2008},
  \bibinfo{journal}{Appl. Phys. Lett.} \textbf{\bibinfo{volume}{93}},
  \bibinfo{pages}{242109}.

\bibitem[{\citenamefont{Tarucha} \emph{et~al.}(1996)\citenamefont{Tarucha,
  Austing, Honda, van~der Hage, and Kouwenhoven}}]{Tarucha1996}
\bibinfo{author}{\bibnamefont{Tarucha}, \bibfnamefont{S.}},
  \bibinfo{author}{\bibfnamefont{D.~G.} \bibnamefont{Austing}},
  \bibinfo{author}{\bibfnamefont{T.}~\bibnamefont{Honda}},
  \bibinfo{author}{\bibfnamefont{R.~J.} \bibnamefont{van~der Hage}}, and
  \bibinfo{author}{\bibfnamefont{L.~P.} \bibnamefont{Kouwenhoven}},
  \bibinfo{year}{1996}, \bibinfo{journal}{Phys. Rev. Lett.}
  \textbf{\bibinfo{volume}{77}}, \bibinfo{pages}{3613}.

\bibitem[{\citenamefont{Thompson and Lampard}(1956)}]{Thompson1956}
\bibinfo{author}{\bibnamefont{Thompson}, \bibfnamefont{A.~M.}}, and
  \bibinfo{author}{\bibfnamefont{D.~G.} \bibnamefont{Lampard}},
  \bibinfo{year}{1956}, \bibinfo{journal}{Nature}
  \textbf{\bibinfo{volume}{177}}, \bibinfo{pages}{888}.

\bibitem[{\citenamefont{Thornton} \emph{et~al.}(1986)\citenamefont{Thornton,
  Pepper, Ahmed, Andrews, and Davies}}]{Thornton1986}
\bibinfo{author}{\bibnamefont{Thornton}, \bibfnamefont{T.}},
  \bibinfo{author}{\bibfnamefont{M.}~\bibnamefont{Pepper}},
  \bibinfo{author}{\bibfnamefont{H.}~\bibnamefont{Ahmed}},
  \bibinfo{author}{\bibfnamefont{D.}~\bibnamefont{Andrews}}, and
  \bibinfo{author}{\bibfnamefont{G.}~\bibnamefont{Davies}},
  \bibinfo{year}{1986}, \bibinfo{journal}{Physical Review Letters}
  \textbf{\bibinfo{volume}{56}}, \bibinfo{pages}{1198}.

\bibitem[{\citenamefont{Thouless} \emph{et~al.}(1982)\citenamefont{Thouless,
  Kohmoto, Nightingale, and den Nijs}}]{Thouless1982}
\bibinfo{author}{\bibnamefont{Thouless}, \bibfnamefont{D.~J.}},
  \bibinfo{author}{\bibfnamefont{M.}~\bibnamefont{Kohmoto}},
  \bibinfo{author}{\bibfnamefont{M.~P.} \bibnamefont{Nightingale}}, and
  \bibinfo{author}{\bibfnamefont{M.}~\bibnamefont{den Nijs}},
  \bibinfo{year}{1982}, \bibinfo{journal}{Phys. Rev. Lett.}
  \textbf{\bibinfo{volume}{49}}, \bibinfo{pages}{405}.

\bibitem[{\citenamefont{Timofeev}
  \emph{et~al.}(2009{\natexlab{a}})\citenamefont{Timofeev, Garcia, Kopnin,
  Savin, Meschke, Giazotto, and Pekola}}]{Timofeev2009}
\bibinfo{author}{\bibnamefont{Timofeev}, \bibfnamefont{A.~V.}},
  \bibinfo{author}{\bibfnamefont{C.~P.} \bibnamefont{Garcia}},
  \bibinfo{author}{\bibfnamefont{N.~B.} \bibnamefont{Kopnin}},
  \bibinfo{author}{\bibfnamefont{A.~M.} \bibnamefont{Savin}},
  \bibinfo{author}{\bibfnamefont{M.}~\bibnamefont{Meschke}},
  \bibinfo{author}{\bibfnamefont{F.}~\bibnamefont{Giazotto}}, and
  \bibinfo{author}{\bibfnamefont{J.~P.} \bibnamefont{Pekola}},
  \bibinfo{year}{2009}{\natexlab{a}}, \bibinfo{journal}{Phys. Rev. Lett.}
  \textbf{\bibinfo{volume}{102}}, \bibinfo{pages}{017003}.

\bibitem[{\citenamefont{Timofeev}
  \emph{et~al.}(2009{\natexlab{b}})\citenamefont{Timofeev, Helle, Meschke,
  M\"ott\"onen, and Pekola}}]{Timofeev2009a}
\bibinfo{author}{\bibnamefont{Timofeev}, \bibfnamefont{A.~V.}},
  \bibinfo{author}{\bibfnamefont{M.}~\bibnamefont{Helle}},
  \bibinfo{author}{\bibfnamefont{M.}~\bibnamefont{Meschke}},
  \bibinfo{author}{\bibfnamefont{M.}~\bibnamefont{M\"ott\"onen}}, and
  \bibinfo{author}{\bibfnamefont{J.~P.} \bibnamefont{Pekola}},
  \bibinfo{year}{2009}{\natexlab{b}}, \bibinfo{journal}{Phys. Rev. Lett.}
  \textbf{\bibinfo{volume}{102}}, \bibinfo{pages}{200801}.

\bibitem[{\citenamefont{Tinkham}(1996)}]{Tinkham}
\bibinfo{author}{\bibnamefont{Tinkham}, \bibfnamefont{M.}},
  \bibinfo{year}{1996}, \emph{\bibinfo{title}{Introduction to
  superconductivity}} (\bibinfo{publisher}{McGraw-Hill}, \bibinfo{address}{New
  York}), \bibinfo{edition}{2} edition.

\bibitem[{\citenamefont{Toppari} \emph{et~al.}(2004)\citenamefont{Toppari,
  Kivioja, Pekola, and Savolainen}}]{Toppari2004}
\bibinfo{author}{\bibnamefont{Toppari}, \bibfnamefont{J.~J.}},
  \bibinfo{author}{\bibfnamefont{J.~M.} \bibnamefont{Kivioja}},
  \bibinfo{author}{\bibfnamefont{J.~P.} \bibnamefont{Pekola}}, and
  \bibinfo{author}{\bibfnamefont{M.~T.} \bibnamefont{Savolainen}},
  \bibinfo{year}{2004}, \bibinfo{journal}{J. Low Temp. Phys.}
  \textbf{\bibinfo{volume}{136}}, \bibinfo{pages}{57}.

\bibitem[{\citenamefont{Tsai} \emph{et~al.}(1983)\citenamefont{Tsai, Jain, and
  Lukens}}]{Tsai1983}
\bibinfo{author}{\bibnamefont{Tsai}, \bibfnamefont{J.-S.}},
  \bibinfo{author}{\bibfnamefont{A.~K.} \bibnamefont{Jain}}, and
  \bibinfo{author}{\bibfnamefont{J.~E.} \bibnamefont{Lukens}},
  \bibinfo{year}{1983}, \bibinfo{journal}{Phys. Rev. Lett.}
  \textbf{\bibinfo{volume}{51}}, \bibinfo{pages}{1109}.

\bibitem[{\citenamefont{Tsukagoshi}
  \emph{et~al.}(1998)\citenamefont{Tsukagoshi, Alphenaar, and
  Nakazato}}]{Tsukagoshi1998}
\bibinfo{author}{\bibnamefont{Tsukagoshi}, \bibfnamefont{K.}},
  \bibinfo{author}{\bibfnamefont{B.~W.} \bibnamefont{Alphenaar}}, and
  \bibinfo{author}{\bibfnamefont{K.}~\bibnamefont{Nakazato}},
  \bibinfo{year}{1998}, \bibinfo{journal}{Appl. Phys. Lett.}
  \textbf{\bibinfo{volume}{73}}, \bibinfo{pages}{2515}.

\bibitem[{\citenamefont{Tsukagoshi}
  \emph{et~al.}(1997)\citenamefont{Tsukagoshi, Nakazato, Ahmed, and
  Gamo}}]{Tsukagoshi1997}
\bibinfo{author}{\bibnamefont{Tsukagoshi}, \bibfnamefont{K.}},
  \bibinfo{author}{\bibfnamefont{K.}~\bibnamefont{Nakazato}},
  \bibinfo{author}{\bibfnamefont{H.}~\bibnamefont{Ahmed}}, and
  \bibinfo{author}{\bibfnamefont{K.}~\bibnamefont{Gamo}}, \bibinfo{year}{1997},
  \bibinfo{journal}{Phys. Rev. B} \textbf{\bibinfo{volume}{56}},
  \bibinfo{pages}{3972}.

\bibitem[{\citenamefont{Tuominen} \emph{et~al.}(1992)\citenamefont{Tuominen,
  Hergenrother, Tighe, and Tinkham}}]{Tuominen1992}
\bibinfo{author}{\bibnamefont{Tuominen}, \bibfnamefont{M.~T.}},
  \bibinfo{author}{\bibfnamefont{J.~M.} \bibnamefont{Hergenrother}},
  \bibinfo{author}{\bibfnamefont{T.~S.} \bibnamefont{Tighe}}, and
  \bibinfo{author}{\bibfnamefont{M.}~\bibnamefont{Tinkham}},
  \bibinfo{year}{1992}, \bibinfo{journal}{Phys. Rev. Lett.}
  \textbf{\bibinfo{volume}{69}}, \bibinfo{pages}{1997}.

\bibitem[{\citenamefont{Tuominen} \emph{et~al.}(1999)\citenamefont{Tuominen,
  Krotkov, and Breuer}}]{Tuominen1999}
\bibinfo{author}{\bibnamefont{Tuominen}, \bibfnamefont{M.~T.}},
  \bibinfo{author}{\bibfnamefont{R.~V.} \bibnamefont{Krotkov}}, and
  \bibinfo{author}{\bibfnamefont{M.~I.} \bibnamefont{Breuer}},
  \bibinfo{year}{1999}, \bibinfo{journal}{Phys. Rev. Lett.}
  \textbf{\bibinfo{volume}{83}}, \bibinfo{pages}{3025}.

\bibitem[{\citenamefont{Turek} \emph{et~al.}(2005)\citenamefont{Turek, Lehnert,
  Clerk, Gunnarsson, Bladh, Delsing, and Schoelkopf}}]{Turek2005}
\bibinfo{author}{\bibnamefont{Turek}, \bibfnamefont{B.~A.}},
  \bibinfo{author}{\bibfnamefont{K.~W.} \bibnamefont{Lehnert}},
  \bibinfo{author}{\bibfnamefont{A.}~\bibnamefont{Clerk}},
  \bibinfo{author}{\bibfnamefont{D.}~\bibnamefont{Gunnarsson}},
  \bibinfo{author}{\bibfnamefont{K.}~\bibnamefont{Bladh}},
  \bibinfo{author}{\bibfnamefont{P.}~\bibnamefont{Delsing}}, and
  \bibinfo{author}{\bibfnamefont{R.~J.} \bibnamefont{Schoelkopf}},
  \bibinfo{year}{2005}, \bibinfo{journal}{Physical Review B}
  \textbf{\bibinfo{volume}{71}}, \bibinfo{pages}{193304}.

\bibitem[{\citenamefont{Tzalenchuk}
  \emph{et~al.}(2010)\citenamefont{Tzalenchuk, Lara-Avila, Kalaboukhov,
  Paolillo, Syvajarvi, Yakimova, Kazakova, M., Fal'ko, and
  Kubatkin}}]{Tzalenchuk2010}
\bibinfo{author}{\bibnamefont{Tzalenchuk}, \bibfnamefont{A.}},
  \bibinfo{author}{\bibfnamefont{S.}~\bibnamefont{Lara-Avila}},
  \bibinfo{author}{\bibfnamefont{A.}~\bibnamefont{Kalaboukhov}},
  \bibinfo{author}{\bibfnamefont{S.}~\bibnamefont{Paolillo}},
  \bibinfo{author}{\bibfnamefont{M.}~\bibnamefont{Syvajarvi}},
  \bibinfo{author}{\bibfnamefont{R.}~\bibnamefont{Yakimova}},
  \bibinfo{author}{\bibfnamefont{O.}~\bibnamefont{Kazakova}},
  \bibinfo{author}{\bibfnamefont{J.~J.~B.} \bibnamefont{M.}},
  \bibinfo{author}{\bibfnamefont{V.}~\bibnamefont{Fal'ko}}, and
  \bibinfo{author}{\bibfnamefont{S.}~\bibnamefont{Kubatkin}},
  \bibinfo{year}{2010}, \bibinfo{journal}{Nat. Nanotech.}
  \textbf{\bibinfo{volume}{5}}, \bibinfo{pages}{186}.

\bibitem[{\citenamefont{Ullom} \emph{et~al.}(1998)\citenamefont{Ullom, Fisher,
  and Nahum}}]{Ullom1998}
\bibinfo{author}{\bibnamefont{Ullom}, \bibfnamefont{J.~N.}},
  \bibinfo{author}{\bibfnamefont{P.~A.} \bibnamefont{Fisher}}, and
  \bibinfo{author}{\bibfnamefont{M.}~\bibnamefont{Nahum}},
  \bibinfo{year}{1998}, \bibinfo{journal}{Phys. Rev. B}
  \textbf{\bibinfo{volume}{58}}, \bibinfo{pages}{8225}.

\bibitem[{\citenamefont{Utko} \emph{et~al.}(2006)\citenamefont{Utko, Lidelof,
  and Gloos}}]{Utko2006}
\bibinfo{author}{\bibnamefont{Utko}, \bibfnamefont{P.}},
  \bibinfo{author}{\bibfnamefont{P.~E.} \bibnamefont{Lidelof}}, and
  \bibinfo{author}{\bibfnamefont{K.}~\bibnamefont{Gloos}},
  \bibinfo{year}{2006}, \bibinfo{journal}{Appl. Phys. Lett.}
  \textbf{\bibinfo{volume}{88}}, \bibinfo{pages}{202113}.

\bibitem[{\citenamefont{Vandersypen}
  \emph{et~al.}(2004)\citenamefont{Vandersypen, Elzerman, Schouten, {Willems
  van Beveren}, Hanson, and Kouwenhoven}}]{Vandersypen2004}
\bibinfo{author}{\bibnamefont{Vandersypen}, \bibfnamefont{L.~M.~K.}},
  \bibinfo{author}{\bibfnamefont{J.~M.} \bibnamefont{Elzerman}},
  \bibinfo{author}{\bibfnamefont{R.~N.} \bibnamefont{Schouten}},
  \bibinfo{author}{\bibfnamefont{L.~H.} \bibnamefont{{Willems van Beveren}}},
  \bibinfo{author}{\bibfnamefont{R.}~\bibnamefont{Hanson}}, and
  \bibinfo{author}{\bibfnamefont{L.~P.} \bibnamefont{Kouwenhoven}},
  \bibinfo{year}{2004}, \bibinfo{journal}{Applied Physics Letters}
  \textbf{\bibinfo{volume}{85}}, \bibinfo{pages}{4394}.

\bibitem[{\citenamefont{Vartiainen}
  \emph{et~al.}(2007)\citenamefont{Vartiainen, M\"{o}tt\"{o}nen, Pekola, and
  Kemppinen}}]{Vartiainen2007}
\bibinfo{author}{\bibnamefont{Vartiainen}, \bibfnamefont{J.~J.}},
  \bibinfo{author}{\bibfnamefont{M.}~\bibnamefont{M\"{o}tt\"{o}nen}},
  \bibinfo{author}{\bibfnamefont{J.~P.} \bibnamefont{Pekola}}, and
  \bibinfo{author}{\bibfnamefont{A.}~\bibnamefont{Kemppinen}},
  \bibinfo{year}{2007}, \bibinfo{journal}{Appl. Phys. Lett.}
  \textbf{\bibinfo{volume}{90}}, \bibinfo{pages}{082102}.

\bibitem[{\citenamefont{Visscher} \emph{et~al.}(1996)\citenamefont{Visscher,
  Lindeman, Verbrugh, Hadley, Mooij, and van~der Vleuten}}]{Visscher1996}
\bibinfo{author}{\bibnamefont{Visscher}, \bibfnamefont{E.~H.}},
  \bibinfo{author}{\bibfnamefont{J.}~\bibnamefont{Lindeman}},
  \bibinfo{author}{\bibfnamefont{S.~M.} \bibnamefont{Verbrugh}},
  \bibinfo{author}{\bibfnamefont{P.}~\bibnamefont{Hadley}},
  \bibinfo{author}{\bibfnamefont{J.~E.} \bibnamefont{Mooij}}, and
  \bibinfo{author}{\bibfnamefont{W.}~\bibnamefont{van~der Vleuten}},
  \bibinfo{year}{1996}, \bibinfo{journal}{Applied Physics Letters}
  \textbf{\bibinfo{volume}{68}}, \bibinfo{pages}{2014}.

\bibitem[{\citenamefont{Wei and Chandrasekhar}(2010)}]{Wei2010}
\bibinfo{author}{\bibnamefont{Wei}, \bibfnamefont{J.}}, and
  \bibinfo{author}{\bibfnamefont{V.}~\bibnamefont{Chandrasekhar}},
  \bibinfo{year}{2010}, \bibinfo{journal}{Nature Phys.}
  \textbf{\bibinfo{volume}{6}}, \bibinfo{pages}{494}.

\bibitem[{\citenamefont{Weis and von Klitzing}(2011)}]{Weis2011}
\bibinfo{author}{\bibnamefont{Weis}, \bibfnamefont{J.}}, and
  \bibinfo{author}{\bibfnamefont{K.}~\bibnamefont{von Klitzing}},
  \bibinfo{year}{2011}, \bibinfo{journal}{Phil. Trans. Royal Soc. A}
  \textbf{\bibinfo{volume}{369}}, \bibinfo{pages}{3954}.

\bibitem[{\citenamefont{Weiss and Zwerger}(1999)}]{Weiss1999}
\bibinfo{author}{\bibnamefont{Weiss}, \bibfnamefont{C.}}, and
  \bibinfo{author}{\bibfnamefont{W.}~\bibnamefont{Zwerger}},
  \bibinfo{year}{1999}, \bibinfo{journal}{Europhys. Lett.}
  \textbf{\bibinfo{volume}{47}}, \bibinfo{pages}{97}.

\bibitem[{\citenamefont{Wellstood} \emph{et~al.}(1994)\citenamefont{Wellstood,
  Urbina, and Clarke}}]{Wellstood1994}
\bibinfo{author}{\bibnamefont{Wellstood}, \bibfnamefont{F.~C.}},
  \bibinfo{author}{\bibfnamefont{C.}~\bibnamefont{Urbina}}, and
  \bibinfo{author}{\bibfnamefont{J.}~\bibnamefont{Clarke}},
  \bibinfo{year}{1994}, \bibinfo{journal}{Phys. Rev. B}
  \textbf{\bibinfo{volume}{49}}, \bibinfo{pages}{5942}.

\bibitem[{\citenamefont{White and Wagner}(1993)}]{White1993}
\bibinfo{author}{\bibnamefont{White}, \bibfnamefont{J.~D.}}, and
  \bibinfo{author}{\bibfnamefont{M.}~\bibnamefont{Wagner}},
  \bibinfo{year}{1993}, \bibinfo{journal}{Phys. Rev. B}
  \textbf{\bibinfo{volume}{48}}, \bibinfo{pages}{2799}.

\bibitem[{\citenamefont{van~der Wiel} \emph{et~al.}(2002)\citenamefont{van~der
  Wiel, Franceschi, Elzerman, Fujisawa, Tarucha, and
  Kouwenhoven}}]{vanderWiel2002}
\bibinfo{author}{\bibnamefont{van~der Wiel}, \bibfnamefont{W.~G.}},
  \bibinfo{author}{\bibfnamefont{S.~D.} \bibnamefont{Franceschi}},
  \bibinfo{author}{\bibfnamefont{J.~M.} \bibnamefont{Elzerman}},
  \bibinfo{author}{\bibfnamefont{T.}~\bibnamefont{Fujisawa}},
  \bibinfo{author}{\bibfnamefont{S.}~\bibnamefont{Tarucha}}, and
  \bibinfo{author}{\bibfnamefont{L.}~\bibnamefont{Kouwenhoven}},
  \bibinfo{year}{2002}, \bibinfo{journal}{Rev. Mod. Phys.}
  \textbf{\bibinfo{volume}{75}}(\bibinfo{number}{1}).

\bibitem[{\citenamefont{Williams} \emph{et~al.}(1992)\citenamefont{Williams,
  Ghosh, and Martinis}}]{Williams1992}
\bibinfo{author}{\bibnamefont{Williams}, \bibfnamefont{E.~R.}},
  \bibinfo{author}{\bibfnamefont{R.~N.} \bibnamefont{Ghosh}}, and
  \bibinfo{author}{\bibfnamefont{J.~M.} \bibnamefont{Martinis}},
  \bibinfo{year}{1992}, \bibinfo{journal}{J. Res. Natl. Inst. Stand. Technol.}
  \textbf{\bibinfo{volume}{1992}}, \bibinfo{pages}{299}.

\bibitem[{\citenamefont{Wood and Solve}(2009)}]{Wood2009}
\bibinfo{author}{\bibnamefont{Wood}, \bibfnamefont{B.~M.}}, and
  \bibinfo{author}{\bibfnamefont{S.}~\bibnamefont{Solve}},
  \bibinfo{year}{2009}, \bibinfo{journal}{Metrologia}
  \textbf{\bibinfo{volume}{46}}, \bibinfo{pages}{R13}.

\bibitem[{\citenamefont{Wright} \emph{et~al.}(2008)\citenamefont{Wright,
  Blumenthal, Gumbs, Thorn, Pepper, Janssen, Holmes, Anderson, Jones, Nicoll,
  and Ritchie}}]{Wright2008}
\bibinfo{author}{\bibnamefont{Wright}, \bibfnamefont{S.~J.}},
  \bibinfo{author}{\bibfnamefont{M.~D.} \bibnamefont{Blumenthal}},
  \bibinfo{author}{\bibfnamefont{G.}~\bibnamefont{Gumbs}},
  \bibinfo{author}{\bibfnamefont{A.~L.} \bibnamefont{Thorn}},
  \bibinfo{author}{\bibfnamefont{M.}~\bibnamefont{Pepper}},
  \bibinfo{author}{\bibfnamefont{T.~J. B.~M.} \bibnamefont{Janssen}},
  \bibinfo{author}{\bibfnamefont{S.~N.} \bibnamefont{Holmes}},
  \bibinfo{author}{\bibfnamefont{D.}~\bibnamefont{Anderson}},
  \bibinfo{author}{\bibfnamefont{G.~A.~C.} \bibnamefont{Jones}},
  \bibinfo{author}{\bibfnamefont{C.~A.} \bibnamefont{Nicoll}}, and
  \bibinfo{author}{\bibfnamefont{D.~A.} \bibnamefont{Ritchie}},
  \bibinfo{year}{2008}, \bibinfo{journal}{Phys. Rev. B}
  \textbf{\bibinfo{volume}{78}}, \bibinfo{pages}{233311}.

\bibitem[{\citenamefont{Wright} \emph{et~al.}(2009)\citenamefont{Wright,
  Blumenthal, Pepper, Anderson, Jones, Nicoll, and Ritchie}}]{Wright2009}
\bibinfo{author}{\bibnamefont{Wright}, \bibfnamefont{S.~J.}},
  \bibinfo{author}{\bibfnamefont{M.~D.} \bibnamefont{Blumenthal}},
  \bibinfo{author}{\bibfnamefont{M.}~\bibnamefont{Pepper}},
  \bibinfo{author}{\bibfnamefont{D.}~\bibnamefont{Anderson}},
  \bibinfo{author}{\bibfnamefont{G.~A.~C.} \bibnamefont{Jones}},
  \bibinfo{author}{\bibfnamefont{C.~A.} \bibnamefont{Nicoll}}, and
  \bibinfo{author}{\bibfnamefont{D.~A.} \bibnamefont{Ritchie}},
  \bibinfo{year}{2009}, \bibinfo{journal}{Phys. Rev. B}
  \textbf{\bibinfo{volume}{80}}, \bibinfo{pages}{113303}.

\bibitem[{\citenamefont{Wulf and Zorin}(2008)}]{Wulf2008}
\bibinfo{author}{\bibnamefont{Wulf}, \bibfnamefont{M.}}, and
  \bibinfo{author}{\bibfnamefont{A.~B.} \bibnamefont{Zorin}},
  \bibinfo{year}{2008}, \bibinfo{journal}{arXiv:0811.3927} .

\bibitem[{\citenamefont{Xue} \emph{et~al.}(2009)\citenamefont{Xue, Ji, Pan,
  Stettenheim, Blencowe, and Rimberg}}]{Xue2009}
\bibinfo{author}{\bibnamefont{Xue}, \bibfnamefont{W.~W.}},
  \bibinfo{author}{\bibfnamefont{Z.}~\bibnamefont{Ji}},
  \bibinfo{author}{\bibfnamefont{F.}~\bibnamefont{Pan}},
  \bibinfo{author}{\bibfnamefont{J.}~\bibnamefont{Stettenheim}},
  \bibinfo{author}{\bibfnamefont{M.~P.} \bibnamefont{Blencowe}}, and
  \bibinfo{author}{\bibfnamefont{A.~J.} \bibnamefont{Rimberg}},
  \bibinfo{year}{2009}, \bibinfo{journal}{Nature Physics}
  \textbf{\bibinfo{volume}{5}}, \bibinfo{pages}{660}.

\bibitem[{\citenamefont{Yamahata} \emph{et~al.}(2011)\citenamefont{Yamahata,
  Nishiguchi, and Fujiwara}}]{Yamahata2011}
\bibinfo{author}{\bibnamefont{Yamahata}, \bibfnamefont{G.}},
  \bibinfo{author}{\bibfnamefont{K.}~\bibnamefont{Nishiguchi}}, and
  \bibinfo{author}{\bibfnamefont{A.}~\bibnamefont{Fujiwara}},
  \bibinfo{year}{2011}, \bibinfo{journal}{Appl. Phys. Lett.}
  \textbf{\bibinfo{volume}{98}}, \bibinfo{pages}{222104}.

\bibitem[{\citenamefont{Yang} \emph{et~al.}(2012)\citenamefont{Yang, Mester,
  Sturgeon, and Meija}}]{Yang2012}
\bibinfo{author}{\bibnamefont{Yang}, \bibfnamefont{L.}},
  \bibinfo{author}{\bibfnamefont{Z.}~\bibnamefont{Mester}},
  \bibinfo{author}{\bibfnamefont{R.~E.} \bibnamefont{Sturgeon}}, and
  \bibinfo{author}{\bibfnamefont{J.}~\bibnamefont{Meija}},
  \bibinfo{year}{2012}, \bibinfo{journal}{Anal. Chem.}
  \textbf{\bibinfo{volume}{84}}, \bibinfo{pages}{2321}.

\bibitem[{\citenamefont{Young and Clerk}(2010)}]{Young2010}
\bibinfo{author}{\bibnamefont{Young}, \bibfnamefont{C.~E.}}, and
  \bibinfo{author}{\bibfnamefont{A.~A.} \bibnamefont{Clerk}},
  \bibinfo{year}{2010}, \bibinfo{journal}{Phys. Rev. Lett.}
  \textbf{\bibinfo{volume}{104}}, \bibinfo{pages}{186803}.

\bibitem[{\citenamefont{Yuan} \emph{et~al.}(2011)\citenamefont{Yuan, Pan, Yang,
  Gilheart, Chen, Savage, Lagally, Eriksson, and Rimberg}}]{Yuan2011}
\bibinfo{author}{\bibnamefont{Yuan}, \bibfnamefont{M.}},
  \bibinfo{author}{\bibfnamefont{F.}~\bibnamefont{Pan}},
  \bibinfo{author}{\bibfnamefont{Z.}~\bibnamefont{Yang}},
  \bibinfo{author}{\bibfnamefont{T.~J.} \bibnamefont{Gilheart}},
  \bibinfo{author}{\bibfnamefont{F.}~\bibnamefont{Chen}},
  \bibinfo{author}{\bibfnamefont{D.~E.} \bibnamefont{Savage}},
  \bibinfo{author}{\bibfnamefont{M.~G.} \bibnamefont{Lagally}},
  \bibinfo{author}{\bibfnamefont{M.~A.} \bibnamefont{Eriksson}}, and
  \bibinfo{author}{\bibfnamefont{A.~J.} \bibnamefont{Rimberg}},
  \bibinfo{year}{2011}, \bibinfo{journal}{Applied Physics Letters}
  \textbf{\bibinfo{volume}{98}}, \bibinfo{pages}{4}.

\bibitem[{\citenamefont{Zaikin}(1994)}]{Zaikin1994}
\bibinfo{author}{\bibnamefont{Zaikin}, \bibfnamefont{A.~D.}},
  \bibinfo{year}{1994}, \bibinfo{journal}{Physica B: Condensed Matter}
  \textbf{\bibinfo{volume}{203}}, \bibinfo{pages}{255}.

\bibitem[{\citenamefont{Zanardi and Rasetti}(1999)}]{Zanardi1999}
\bibinfo{author}{\bibnamefont{Zanardi}, \bibfnamefont{P.}}, and
  \bibinfo{author}{\bibfnamefont{M.}~\bibnamefont{Rasetti}},
  \bibinfo{year}{1999}, \bibinfo{journal}{Phys. Lett. A}
  \textbf{\bibinfo{volume}{264}}, \bibinfo{pages}{94}.

\bibitem[{\citenamefont{Zhang} \emph{et~al.}(2005)\citenamefont{Zhang, Tan,
  Stormer, and Kim}}]{Zhang2005}
\bibinfo{author}{\bibnamefont{Zhang}, \bibfnamefont{Y.}},
  \bibinfo{author}{\bibfnamefont{Y.-W.} \bibnamefont{Tan}},
  \bibinfo{author}{\bibfnamefont{H.~L.} \bibnamefont{Stormer}}, and
  \bibinfo{author}{\bibfnamefont{P.}~\bibnamefont{Kim}}, \bibinfo{year}{2005},
  \bibinfo{journal}{Nature} \textbf{\bibinfo{volume}{438}},
  \bibinfo{pages}{201}.

\bibitem[{\citenamefont{Zimmerli} \emph{et~al.}(1992)\citenamefont{Zimmerli,
  Eiles, Kautz, and Martinis}}]{Zimmerli1992}
\bibinfo{author}{\bibnamefont{Zimmerli}, \bibfnamefont{G.}},
  \bibinfo{author}{\bibfnamefont{T.~M.} \bibnamefont{Eiles}},
  \bibinfo{author}{\bibfnamefont{R.~L.} \bibnamefont{Kautz}}, and
  \bibinfo{author}{\bibfnamefont{J.~M.} \bibnamefont{Martinis}},
  \bibinfo{year}{1992}, \bibinfo{journal}{Applied Physics Letters}
  \textbf{\bibinfo{volume}{61}}, \bibinfo{pages}{237}.

\bibitem[{\citenamefont{Zimmerman} \emph{et~al.}(2004)\citenamefont{Zimmerman,
  Hourdakis, Ono, Fujiwara, and Takahashi}}]{Zimmerman2004}
\bibinfo{author}{\bibnamefont{Zimmerman}, \bibfnamefont{N.~M.}},
  \bibinfo{author}{\bibfnamefont{E.}~\bibnamefont{Hourdakis}},
  \bibinfo{author}{\bibfnamefont{Y.}~\bibnamefont{Ono}},
  \bibinfo{author}{\bibfnamefont{A.}~\bibnamefont{Fujiwara}}, and
  \bibinfo{author}{\bibfnamefont{Y.}~\bibnamefont{Takahashi}},
  \bibinfo{year}{2004}, \bibinfo{journal}{Journal of Applied Physics}
  \textbf{\bibinfo{volume}{96}}, \bibinfo{pages}{5254}.

\bibitem[{\citenamefont{Zimmerman and Keller}(2003)}]{Zimmerman2003}
\bibinfo{author}{\bibnamefont{Zimmerman}, \bibfnamefont{N.~M.}}, and
  \bibinfo{author}{\bibfnamefont{M.~W.} \bibnamefont{Keller}},
  \bibinfo{year}{2003}, \bibinfo{journal}{Meas. Sci. Technol.}
  \textbf{\bibinfo{volume}{14}}, \bibinfo{pages}{1237}.

\bibitem[{\citenamefont{Zimmerman} \emph{et~al.}(2006)\citenamefont{Zimmerman,
  Simonds, and Wang}}]{Zimmerman2006}
\bibinfo{author}{\bibnamefont{Zimmerman}, \bibfnamefont{N.~M.}},
  \bibinfo{author}{\bibfnamefont{B.~J.} \bibnamefont{Simonds}}, and
  \bibinfo{author}{\bibfnamefont{Y.}~\bibnamefont{Wang}}, \bibinfo{year}{2006},
  \bibinfo{journal}{Metrologia} \textbf{\bibinfo{volume}{43}},
  \bibinfo{pages}{383}.

\bibitem[{\citenamefont{Zorin}(1996)}]{Zorin1996}
\bibinfo{author}{\bibnamefont{Zorin}, \bibfnamefont{A.~B.}},
  \bibinfo{year}{1996}, \bibinfo{journal}{Physical Review Letters}
  \textbf{\bibinfo{volume}{76}}, \bibinfo{pages}{4408}.

\bibitem[{\citenamefont{Zorin}(2001)}]{Zorin2001}
\bibinfo{author}{\bibnamefont{Zorin}, \bibfnamefont{A.~B.}},
  \bibinfo{year}{2001}, \bibinfo{journal}{Physical Review Letters}
  \textbf{\bibinfo{volume}{86}}, \bibinfo{pages}{3388}.

\bibitem[{\citenamefont{Zorin} \emph{et~al.}(2000)\citenamefont{Zorin, Lotkhov,
  Zangerle, and Niemeyer}}]{Zorin2000}
\bibinfo{author}{\bibnamefont{Zorin}, \bibfnamefont{A.~B.}},
  \bibinfo{author}{\bibfnamefont{S.~V.} \bibnamefont{Lotkhov}},
  \bibinfo{author}{\bibfnamefont{H.}~\bibnamefont{Zangerle}}, and
  \bibinfo{author}{\bibfnamefont{J.}~\bibnamefont{Niemeyer}},
  \bibinfo{year}{2000}, \bibinfo{journal}{J. Appl. Phys.}
  \textbf{\bibinfo{volume}{88}}, \bibinfo{pages}{2665}.

\end{thebibliography}

\end{document}